\documentclass[longauth]{aa}

\usepackage{graphicx}
\usepackage{txfonts}
\usepackage[allcolors=blue]{hyperref}

\usepackage{upquote} 

\usepackage{hyperref}
\usepackage{makecell}
\newcommand{\orcit}[1]{\protect\href{https://orcid.org/#1}{\protect\includegraphics[width=8pt]{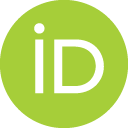}}}

\usepackage{placeins}

\usepackage{soul}

\newcolumntype{L}[1]{>{\raggedright\let\newline\\\arraybackslash\hspace{0pt}}m{#1}}
\newcolumntype{C}[1]{>{\centering\let\newline\\\arraybackslash\hspace{0pt}}m{#1}}
\newcolumntype{R}[1]{>{\raggedleft\let\newline\\\arraybackslash\hspace{0pt}}m{#1}}

\def\deg{\ensuremath{^\circ}}
\def\arcmin{\ensuremath{'}}
\def\arcsec{\ensuremath{''}}

\newcommand\gdr[1]{\gaia~DR#1}

\newcommand{\gaia}{\textit{Gaia}\xspace}

\def\gmag{\ensuremath{G}\xspace}
\def\gbp{\ensuremath{G_{\rm BP}}\xspace}
\def\grp{\ensuremath{G_{\rm RP}}\xspace}
\def\grvs{\ensuremath{G_{\rm RVS}}\xspace}

\newcommand{\cxs}{\ensuremath{C^{\ast}}\xspace}

\def\rv{\ensuremath{RV}\xspace}
\newcommand{\bpminrp}{\ensuremath{G_\mathrm{BP}-G_\mathrm{RP}}\xspace}

\def\a0{\ensuremath{A_{\rm 0}}\xspace}

\def\ipdGofHarmAmpl{\texttt{ipd\_gof\_harmonic\_amplitude}\xspace}
\def\ipdGofHarmPhase{\texttt{ipd\_gof\_harmonic\_phase}\xspace}
\def\ipdFracMultiPeak{\texttt{ipd\_frac\_multi\_peak}\xspace}
\def\ipdFracOddWin{\texttt{ipd\_frac\_odd\_win}\xspace}

\def\rIpd{$r_\text{ipd}$\xspace}
\def\rExf{$r_\text{exf}$\xspace}
\def\rExfG{$r_{\text{exf,$G$}}$\xspace}
\def\rExfBp{$r_{\text{exf,$BP$}}$\xspace}
\def\rExfRp{$r_{\text{exf,$RP$}}$\xspace}
\def\rIpdG{$r_{\text{ipd,$G$}}$\xspace}
\def\rIpdBp{$r_{\text{ipd,$BP$}}$\xspace}
\def\rIpdRp{$r_{\text{ipd,$RP$}}$\xspace}

\def\numAllBands{$N_\text{3band}$\xspace}
\def\numObsExclEpslG{$N_\text{noEpsl,$G$}$\xspace}
\def\numObsExclEpslBp{$N_\text{noEpsl,$BP$}$\xspace}
\def\numObsExclEpslRp{$N_\text{noEpsl,$RP$}$\xspace}
\def\numObsExclEpslGBpRp{$N_\text{noEpsl,$G$/$BP$/$RP$}$\xspace}

\def\redChiSqG{$\chi^2_\text{red,$G$}$\xspace}
\def\redChiSqBp{$\chi^2_\text{red,$BP$}$\xspace}
\def\redChiSqRp{$\chi^2_\text{red,$RP$}$\xspace}
\def\redChiSq{$\chi^2_\text{red}$\xspace}

\def\aG{$a_G$\xspace}
\def\thetaG{$\theta_G$\xspace}

\def\tabGaiaSource{\texttt{gaia\_source}\xspace}
\def\tabGalaxy{\texttt{galaxy}\xspace}

\usepackage{xcolor}

\graphicspath{{./images/}}
\usepackage[normalem]{ulem}

\begin{document} 

    \title{\gaia Data Release 3}
    \subtitle{\gaia scan-angle-dependent signals and spurious periods\thanks{Table \texttt{gaia\_dr3.vari\_spurious\_signals} is available at the \gaia archive via \url{https://gea.esac.esa.int/archive/} and will soon also be available on the CDS.}}

    \titlerunning{\gaia~DR3: Scan-angle-dependent signals and spurious periods}

   \author{
B.~                          Holl\orcit{0000-0001-6220-3266}\inst{\ref{inst:0003},\ref{inst:0006}}\fnmsep\thanks{Corresponding author: B. Holl (\href{mailto:berry.holl@unige.ch}{\tt berry.holl@unige.ch})}
\and C.~ Fabricius\orcit{0000-0003-2639-1372}\inst{\ref{inst:0073},\ref{inst:0074}}
\and J.~ Portell\orcit{0000-0002-8886-8925}\inst{\ref{inst:0074},\ref{inst:0073}}
\and L.~ Lindegren\orcit{0000-0002-5443-3026}\inst{\ref{inst:10016}}
\and P.~ Panuzzo\orcit{0000-0002-0016-8271}\inst{\ref{inst:10006}}
\and M.~Bernet\orcit{0000-0001-7503-1010}\inst{\ref{inst:0074},\ref{inst:0073}}
\and J.~Casta{\~n}eda\orcit{0000-0001-7820-946X}\inst{\ref{inst:0074},\ref{inst:0073}}
\and G.~ Jevardat de Fombelle\inst{\ref{inst:0003}}
\and M.~Audard\orcit{0000-0003-4721-034X}\inst{\ref{inst:0003},\ref{inst:0006}}
\and C.~ Ducourant\orcit{0000-0003-4843-8979}\inst{\ref{inst:0075}}
\and D.L.~Harrison\orcit{0000-0001-8687-6588}\inst{\ref{inst:0001},\ref{inst:0001a}}
\and D.W.~Evans\orcit{0000-0002-6685-5998}\inst{\ref{inst:0001}}
\and G.~Busso\orcit{0000-0003-0937-9849}\inst{\ref{inst:0001}}
\and A.~Sozzetti\orcit{0000-0002-7504-365X}\inst{\ref{inst:13}}
\and E.~Gosset\inst{\ref{inst:10079},\ref{inst:10021}}
\and F.~Arenou\orcit{0000-0003-2837-3899}\inst{\ref{inst:10006}}
\and F.~De Angeli\orcit{0000-0003-1879-0488}\inst{\ref{inst:0001}}
\and M.~Riello\orcit{0000-0002-3134-0935}\inst{\ref{inst:0001}}
\and L.~Eyer\orcit{0000-0002-0182-8040}\inst{\ref{inst:0003}}
\and         L.~                     Rimoldini\orcit{0000-0002-0306-585X}\inst{\ref{inst:0006}}
\and         P.~                        Gavras\orcit{0000-0002-4383-4836}\inst{\ref{inst:0022}}
\and         N.~                       Mowlavi\orcit{0000-0003-1578-6993}\inst{\ref{inst:0003}}
\and         K.~                  Nienartowicz\orcit{0000-0001-5415-0547}\inst{\ref{inst:0005},\ref{inst:0006}}
\and         I.~                 Lecoeur-Ta\"ibi\orcit{0000-0003-0029-8575}\inst{\ref{inst:0006}}
\and         P.~              Garc\'{i}a-Lario\orcit{0000-0003-4039-8212}\inst{\ref{inst:0013}}
\and         D.~              Pourbaix$^\dagger$\orcit{0000-0002-3020-1837}\inst{\ref{inst:10020},\ref{inst:10021}}
}
\institute{
Department of Astronomy, University of Geneva, Chemin Pegasi 51, 1290 Versoix, Switzerland\relax
\label{inst:0003}
\and Department of Astronomy, University of Geneva, Chemin d'Ecogia 16, 1290 Versoix, Switzerland\relax
\label{inst:0006}
\and Institut d'Estudis Espacials de Catalunya (IEEC), c.\ Gran Capit\`a, 2-4, 08034 Barcelona, Spain\relax                                                                                               \label{inst:0073}
\and  Institut de Ciències del Cosmos (ICCUB), Universitat de Barcelona (UB), c.\ Mart\'i i Franqu\`es, 1, 08028 Barcelona, Spain\relax                                                                
\label{inst:0074}
\and Lund Observatory, Department of Astronomy and Theoretical Physics, Lund University, Box 43, 22100 Lund, Sweden\relax                                                                                        \label{inst:10016}
\and GEPI, Observatoire de Paris, Universit\'{e} PSL, CNRS, 5 Place Jules Janssen, 92190 Meudon, France\relax                                                                                              \label{inst:10006}
\and Laboratoire d'Astrophysique de Bordeaux, Univ. Bordeaux, CNRS, B18N, all{\'e}e Geoffroy Saint-Hilaire, 33615 Pessac, France\relax
\label{inst:0075}
\and Institute of Astronomy, University of Cambridge, Madingley Road, Cambridge CB3 0HA, United Kingdom\relax                                                                                             \label{inst:0001}
\and Kavli Institute for Cosmology, Institute of Astronomy, Madingley Road, Cambridge, CB3 0HA, UK\relax  \label{inst:0001a}
\and INAF - Osservatorio Astrofisico di Torino, Via Osservatorio 20, I- 10025 Pino Torinese, Italy\relax
\label{inst:13}
\and Institut d'Astrophysique et de G\'{e}ophysique, Universit\'{e} de Li\`{e}ge, 19c, All\'{e}e du 6 Ao\^{u}t, B-4000 Li\`{e}ge, Belgium\relax                                                                 \label{inst:10079}
\and F.R.S.-FNRS, Rue d'Egmont 5, 1000 Brussels, Belgium\relax                                            \label{inst:10021}
\and RHEA for European Space Agency (ESA), Camino bajo del Castillo, s/n, Urbanizacion Villafranca del Castillo, Villanueva de la Ca\~{n}ada, 28692 Madrid, Spain\relax                                          \label{inst:0022}
\and Sednai S\`{a}rl, 4 Rue des Marbiers, 1204, Geneva, Switzerland\relax                                                                  \label{inst:0005}
\and European Space Agency (ESA), European Space Astronomy Centre (ESAC), Camino bajo del Castillo, s/n, Urbanizacion Villafranca del Castillo, Villanueva de la Ca\~{n}ada, 28692 Madrid, Spain\relax             \label{inst:0013}
\and Institut d'Astronomie et d'Astrophysique, Universit\'{e} Libre de Bruxelles CP 226, Boulevard du Triomphe, 1050 Brussels, Belgium\relax                                                                    \label{inst:10020}                                      
}

    \authorrunning{B.~Holl, C.~ Fabricius, J.~ Portell, et al.}

   \date{Received ?; accepted ?}

 
   \abstract
{
\gaia Data Release 3 (\gdr{3}) time series data may contain spurious signals related to the time-dependent scan angle.
}
{
We aim to explain the origin of scan-angle-dependent signals and how they can lead to spurious periods, provide statistics to identify them in the data, and suggest how to deal with them in \gdr{3} data and in future releases.
}
{
Using real \gaia (DR3) data alongside numerical and analytical models, we visualise and explain the features observed in the data. 
}
{
We demonstrated with \gaia (DR3) data that source structure (multiplicity or extendedness) or pollution from close-by bright objects can cause biases in the image parameter determination from which photometric, astrometric, and (indirectly) radial velocity time series are derived. These biases are a function of the time-dependent scan direction of the instrument and thus can introduce scan-angle-dependent signals, which due to the scanning-law-induced sampling of \gaia can result in specific spurious periodic signals. 
Numerical simulations in which a period search is performed on \gaia time series with a scan-angle-dependent signal qualitatively reproduce the general structure observed in the spurious period distribution of photometry and astrometry, and the associated spatial distributions on the sky. A variety of statistics allows for the deeper understanding and identification of affected sources.
}
{
The origin of the scan-angle-dependent signals and subsequent spurious periods is well understood and is mostly caused by fixed-orientation optical pairs with a separation $<0.5$\arcsec (including binaries with $P\gg 5$y) and (cores of) distant galaxies.  
Although most of the sources with affected derived parameters have been filtered out from the \gaia archive  \texttt{nss\_two\_body\_orbit} and several \texttt{vari}-tables, \gdr{3} data remain that should be treated with care (no sources were filtered from \texttt{gaia\_source}).
Finally, the various statistics discussed in the paper can be used to identify and filter affected sources and also reveal new information about them that is not available through other means, especially in terms of binarity on sub-arcsecond scale.
}


   \maketitle
%

\section{Introduction} \label{sec:introduction}

The ongoing processing and analyses of \gaia data by the Data Processing Analysis Consortium (DPAC) and scientific community is leading to an increasingly more detailed and refined understanding of the instrument responses and of the data properties. 
This paper is mainly dedicated to so-called 
scan-angle-dependent signals in the \gaia data, which is a product of the on-sky source structure (mainly multiplicity or extendedness), \gaia scanning law, the on-board sampling and windowing observation strategy, and on-ground observation modelling. 
These signals 
can lead to the emergence of biases in the derived parameters such as the periodicity, giving rise to specific spurious periods.

A quick overview of the paper is given in the discussion in
Sect.~\ref{sec:discussion}, where the whole paper is condensed around several relevant topics and questions that point out the relevant sections for further reading.

To properly understand and explain the mentioned effects, we structured the paper in the following way. 
First, the basic \gaia observation mode and its properties are explained in Sect.~\ref{sec:gaiaObserving}.
Then Sect.~\ref{sec:instrumentCal} discusses and demonstrates the relevant scan-angle-related modelling errors 
for each \gaia
instrument that can be introduced in the derived data.
Examples and interpretation of observed spurious period distributions are then discussed in Sect.~\ref{sec:spuriousPeriodsInGaiaData}.
In Sect.~\ref{sec:responseToSaSignal} we introduce a photometric and astrometric scan-angle-dependent bias signal model 
and demonstrate through simulations how it qualitatively reproduces the observed spurious periods.
Section~\ref{sec:detectSaSignals} then focuses on statistics that can detect scan-angle-dependent signals and several other relevant features.  Sect.~\ref{sec:discussion} contains condensed discussions around the subjects related to this paper, which is followed by our concluding remarks in Sect.~\ref{sec:conclusions}.

In Appendix~\ref{sec:auxTablesStats} we describe the \gaia archive table data that are published with this paper for all sources with published time series in \gaia Data Release 3 (\gdr{3}), containing the statistical parameters of Sect.~\ref{sec:detectSaSignals}. Appendix~\ref{sec:appendix-bcn-full-data} contains additional examples of sources that are affected by the scan-angle signal. In Appendix~\ref{sec:simPeaksExamples} we show the sky distribution of specific spurious peaks as identified in Sect.~\ref{ssec:propSimSignals}. Finally, Appendix~\ref{sec:eclScanAngle} describes the conversion between equatorial and ecliptic scan position angles.


\begin{figure}[t]
  \includegraphics[width=0.5\textwidth]{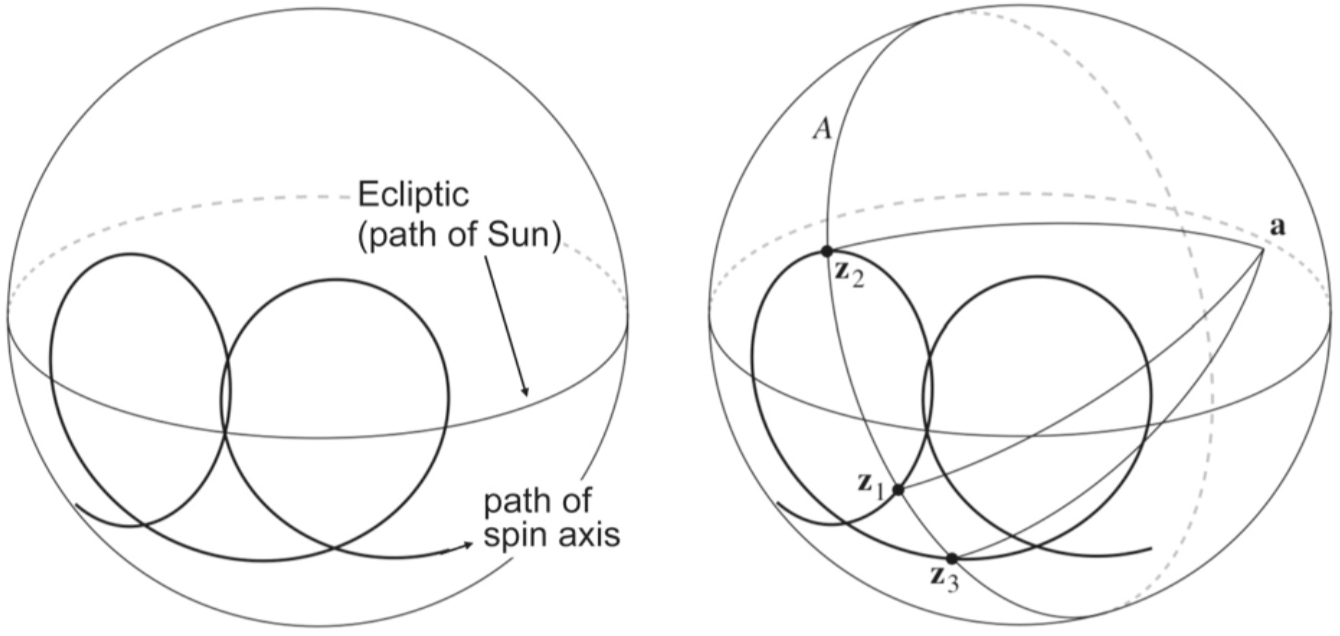}                 
\caption{Overview of the Gaia scanning law. Left:
 During the nominal scanning law, the spin axis $\vec{\rm z}$ makes overlapping loops around the Sun at a separation of $45\degr$ and rate of 5.8~cycles yr$^{-1}$. Right: One source at point $\vec{\rm a}$ may be scanned whenever $\vec{\rm z}$ is $90\degr$ from $\vec{\rm a}$, that is, on the great circle $A$ at $\vec{\rm z}_1$, $\vec{\rm z}_2$, $\vec{\rm z}_3$, etc. Reproduction with permission of fig.~7 in \cite{2016A&A...595A...1G}.
}
\label{fig:saPrustiFig}
\end{figure}

\section{How \gaia observes the sky} \label{sec:gaiaObserving}

\begin{figure}[t!]
 \includegraphics[width=0.5\textwidth]{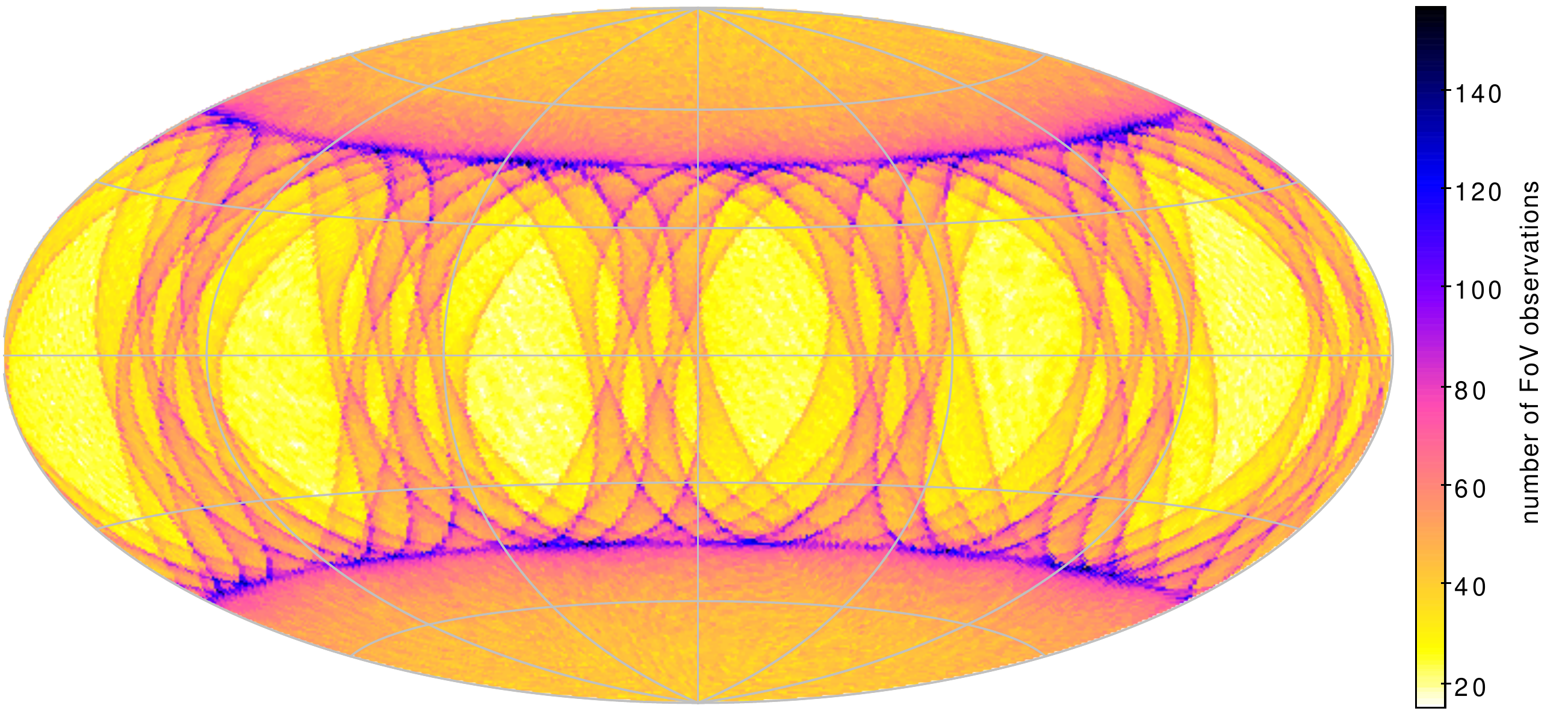}         
\includegraphics[width=0.5\textwidth]{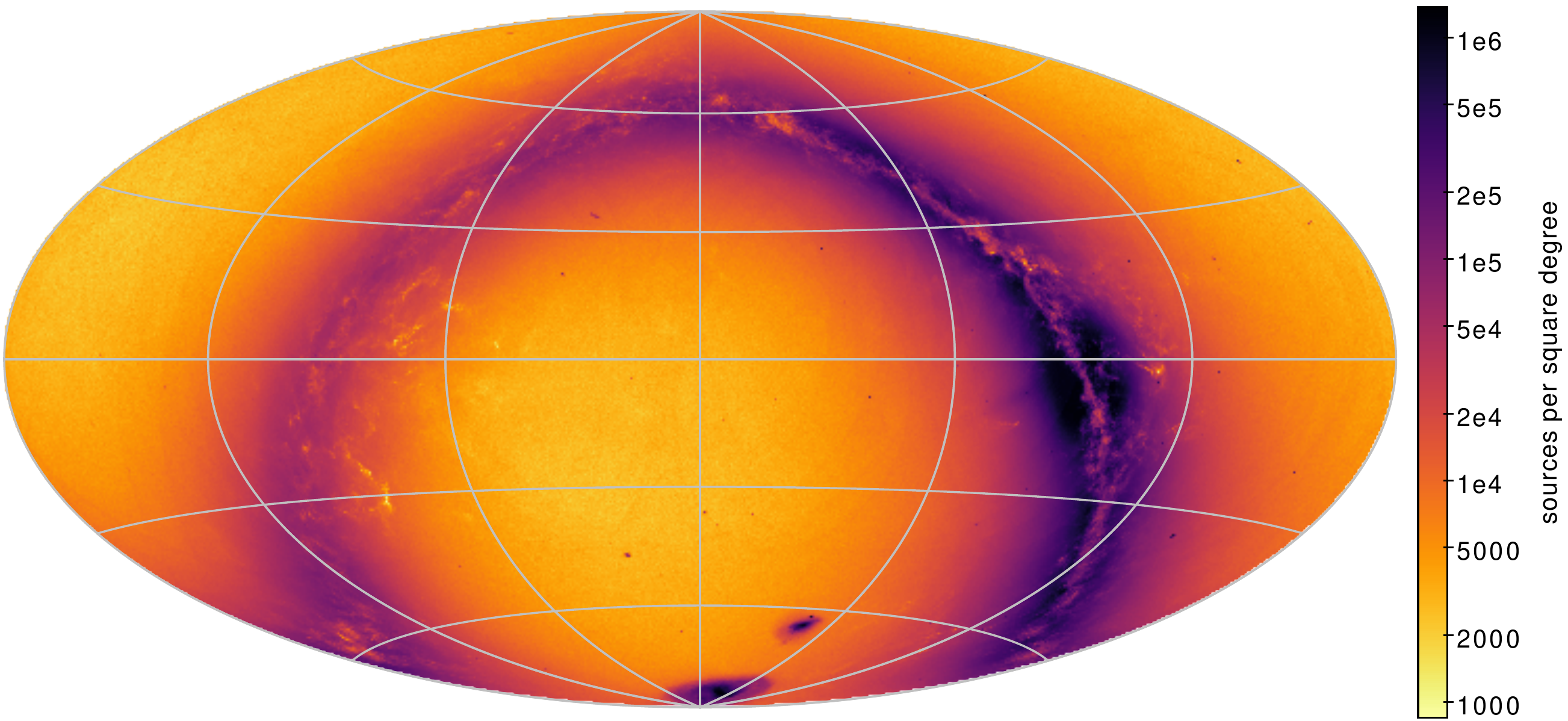} 
\caption{Ecliptic coordinate plots with longitude zero at the centre and increasing to the left. Top panel: Simulated number of field-of-view observations during the nominal scanning law phase of the \gdr{3} time range. Bottom panel: Sky density of the published \gdr{3} sources.
}
\label{fig:nslNumObsAndDensity}
\end{figure}

We start with a brief overview of the \gaia scanning-law properties that are relevant for this study (for more details, see
\cite{2016A&A...595A...1G, 2010EAS....45..109L, 2010IAUS..261..331D}). We only consider operations under the nominal scanning law (NSL) and ignored other non-nominal modes because they do not affect the majority of the data significantly and are not essential for the understanding of the discussed features. The NSL dictates the way in which the \gaia spacecraft scans the sky; its two fields of view are separated by 106.5\deg, and it rotates in a plane orthogonal to the spacecraft spin axis with a period of 6~hours. Each field of view has an instantaneous coverage of about 0.5~deg$^2$ ($0.72\degr \times 0.69\degr$), and a source is typically observed sequentially 
by at least one pair of the {preceding} and {following} field of view, with decreasing frequency of longer sequences of recurring observations due to the slow and non-constant precession rate of the spin axis \citep[see for example][for these all-sky sequence statistics]{2017arXiv170203295E}.
For observations around a certain time at a specific sky location, a low or high AC-scan velocity (see Sect.~\ref{ssec:acMotion}) will produce more or fewer sequences of recurring observations, respectively.
If the spin axis had a fixed orientation in space, a single great circle alone would be scanned on the sky. 
In reality, the spacecraft orbits the second Lagrangian point (L2) of the Earth-Sun system, and thus, the spacecraft has to rotate its spin axis with a yearly cycle to keep the instrumentation behind the solar shield. To be able to acquire useful astrometric measurements throughout the sky (in terms of temporal sampling and required instrument orientation), 
the spin axis is made to precess at a 45\deg\ angle around the direction towards the Sun with a frequency of 5.8 cycles yr$^{-1}$, which is about 63.0~d per cycle (see the left panel of Fig.~\ref{fig:saPrustiFig}).
To be precise, this precession is around a fictitious nominal Sun direction as seen from L2 
(that is, along the Earth-Sun vector), and not from \gaia orbiting L2, although the offset is always less than 0.15\degr \ \citep[see][]{2016A&A...595A...1G}.
This gives rise to the specific observation distribution, as illustrated in the top panel of Fig.\,\ref{fig:nslNumObsAndDensity}, along with the published \gdr{3} source sky density in the bottom panel for comparison.

Because of the approximately 3:1 aspect ratio of the \gaia primary mirrors \citep{2016A&A...595A...1G} and matching 1:3 pixel aspect ratio (to achieve diffraction-limited sampling), the highest image sampling resolution of 58.9~mas/pixel is achieved in the so-called along-scan (AL) direction. This is the direction in which a field of view passes over a particular source due to the spinning motion of the spacecraft. Its direction is indicated by the time-dependent scan angle $\psi$ that is illustrated in Fig.~\ref{fig:AfAngularSepIllustration}. The direction orthogonal to AL is called across-scan (AC), and it is sampled with a resolution of 176.8~mas/pixel. Depending on the magnitude of a detected source and the instrument, 
the details of the data acquisition vary, as described in Sect.~\ref{sec:instrumentCal}.

The most important information in this section is that the vast majority of \gaia information is encoded and contained in the AL-scan measurement, which is taken in the direction of the scan angle over a source at a particular time.

\subsection{Scan-angle distribution of source observations\label{ssec:obsDistrSa}} 

\begin{figure}[t]
  \includegraphics[width=0.5\textwidth]{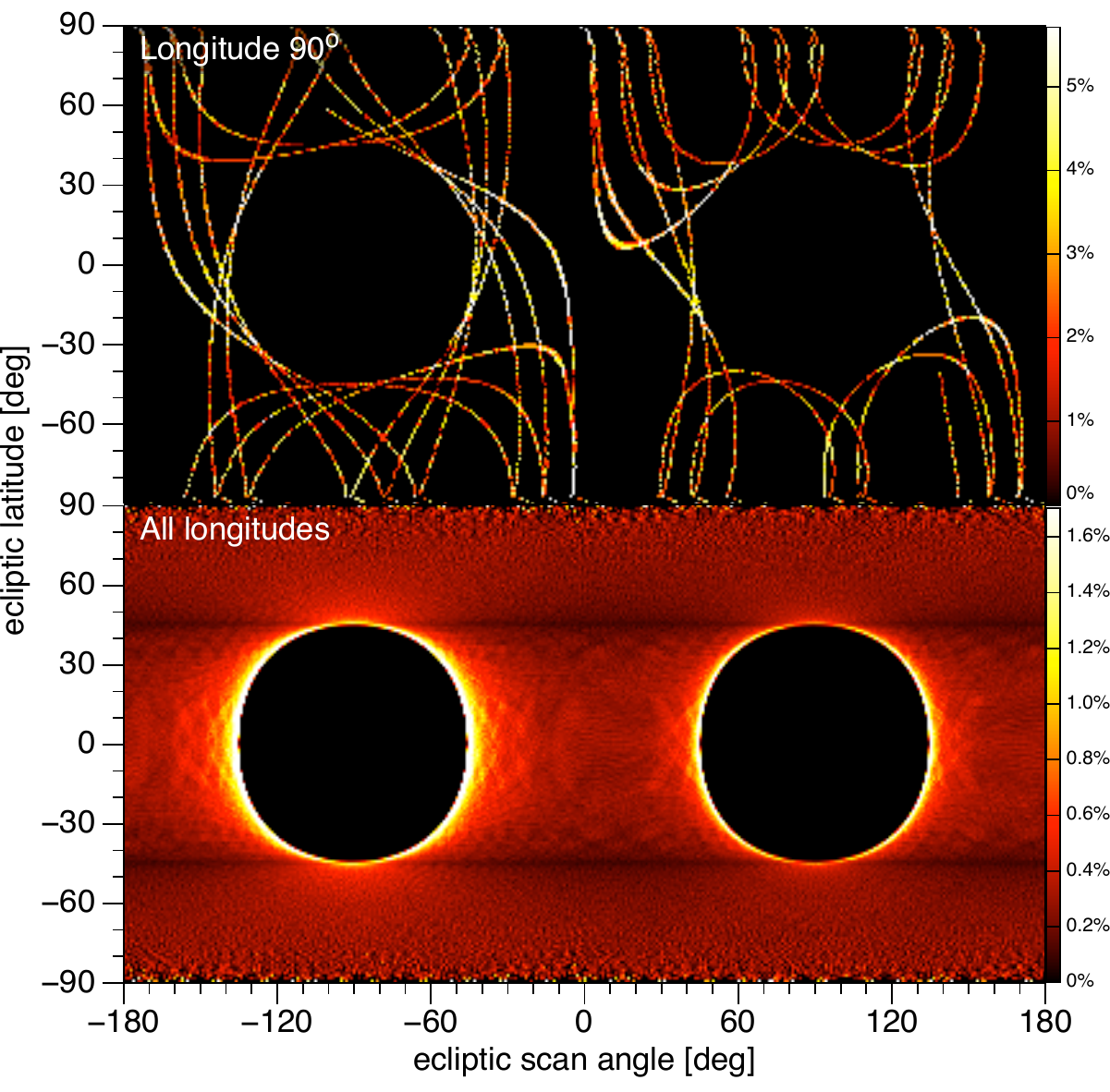}                         
\caption{Ecliptic scan-angle distribution for the nominal scanning law during the \gdr{3} time range. For a certain ecliptic latitude (horizontal slice), the colour represents the occupancy percentage per 1\degr \ scan-angle bin (summing up to 100\% over all scan angles) to highlight non-uniformities in the scan-angle distribution at different ecliptic latitudes. Top panel: Distribution for sources along a half-circle slice at ecliptic longitude $\lambda=90\deg$. Bottom panel: 
Same as top panel, but for an all-sky uniform HEALPix grid of sources (that is, all ecliptic longitudes for a given latitude). 
The strong imbalance of scan angles for sources $|\beta|\leq45^\circ$ has a strong impact on the propagation strength of certain scan-angle-dependent signals; see text for details.
}
\label{fig:saDistr}
\end{figure}

\begin{figure}[t]
  \includegraphics[width=0.5\textwidth]{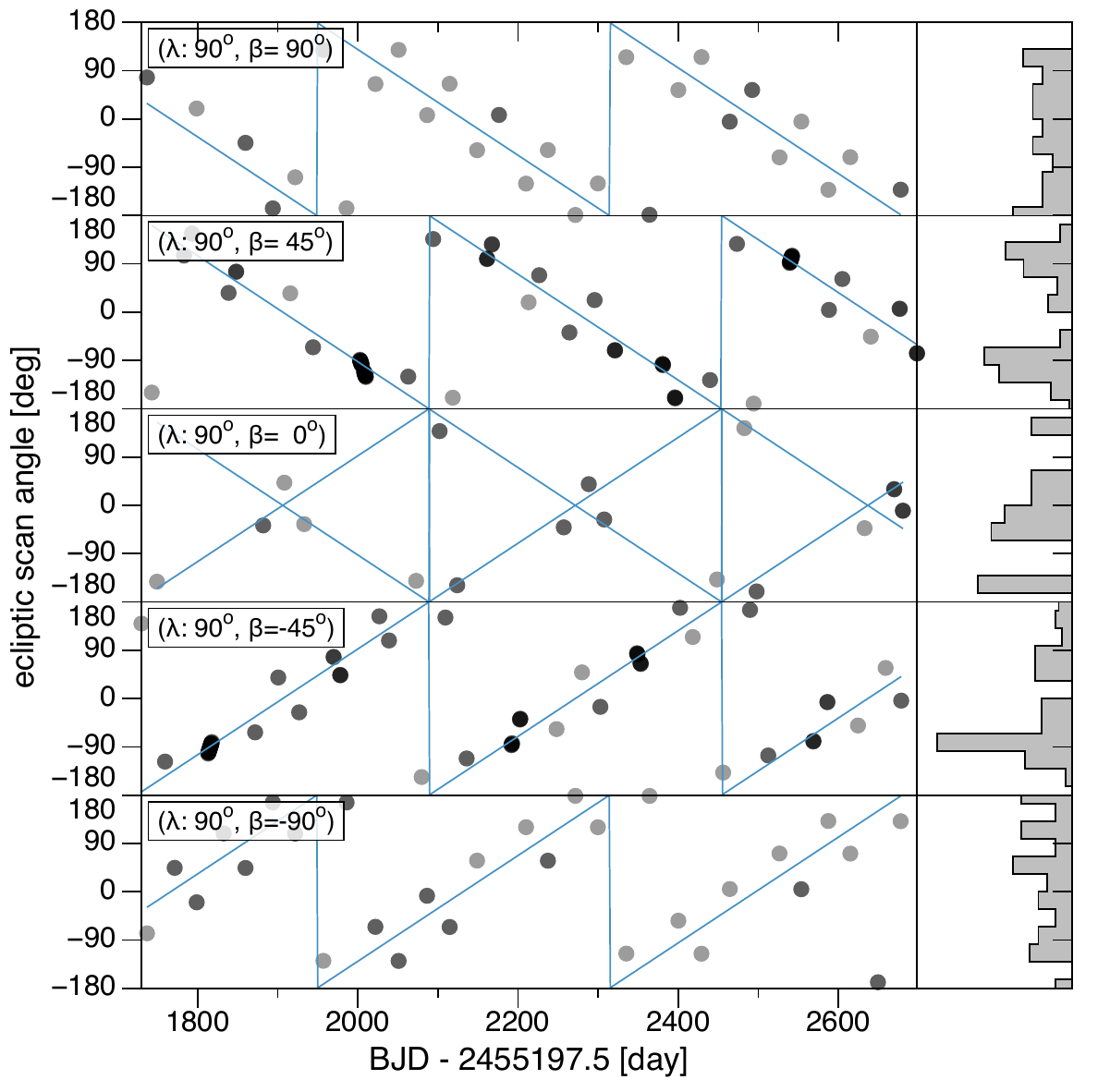}                 
\caption{Time series of the ecliptic scan-angle distribution during the \gdr{3} NSL time range for five ecliptic latitudes along the half-circle slice with ecliptic longitude $\lambda=90\deg$ (same as the top panel of Fig.~\ref{fig:saDistr}). Each point is semi-transparent, so that a darker colour means more observations. Blue cyclic lines illustrate the slopes due to the yearly rotation around the Sun. The histogram on the right side has a bin size of 32.7\degr (360/11) and shows the relative distribution of scan angles, corresponding to the top panel of Fig.~\ref{fig:saDistr} for the specified ecliptic latitudes. 
}
\label{fig:tsSaDistr}
\end{figure}

The nominal scanning law not only dictates the cadence and thus total number of observations for each position on the sky (as shown in the top panel of Fig.~\ref{fig:nslNumObsAndDensity}), but also the associated observation scan angles. 
The scan angle $\psi$ in Fig.~\ref{fig:AfAngularSepIllustration} at a certain sky position and time is zero when pointing toward the local equatorial north  and $90\degr$ when pointing towards the local equatorial east direction. 
To illustrate the all-sky scan-angle distribution in the bottom panel of Fig.~\ref{fig:saDistr}, we collapsed all sky positions along the {ecliptic} longitude because the nominal scanning law induces the most distinctive scan-angle variations as a function of ecliptic latitude, as also seen in the observation counts of Fig.~\ref{fig:saDistr}. We use the hierarchical equal area isolatitude pixelation (HEALPix) of the celestial sphere \citep{2002ASPC..281..107G}. The normal (equatorial-based) scan-angle would cause a sky-position-dependent offset of the scan-angles of a source due to the offset between the equatorial and ecliptic reference frame, however, thus blurring the image.
To circumvent this issue, we thus introduce the {ecliptic} scan angle, $\psi_\text{ecl}$, which is defined with respect to the ecliptic local north and east directions. It effectively is the (equatorial) scan angle plus an offset that depends on sky position, as given by Eq.~\ref{eq:scanAngleE8} of Appendix~\ref{sec:eclScanAngle}. 

The top panel of Fig.~\ref{fig:saDistr} shows the ecliptic scan-angle distribution for sources along a half-circle slice with ecliptic longitude $\lambda$\,=\,90\deg, starting at the north ecliptic pole (NEP; at ecliptic latitude $\beta$\,=\,90\deg) and extending to the South Ecliptic Pole (SEP, at $\beta$\,=\,--90\deg). 
The specific choice of $\lambda$\,=\,90\deg\, was made because it intersects the equatorial north pole, causing the equatorial and ecliptic scan angles to be identical for $\beta<66.6\degr$ and $180\degr$ offset above.
We normalised the distribution of observation scan-angles over each ecliptic latitude bin with per-source normalised observation weights to compensate for the different numbers of observations of each source position. Then we colour-coded this 
to highlight non-uniformities in the distribution of scan angles of sources at different ecliptic latitudes: yellow means a high concentration of scan angles at the particular scan-angle bin (bin width 1\deg), and dark red means that it was only sampled once or twice.

Although the distribution is approximately spread out evenly towards the ecliptic poles, it becomes much tighter and imbalanced for $|\beta|\leq45^\circ$.
These asymmetries become even more apparent when we combine the scan-angle distribution of sources that are uniformly distributed over the sky on a HEALPix grid level 5, 
as shown in the bottom panel of Fig.~\ref{fig:saDistr}. The sources close to the ecliptic poles have indeed very similarly regularly spread scan angles (red). The feature resulting from the geometric constraints of the NSL is now very clear: a circle of avoidance for $|\beta|<45\degr$ centred on $(\psi_\text{ecl},\beta)=(-90\degr, 0\degr)$ and $(90\degr, 0\degr)$, with an overabundance of observations at the very specific scan angles close to the border of these circles (yellow). 
Additionally, for sources located at $|\beta|\sim 45\degr$ , the scan angles are very clustered at ecliptic scan angles $\psi_\text{ecl} = \pm90\degr$ (upper and lower parts of the yellow circles) with hardly any observations at other scan angles, that is, the dark horizontal zones. 
The very specific clusters in scan angle for sources with $|\beta|\leq45\degr$ have important implications for the selection function of signals that have a strong dependence on scan angle, such as astrometric orbits and the scan-angle-dependent signals we discuss here: depending on the phasing of this signal, it might or might not be detectable. For example, a signal that peaks in the circle of avoidance  might be completely undetected.

Continuing with the properties of the nominal scanning law, we noted earlier that the spacecraft rotation around the Sun will induce a yearly rotation of its spin axis. In Fig.~\ref{fig:tsSaDistr} we show the temporal distribution of the scan angles of five positions along the ecliptic half-circle used in the top panel of Fig.~\ref{fig:saDistr}. The figure shows that this yearly rotation clearly dominates the temporal distribution of the scan angles of the sources. To guide the eye, we added the slope of this yearly cyclic rotation with a blue line. Depending on the ecliptic hemisphere, this gives a negative or positive slope.

In addition to the yearly rotation, we have additional modulations due to the spin axis rotation around the nominal Sun direction, as is illustrated in the right panel of Fig.~\ref{fig:saPrustiFig} for a source at an arbitrary position $\vec{\rm a}$. For simplicity, we study a source located at the north ecliptic pole 
$(\beta = 90\degr)  $ in more detail, that is, the top panel of Fig.~\ref{fig:tsSaDistr}. In this case, all observations are generated when the spin axis crosses the ecliptic equator, which occurs at a rate of about twice the spin-axis precession rate, that is, 11.6 cycles yr$^{-1}$, or approximately 31.5~d intervals. Alternating with these crossings are upwards/ahead or downwards/trailing the Sun nominal direction, which are therefore offset vertically, as is clearly visible in the distribution of data points in Fig.~\ref{fig:tsSaDistr}. However, to be precise, 
the precession rate is not constant during its 63~d period, 
nor during a one-year cycle \citep[see eq.1 of][]{2016A&A...595A...1G}, thus the interval between up- and downward cycles is not as symmetric as we suggested just now. This already indicates one of the reasons for the complexity and broadness of the spurious period peak distributions discussed later in Sects.~\ref{sec:spuriousPeriodsInGaiaData} and \ref{sec:responseToSaSignal}.

The NSL scanning law observations in this section were generated by a reduced version of the astrometric global iterative solution, AGISLab \citep{Holl:2012fk}, but the same data can be generated with the public Gaia Observation Forecast Tool (GOST) at \url{https://gaia.esac.esa.int/gost/}.

\begin{figure}[t]
    \centering
    \includegraphics[width=0.49\textwidth]{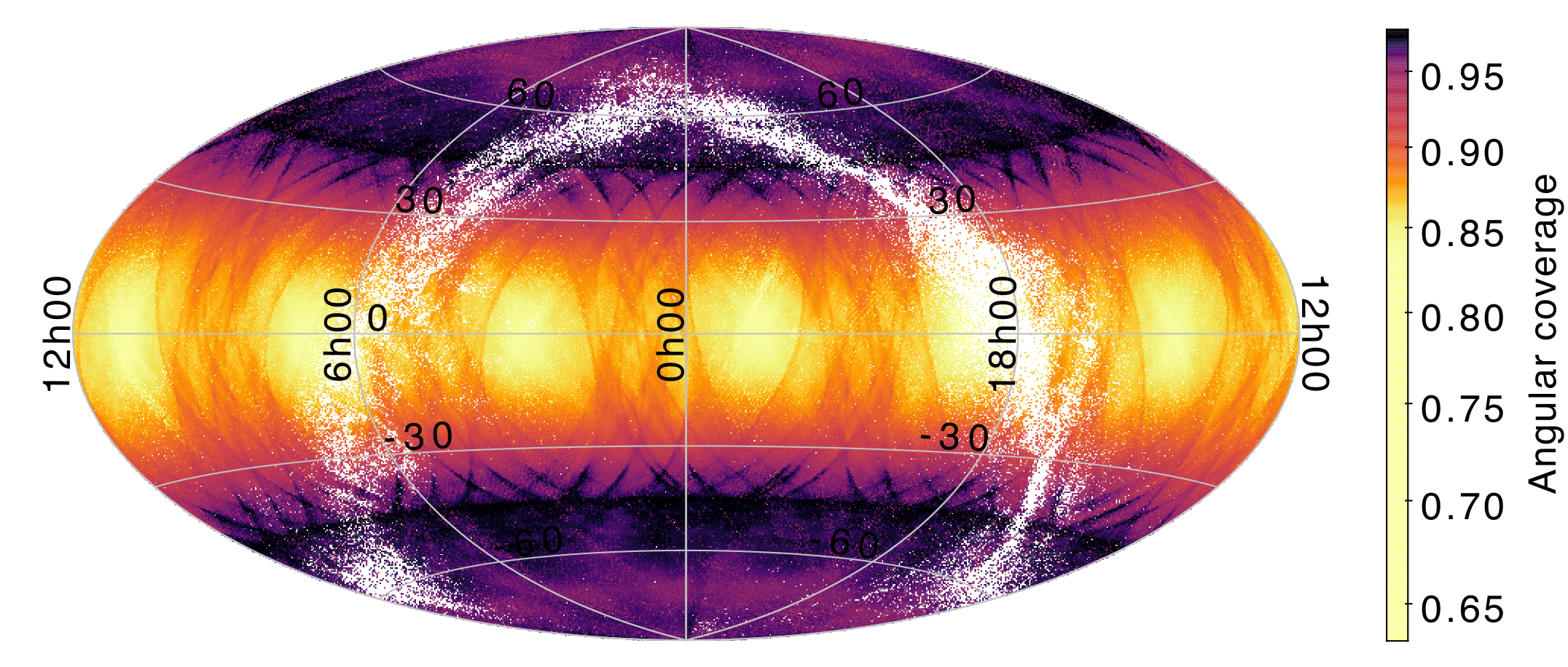}
    \caption{Sky distribution in ecliptic coordinates of extragalactic sources  analysed in terms of surface brightness profile in \gaia\ DR3, colour-coded with the angular coverage of the sources. The non-linear shader table reveals the NSL pattern.}
    \label{fig:sky_coord_angular_coverage}
\end{figure}

\subsection{Angular coverage of extended sources}

Extended objects are directly concerned by the dependence of the NSL on the ecliptic latitude. 
For these objects with a spatial extension, the variety of scan angles is crucial for reconstructing their structures.
We define the angular coverage as the fraction of the sky area that is covered by the union of the observation windows for a particular source, relative to the ideal case of a uniform distribution in scan angles 
\citep[see fig.~3 in][for an illustration, and
 \cite{garce2011}]{2022arXiv220614491D}. This quantity is mainly dependent on how the scan angles are spread over the source. Preferential scan directions will result in lower angular coverage. Figure \ref{fig:sky_coord_angular_coverage} presents the distribution of $\sim$1.3 million extragalactic sources on the sky in ecliptic coordinates, colour-coded with their angular coverage. The surface coverage of sources with $|\beta|<$30$\deg$ is  frequently lower than 85\%, as is well understood from the circle of avoidance in scan angles shown in Fig.~\ref{fig:saDistr}. 
Only for sources with a coverage larger than 85\% is the morphology of extended sources provided in \gdr{3} data.

\subsection{Across-scan velocity and scan phase\label{ssec:acMotion}}
As mentioned in sect.~\ref{sec:gaiaObserving}, there are variations in the AC velocity of  sources transiting the focal plane. The AC-scan velocity varies sinusoidally with time, with the nominal satellite rotation period of 6~h and an amplitude of 173~mas~s$^{-1}$ \citep{2016A&A...595A...1G}. This means that the AC-scan velocity varies along the great circle that is scanned on the sky by each field of view (FoV) over this 6~h rotation. The phase of this oscillation is determined by the scan {phase} (not to be confused with the previously discussed scan {angle}), which is defined as the angle between the plane containing the Sun and the spin axis and the plane spanned by the spin axis and the vector pointing exactly in between the two fields of view; see sect.~5.2 of \cite{2016A&A...595A...1G} and fig.~2 of \cite{2012A&A...538A..78L} for details about the scan phase and its definition, and Sect.~\ref{ssec:afInstr} for a calibration related effect. The reference point of this scan phase slowly drifts over time due to the 63~d precession of the spin axis. For several-day stretches of time, this means that for particularly narrow localised regions on the sky, the AC-scan velocity will vary slowly. When it is close to zero, it means that the succeeding great-circle passes will be able to scan this position more times, while a high AC-scan velocity will cause the source to drift out of the FoV before many passes can be made. At the time of the sinusoidal crossing of zero AC-scan velocity, some sky positions along the great circle experience a reverse of the AC-scan velocity in a period of a few days, causing them to stay within the across-scan bounds of the transiting FoVs. At some specific moments, this can cause the accumulation of very many (near-) continuous transits along a small band on the sky called a cusp, for example, 28~FoVs during $\sim 3.5$~d. These cusps usually occur in regions on the sky with $|\beta|\leq45\degr$.


\section{Scan-angle-dependent instrument calibration features \label{sec:instrumentCal}}

The several instruments on board \gaia each have their own specific way of collecting data, which are processed differently to extract the most relevant science data (see \cite{2016A&A...595A...1G} for a general overview and the references below for full details). In this section, we concentrate on the aspects of the data collection for each instrument and on the processing that is relevant to the introduction of 
spurious signals that are dependent on the scan angle.

Examining each instrument separately, we start with the astrometric field in Sect.~\ref{ssec:afInstr}, followed by the blue and red photometer instrument in Sect.~\ref{ssec:xpInstr},  followed in turn by the radial velocity spectrometer instrument in Sect.~\ref{ssec:rvsInstr}.


\subsection{Astrometric Field instrument\label{ssec:afInstr}}

The Astrometric Field (AF) instrument consists of a grid of 62~charge-coupled devices (CCDs) that are used to extract astrometric transit time information and photometric G-band flux measurements. It does so by reading out windows of a particular size around each star depending on the on-board determined magnitude on the Sky Mapper (SM) CCDs. Specifically, the typical sizes are $12\times12$ pixels ($0.7\arcsec\times2.1\arcsec$) for $\gmag \geq16$, and $18\times12$ pixels ($1.1\arcsec\times2.1\arcsec$) for $\gmag < 16$ \citep[][sect.\ 1.1.3]{2022gdr3.reptE...1D}.

In order to reduce readout noise, the pixels are collapsed in the across-scan
direction during the reading process, leaving only a one-dimensional set of 12 or 18 samples containing along-scan astrometric information, and G-band photometric information. For the brightest sources ($\gmag<13$\,mag), this would lead to saturation, and for these sources, the full two-dimensional window is therefore read.
These windows provide the best angular resolution achievable by \gaia, which is 59~mas $\times$ 177~mas, that is, the nominal angular size of its CCD pixels.
Additionally, for sources with $\gmag \lesssim 12,$ a gating scheme reduces the integration time to prevent these very bright stars from obtaining too many saturated pixels. The proper calibration of the various gating and windowing regimes is extremely non-trivial, causing residual calibration effects to be enhanced in sources that have observations taken in multiple calibration regimes (see Sect.~\ref{ssec:corExFlux} for a statistic that can help to identify affected sources). Typically, the mean magnitudes of these are sources are close to a regime-changing magnitude, or they are variable stars with large amplitudes.

\begin{table*}[t]
\caption{\label{tab:ipdSignalsList} Overview of source examples in this work that have scan-angle-dependent signals, along with diagnostic statistics and fitted parameters. }
\centering
\begin{scriptsize}
\begin{tabular}{
 m{0.12\textwidth} 
 m{0.125\textwidth}
|
 >{\raggedleft}m{0.02\textwidth}
 >{\raggedleft}m{0.02\textwidth}
|
 >{\raggedleft}m{0.02\textwidth}
  >{\raggedleft}m{0.02\textwidth}
   >{\raggedleft}m{0.02\textwidth}
   |
   >{\raggedleft}m{0.02\textwidth}
   >{\raggedleft}m{0.02\textwidth}
   >{\raggedleft}m{0.02\textwidth}
   >{\raggedleft}m{0.02\textwidth}
   |
   >{\raggedleft}m{0.02\textwidth}
   >{\raggedleft}m{0.02\textwidth}
   >{\raggedleft}m{0.02\textwidth}
   >{\raggedleft}m{0.02\textwidth}
   >{\raggedleft}m{0.02\textwidth}
   |
 m{0.055\textwidth}
   }

\hline\hline
SourceID & Case & \rExfG & \rIpdG &  $a_\text{ipd}$ & $\varphi_\text{ipd}$ & $\kappa$ &  $c_0$ & \aG & \thetaG & \redChiSqG  &  $G_p$ & $G_s$ & $\rho$ & $\theta$ & \redChiSqG & Fig(s). \\
\hline

\multicolumn{17}{l}{DR3 pairs, triplets and bright neighbours, with angular separation and position angle:} \\ \hline
382074694311961856$^a$  & 0.36\arcsec, 19\degr          & 0.92 & 0.17 & 0.21 & 95\degr & 84\% & 13.1 & 0.21 & 8\degr & 68.5 & 13.3 & 13.9 & 0.33\arcsec & 15\degr & 1.9 & \ref{fig:closePair371mas} \\
379163256239241216$^a$  & 0.95\arcsec, 27\degr          & 0.76 & 0.21 & 0.08 & 90\degr & 5\% & 17.1 & 0.01 & 29\degr & 4.7 & 17.1 & 21.6 & 0.36\arcsec & 99\degr & 4.6 & \ref{fig:ep-sa-379163256239241216} \\
388877407113709056$^a$  & 0.56\arcsec, 87\degr + 1.04\arcsec, 69\degr  & 0.36 & 0.09 & 0.05 & 131\degr & 19\% & 15.9 & 0.00 & 92\degr & 9.8 & 15.9 & 19.9 & 1.19\arcsec & 86\degr & 9.4 & \ref{fig:ep-sa-388877407113709056} \\
385771836519005184$^a$  & bright, 3.1\arcsec, 24\degr    & 0.00 & 0.07 & 0.05 & 50\degr & 20\% & 13.1 & 0.00 & 19\degr & 8.3 & 13.1 & 18.0 & 0.66\arcsec & 104\degr & 8.0 & \ref{fig:ep-sa-385771836519005184} \\
\hline

\multicolumn{17}{l}{Blind search in DR3 looking for multi-peak indicators:} \\ \hline
367388551858425344$^a$  & Pre-DR4 split in two  & 0.99 & 0.85 & 0.16 &  10\degr & 40\% & 12.4 & 0.09 &  98\degr & 13.6 & 12.4 & 13.8 & 0.31\arcsec & 18\degr & 2.3 & \ref{fig:ep-sa-367388551858425344}, \ref{fig:foldedPeriodGExamples1} \\
383556286230747520$^{av}$  & Variable flag set & 0.99 & 0.95 & 0.40 & 61\degr & 39\% & 16.1 & 0.42 & 150\degr & 15.1 & 19.5 & 15.7 & 0.07\arcsec & 150\degr & 15.2 & \ref{fig:ep-sa-383556286230747520}, \ref{fig:foldedPeriodGExamples2} \\
380538569192874112$^a$  & 0.49\arcsec, 112\degr  & 0.91 & 0.86 & 0.47 & 42\degr & 37\% & 16.5 & 0.46 & 127\degr & 208.6 & 16.8 & 16.9 & 0.48\arcsec & 113\degr & 13.4 & \ref{fig:ep-sa-380538569192874112}, \ref{fig:foldedPeriodGExamples1} \\
\hline

\multicolumn{17}{l}{DR3 sources identified as galaxy candidates, with De Vaucouleurs radius, ellipticity and position angle:} \\ \hline
373852271480563968$^a$  & 0.99\arcsec, 0.10, 7\degr    & 0.13 & 0.42 & 0.06 &  92\degr &  0\% & 19.8 & 0.06 & 162\degr & 26.1 & 19.8 & 22.3 & 0.35\arcsec & 70\degr & 26.8 & \ref{fig:ep-sa-373852271480563968} \\
366951667785042688$^a$  & 1.65\arcsec, 0.35, 63\degr   & 0.63 & 0.95 & 0.26 & 150\degr &  0\% & 19.9 & 0.29 & 61\degr &  1.3 & 15.7 & 15.7 &  0.05\arcsec & 151\degr &  1.4 & \ref{fig:galaxymid}, \ref{fig:foldedPeriodGExamples2} \\
364175332206026368$^a$  & 1.95\arcsec, 0.56, 85\degr   & 0.95 & 0.96 & 0.67 & 173\degr &  0\% & 20.3 & 0.77 & 82\degr &  5.8 & 21.1 & 19.7 & 0.18\arcsec & 85\degr &  3.8 & \ref{fig:ep-sa-364175332206026368} \\
\hline

\multicolumn{17}{l}{DR3 Spectroscopic Binaries (SB1), probably spurious (63-day period), furthermore flagged as Variable:} \\ \hline
415146526611154176$^v$  & 0.27\arcsec, 63\degr  & 0.99 & 0.72 & 0.05 & 157\degr & 64\% & 9.9 & 0.04 & 58\degr & 10.8 & 9.8 & 12.4 & 0.19\arcsec & 149\degr & 10.0 & \ref{fig:ep-sa-415146526611154176} \\
5815369024263284352$^v$ & -             & 0.94 & 0.24 & 0.02 & 47\degr & 0\% & 8.5 & 0.01 & 115\degr & 3.5 & 8.5 & 13.3 & 0.19\arcsec & 118\degr  & 3.5 & \ref{fig:ep-sa-5815369024263284352} \\
\hline

\multicolumn{17}{l}{Pre-DR4 close pairs, with their angular separation and position angle (if available):} \\ \hline
389636619892245248$^a$  & 0.13\arcsec               & 0.98 & 0.96 & 0.23 & 122\degr & 29\% & 15.4 & 0.17 & 28\degr & 34.1 & 15.2 & 16.4 & 0.21\arcsec & 118\degr & 8.0 & \ref{fig:closePair130mas} \\
378810450446502400$^a$  & 0.14\arcsec, 30°           & 0.99 & 0.92 & 0.25 & 129\degr & 36\% & 11.7 & 0.09 & 34\degr & 25.6 & 11.6 & 12.8 & 0.40\arcsec & 111\degr & 6.1 & \ref{fig:ep-sa-378810450446502400} \\
388466602081536640$^a$  & 0.17\arcsec               & 0.94 & 0.94 & 0.23 & 173\degr & 26\% & 18.2 & 0.24 & 82\degr & 14.8 & 18.6 & 18.8 & 0.15\arcsec & 84\degr & 13.6 & \ref{fig:ep-sa-388466602081536640}, \ref{fig:foldedPeriodGExamples2} \\
376045247423005184$^a$  & 0.22\arcsec               & 0.94 & 0.85 & 0.16 &  28\degr & 38\% & 17.2 & 0.16 & 106\degr &  8.4 & 17.3 & 18.3 & 0.27\arcsec & 104\degr &  2.0 & \ref{fig:ep-sa-376045247423005184} \\
382159975182923264$^a$  & 0.25\arcsec, 82°           & 0.99 & 0.93 & 0.48 & 154\degr & 46\% & 17.5 & 0.50 & 66\degr & 26.7 & 17.8 & 17.7 & 0.26\arcsec & 75\degr & 12.1 & \ref{fig:ep-sa-382159975182923264}, \ref{fig:foldedPeriodGExamples2} \\
385844060692409344$^a$  & 0.35\arcsec               & 0.98 & 0.77 & 0.25 & 35\degr & 43\% & 16.1 & 0.28 & 121\degr & 259 & 16.3 & 16.9 & 0.37\arcsec & 109\degr & 161 & \ref{fig:ep-sa-385844060692409344}, \ref{fig:foldedPeriodGExamples1} \\
385804856230839552$^a$  & 0.35\arcsec               & 0.99 & 0.64 & 0.08 & 17\degr & 44\% & 9.1 & 0.14 & 97\degr & 43.7 & 9.2 & 10.5 & 0.28\arcsec & 96\degr  & 13.1 & \ref{fig:ep-sa-385804856230839552} \\
387325652606842368$^a$  & 0.63\arcsec, 101\degr & 0.45 & 0.22 & 0.03 & 82\degr & 24\% & 18.8 & 0.01 & 178\degr & 26.9 & 18.8 & 17.9 & 1.91\arcsec & 97\degr & 7.1 & \ref{fig:ep-sa-387325652606842368} \\
385367010081173504$^a$  & 0.69\arcsec, 133\degr & 0.43 & 0.41 & 0.02 & 46\degr & 23\% & 18.1 & 0.02 & 125\degr & 4.2 & 18.2 & 18.5 & 0.60\arcsec & 140\degr & 2.0 & \ref{fig:ep-sa-385367010081173504} \\
\hline

\end{tabular}
\tablefoot{
For each case, we list the Spearman correlations (\rExfG and \rIpdG, Sect.~\ref{sec:detectSaSignals}), the \ipdGofHarmAmpl, \ipdGofHarmPhase, and \ipdFracMultiPeak ($a_\text{ipd}$, $\varphi_\text{ipd}$ and $\kappa$, Sect.~\ref{sssec:ipdStats}), the results of a sinusoidal or small-separation fit to the $G$ photometric scan-angle signal ($c_0$, \aG, \thetaG , and the \redChiSqG of the fit, Eq.~\ref{eq:simu2}), and finally, the results of a source-pair fit to the G-band photometric scan-angle signal ($G_p$, $G_s$, $\rho$, $\theta$ , and the $\chi^2$ of the fit, Eq.~\ref{eq:simu}).}
\tablebib{
$^a$~source in the \gaia Andromeda photometric survey (GAPS);
$^v$~source in the variability catalogue; in both cases, the photometric time series and a subset of diagnostic statistics 
 and fitted parameters are published; see Sect.~\ref{sec:auxTablesStats} for details.
}
\end{scriptsize}

\end{table*}

The instruments based on non-dispersive optics (that is, SM and AF) define the detection capabilities of \gaia. 
From these, we can distinguish the \gaia sources as single point-like sources, multiple point-like sources, or extended sources, although this is not yet systematically done in DR3.
Multiple sources, such as pairs of stars separated by a small apparent angle (or close pairs), are especially interesting for this work.
Depending on their angular separation and the sampling scheme used by \gaia, a close pair (or a multiple source, in general) can be resolved, partially resolved, or unresolved. This classification indicates the capability of \gaia or DPAC to detect the source multiplicity in all, a few, or none of the scans.
Multiple sources that are resolved or partially resolved will typically lead to different source entries in the catalogue. We should note that sources can either be resolved on board, leading to different windows for each of the detected sources, or on ground, eventually leading to separate source entries for sources sharing the same acquisition windows. As of \gdr{3}, DPAC has only considered sources that were resolved (or partially resolved) on board.

For completeness, we would like to point out that the point spread function (PSF) models at close to zero AC-scan velocities (Sect.~\ref{ssec:acMotion}) were lacking accuracy in the \gdr{3} data, causing systematics (at a level of up to several mmag) in the recovered fluxes that are biased as a function of the AC-scan velocity of the field of view. 
These scan-phase flux dependences are not included in the current study. However, because scan phase and scan angle are correlated, there is a potential for interaction between scan-angle and scan-phase dependences for two-dimensional images of which users should be aware.

\subsubsection{IPD modelling error of non-point-like sources\label{sssec:ipdModErrNonPointLike}}

One of the main steps in the extraction of science data from the individual AF CCD observations is the image parameter determination, IPD \citep[][sect.\ 3.3.6]{2022gdr3.reptE...3C}. 
In the IPD procedure, a two-dimensional PSF or one-dimensional line spread function (LSF) model is fitted with a maximum likelihood procedure to the two-dimensional ($\gmag\lesssim 13$) or one-dimensional counts in the window around the source to estimate the position and flux of a presumed single point-like source in the window, along with the background level \citep[see][]{EDR3-PSF-Rowell}.
In reality, this procedure involves more complex interactions with the astrometric global iterative solution (AGIS) and 
photometric processing
to estimate the effective wave number and/or colour, and the precise calibration of the PSF and LSF profile as a function of time, focal-plane location, CCD transit position, applied gates, and time since charge injection. 

Consider now that \gaia observes an asymmetric (non-point-like) object on the sky, such as a close pair or the core of a galaxy. The different observations will be sampled by the \gaia instruments with a variety of scan angles (or lack thereof), as discussed in Sect.~\ref{ssec:obsDistrSa}. 
Because the LSF or PSF profiles are calibrated on point sources and do not account for the additional source structure resolved in the specific scan direction, this is likely to result in a bias of some sort in the estimated position or total flux. Any asymmetry in the source structure will bias the estimate differently, depending on the direction in which \gaia scans over the object, introducing a scan-angle-dependent bias signal.
This effect is (partly) mitigated when a secondary peak is detected in the window data. The affected samples (pixels) are then excluded from the PSF or LSF fitting \citep[][sect.\ 3.3.6]{2022gdr3.reptE...3C}, 
thereby diminishing the bias in position and flux. For future \gaia data releases, a more detailed image analysis is planned.

A discussion of how the source structure and environment around each of the billion \gaia sources might be estimated can be found in Sect.~\ref{ssec:sourceEnv}.

\subsubsection{IPD model error statistics and scan-angle model \label{sssec:ipdStats}}

Although the IPD procedure used in \gdr{3} does not fit for multiple peaks or non-point-like source structure, it does populate several useful statistics in the \tabGaiaSource table that give information about possible perturbations \citep[][sect.\ 5]{2021A&A...649A...2L}. \gaia early data release 3 (EDR3) and DR3 include the following four IPD-related statistics:
(1)~
\href{https://gea.esac.esa.int/archive/documentation/GDR3/Gaia_archive/chap_datamodel/sec_dm_main_source_catalogue/ssec_dm_gaia_source.html#gaia_source-ipd_frac_multi_peak}{\ipdFracMultiPeak}: the percentage of windows, $\kappa$, for which the IPD algorithm has identified more than one peak, computed for all transits in which the IPD was successful. When processing each window, the IPD masks (suppresses) these secondary peaks, which typically allows for a better fit to the main peak.
(2)~\href{https://gea.esac.esa.int/archive/documentation/GDR3/Gaia_archive/chap_datamodel/sec_dm_main_source_catalogue/ssec_dm_gaia_source.html#gaia_source-ipd_frac_odd_win}{\ipdFracOddWin}: the percentage of transits with truncated windows or multiple gate, computed for all transits in which the IPD was successful. This means that the target is likely disturbed by a brighter source close by.      
(3)~\href{https://gea.esac.esa.int/archive/documentation/GDR3/Gaia_archive/chap_datamodel/sec_dm_main_source_catalogue/ssec_dm_gaia_source.html#gaia_source-ipd_gof_harmonic_amplitude}{\ipdGofHarmAmpl}, measuring the amplitude, $a_\text{ipd}$, of a model of the IPD goodness of fit (GoF, $\chi^2_{\mathrm{red}}$) as a function of the position angle of the scan direction; see Eq.~\ref{eq:ipdScanAngleModel} below.
(4)~\href{https://gea.esac.esa.int/archive/documentation/GDR3/Gaia_archive/chap_datamodel/sec_dm_main_source_catalogue/ssec_dm_gaia_source.html#gaia_source-ipd_gof_harmonic_phase}{\ipdGofHarmPhase}, measuring the phase, $\varphi_\text{ipd}$, of the variation of the IPD GoF ($\chi^2_{\mathrm{red}}$) as a function of the position angle of the scan direction; see Eq.~\ref{eq:ipdScanAngleModel} below.

As described in the \gdr{3} archive documentation \cite[sect~20.1.1 of][]{2022gdr3.reptE....V},
these two last parameters relate to a sinusoidal model fit to the natural logarithm of the IPD reduced $\chi^2$ (determined for each CCD observation) as function of scan angle, $\psi$, for the observations used in the astrometric solution, 
\begin{align}
&\ln(\chi^2_{\mathrm{red}}) = M_\text{ipd}( \psi ) =  c_0  +  c_2 \, \cos(2\psi) + s_2 \, \sin(2\psi)   \label{eq:ipdScanAngleModel}  \\
&a_\text{ipd} = \ipdGofHarmAmpl =\sqrt{ c_2^2 + s_2^2 } \nonumber \\
&\varphi_\text{ipd} = \ipdGofHarmPhase =\frac{1}{2} \mathrm{atan2}( s_2, c_2) \ \  (+ 180\deg) \, \, \, . 
\nonumber
\end{align}
As explained in Sect.~\ref{sssec:ipdModErrNonPointLike}, the $\chi^2_{\mathrm{red}}$ for brighter sources ($G\lesssim 13$) relates to a fitting using a two-dimensional PSF, and for fainter sources, it leads to a one-dimensional LSF fitting. 
When two sources blend, we obtain biased image parameters and a high value for the goodness of fit. This happens more easily for LSF fitting where we cannot benefit from the separation of the sources in the across-scan direction.

The main assumption in this model is that the source image to first order is axis-symmetric with respect to a certain line on the sky, parametrised by \ipdGofHarmPhase, which can be interpreted as the scan angle ($\pm$ 180\deg) corresponding to the worst fit. The interpretation of this direction is not straightforward. For a close binary, not detected as such by the IPD (and thus unresolved, following the nomenclature introduced at the end of Sect.~\ref{ssec:afInstr}), the worst fit will be along the line joining the two sources, known as the {position angle}. For these unresolved pairs, \ipdFracMultiPeak will be small (typically, a secondary peak is detected in  fewer than 10\% of the windows). This situation will happen for separations $\lesssim 0.1\arcsec$. On the other hand, for a somewhat wider binary (partially resolved), the two peaks are best detected when scanning along this line, and because the secondary signal is then suppressed, this is where we obtain the best fit. For these pairs, and especially for resolved pairs, \ipdFracMultiPeak will be high (around 30 to 50\%, perhaps even approaching 100\% if the secondary peak is detected in nearly all scans). In this case, \ipdGofHarmPhase differs from the position angle of the binary (modulo 180$\deg$) by approximately $\pm$90$\deg$. For galaxies, the angular extent is also important, but typically, the disk will be interpreted as high background by the IPD when scanning along the major axis, and this will then be the direction of the best fit. Here, \ipdGofHarmPhase will also differ from the position angle of the major axis (modulo 180$\deg$) by approximately~$\pm$90$\deg$.

The exact scan-angle dependence on the logarithm of $\chi^2_{\mathrm{red}}$  will obviously not always be well-characterised by a sinusoidal model, but it nonetheless provides us with a very useful first-order model that allows us to know the amplitude and direction (phase) of the distortion. The reference level $c_0$ of Eq.~\ref{eq:ipdScanAngleModel} 
is not published.

\cite{2021A&A...649A...5F} and \cite{DR3-DPACP-100}
showed that the \ipdGofHarmAmpl and \ipdGofHarmPhase are useful for identifying spurious solutions of resolved doubles. As of \gaia DR3, these are not yet correctly handled in the \gaia\ astrometric processing. Even though IPD is able to detect secondary peaks in the windows, no PSF or LSF fitting is attempted for them, which also means that they are not cross-matched to any source.
In Sect.~\ref{ssec:ipdParamAsDetectionMethod} we discuss the values of the IPD statistics (and others) in more detail that might be significant in relation to scan-angle-dependent signals.

We concentrated on explicit scan-angle-dependent biases from the IPD. These biases will lead to poorer astrometric and photometric solutions and will to some extent be reflected in the various astrometric and photometric quality indicators in \tabGaiaSource.

\subsubsection{Demonstration of scan-angle-dependent signals resulting from IPD outputs \label{sssec:demoSaSignalsPointSrc}}

\begin{figure}[h]
\centering
  \includegraphics[width=0.49\textwidth]{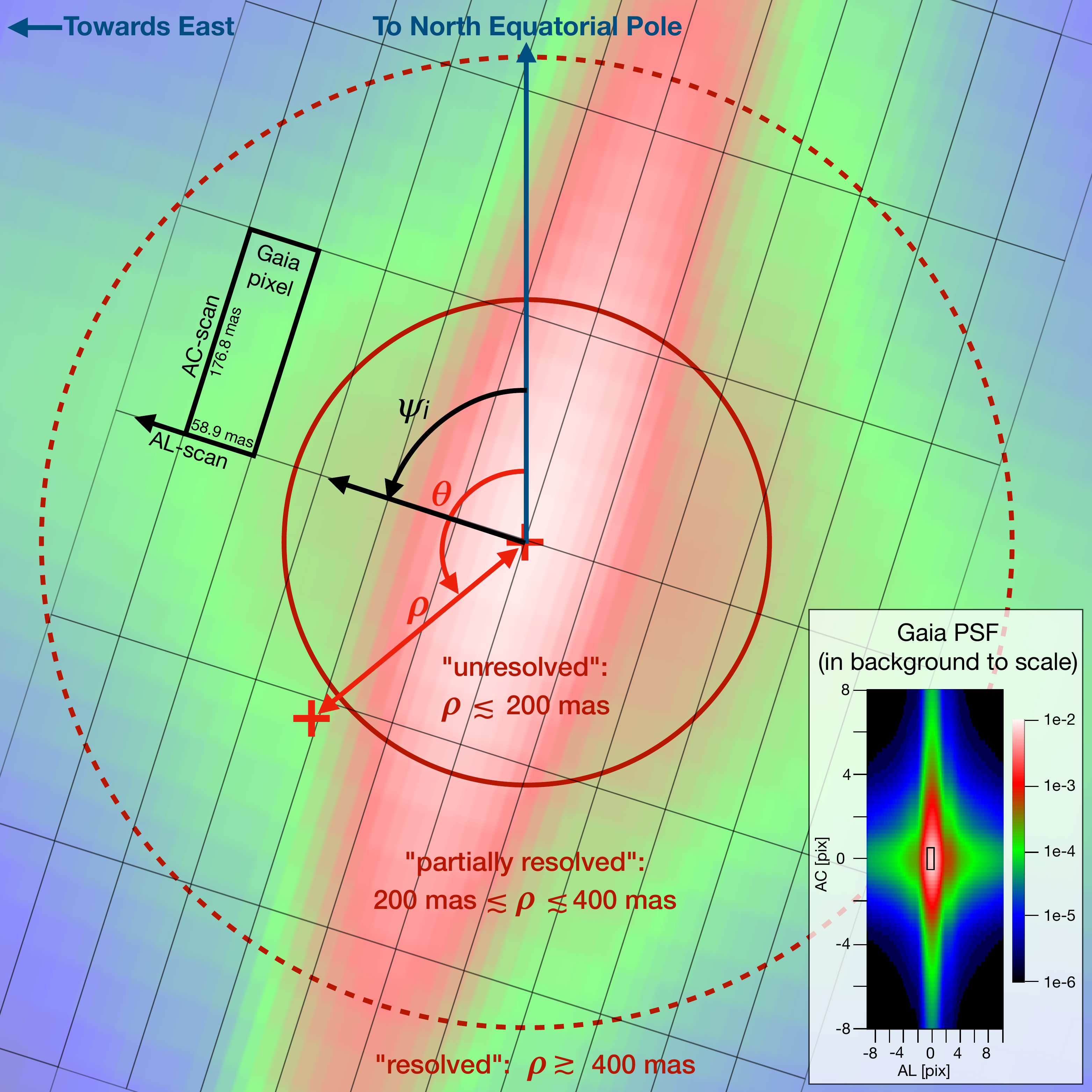}                 
\caption{\label{fig:AfAngularSepIllustration}Sky-projected illustration of the rough zones in which equal-brightness optical binary stars (two red crosses) at angular separation~$\rho$ and position angle $\theta$ can be resolved by \gaia. In the partially resolved region, the stars are resolved into two components depending on the scan angle $\psi_i$ of the observation $i$, because of the asymmetric PSF, which has the highest spatial resolution in the along-scan direction. The bottom right inset shows a typical PSF profile  \citep{2016A&A...595A...3F} that is rotated and scaled in the background image to represent the expected PSF of the upper right component of the binary star for the given scan angle. East direction (increasing RA) is towards the left.}
\end{figure}

As explained in Sect.~\ref{sssec:ipdModErrNonPointLike}, IPD model errors can arise from multi-peak or non-point-like (extended) sources. The latter is also clearly demonstrated in \cite{DR3-DPACP-101} for galaxies that are extended by definition. Figure~\ref{fig:AfAngularSepIllustration} illustrates the main concepts involved here, such as the pixel size and proportions, the PSF, the scan angle, and the separation between the two components of an optical binary star.
The \gaia scan angle is defined as $\psi=0\degr$ when the field of view is moving towards local north, and $\psi=90\degr$ towards local east, which is different from that used for Hipparcos \citep[for example][]{Leeuwen:2007kx}.
In the following, we illustrate the main three cases described in the detailed description of \ipdGofHarmPhase \footnote{See \gaia archive documentation for \href{https://gea.esac.esa.int/archive/documentation/GDR3/Gaia_archive/chap_datamodel/sec_dm_main_source_catalogue/ssec_dm_gaia_source.html\#gaia_source-ipd_gof_harmonic_phase}{\ipdGofHarmPhase}.}. All examples shown in this study are also summarised in Table~\ref{tab:ipdSignalsList}. 

\begin{figure}[h]
\centering
  \includegraphics[width=0.49\textwidth]{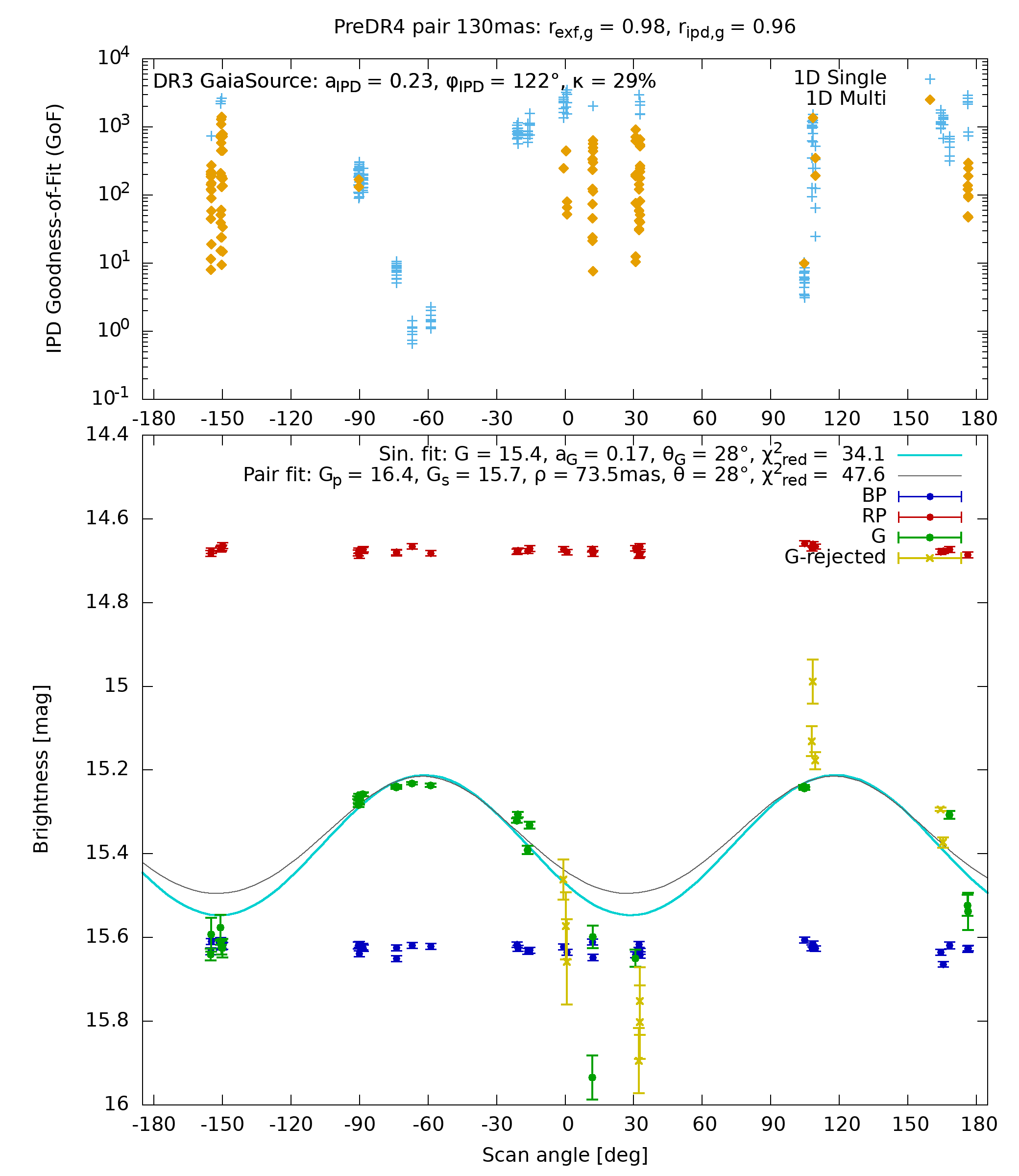}
  \includegraphics[width=0.45\textwidth]{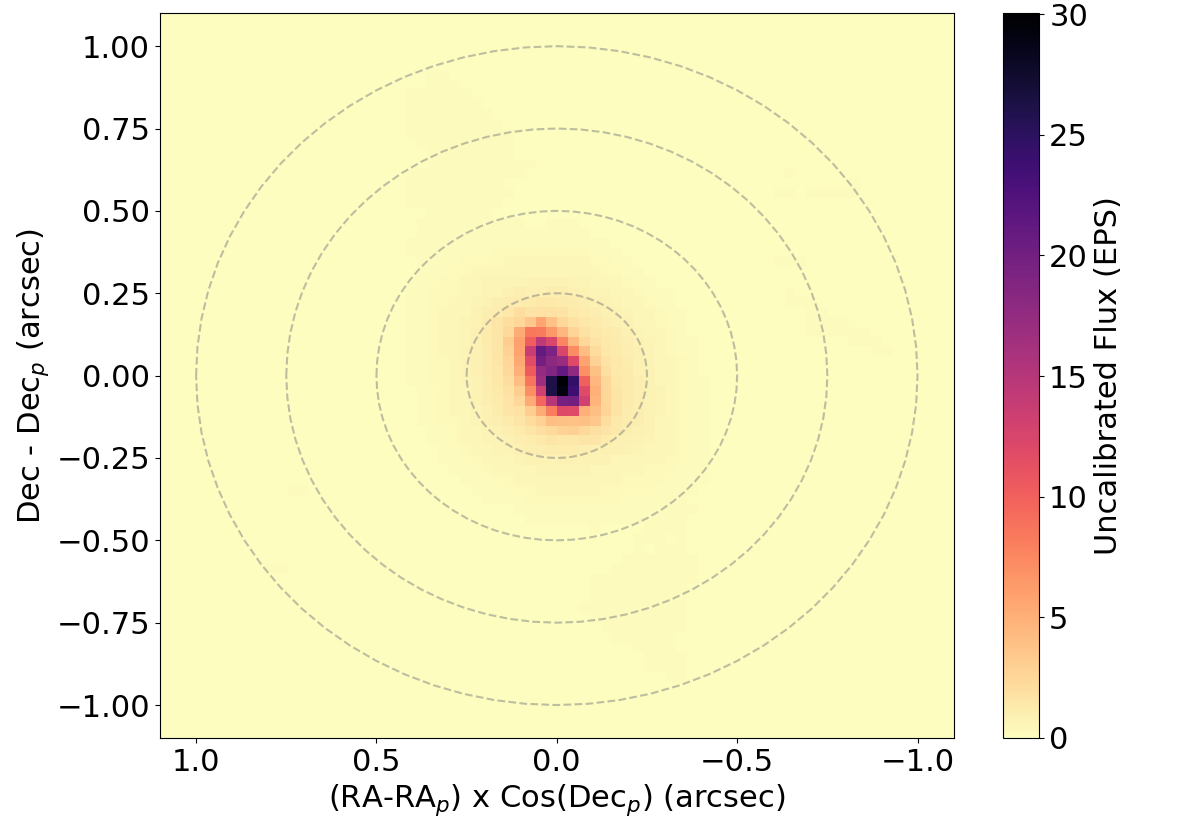}
\caption{\label{fig:closePair130mas}SourceID 389636619892245248: Scan-angle signatures from a partially resolved double star with similar magnitudes and a separation of about 130~mas between the two components (determined by IDU for DR4).
The top panel shows the unpublished IPD epoch GoF values determined by IDU in DR3, where we indicate the scans for which the IPD detected multiple peaks.
The central panel shows the brightness in the G band and in the BP and RP photometry as provided by the epoch-photometry table published in DR3, to illustrate the differences in that instrument. It also includes the fits to $G$ using Eq.~\ref{eq:simu} (pair model) and Eq.~\ref{eq:simu2} (sinusoidal model).
The bottom panel shows the image reconstructed by SEAPipe, with dashed grey circles at increasing radii in steps of 250~mas from the image centre. See text for further details.}
\end{figure}

    \noindent \textbf{Case 1: A double star with separation $\lesssim 0.1 \arcsec$}, where the GoF is expected to be higher (worse) when the scan is along the arc joining the components than in the perpendicular direction, and the \ipdFracMultiPeak should be small.
    Figure~\ref{fig:closePair130mas} shows an example of a partially resolved binary with the scan angle, IPD, and photometric signals. We make use of some internal information provided by the intermediate data updating system \citep[IDU; see][]{2016A&A...595A...3F} during its initial runs for the \gaia data release 4 (DR4), and some unpublished data of DR3. The $r$ values in the title are Spearman correlations introduced in Sect.~\ref{sec:detectSaSignals}, which help determine whether this is a scan-angle-dependent signal, which in this case is very likely given that both correlations are close to 1.
    The top panel lists the published \ipdGofHarmAmpl, \ipdGofHarmPhase, and \ipdFracMultiPeak as $a_\text{ipd}$, $\varphi_\text{ipd}$, and $\kappa$, together with the time series of the IPD goodness of fit (per CCD observation) differentiated between observations detected as single peak or multiple peaks.
    The central panel shows the derived photometry (per field-of-view transit) in \gmag, \gbp , and \grp. For \gmag we include two fits to the data that are detailed in Sect.~\ref{sec:responseToSaSignal}: (1) 
    a Pair fit (Eq.~\ref{eq:simu}), which is a generalisation of Eq.~\ref{eq:ipdScanAngleModel}, leading to magnitude estimates for the primary and secondary components ($G_p$ and $G_s$), their separation $\rho,$ and their position angle $\theta$; and (2) a small-separation simpler Sine fit (Eq.~\ref{eq:simu2}) that has the same parametrisation as Eq.~\ref{eq:ipdScanAngleModel} and whose amplitude \aG is comparable to that of $a_\text{ipd}$ (although with a different unit); see also Fig.~\ref{fig:samAmplVsIpdAmpl} of Appendix~\ref{sec:auxTablesStats}. Depending on the separation, the phase \thetaG is $\pm 90\degr$ offset to $\varphi_\text{ipd}$ (as is the case here) or similar in value. 
    This second simpler model can perform better than Eq.~\ref{eq:simu} for small separations, especially when the secondary peak is never resolved, and it is provided for all photometric sources with available time series in Appendix~\ref{sec:auxTablesStats}. The goodness of both fits is indicated as \redChiSq. 
    The bottom panel shows an approximate reconstruction of the source environment using the source environment analysis pipeline, SEAPipe \citep[][see also Sect.~\ref{ssec:sourceEnv}]{harrison11}.
    In this example, as in many other cases with partially resolved pairs, AGIS was unable to determine a full solution, and therefore only a two-parameter solution is available in DR3. As can be seen, the G-band photometry strongly depends on the scan angle, with fainter values for the scans in which IPD was able to detect and mask the secondary peak (labelled ``{\tt Multi}'' in the top panel). In the unresolved scans (``{\tt Single}''), both peaks are combined in the IPD fitting, leading to an artificially brighter value. In this case, the separation is large enough to allow for a 
    rather high $\kappa$ of 29\% and a good fit with the pair model. 
    Photometry from the blue and red photometer instruments (BP and RP, respectively) is mostly constant because of the larger windows used there. Finally, the lower (better) epoch GoF values are found in the scans in which the two peaks are not resolved, as expected. For completeness, the central panel also includes the G-band observations that were rejected during variability processing and excluded from our fitting procedure, that is, with \href{https://gea.esac.esa.int/archive/documentation/GDR3/Gaia_archive/chap_datamodel/sec_dm_photometry/ssec_dm_epoch_photometry.html#epoch_photometry-variability_flag_g_reject}{\tt variability\_flag\_g\_reject}\texttt{=true}. 
    
\begin{figure}[h]
\centering
  \includegraphics[width=0.49\textwidth]{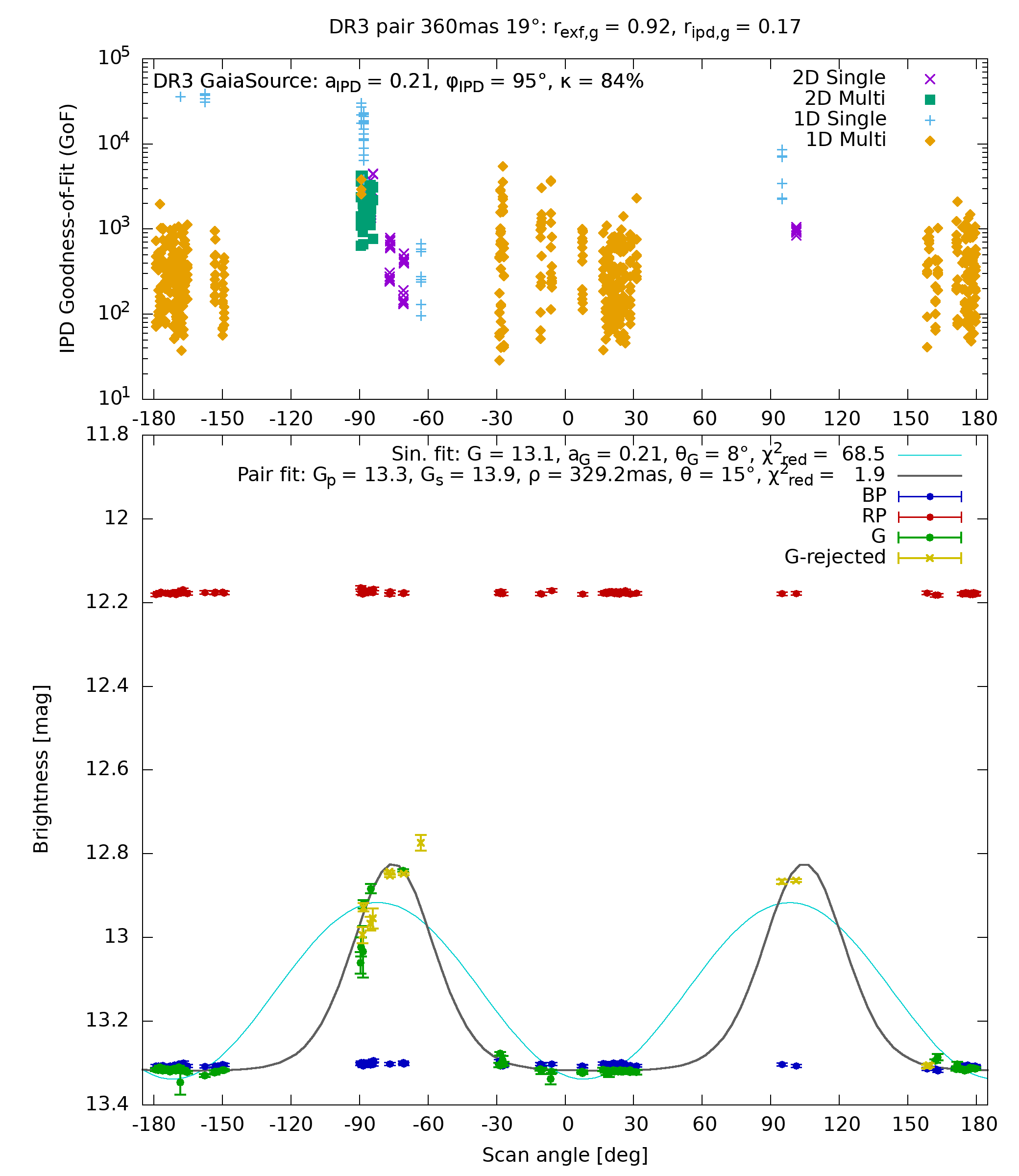}
  \includegraphics[width=0.45\textwidth]{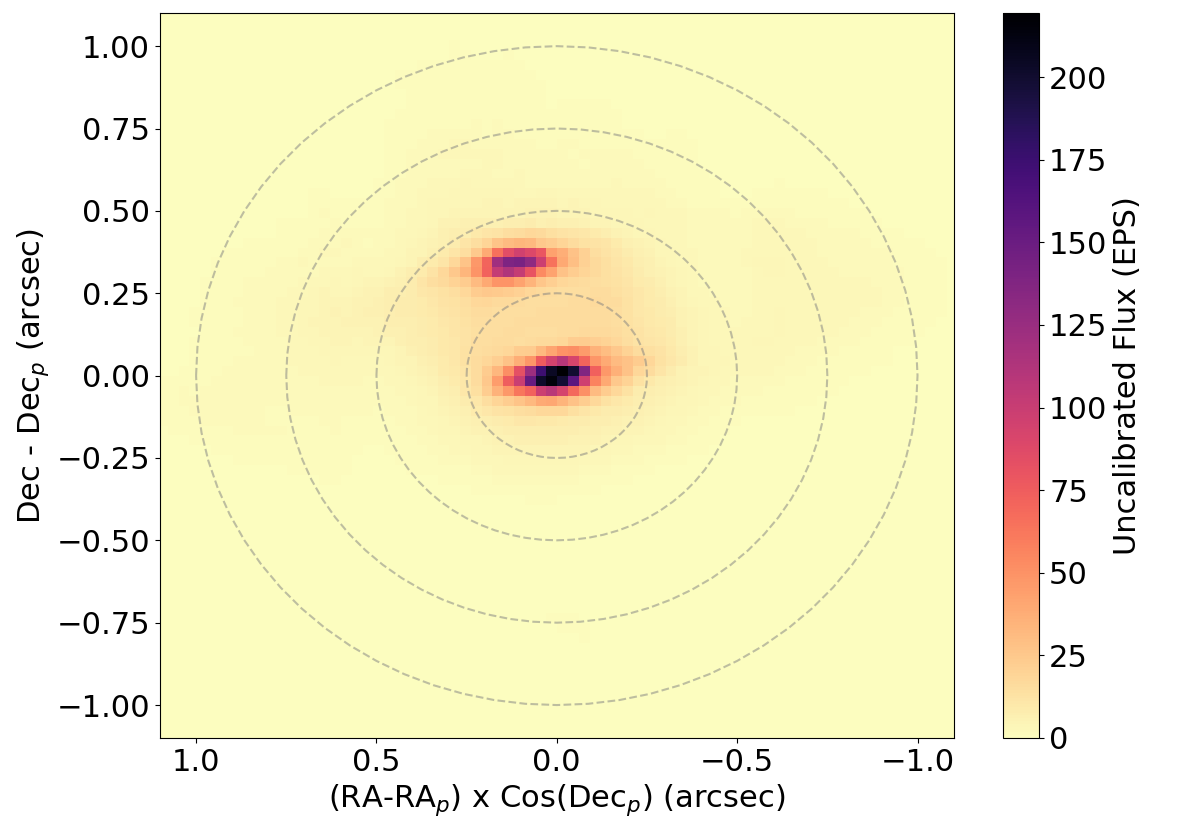}
\caption{\label{fig:closePair371mas}SourceID 382074694311961856: Scan-angle signatures (top and central panels) and image reconstructed by SEAPipe (bottom panel) from a resolved double star with a separation of about 360~mas between the two components, available as two separate sources in DR3 (only one of the two sources is shown in the top and central panels). See text and Fig.~\ref{fig:closePair130mas} for further details.}
\end{figure}

   \noindent \textbf{Case 2: A resolved binary}, in which the GoF is expected to be smaller (better) when the scan is along the arc joining the two components (along the position angle), and the \ipdFracMultiPeak value should be high.
    Figure~\ref{fig:closePair371mas} shows an example that again shows strong variation in G-band photometry with the scan angle. With this larger separation, Eq.~\ref{eq:simu} fits  the signal much better than Eq.~\ref{eq:simu2}. In this case, depending on the epoch, this source is assigned one-dimensional or two-dimensional windows because its magnitude is close to 13. The G-band photometry again becomes brighter, especially when the IPD is unable to detect the secondary peak, and vice versa. The separation is still too small to cause any significant variations in the larger BP and RP windows, meaning that these bands will contain the contribution from both sources. For the epoch IPD GoF, better fits (lower values) are obtained when the IPD detects and masks the secondary source, as expected. This typically occurs when the scan is made along the arc joining the two components. The value of $\kappa$ (84\%) is very high.

\begin{figure}[h]
\centering
  \includegraphics[width=0.49\textwidth]{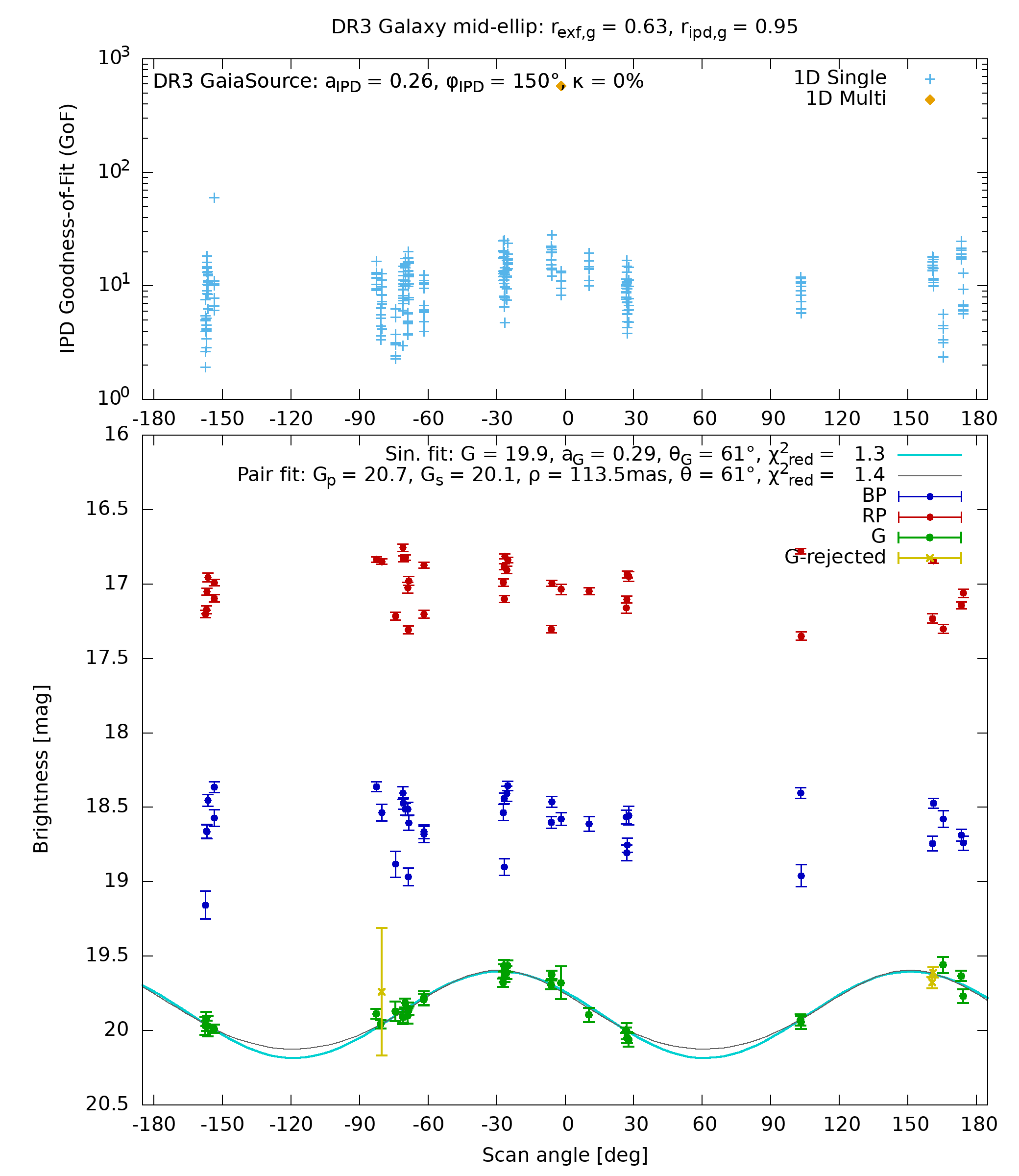}
  \includegraphics[width=0.45\textwidth]{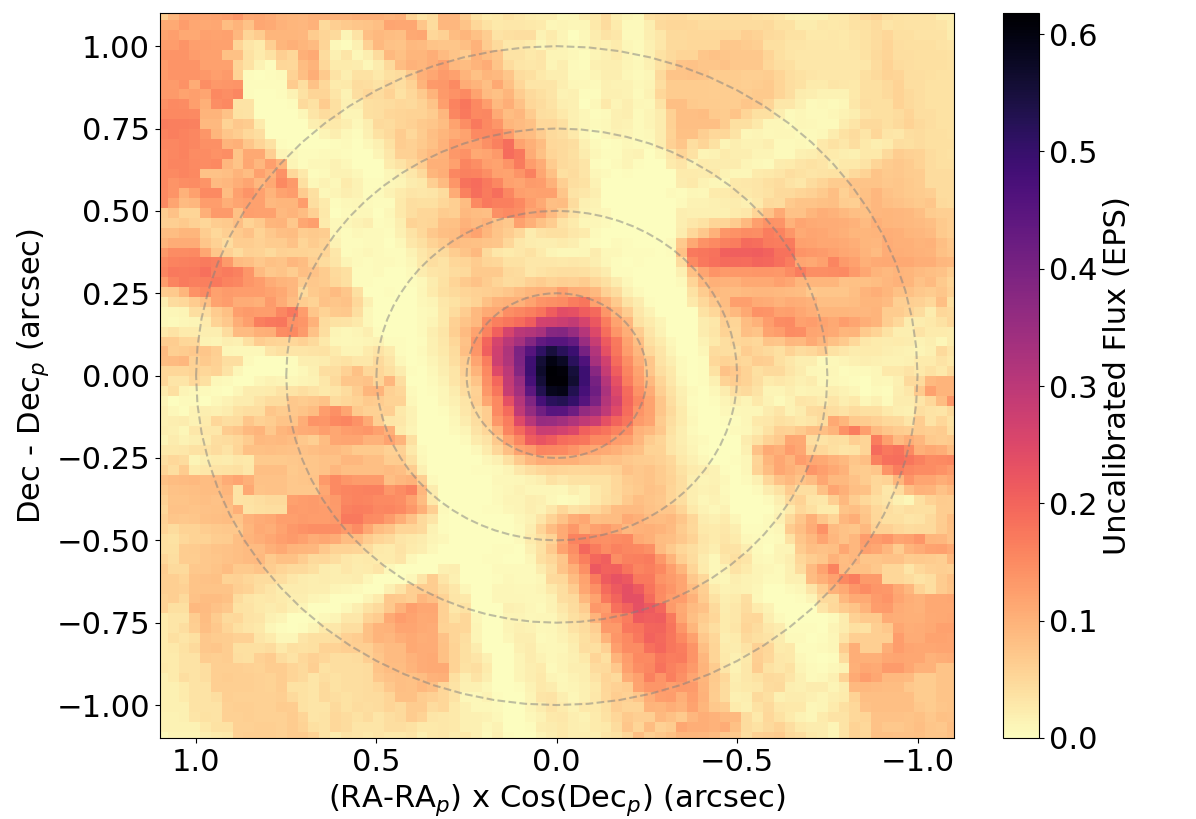}
\caption{\label{fig:galaxymid}Galaxy LEDA~2112767 \citep{2003A&A...412...45P}, sourceID 366951667785042688: Scan-angle signatures (top and central panels) and image reconstructed by SEAPipe (bottom panel).
This galaxy was published in \gdr{3} with a moderate ellipticity.
See text and Fig.~\ref{fig:closePair130mas} for further details. }
\end{figure}

    \noindent \textbf{Case 3: A galaxy with elongated intensity distribution}, for which a smaller GoF is expected when the scan is along the major axis of the image.
    Figure~\ref{fig:galaxymid} shows an example for a galaxy candidate with G-band magnitude around 20, two-parameter AGIS solution,
    and the following {\em de Vaucouleurs} fitted parameters \citep{2022arXiv220614491D}: radius $1.65\arcsec$, ellipticity 0.35, and position angle $63.4\deg$.  \ipdGofHarmPhase takes a value of about 150$\deg$. The predicted difference is nearly 90$\deg$ with respect to the correctly fitted position angle \thetaG. 
    This time, the variations in \gbp and \grp photometry are significant because the source is very extended, as further explained in Sect.~\ref{ssec:xpInstr}. This also causes the \gbp and \grp to be brighter than the \gmag. On the other hand, $\kappa$ is zero, as 
    expected, since this smooth extension of the source cannot be identified as secondary peaks by the IPD (except in one single transit, which was probably a spurious detection). The pair model is obviously not applicable here: after the maximum number of iterations allowed is exceeded, the best fit indicates an unrealistically low $\rho$ value of 46~mas.


Appendix~\ref{sec:appendix-bcn-full-data} provides additional examples for the different cases. In general, the G-band photometric signal in magnitude is well modelled by 
Eq.\,\ref{eq:simu2} (which is identical to the IPD model of Eq.\,\ref{eq:ipdScanAngleModel}), although there can be exceptions where Eq.~\ref{eq:simu} performs significantly better. These models seems to provide better fits on photometry than on the IPD GoF, although \ipdGofHarmAmpl typically provides a reliable indication of scan-angle-dependent astrometric signals for the source. Combined with \ipdFracMultiPeak, it is a quite powerful tool for detecting extended sources or multiple point-like sources.
In addition to these published quantities, Eq.~\ref{eq:simu} seems to provide very interesting fits in case of moderate separations, even allowing us to localise a neighbouring source with quite some reliability.

\begin{figure}
    \centering
    \includegraphics[width=0.49\textwidth]{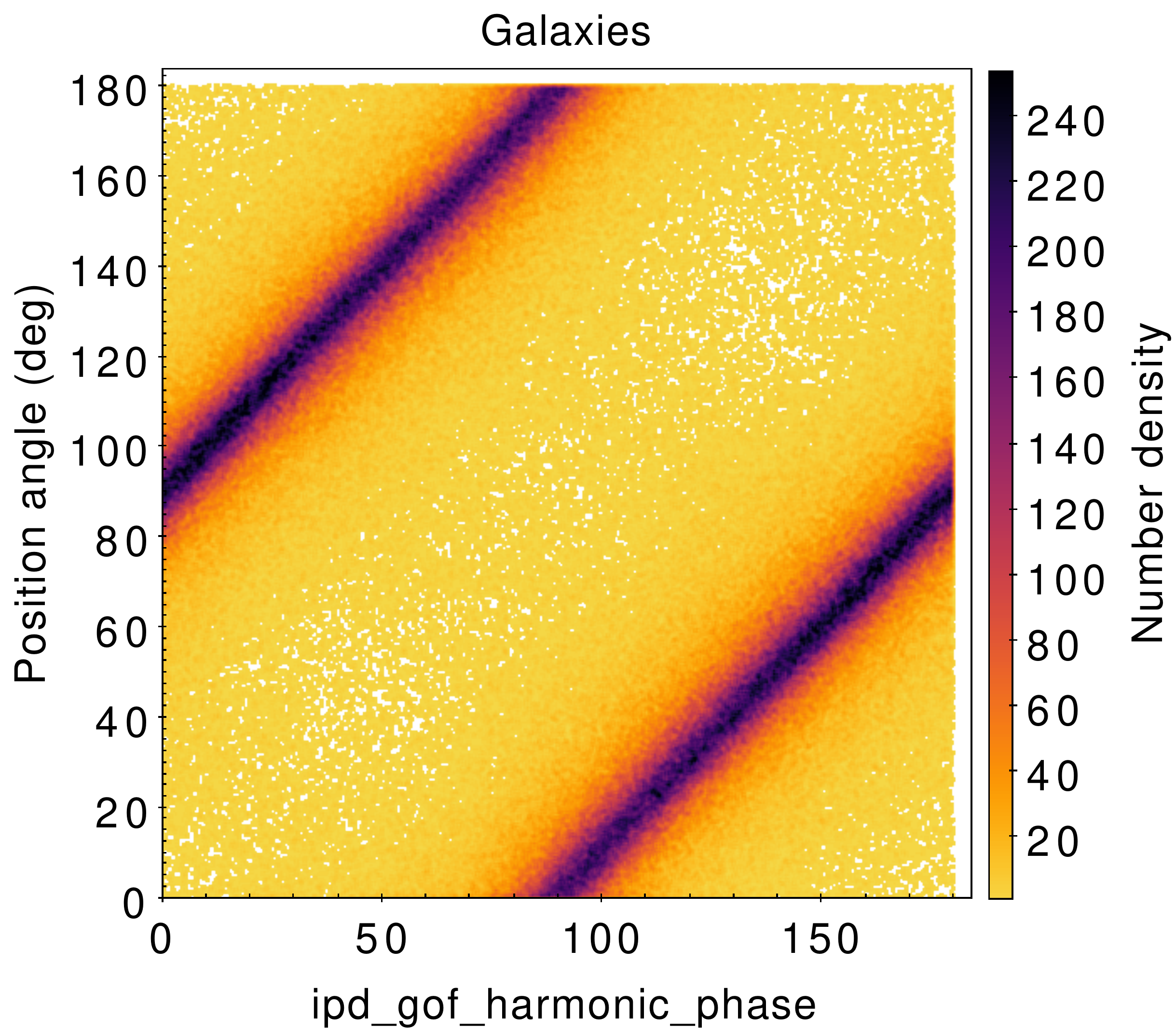} 
    \caption{Comparison of the position angle of extended galaxies measured with \gaia data with the \ipdGofHarmPhase parameter.}
    \label{pa_ipdPhaseGalaxies}
\end{figure}

To further illustrate the usefulness of these IPD-related published parameters, Figure~\ref{pa_ipdPhaseGalaxies} presents the density plot of the comparison of \ipdGofHarmPhase with the position angle for $\sim$914\,000 galaxies measured by the \gaia surface-brightness profile fitting pipeline of the extended objects processing in the fourth coordination unit, CU4-EO \citep{DR3-DPACP-153, DR3-DPACP-101}. The two parameters agree very well (with a 90$\deg$ shift, as previously explained). Sources that depart from the dense lines are quasi-circular galaxies for which the position angle is meaningless.

All \gmag, \gbp, and \grp photometry shown in this study is as published in \gdr{3} \citep{2021A&A...649A...3R,DR3-DPACP-142} (unless otherwise stated). The observations flagged by the variability analyses of \cite{DR3-DPACP-162} were rejected, that is, observations with 
\href{https://gea.esac.esa.int/archive/documentation/GDR3/Gaia_archive/chap_datamodel/sec_dm_photometry/ssec_dm_epoch_photometry.html#epoch_photometry-variability_flag_g_reject}{\tt variability\_flag\_g\_reject}\texttt{=true} in the epoch-photometry time-series data.

\subsection{Scan-angle-dependent signals in the blue and red photometer instruments \label{ssec:xpInstr}}

\begin{figure}
    \centering
    \includegraphics[width=0.49\textwidth]{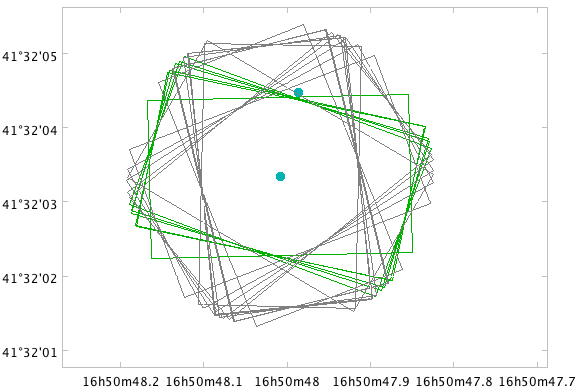} 
    \caption{
    Example of two sources causing blended spectra in some of the observations. The rectangular shapes show the footprint of the observing window for real transits over one of the two sources. In the transits highlighted in green, the secondary source is located beyond the window, while both sources are inside the window for grey transits. The dispersion direction is along the major side of the observing window.} 
    \label{Fig:xpScan}
\end{figure}

The BP and RP instruments 
measure a low-resolution spectrum ($R\sim 60$) in the blue [300-700] nm and red [600-1100] nm part of the spectrum. For a detailed description of the instrument, we refer to sect.~3.3.6 of \cite{2016A&A...595A...1G}. 
The integration of the flux within the window (aperture photometry) in the two photometers generates the \gbp and \grp magnitudes. Because no LSF or PSF fitting is involved to process the data, it is less likely to introduce scan-angle-dependent model errors due to subtle LSF or PSF mismatches. However, scan-angle-dependent model errors can still be introduced through blending because the BP and RP spectra are acquired with windows that are much wider than those used for the AF observations in the AL direction, with a length of 60 pixels, corresponding to $\sim 3.5\arcsec$. 
This means that more blending is expected to occur in BP and RP than in $G$, especially in crowded regions. How strong the crowding effect is depends on the separation between sources, but also on the scan angle, as the amount of flux from the blending source will vary depending on the mutual position of the sources and the scanning direction.
An example is shown in Fig.~\ref{Fig:xpScan}. 

\begin{figure}
    \centering
    \includegraphics[width=0.49\textwidth]{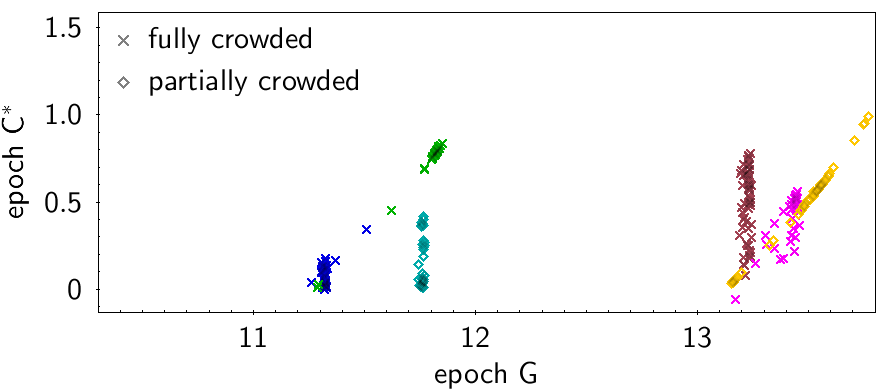} 
    \includegraphics[width=0.49\textwidth]{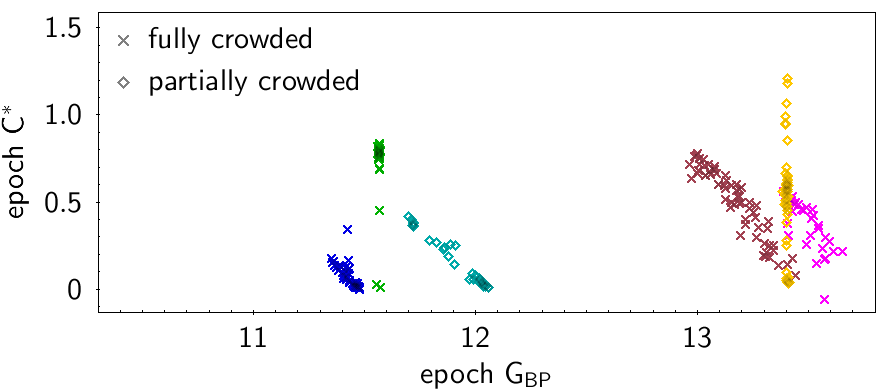} 
    \includegraphics[width=0.49\textwidth]{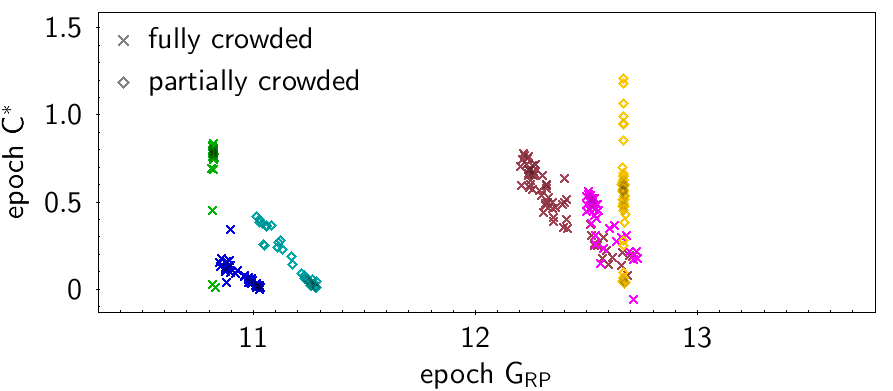} 
    \caption{
    Examples of crowding effects on the photometry for six sources, each with a different colour.  From top to bottom, we show the epoch-corrected flux excess \cxs as a function of epoch \gmag, \gbp, and \grp. 
    Sources shown as crosses were estimated as crowded in every BP/RP transit. Sources shown as diamonds were estimated as crowded in only a few transits. 
    } 
    \label{Fig:xpCstar}
\end{figure}

The \gbp and \grp magnitudes are calculated by integrating the respective spectra, and because no deblending correction has been applied in DR3 yet, they can be affected by crowding. 
In sect.~6 of \cite{2021A&A...649A...3R}, the corrected BP and RP flux excess \cxs for the photometry was defined. \cxs is a consistency metric for the mean photometry, and it depends on the \gmag, \gbp, and \grp fluxes and the colour. 
We have computed the equivalent \cxs from epoch photometry.
Figure~\ref{Fig:xpCstar} shows some examples of the variation in epoch-flux excess  \cxs as a function of the epoch photometry. 
In some cases (green crosses and yellow diamonds), \cxs correlates with \gmag, but it does not depend on \gbp and \grp: here the two sources are close enough that every BP and RP transit contains the flux of both sources, while the amount of flux in the AF varies with the scan angle. 
In the other cases, \gbp and \grp are instead correlated with \cxs, indicating that the amount of flux in BP/RP varies with the scan angle, while in \gmag, the two sources are distant enough to prevent contamination, and \gmag is not affected or it is affected only negligibly. 
The crowding evaluation was carried out in the BP/RP processing, and the crowding status in the plots is relevant only for those instruments. 
While the examples represent only a handful of cases, the bottom panel of fig.~19 in  \cite{2021A&A...649A...3R} shows this effect in a more global way: The corrected BP and RP flux excess with the colour of the sources colour-coded by blend probability clearly shows that \cxs is closer to zero (indicating good and consistent photometry) when the blend probability is lower than 20\%.

The examples in  Fig.~\ref{Fig:xpCstar} show that sources can be strongly correlated between the epoch-corrected flux excess \cxs and photometric \gmag, \gbp, and \grp. We quantify this correlation using three Spearman correlations: \rExfG, \rExfBp, and \rExfRp, as detailed in Sect.~\ref{ssec:corExFlux} and published with this paper
for all sources with published photometric time series in \gdr{3} as described in Appendix~\ref{sec:auxTablesStats}.

\subsection{Radial Velocity Spectrometer\label{ssec:rvsInstr}}

The Radial Velocity Spectrometer (RVS) \citep{2018A&A...616A...5C} produces high-resolution spectra ($R\sim 11,500$) of sources between 845 and 872 nm. The light of sources observed by the RVS is dispersed (0.0245 nm pixel$^{-1}$) by a grating plate, resulting in a spectrum that spreads over 
about 1100 pixels in the AL direction, corresponding to 65 arcsec. The RVS focal plane is composed of 12 CCDs arranged in four rows. 

The wavelength range of RVS spectra contains, amongst other possible lines, the calcium triplet, which allows measuring the radial velocity for a wide range of stellar types. The RVS instrument is aimed to estimate the mean radial velocity (\rv) at the end of the mission for basically all stars up until $\grvs\sim\,$16, and to provide time-series \rv data up until $\grvs \lesssim\,$13.

\subsubsection{Scan-angle-dependent signals in RVS data of astrometric binaries\label{sssec:rvSignalAstrometric}}

The RVS does not have any internal calibration source to perform a wavelength calibration of individual spectra.
The wavelength values are associated with each sample of a spectrum, assuming that the wavelength is a polynomial function of the field angles ($\eta$,\, $\zeta$) of the source, that is, the position of the source in the FoV reference system \citep[FoVRS; see][for the definition of this reference system]{2012A&A...538A..78L}, at the time at which the sample crossed the fiducial line of the CCD. The coefficients of the above function are quantities that evolve slowly in time, and they are determined using the observations of bright stars with known radial velocity \citep[see][for more details]{2018A&A...616A...6S}.

In the RVS DR3 processing, the field angles of the observed source are computed from the single-star astrometric parameters determined by AGIS \citep{2012A&A...538A..78L}. If the real position of the source in the FoVRS is different from the predicted position, the wavelength associated with the spectrum samples will be incorrect. This mismatch can be due either to a problem in the AGIS astrometric parameters of the source or to the astrometric motion along the Keplerian orbit of the star in the case of an astrometric binary.
Because the RVS dispersion occurs in the AL direction, the effect of a mismatch in the position is at first order proportional to $\Delta\eta$, that is, the displacement projected on the AL direction. The effect on the epoch radial velocity (that is, the difference between the measured and the real \rv ) will be $\Delta RV=-0.146\cdot \Delta\eta$, with $\Delta RV$ in km s$^{-1}$, and $\Delta \eta$ (in mas) being the difference between the real and the assumed $\eta$ value. 

\begin{figure}[h]
  \includegraphics[width=0.45\textwidth]{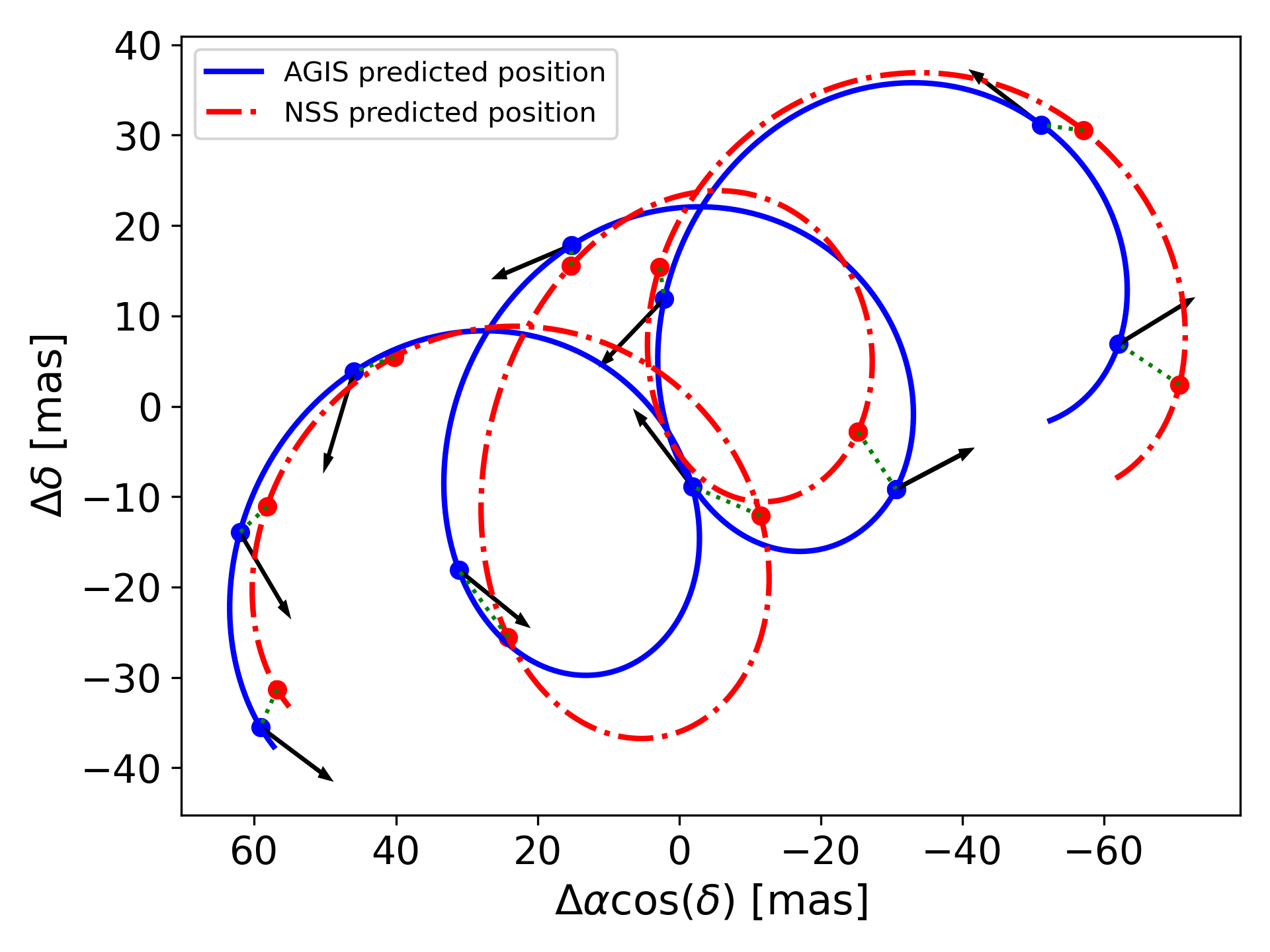}                         
  \includegraphics[width=0.45\textwidth]{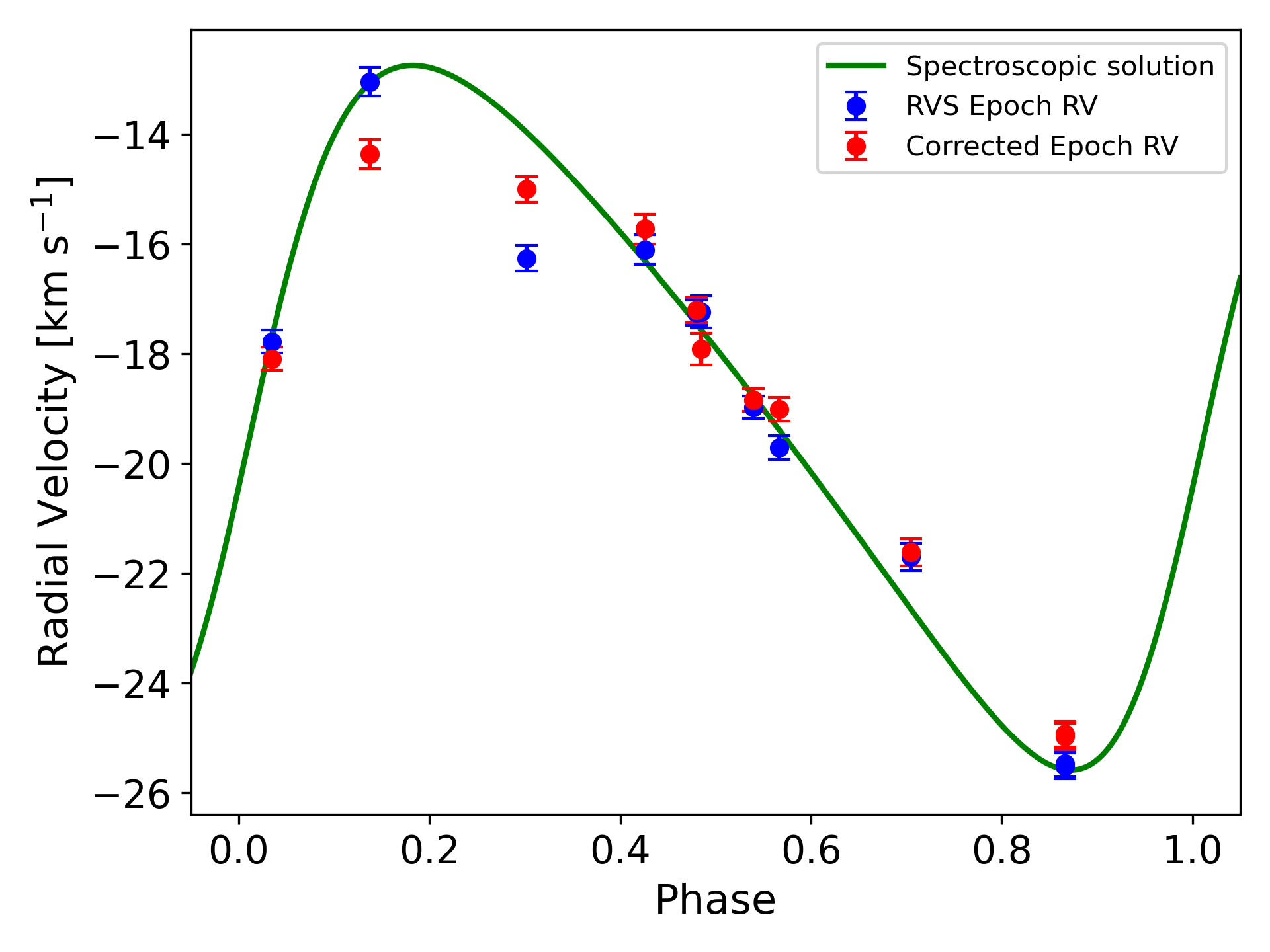}                         
\caption{Example demonstrating how insufficient astrometric modelling leads to incorrect RV determinations.
Top panel: Motion of \gaia\ DR3 6631710606341412096 on the sky with respect to its reference position in the reference system moving with \gaia, as predicted from the five-parameter single-star AGIS solution (solid blue line), compared with the position predicted by the NSS \texttt{AstroSpectroSB1} solution (dot-dashed red line), which included the Keplerian orbit. Circles show the positions at which the epoch \rv were measured, and the arrows show the scanning direction at the epoch. Bottom panel: RV data, folded in phase, as provided by the DR3 pipeline (blue dots) and the data corrected for the displacement (in red), compared with the radial velocity predicted by the \texttt{AstroSpectroSB1} solution (green line).
}
\label{fig:rvs_orbital}
\end{figure}

An example of the effect of the astrometric orbit on the epoch \rv of an astrometric binary, \gaia\ DR3 6631710606341412096, is shown in Fig.~\ref{fig:rvs_orbital}. The \texttt{AstroSpectroSB1} solution of this source has an orbit with period of 937.0 days, a semi-axis $a_0=11.79$ mas, and a parallax of 25.98 mas.
See \citet{DR3-DPACP-100} for the description of non--single--star (NSS) solutions.
In the top panel of Fig.~\ref{fig:rvs_orbital}, we show the position of the source on the sky in the reference system moving with \gaia, as predicted from the AGIS single-star astrometric solution, compared with the position predicted when the astrometric orbit is included. The bottom panel shows the epoch \rv data, folded in phase, as provided by the DR3 pipeline (blue dots), and the data corrected for the displacement (in red), compared with the \texttt{AstroSpectroSB1} solution.

The bottom panel of Fig.~\ref{fig:rvs_orbital} shows that the effect of the astrometric orbit on the epoch \rv~(not published in \gdr3) can, as a consequence, produce an NSS solution that is not fully correct. In the case of \gaia\ DR3 6631710606341412096, which was chosen from those in which the effect is strongest, the semi-amplitude of the \rv curve is certainly affected, but not dramatically so. The semi-amplitude is the most affected spectroscopic orbital parameter, but the eccentricity (and the argument of periastron) might also be sensitive. However, the spectroscopic values are certainly not predominant
in the combined orbital solution; the dominant constraints always come from the astrometry.

It should be noted that this effect is weaker than the epoch \rv errors for the vast majority of the astrometric binaries detected by \gaia, and it is relevant only for nearby and bright astrometric binaries with a large semi-axis orbit. Using the orbit semi-axis from the published non-single star (NSS) \texttt{Orbital} and \texttt{AstroSpectroSB1} solutions as estimate of the maximum expected $\Delta \eta$, we found that 1876 sources with \texttt{AstroSpectroSB1} and 1024 sources with \texttt{Orbital} solutions have an effect on the epoch radial velocity that is stronger than the mean of epoch \rv errors. The means of epoch \rv errors are not published in \gdr3. A correction of this problem is planned for the next release.

\subsubsection{Scan-angle-dependent signals in RVS data of non-point-like sources\label{sssec:rvSignalResolved}}

The second situation in which the epoch radial velocities of a source receive a spurious signal that depends on the scan angle is when the source is a resolved or partially resolved binary or double star. The meaning of resolved or partially resolved is discussed in Sect. \ref{sssec:ipdStats}.

If the onboard software is not able to distinguish the two stars composing the source, a single window is generated when the source is observed. In this case, the RVS pipeline is not able to deblend the overlapping spectra of the two components, and the spectra will be processed as if it were a single star.

The absorption lines of the two components in the processed spectra will appear shifted in wavelength with respect to their expected position, proportionally to the displacement with respect to the predicted position in the FoVRS, projected on the AL direction, according to the relation $\Delta RV=-0.146\cdot \Delta\eta$ described in the previous section. 

The RVS pipeline includes an implementation of an algorithm similar to the two-dimensional correlation technique, \texttt{TodCor} \citep[][]{1994ApJ...420..806Z,DR3-DPACP-161} to identify double-lined spectra. As described in \citet{DR3-DPACP-161}, the algorithm has limits: When the source is fainter than $G_{\mathrm{RVS}} = 11$, or when the faintest component is more than five times fainter than the primary, or when the \rv separation between the lines of the two components is below 15 km s$^{-1}$, the RVS pipeline is unable to identify the spectrum as double lined.

When the blended spectra are not detected as double lined and the two components have similar radial velocities (for example, as expected in wide binary systems), the blending will result in a shift of the position of the centroid of the absorption lines. This will generate a radial velocity signal that is proportional to the separation projected on the AL direction. The spurious signal will be
\begin{equation}
\label{eq:rv_scan_pos}
    \Delta RV\sim K\cos(\psi-\theta)
,\end{equation}
where the semi-amplitude $K$ depends on the separation, the luminosity ratio, and the respective spectral types, while $\psi$ and $\theta$ are the scan angle and the position angle of the secondary, respectively.
Because the scanning angle has the same periodicity as the spacecraft precession, this will generate spurious NSS SB1 solutions with a similar period, as noted in \citet{DR3-DPACP-100}.

\begin{figure}[h]
  \includegraphics[width=0.45\textwidth]{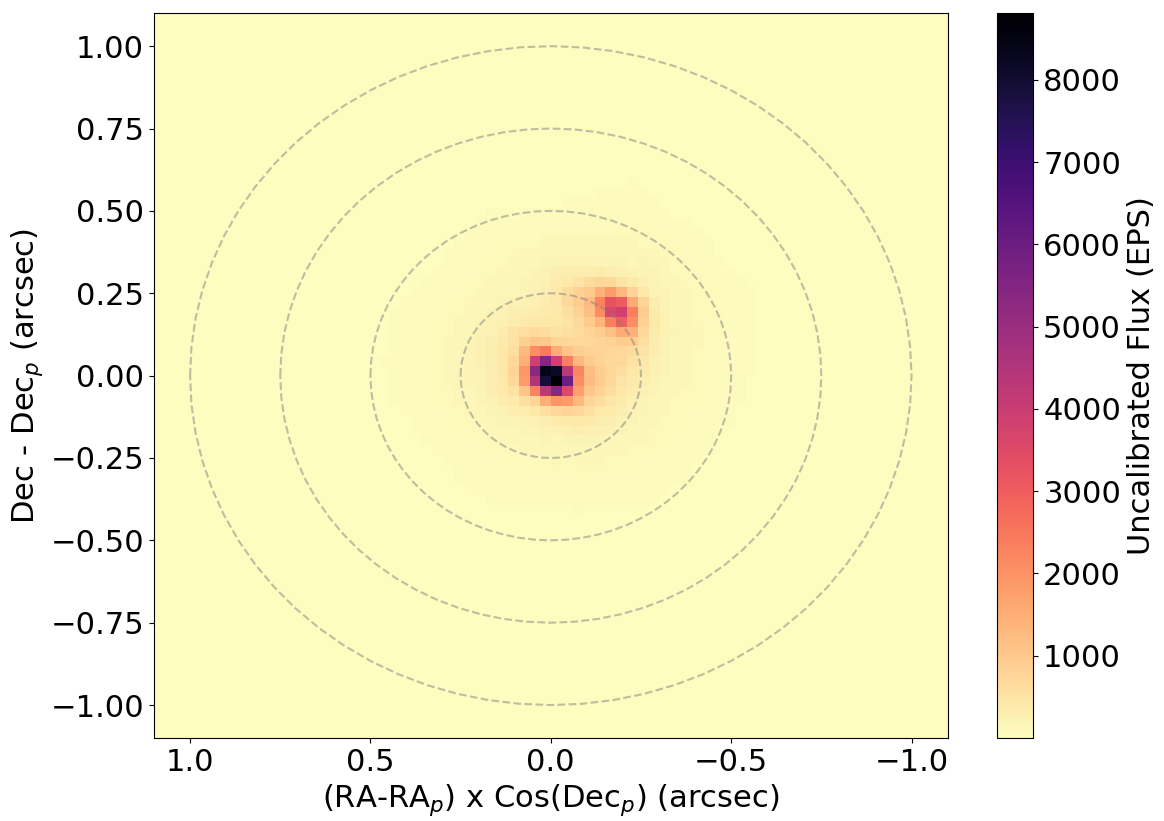}                         
  \includegraphics[width=0.45\textwidth]{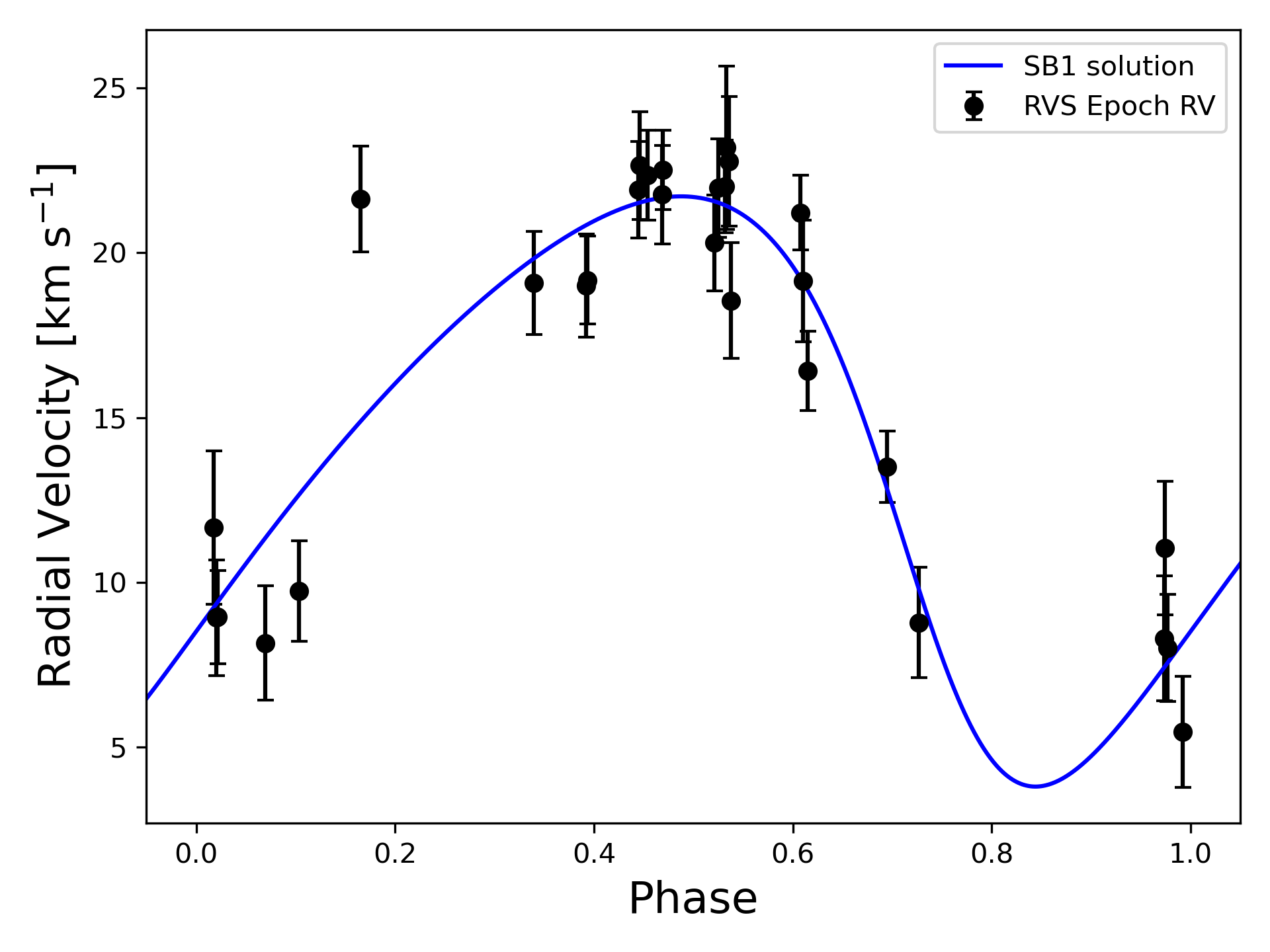}                         
  \includegraphics[width=0.45\textwidth]{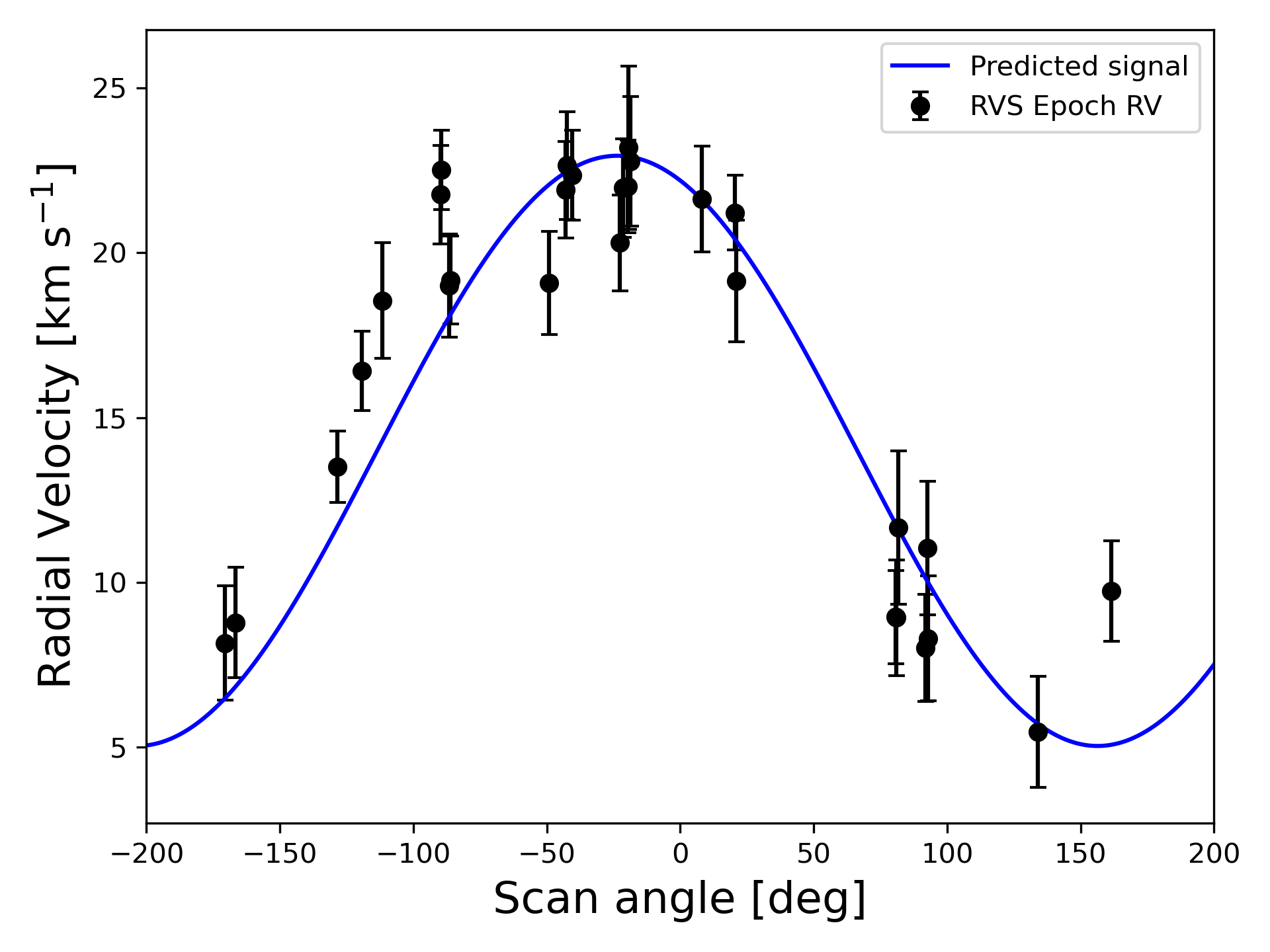}                         
\caption{Top panel: Image of the source \gaia\ DR3 5648209549925093504 produced by SEAPipe. Middle panel:  RV data of \gaia\ DR3 5648209549925093504, folded in phase, as provided by the DR3 pipeline (black dots), compared with the SB1 solution provided in DR3 (blue line). Bottom panel: RV data as a function of the scan angle $\psi$, compared with a sinusoidal signal as predicted by equation \ref{eq:rv_scan_pos}.
}
\label{fig:rvs_resolved_binary}
\end{figure}

An example of a spurious SB1 solution due to a resolved binary is \gaia\ DR3 5648209549925093504. This source, as revealed by SEAPipe preliminary results shown in the top panel of Fig.~\ref{fig:rvs_resolved_binary}, is composed of two stars 
that are separated by about 300 mas. This source has $\ipdFracMultiPeak=90$ and $\ipdGofHarmPhase=66.4\deg$.  As explained in Sect.~\ref{sssec:demoSaSignalsPointSrc}, we obtain a position angle $\theta \sim  \ipdGofHarmPhase-90\deg=336.4\deg$, which agrees well with what is seen in the SEAPipe image.

The middle panel shows the epoch \rv data, folded in phase, compared with the published SB1 solution. In the bottom panel, the epoch \rv data, folded with the scan angle, are compared with the \rv{} predicted by equation \ref{eq:rv_scan_pos}, and with $K$ equal to the semi-amplitude of the SB1 solution.
The measured \rv{} variability is well reproduced by the scan-angle effect. This proves that this is a spurious SB1 solution. An algorithm that can identify spurious solutions like this is planned for the next release.

When the \texttt{TodCor} algorithm identifies the spectra as double lined, the source might be identified as a double-lined spectroscopic binary (SB2) by the NSS pipeline, with a period near the precession period. We found no spurious SB2 solution in the DR3 data.

\subsubsection{Scan-angle-dependent signals in RVS data due to contamination\label{sssec:rvSignalContam}}

A third type of scan-angle-dependent signal is introduced by the contamination of a spectrum with the light from a nearby source \citep[see][]{2021A&A...653A.160S,2019MNRAS.486.2618B}. 
The RVS pipeline for DR3 \citep{DR3-DPACP-159} contains a deblending algorithm \citep[described in][]{2021A&A...653A.160S} that treats the case of overlap of windows of two (or more) sources. 
When a source is bright enough, however, its light can contaminate the spectra of nearby fainter stars even if their windows do not overlap.
As discussed in more detail in \citet{DR3-DPACP-159}, the RVS pipeline for DR3 is not able to identify such cases. 
During the DR3 validation phase, a method was identified to filter out these cases, although it was only applied to the mean radial velocity. The epoch $RV$s of some contaminated stars were instead processed in the NSS pipeline, generating occasionally spurious SB1 solutions. One example is \gaia DR3 2006840790676091776, which is a $G=11.18$ source that is contaminated by \gaia\ DR3 2006840790679122688 ($G=3.86$) at 31.9 arcsec.

\begin{figure}
    \centering
    \includegraphics[width=0.45\textwidth]{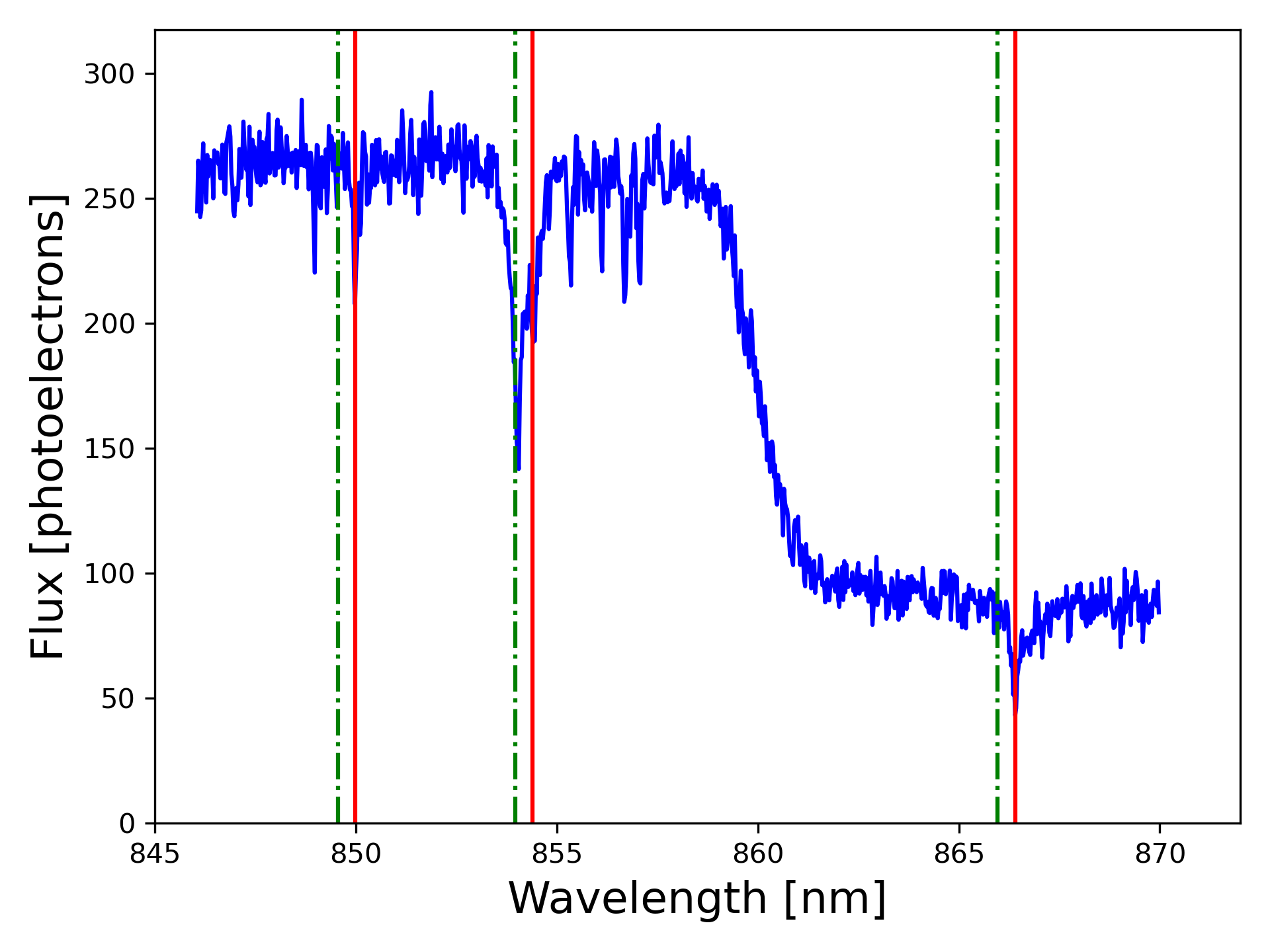}
    \caption{Spectrum of \gaia\ DR3 2006840790676091776, contaminated by the nearby source \gaia\ DR3 2006840790679122688 recorded in a transit. The solid vertical red lines show the real position of the \ion{Ca}{ii} triplet lines of \gaia DR3 2006840790676091776, and the dot-dashed green lines show the position of the same lines as found by the pipeline.}
    \label{fig:rvsSpecCont}
\end{figure}

In Fig.~\ref{fig:rvsSpecCont} we show the spectrum of \gaia\ DR3 2006840790676091776, recorded in one of the contaminated transits. At wavelengths shorter than 859\,nm, the spectrum is dominated by the light of the contaminating bright source \gaia\ DR3 2006840790679122688. The shoulder at 859\,nm corresponds to the red limit of the transmission band
associated with the contaminating source (and thus shifted by some 12\,nm here).
The solid vertical red lines show the real position of the \ion{Ca}{ii} triplet 
lines of \gaia\ DR3 2006840790676091776. The presence of the \ion{Ca}{ii} 866.452\,nm from the bright contaminating source (thus also shifted) near the \ion{Ca}{ii} 854.444 nm line of the contaminated source produces a peak in the cross-correlation function when the spectrum is compared with the template \citep[see][for details about the \rv derivation]{2018A&A...616A...6S}, resulting in an incorrect \rv.

\section{Spurious periods in \gaia data \label{sec:spuriousPeriodsInGaiaData}} 

\subsection{Observed period structure\label{ssec:obsPerDistr}}

\begin{figure*}[t!]
    \centering
    \includegraphics[width=\textwidth]{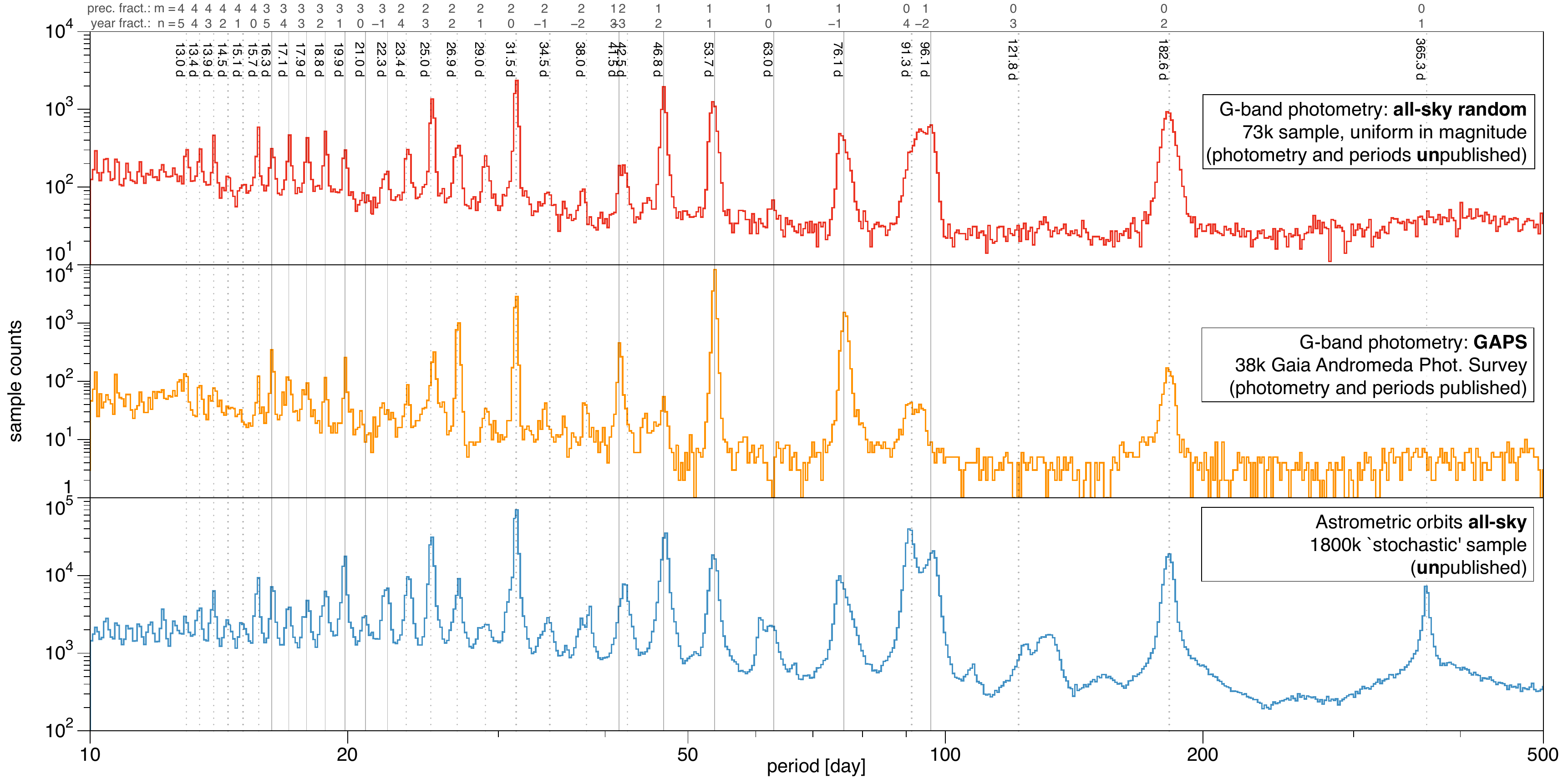}
    \caption{Period distributions of (largely unpublished) \gaia data to show the diversity and (dis)similarities of various peak locations and amplitudes. See also Figs.~\ref{fig:simCompareToPhotAllSky}, \ref{fig:simCompareToPhotGaps}, and \ref{fig:astrSimP} for comparison with period search results on simulated scan-angle signals that qualitatively reproduce these peaks.}
    \label{fig:observedPhotPeriod}
\end{figure*}

\begin{figure*}[t!]
    \centering
    \includegraphics[width=\textwidth]{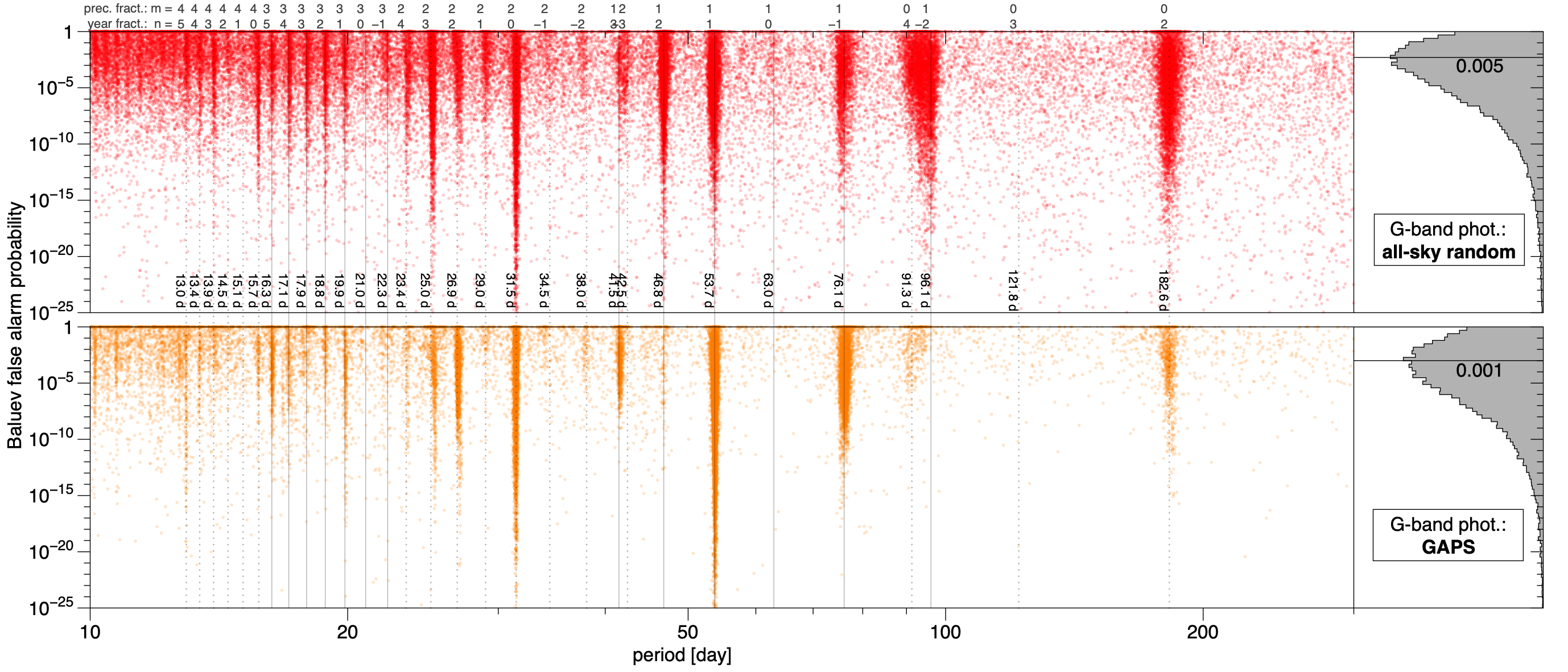}
    \caption{Distribution of false-alarm probabilities of the two photometric samples shown in Fig.~\ref{fig:observedPhotPeriod}, illustrating the highly significant nature of most of the spurious periods.}
    \label{fig:observedPhotPeriodFaps}
\end{figure*}

A clear feature observed during the data processing for \gdr{3} is that specific periods are identified much more frequently than others. This is illustrated with several public and non-public data sets in Fig.~\ref{fig:observedPhotPeriod}.
The first set (top panel) is drawn from a (unpublished) sample of about 1.6~million sources that were selected by randomly sampling from the full range of magnitude in the G-band photometric data, with an upper limit of 6000~objects per 0.05\,mag interval, and then by filtering out sources with fewer than five~FoV transits in the G band and those without any measurement in both \gbp and \grp. They were then processed by the default variability pipeline of \cite{DR3-DPACP-162}, in which only sources were selected that passed a general variability test. An unweighted periodogram was then made using generalised least-squares \citep{1985A&AS...59...63H,CummingMarcyBulter99,ZechmeisterKurster09}
(an extension of the Fourier periodogram on unevenly sampled data that is independent of the mean of the data), 
followed by a refinement of the period with the highest power using an unweighted multi-harmonic modelling step. The periodogram was computed between 25 cycles day$^{-1}$ (about 1~h) and $7 \cdot 10^{-4}$ cycles day$^{-1}$ (1700~d), with a step size of typically $10^{-5}$ cycles day$^{-1}$. We only display the 73~k sources with periods in the range 10 to 500~d in which most of the easily identifiable spurious peaks appear.

The second set (middle panel) is extracted from the public photometry published as part of the \gaia Andromeda photometric survey \citep[GAPS;][]{DR3-DPACP-142}. Periods and false-alarm probabilities are provided in Appendix~\ref{sec:auxTablesStats}. The same processing and selections as for the first set were applied, resulting in 38~k sources with periods in the range 10 to 500~d, as shown. For the first and second dataset, we show in Fig.~\ref{fig:observedPhotPeriodFaps} the Baluev false-alarm probability \citep[FAP;][]{2009MNRAS.395.1541B}. The FAP shows that a significant fraction of the peaks is highly significant. As a result of the initial blind source selection, both sets will contain a mix of truly photometric variable objects and spurious variables (for example galaxies and close pairs) due to induced scan-angle-dependent signals or other disturbances in the \gaia data. Most sources of each data set will not exhibit any (periodic) variability at all, however.

In Fig.~\ref{fig:observedPhotPeriod}, the third set (bottom panel) is a set of 1.8~million unpublished astrometric orbital solutions produced by the exoplanet pipeline on a set of stochastic sources \cite[for details, see section 5.1.1 of][]{DR3-DPACP-176}.

As already illustrated by the vertical period-lines in Figs.~\ref{fig:observedPhotPeriod} and \ref{fig:observedPhotPeriodFaps} and in the figures in following sections, the positions of the main peaks are approximately centred on periods $P$ [d], 
\begin{equation}
\label{eq:periodPeaks}
    365.25/P = m\, 5.8 + n, \quad \text{where $m$ and $n$ are small integers,}
\end{equation}
where 5.8 cycles yr$^{-1}$ (about 63.0~d) is the precession frequency of the spin axis around the Sun during the nominal scanning law discussed in Sect.~\ref{sec:gaiaObserving}. The symbol $n$ marks the number of cycles yr$^{-1}$. In clearly identifiable peaks (in the marked range above 13 days), $n$ varies from about --3 to 5.   $m$ marks the number of cycles per precession period, which starts at 0 and increases towards shorter periods (only illustrated until $m=4,$ but continuing beyond). 

The strength and significance of the peaks significantly depends on the ecliptic latitude, which explains the difference between the top two panels of Fig.~\ref{fig:observedPhotPeriod} and \ref{fig:observedPhotPeriodFaps}. This is explored in detail Sect.~\ref{ssec:propSimSignals} and in the associated sky plots in Appendix~\ref{sec:simPeaksExamples}.

\subsection{Interpreting the period peak structure}

As remarked in \cite{2022arXiv220605745L}, this structure might be interpreted as some sort of aliasing of the combined periodicities. 
However, the term aliasing is misleading because in this case, the signal consists of a frequency in the scan-angle domain that is mapped onto the time-domain through a sky-position-dependent transformation encoded in the NSL, in contrast to the usual aliasing (the sample aliasing), where
a true frequency in a frequencygram is distributed over different frequencies due to a convolution with a specific window function. A full analysis of the origin of Eq.~\ref{eq:periodPeaks} and the resulting prevalence of expected frequencies is beyond the scope of this paper, but it might be thought of as the combined effect of different harmonics of the yearly and spin-axis period of the satellite, where the power in the harmonics comes from the deliberate non-integer fraction of cycles per year of the precession frequency, to randomise both scan-angle orientations and observation times. In addition, lower-order perturbations come from the non-constant precession phase-rate discussed in Sect.~\ref{ssec:obsDistrSa}, which is also further complicated by the slightly elliptical orbit around the Sun.

These same period locations relate to expected variations in the selection function of photometric periodic sources and astrometric orbits and in the derived-parameter biases \cite[see for example][]{LL:LL-136,2022MNRAS.513.2437P}, which undoubtedly also affect the shown samples. They are part of the expected features of the data sampling and adopted source model parametrisations, however, which is not the subject of this particular study and thus is not discussed further in this paper. 
We do not show the period distribution below 10~d (which is only an aesthetic choice to focus on the clearest longer-period peaks), but photometric spurious peaks from scan-angle-dependent signals have been identified down to much shorter periods, and thus higher $m$. For example, about 1000 galaxies that were misclassified as RR Lyrae stars with periods of about 0.3~d were already identified in \gdr{2} data \citep[see table C.1 of][]{2019A&A...622A..60C}.


\section{Simulated scan-angle signals and spurious periods\label{sec:responseToSaSignal}} 
We start in Sect.~\ref{ssec:numSimSAbias} to 
numerically simulate the expected scan-angle-dependent bias signal 
in the photometric magnitudes and astrometric AL-scan centroids of the sources, mimicking those introduced through the mechanisms explained in Sect.~\ref{sec:instrumentCal}. Next we provide analytical expressions for these bias signals in Sects.~\ref{ss:analytPhosBias} and \ref{ss:analytAlBias}, respectively. Finally, in Sect.~\ref{ssec:propSimSignals}, we use harmonic decompositions of these analytical expressions to simulate how they propagate in the derived photometric period and astrometric (orbital) parameters when left unmodelled (as is the case for \gdr{3}), and compare them qualitatively with the observed distributions in \gaia data introduced in Sect.~\ref{ssec:obsPerDistr}.

\subsection{Numerical simulation of the scan-angle bias\label{ssec:numSimSAbias}}

\begin{figure}
 \includegraphics[width=0.5\textwidth]{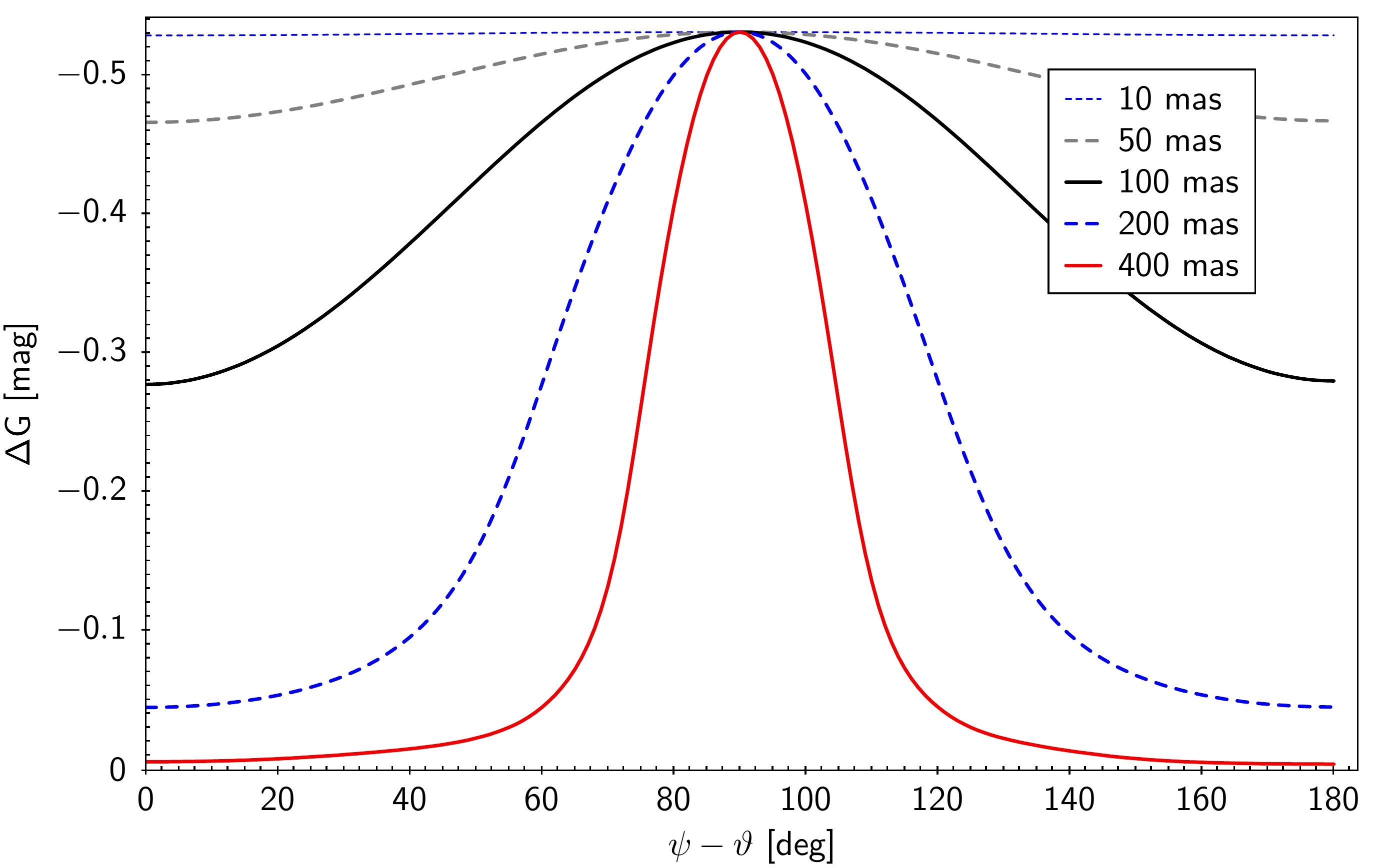}                    
  \includegraphics[width=0.5\textwidth]{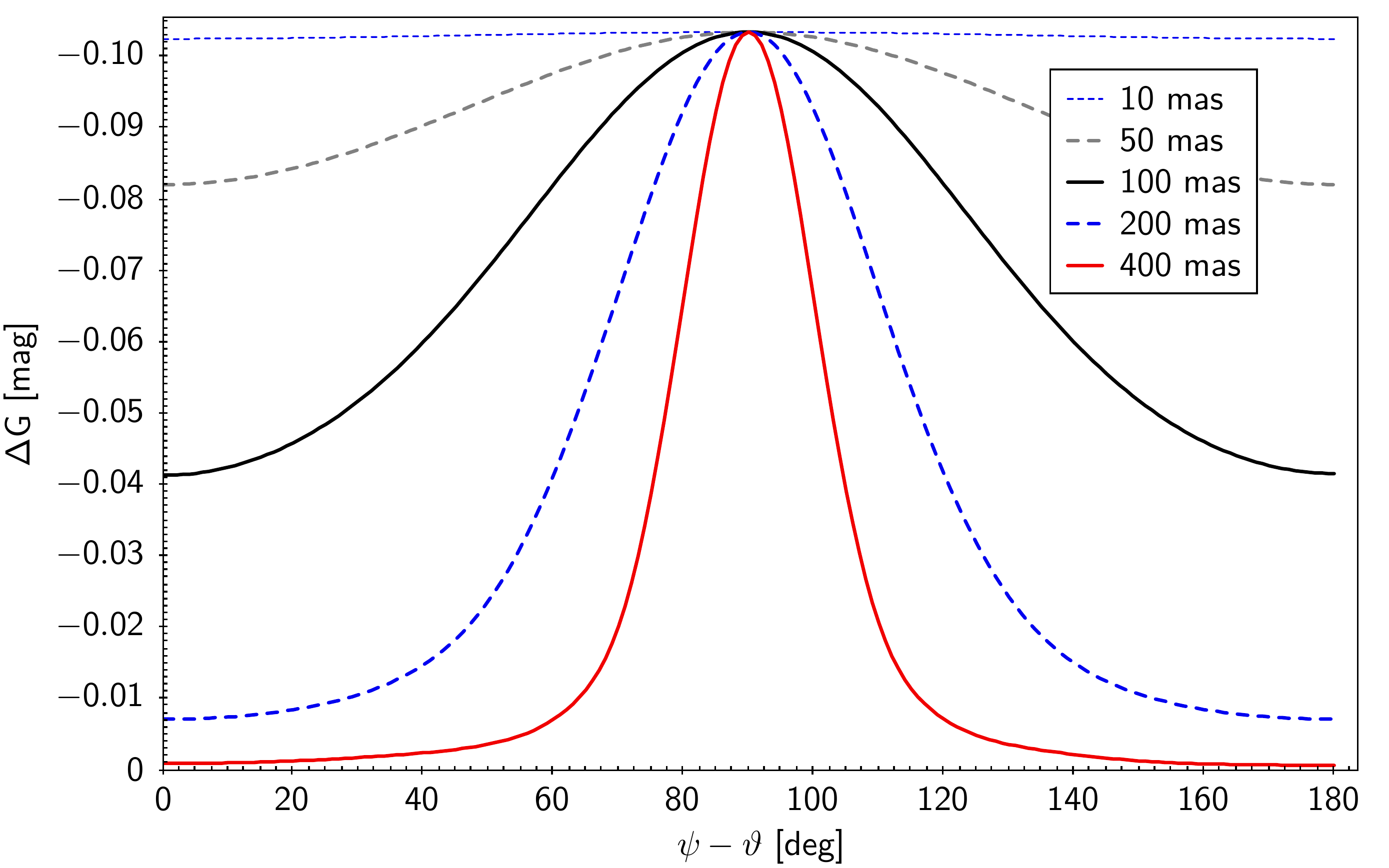}                          
\caption{Simulation of the magnitude bias, $\Delta$\gmag, for the brighter component of a close source pair for five different separations and as a function of the difference between the position angle of the scan and the position angle of the fainter component. The magnitude differences in the top panel are 0.5\,mag and in the bottom panel 2.5\,mag.
}
\label{fig:ipd_dg_simu}
\end{figure}

\begin{figure}
 \includegraphics[width=0.5\textwidth]{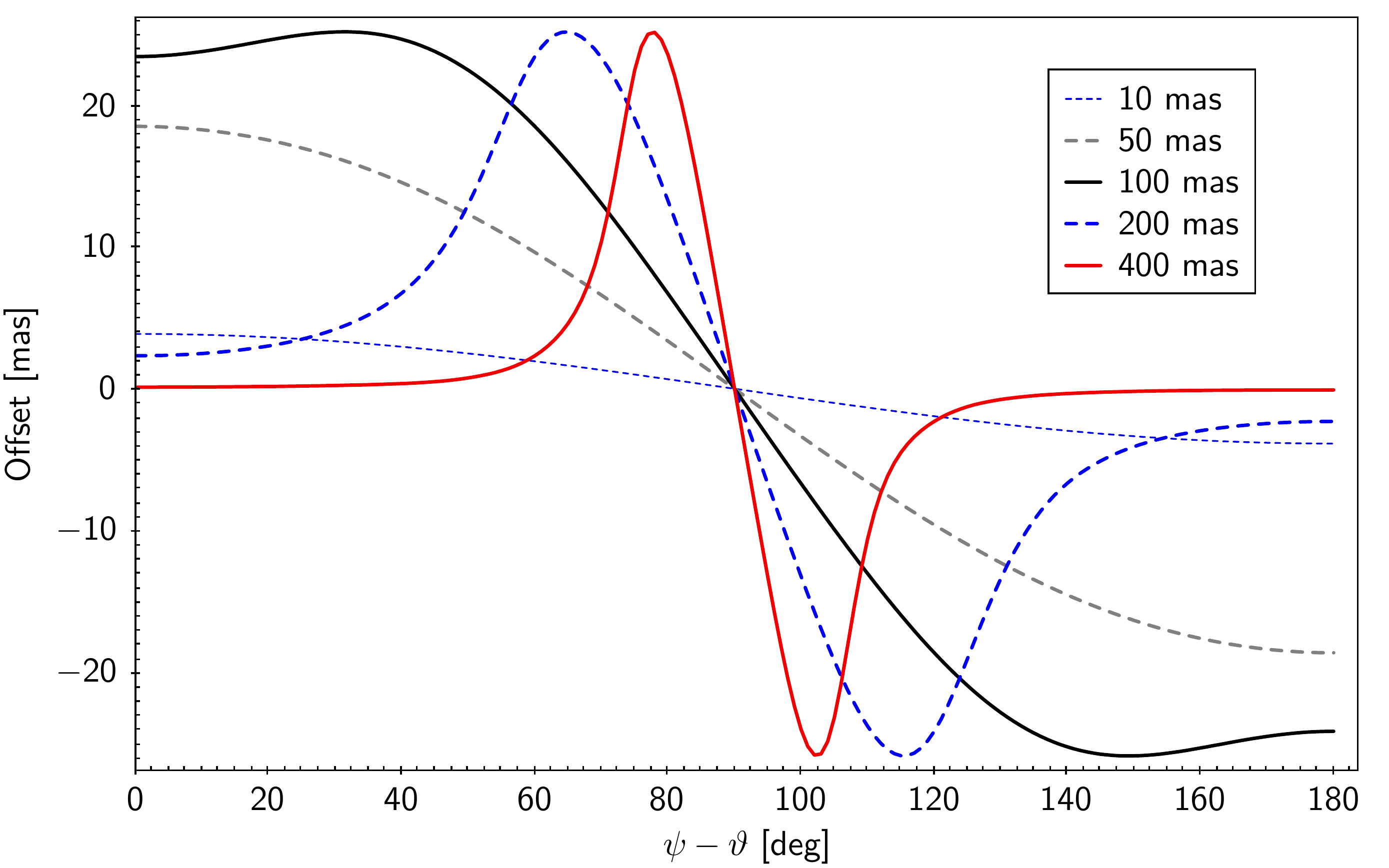}                   
  \includegraphics[width=0.5\textwidth]{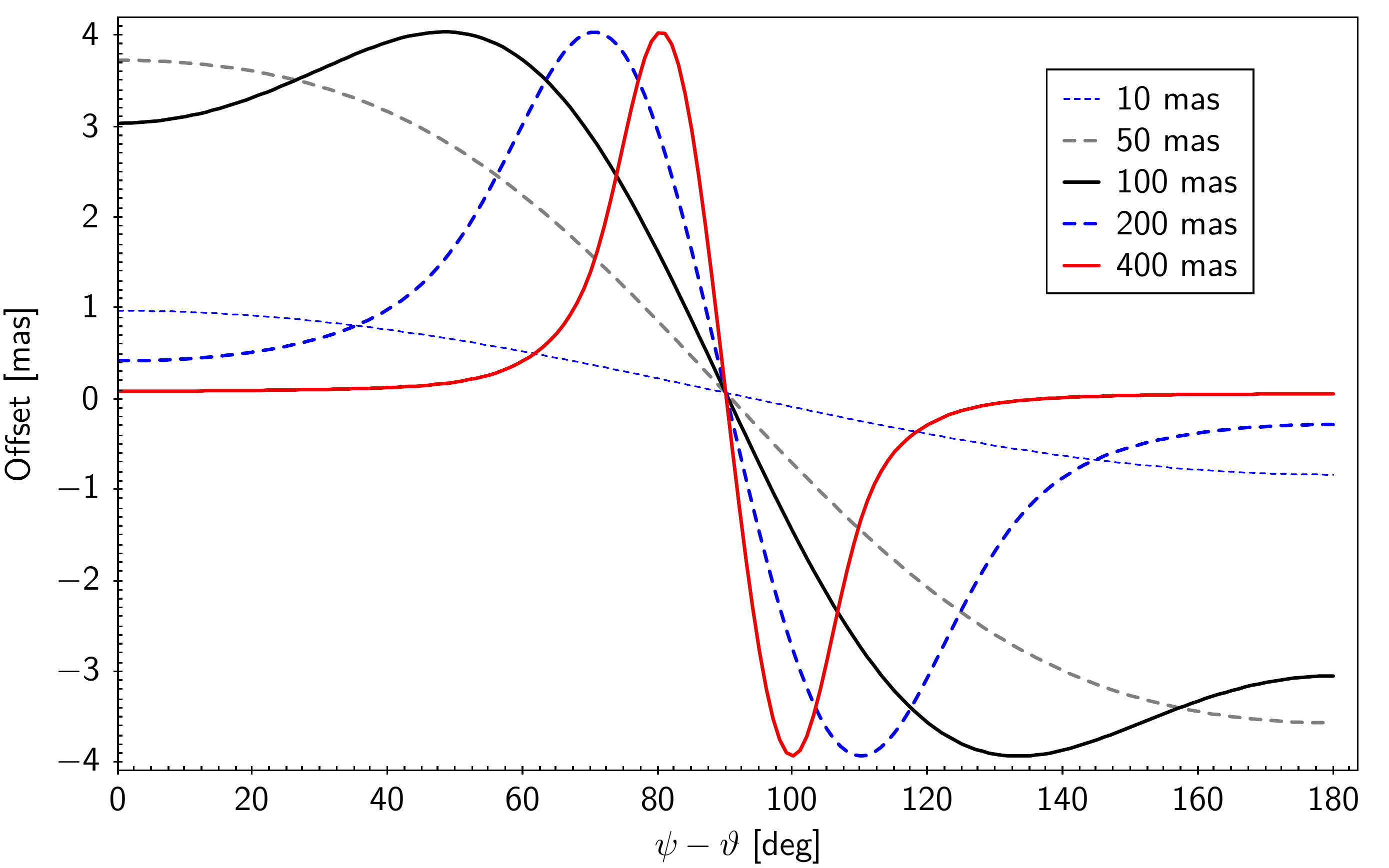}                         
\caption{Simulation of the positional bias for the brighter component of a close source pair for five different separations and as a function of the difference between the position angle of the scan and the position angle of the fainter component. The magnitude differences in the top panel are 0.5\,mag and in the bottom panel 2.5\,mag.
}
\label{fig:ipd_dx_simu}
\end{figure}

We made a simple numerical simulation of the observation of two close sources in different scan directions in order to determine how much the observed position and  magnitude of the brighter source is biased by the presence of the fainter neighbour.  The simulation was noise free and used a realistic LSF, and it assumed the data processing does not suppress the signal from the fainter source.  We simulated five different separations (10, 50, 100, 200, and 400\,mas) and two different magnitude differences (0.5 and 2.5\,mag). The result is shown in Fig.~\ref{fig:ipd_dg_simu} for the observed magnitude and in Fig.~\ref{fig:ipd_dx_simu} for the observed position. We note that the dependence on scan angle has a similar overall appearance for both magnitude differences, but the amplitude strongly depends on this difference. 
 
When scanning at 90\deg\ relative to the position angle of the pair, the two images will overlap and we obtain the brightest flux in photometry and no effect in position. As the scan angle moves away from 90\deg, the effect in flux diminishes and more rapidly so for the larger separations. In position, the effect is strong as long as the two images partly overlap, and it then diminishes. The positional biases shown in Fig.~\ref{fig:ipd_dx_simu} represent the offset in the scan direction from the position of the brighter component. In practice, for a source pair separated by less than about 50\,mas, the observed position will represent the photocentre, and the variation with scan angle in the observed position with respect to the photocentre will be much smaller than the variation with respect to one of the components. For large separations, the data processing will detect the fainter source as soon as the relative scan angle is sufficiently far from 90\deg\ and will reduce its influence. The observed position will therefore not be quite as strongly affected as the figure may suggest. Depending on how observations are distributed in time and in position angle, the biased positions will distort the astrometric solution differently. 
The astrometric residuals with respect to this distorted solution will therefore not have the simple form shown in the figure, and this is further discussed in Sect.~\ref{ss:analytAlBias}. 

\subsection{Analytical expression for photometric bias\label{ss:analytPhosBias}}
It is clear from Fig.~\ref{fig:ipd_dg_simu} that we can expect a star pair to produce a significant variation in the observed \gmag\ magnitude with scan angle. The form and phase of this variation will strongly depend on the separation, the magnitude difference, and the position angle of the source, so that we can even estimate these three parameters from the light curve, except for a 180\deg\ ambiguity in position angle. On the other hand, following Sect.~\ref{ssec:xpInstr}, for \gbp\ and \grp\ , we do not expect a significant scan-angle dependence as long as the separation is small enough for the spectra of the two sources to be contained within the observing window. 

We obtain a crude representation of the simulated variation of the observed magnitude with scan angle from the following expression, which is inspired by the discussion in \cite{LL:LL-136}:
\begin{equation}
G(\psi) = G_{\rm p} + g \exp \left(-\frac{1}{2} \left(\frac{\rho\cos(\psi-\theta)}{b}\right)^2\right)
\label{eq:simu}
,\end{equation}
where \begin{itemize}
\item $G_{\rm p}$ is the magnitude of the primary component,
\item $g = -2.5 \log\bigl(1 + 10^{-0.4 (G_{\rm s}-G_{\rm p})}\bigr)$ ,
\item $G_{\rm s}$ is the magnitude of the secondary component,
\item $\rho$ is the angular separation of the pair,
\item $\theta$ is the position angle of the secondary,
\item $\psi$ is the position angle of the scan, and
\item $b$ is a measure of the width of the LSF.
\end{itemize}

This can only serve as a first approximation. The problematic quantity is the width, $b$, which takes lower values for larger magnitude differences, where the secondary only has a small effect on the image shape. For the simulations shown in Fig.~\ref{fig:ipd_dg_simu}, values of $b = 74\,{\rm mas}$ and $b=90\,{\rm mas}$ are representative for the larger and smaller magnitude difference, but we expect that higher values are needed for the actual observations. We have used $b=100\,{\rm mas}$ in the fits shown in Figs.~\ref{fig:closePair130mas} to~\ref{fig:galaxymid} and tabulated in Table~\ref{tab:ipdSignalsList}.

We note that the fundamental frequency of Eq.~\ref{eq:simu} is at twice the scan angle (as seen next in Eq.~\ref{eq:simu2}) and that higher harmonics are implicitly constrained to even multiples of the 
scan angle.
This is a relevant observation when the propagation of this bias signal into a period-detection algorithm is simulated in Sect.~\ref{ssec:propSimSignals}, where we sample several harmonic components separately. It is important to point out that we adopted the simplified assumption that the position angle does not significantly change over the mission duration and thus can be considered constant. Taking changing position angles into account would cause non-trivial distortions of the induced signal that are beyond the scope of this work.

\subsubsection{Photometric bias model at small separations\label{sec:simu_small}}

For small separations relative to the LSF width, the expression in Eq.~\ref{eq:simu} can be approximated with the sinusoidal expression that was introduced in Eq.~\ref{eq:ipdScanAngleModel} to fit the natural logarithm of the IPD goodness of fit, which is therefore also adequate for modelling the affected \gmag photometric signal,
\begin{equation}
    G(\psi) = c_0  +  c_2 \, \cos 2\psi + s_2 \, \sin 2\psi,
    \label{eq:simu2}
\end{equation}

where \begin{itemize}
    \item $c_0 = G_p + g - \frac{1}{4}g\frac{\rho^2}{b^2}$,
    \item $c_2 = - \frac{1}{4}g\frac{\rho^2}{b^2} \cos 2 \theta$,
    \item $s_2 = - \frac{1}{4}g\frac{\rho^2}{b^2} \sin 2 \theta$.
\end{itemize}

From these coefficients, we find \begin{itemize}
    \item $\theta_\text{G} = \frac{1}{2} \mathrm{atan}( s_2, c_2) \ \  (+ 180\deg)$,
    \item $a_\text{G} = -\frac{1}{4} g \frac{\rho^2}{b^2} =  \sqrt{ c_2^2 + s_2^2 }$,
\end{itemize}
where \aG is the amplitude of the magnitude variation. 

We note that for a given amplitude of the sinusoid, the brightening in magnitude, $g$, from adding the secondary source is inversely proportional to the square of the separation.

\subsection{Analytical expression for astrometric AL-scan bias\label{ss:analytAlBias}}
\begin{figure*}[t]
\includegraphics[width=\textwidth]{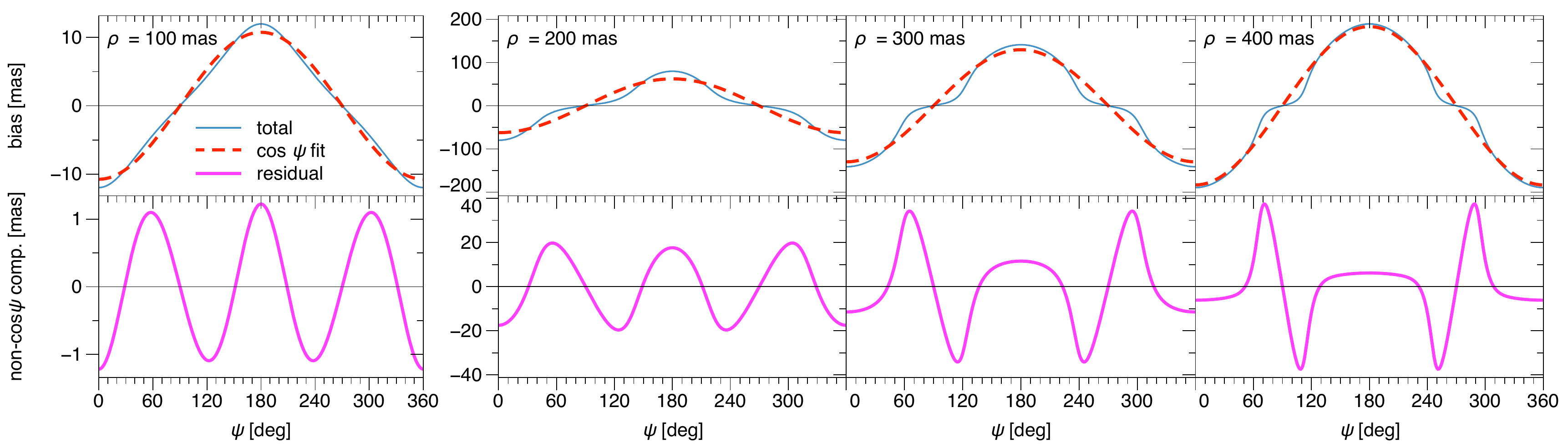}
\caption{Astrometric along-scan bias $\delta\eta$ of Eq.~\ref{eq:alBias} as a function of scan angle $\psi$ for a flux ratio $f = 0.656$ ($\Delta G= 0.46$, mass ratio $q$=0.9), and position angle $\theta$=0$\degr$. Left to right: Source separation $\rho$=100, 200, 300, and 400~mas. Top panels: Total bias $\delta\eta$ (blue line), and a cosine fit (dashed red line) that will propagate into a position bias. Bottom panels: Residuals of the cosine fit (magenta line) that can bias other source parameters. 
}
\label{fig:alBiasExamples}
\end{figure*}

\begin{figure*}[t]
\includegraphics[width=\textwidth]{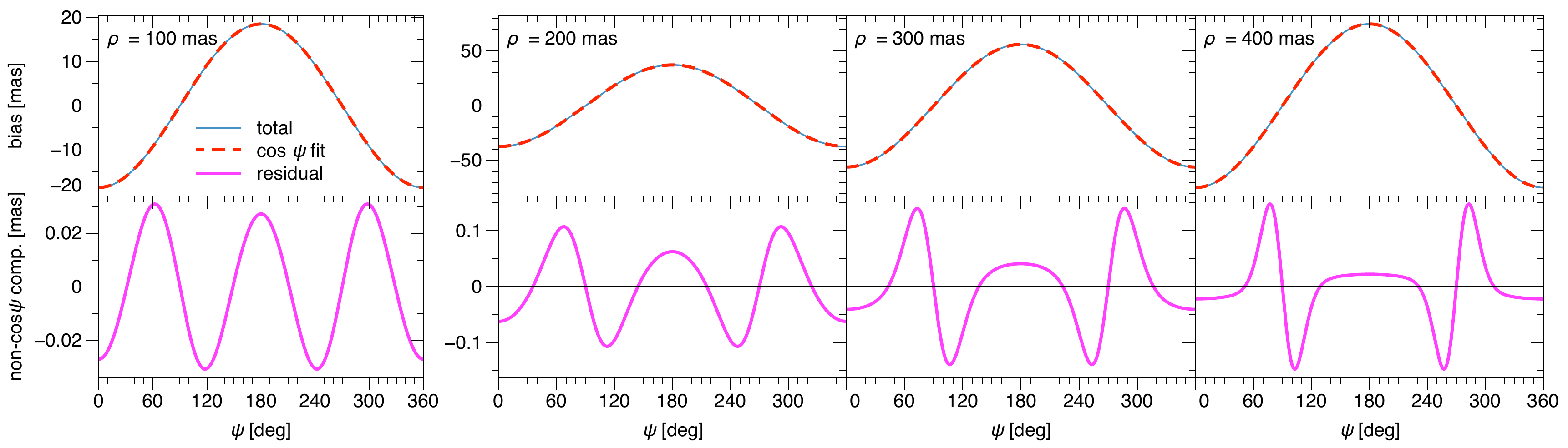}
\caption{Same as Fig.~\ref{fig:alBiasExamples}, but for a flux  ratio $f = 2.8\times 10^{-3}$ ($\Delta G= 6.38$), corresponding to the mass ratio $q\,=\,0.23$ of a typical binary.
}
\label{fig:alBiasExamplesQ0p23}
\end{figure*}

A detailed study of the astrometric bias and its effect on detection of astrometric binaries can be found in \cite{LL:LL-136}, from which we adopt the analytical expression for the astrometric AL-scan bias relative to the barycentre (their eq.~12),
\begin{equation}\label{eq:alBias}
\delta\eta = \begin{cases}
\left({\displaystyle\frac{f}{1+f}}-{\displaystyle\frac{q}{1+q}}\right)
\Delta\eta &\text{if $|\Delta\eta/u|\le 0.1$,}\\[12pt]
uB(f, \Delta\eta/u) - {\displaystyle\frac{q}{1+q}}\,\Delta\eta &\text{if $0.1<|\Delta\eta/u|\le 3-f$,}\\
-{\displaystyle\frac{q}{1+q}}\,\Delta\eta &\text{if $3-f<|\Delta\eta/u|$,}
\end{cases}
\end{equation}
with $f$ and $q$ the flux and mass ratio, respectively, in the sense fainter divided by brighter. 
The bias is primarily a function of the projected AL separation  $\Delta\eta=\rho\cos(\psi-\theta)$, with $\rho$ the binary separation and $\theta$ the position angle of the binary. $B(f,x)$ is the dimensionless anti-symmetric bias function (that is, $B(f,-x)=-B(f,x)$) introduced in their appendix~E, and $u=90$~mas is the resolution unit of the instrument. 
In Fig.~\ref{fig:alBiasExamples} we illustrate the shape of the $\delta\eta$ AL-scan bias as a function of the scan angle for a fixed-orientation binary ($P\gg 5$~y) with $\theta$\,=\,0$\degr$, mass ratio $q\,=\,0.9$, flux ratio $f = 0.656$ ($\Delta G= 0.46$), and separations varying between 100 and 400~mas. We are assuming $f=q^4$ for non-giant binaries \cite[sect. 2.2,][]{LL:LL-136}. Now, we consider the propagation of this signal into the astrometric source parameters. To first order, this signal (shown as the blue line in the top panels) is proportional to $\cos(\psi)$, which will be absorbed as a position bias \citep[see eq. 13 of][]{LL:LL-136}, with the offset being the signal amplitude, and the direction determined by the position angle. The residual of this cosine signal (magenta line in the bottom panels) will be available to propagate into, and thus bias, other astrometric parameters, as we further examine in Sect.~\ref{ssec:propSimSignals}. These residual  signals are usually of much smaller magnitude and are nearly proportional to $\cos(3\psi)$, while even higher odd harmonics appear for greater separations. We note that in Eq.~\ref{eq:alBias}, all harmonics are implicitly constrained to odd multiples of the scan angle.
For completeness, we also show in Fig.~\ref{fig:alBiasExamplesQ0p23} the bias signal induced by a typical binary with a mass ratio of $q\,=\,0.23$ \citep{1991A&A...248..485D} and 
a flux ratio of $f = 2.8\times 10^{-3}$ ($\Delta G= 6.38$).

\subsection{Propagation of bias signals into derived parameters\label{ssec:propSimSignals}}

\begin{table}[t]
\caption{\label{tab:simHarmonics} Simulated harmonics for studying bias propagation in photometric G band and astrometric AL-scan bias signals based on the harmonic decomposition of Eq.~\ref{eq:harmonicBias}.}
\centering                                      
\begin{footnotesize}
\begin{tabular}{l l p{0.7\columnwidth}}
\hline\hline
$k$ & Signal & Notes \\
\hline
1 & Astro & Dominant astrometric bias signal, \newline fully absorbed as position offset. \\
\hline
\textbf{2} & \textbf{G-band} & \textbf{Dominant photometric bias signal} (Eq.\ref{eq:simu2}), \newline fit exported in Appendix~\ref{sec:auxTablesStats} for many sources. \\
\textbf{3} & \textbf{Astro} & \textbf{First harmonic of astrometric bias signal; \newline dominant signal for other parameter biases.} \\
\hline
4 & G-band & First harmonic of photometric bias signal. \\
5 & Astro & Second harmonic of astrometric bias signal. \\
\hline
6 & G-band & Second harmonic of photometric bias signal. \\
7 & Astro & Third harmonic of astrometric bias signal. \\
\hline
\end{tabular}
\end{footnotesize}
\end{table}

In this section, we explore how a photometric bias signal as in Eq.~\ref{eq:simu} and an astrometric AL-scan bias signal as in Eq.~\ref{eq:alBias} propagate into derived parameters such as the derived (orbital) period. Although the two equations look rather different, we have already identified in their originating sections that they can be decomposed into harmonics of even and uneven multiples of the scan angle, respectively, with the highest power usually residing in the lowest harmonic, as listed in Table~\ref{tab:simHarmonics}. A generic expression for this harmonic decomposition is

\begin{eqnarray} \label{eq:harmonicBias}
    b(\psi) &=& c_0  +  \sum_k c_k \, \cos k\psi + s_k \, \sin k\psi,
     \\
    &&\text{with} \quad k =  
    \begin{cases}
    \text{even: photometric bias [mag]}\\
    \text{odd: astrometric AL-scan bias [mas]}\\
\end{cases} \nonumber \\
    && \text{and} \quad a_k = \sqrt{s_k^2 + c_k^2} \nonumber \,\,\, .
\end{eqnarray}

Because a full exploration of the parameter space of the original photometric Eq.~\ref{eq:simu} and astrometric Eq.~\ref{eq:alBias} biases is beyond the scope of this study, we instead assess the properties of their propagated biases by simulating the noiseless scan-angle bias signal represented by each $k$ individually, and qualitatively compare this to observed distributions of Fig.~\ref{fig:observedPhotPeriod}.
We do this by simulating the signal for sources on a uniformly HEALPix 
sky grid for 20 uniformly spread position angles to obtain an impression of the general all-sky response. For photometry, we used a HEALPix grid level of 6 ($\sim49$~k positions with a granularity of about 0.92\arcmin{}), leading to $\text{about one}$~million simulations for each even $k$, and in astrometry level 5 ($\sim12$~k positions with a granularity of about 1.8\degr), leading to $\sim250$~k simulations for each uneven $k$. To simulate the GAPS data, we simply analysed the subset of 113~level 6 HEALPix pixels within the defined 5.5\degr radius around the Andromeda Galaxy (that is, 2260 simulations for each even $k$).

Because we compared our simulations with the observed data samples introduced in Sect.~\ref{ssec:obsPerDistr},  the representability  for the two all-sky samples might be enhanced by weighting the output of our uniform all-sky simulation as function of the expected sky density. We avoided adding this complexity (and choice of prior) because the result already represents the data very well, as shown in the following sections.

In these simulations, we made the important implicit assumption that the (main) source of scan-angle-dependent signals arises because (close) pair stars around the resolving limit of \gaia have a fixed position angle on the sky for the duration of the \gaia mission data. For galaxies, we showed in Sect.~\ref{sssec:demoSaSignalsPointSrc} that their photometric signal is similar to that of unresolved star pairs (with a shift of the phase at which their scan-angle-dependent bias signal peaks), and thus they are equally well included in this simulation. The astrometric signal for galaxies has not been studied yet, but generally, the astrometric signal of galaxies is not well represented by a five-parameter model and ends up as a two-parameter solution.

\subsubsection{Propagation of the photometric bias signal into period\label{sssec:propSimSignalsPhot}}

To derive the most dominant photometric period generated by the noiseless photometric scan-angle bias signal, we used the same generalised least-squares method as discussed in Sect.~\ref{ssec:obsPerDistr}, but now with a lower period limit of 1~d and no further modelling or frequency refinement.
Changing the maximum frequency from 25 cycles d$^{-1}$ to 1 speeds up the computation dramatically, while spurious peaks induced by scan-angle signal below 1~d occur infrequently. We only inspected periods longer than 10~d.
The periodogram is insensitive to any offset because it is fitted as part of the method, and thus $c_0$ of Eq.~\ref{eq:harmonicBias} was set to zero.  This fit remained unweighted because we studied the propagation of a noiseless signal. 

\begin{figure*}[t!]
    \centering
    \includegraphics[width=\textwidth]{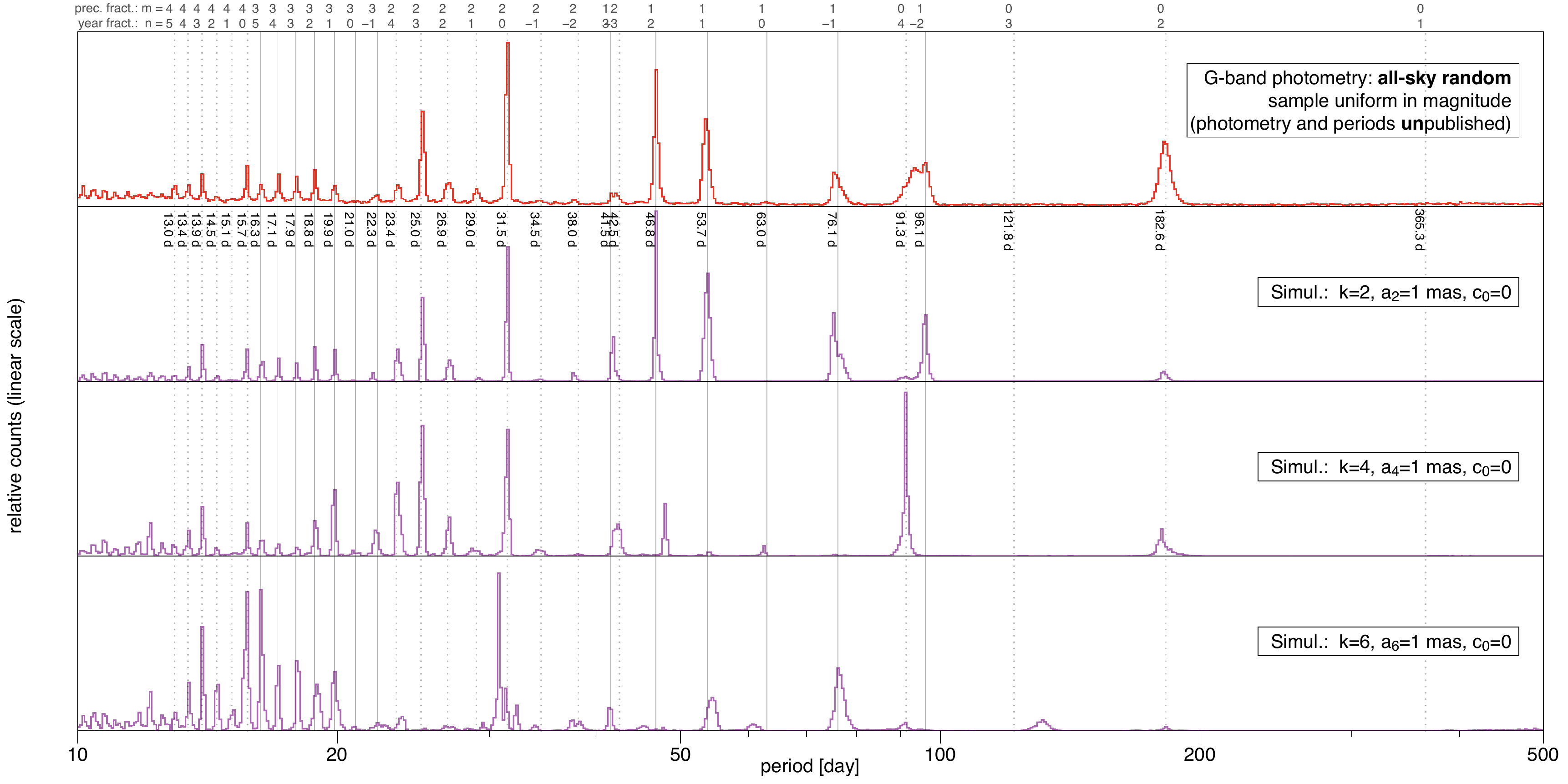}
    \caption{Comparison between the unpublished observed period distribution of an all-sky photometric sample (top panel, red line, same as top panel of Fig.~\ref{fig:observedPhotPeriod}) and that predicted by our noiseless sampled bias model of Eq.~\ref{eq:harmonicBias} for different scan-angle harmonics $k$ (purple lines in following panels). 
    See Figs.~\ref{fig:skyplotDistrPhotAllsky1} and \ref{fig:skyplotDistrPhotAllsky2} for ecliptic sky maps and  Figs.~\ref{fig:foldedPeriodGExamples1} and \ref{fig:foldedPeriodGExamples2} for spurious-period-folded time series of the most prominent peaks.}
    \label{fig:simCompareToPhotAllSky}
\end{figure*}

The top panel of Fig.~\ref{fig:simCompareToPhotAllSky} shows the observed \gaia period distribution  of the unpublished all-sky sample introduced in Sect.~\ref{ssec:obsPerDistr}. This histogram is in linear scale, compared to log scale in Fig.~\ref{fig:observedPhotPeriod}. The three panels below show the period distribution of our simulations in the range 10 to 500~d for $k=$2, 4, and 6, as listed in Table~\ref{tab:simHarmonics}. Although our simulations were not adjusted for the observed sky-distribution differences, the match with the $k=2$ simulation is already strikingly good. We recall that this is (equivalent to) the small separation model of Eq.~\ref{eq:simu2}, \textbf{which suggests that the vast majority of observed photometric spurious periods is due to close pairs with a fixed position angle on the sky}. 
The propagation of the peaks for models with $k=4$ and $6$ shows that specific period peaks are additionally enhanced when the photometric scan-angle bias signal has higher harmonics (especially around 31.5~d), as is the case when the pair is partially resolved (Eq.~\ref{eq:simu}). We note that propagating the harmonics separately is not equivalent to propagating a multi-harmonic model itself, but it will highlight the type of periods that would become dominant in the periodogram if the harmonic itself were to become the dominant component of the bias signal, which is thought to be sufficient for the qualitative comparison we make.

To examine in more detail whether our simulated peaks are related to the observed peaks, we plot the position of some of the peaks of Fig.~\ref{fig:simCompareToPhotAllSky} on the sky and compare them with the observed sky distributions, as provided in Appendix~\ref{sec:simPeaksExamples}. 
In addition to the underlying density variations, the detection regions on the sky agree well. The main exception is probably the observed 182~d peak sky distribution, which appears to be not fully reproduced by our model. It might contain contributions from another calibration effect. The colour-coding also clearly shows that for $|\beta|<45\degr$ , the peak detection generally occurs in a relatively small fraction of the 20~sample phase steps, which is consistent with the predictions of Sect.~\ref{ssec:obsDistrSa} based on the non-uniformities in the scan-angle distribution closer to the ecliptic plane.

\begin{figure*}[t!]
    \centering
    \includegraphics[width=\textwidth]{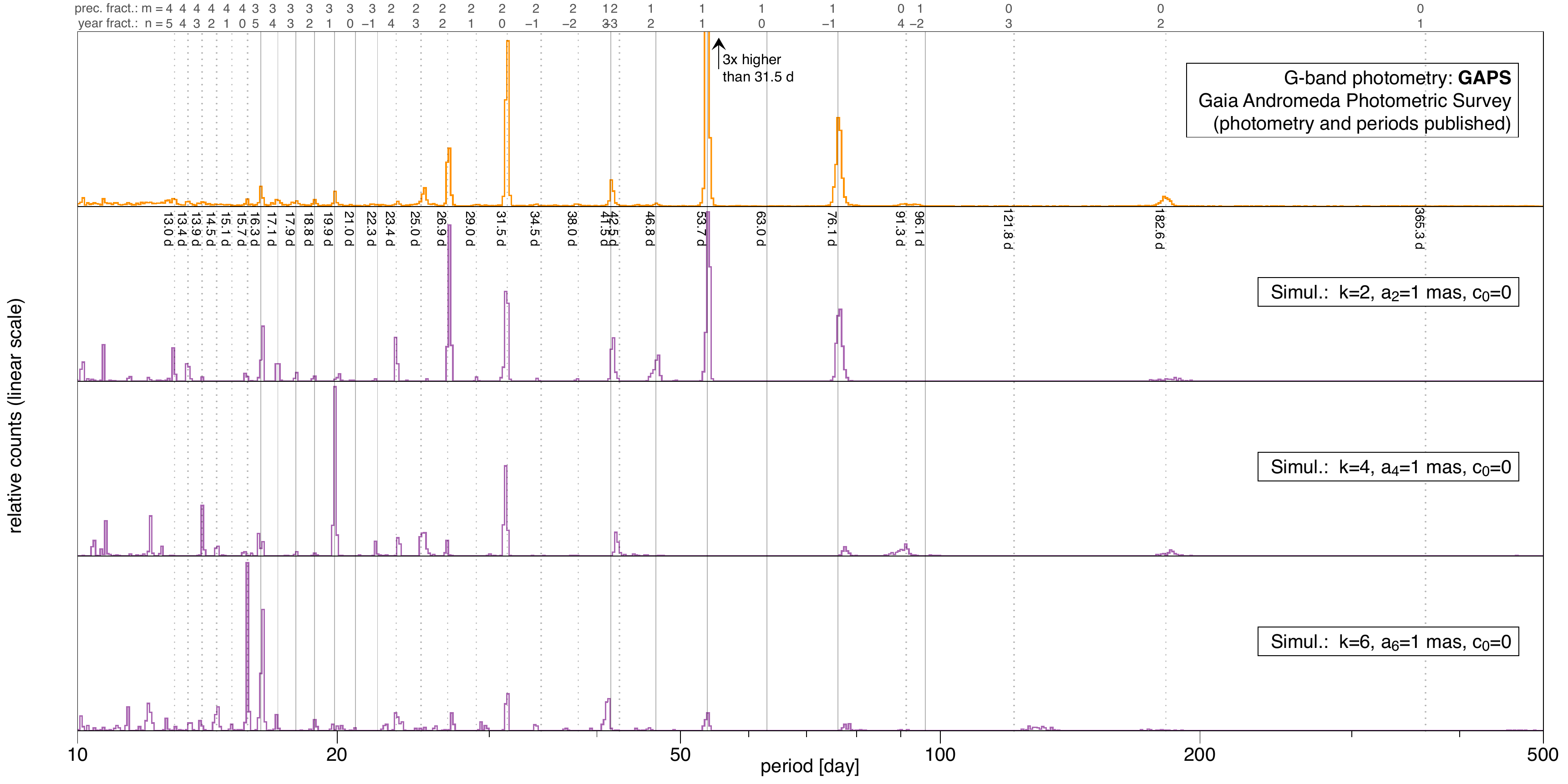}
    \caption{Same as Fig.~\ref{fig:simCompareToPhotAllSky}, but for the published GAPS photometry. The top panel (orange line) shows the same data as in the second panel of Fig.~\ref{fig:observedPhotPeriod}. The green circles in Figs.~\ref{fig:skyplotDistrPhotAllsky1} and \ref{fig:skyplotDistrPhotAllsky2} show ecliptic sky maps of the most prominent peaks.}
    \label{fig:simCompareToPhotGaps}
\end{figure*}

For the public GAPS sample, which was processed in the same way as the all-sky sample, we extracted the simulated data for the 113 HEALPix level 6 pixels within a 5.5\degr\ radius from the Andromeda galaxy \citep[for details about the GAPS sample selection, see][]{DR3-DPACP-142}. The result is shown in Fig.~\ref{fig:simCompareToPhotGaps}. This sample is centred around ecliptic latitude 
33.4\degr\ and thus is well within the region of $|\beta|<45\degr$ , in which the more irregular scan-angle sampling occurs (see Sect.~\ref{ssec:obsDistrSa}). The $k=2$ simulation shows that the same location is generally obtained for the four highest peaks, even though the relative period distribution is not exactly reproduced. A $k=4$ , the 31.5~d peak might additionally be boosted by this harmonic, but we do not seem to predict the same level of boosting of the 53.7~d period as seen in the observations. Our simple simulations clearly cannot account for all detailed features, but even for such a specific sky location, it is satisfying that we detect the dominant features well.

\subsubsection{Propagation of the astrometric bias signal into astrometric orbit\label{sssec:propSimSignalsAstro}}

To derive the most dominant astrometric period generated by a noiseless astrometric scan-angle bias signal 
with an amplitude of
1~mas, we ran the genetic orbit-fitting algorithm described in the \gaia exoplanet pipeline \citep{DR3-DPACP-176} and retrieved the best-fitting Keplerian orbit. Similarly to the photometry, we simulated the response of one  harmonic at a time. As listed in Table~\ref{tab:simHarmonics}, these are the uneven $k$, starting with $k=3,$ and shown here until $k=7$. We note that the adopted model of Eq.~\ref{eq:alBias} does not contain an offset. We therefore set $c_0$\,=\,0 for all simulations.

\begin{figure*}[t!]
    \centering
    \includegraphics[width=\textwidth]{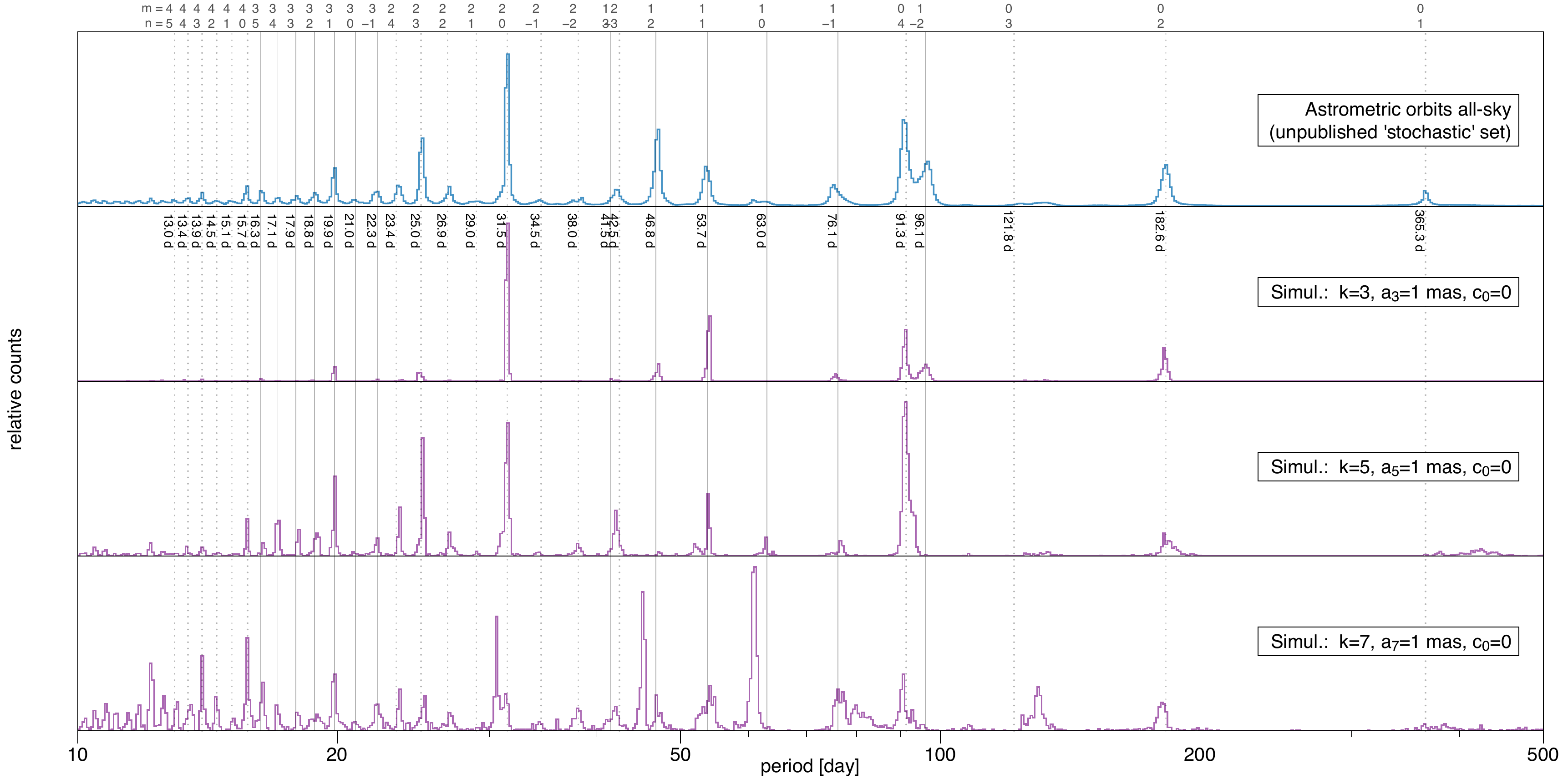}
    \caption{Comparison between the unpublished period distribution of an all-sky astrometric orbit sample (top panel, blue line, same as the bottom panel of Fig.~\ref{fig:observedPhotPeriod}) and that predicted by our fits to the noiseless sampled bias model of Eq.~\ref{eq:harmonicBias} for different scan-angle harmonics $k$ (purple lines in the following panels). See Figs.~\ref{fig:skyplotDistrAstroAllsky1} and \ref{fig:skyplotDistrAstroAllsky2} for ecliptic sky maps of the most prominent peaks. }
    \label{fig:astrSimP}
\end{figure*}

The top panel of Fig.~\ref{fig:astrSimP} shows the observed \gaia period distribution  of the unpublished all-sky stochastic sample introduced in Sect.~\ref{ssec:obsPerDistr}. The three panels below show the period distribution of our simulations in the range 10 to 500~d for $k=$3, 5, and 7, as listed in Table~\ref{tab:simHarmonics}. The $k=1$ simulation results in an orbital amplitude of zero because the signal is fully absorbed in the position offset. Even though our simulations were not adjusted for the observed sky-distribution differences, the match with the $k=3$ simulation is promising. It contains most of the significant peaks seen in the observational data. Adding $k=5$ appears to enhance the correctly identified peaks to better match the observed sample. The $k=7$ simulation appears to enhance peaks that are generally not in the observed sample, however, and it might well be that this harmonic usually has very low power in the astrometric scan-angle bias signal because the companions are sufficiently well separated (and are bright enough) to be resolved individually by \gaia. This relation between higher harmonics and larger companion separation is illustrated in Figs.~\ref{fig:alBiasExamples} and \ref{fig:alBiasExamplesQ0p23}.
In the same way as for the photometry, we plot the position of some of the peaks of Fig.~\ref{fig:astrSimP} on the sky and compare them with the observed sky distributions, as provided in Appendix~\ref{sec:simPeaksExamples}. The observed and predicted location of the specific spurious periods agree well in general.

\begin{figure*}[t!]
    \centering
    \includegraphics[width=\textwidth]{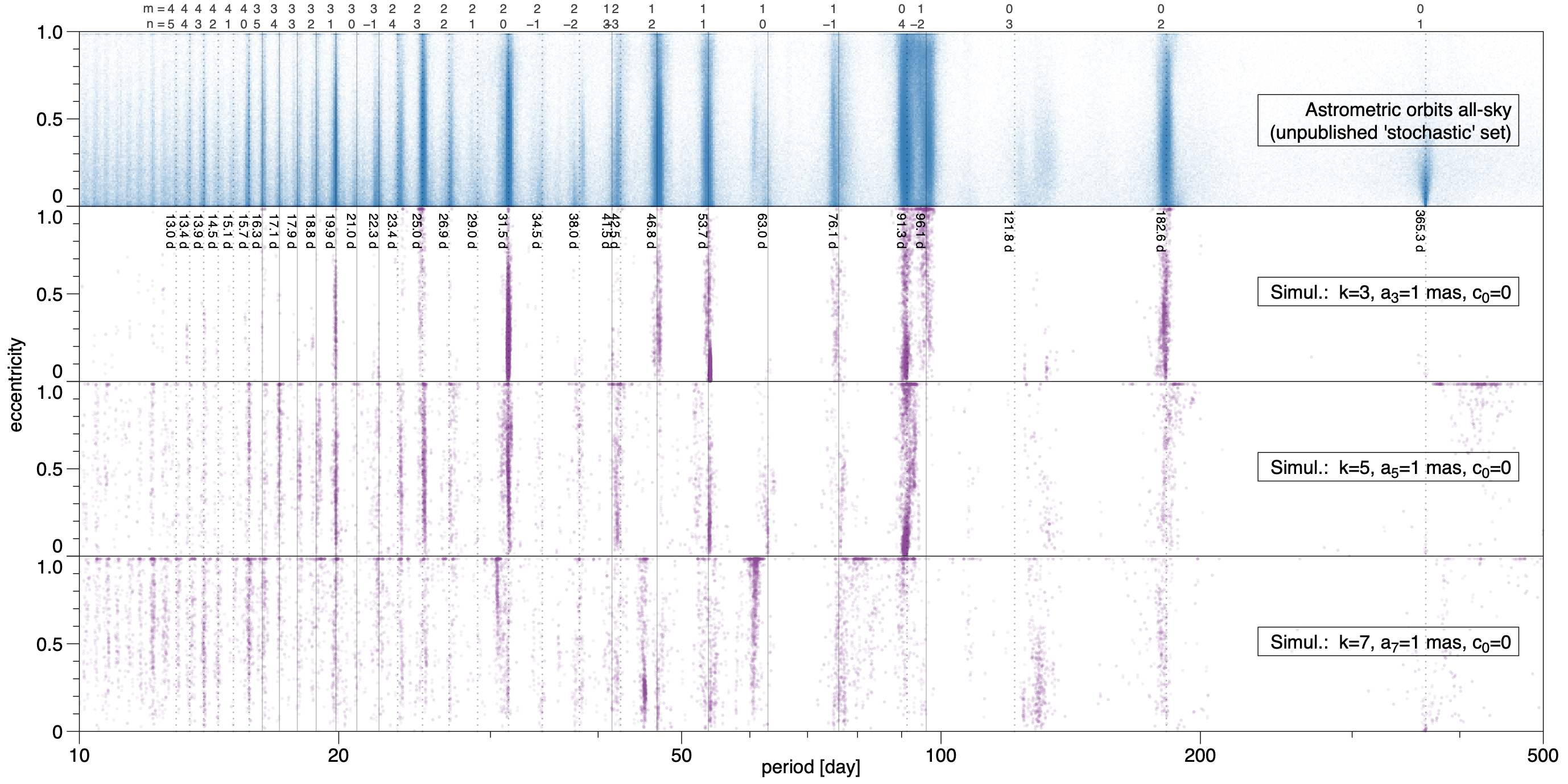}
    \caption{Same unpublished data as Fig.~\ref{fig:astrSimP}, now showing the period-eccentricity relation.}
    \label{fig:astrSimPEcc}
\end{figure*}

To further assess the qualitative similarity of the simulations with the observed data, we additionally plot the period-eccentricity diagram in Fig.~\ref{fig:astrSimPEcc} \citep[which has previously been shown in fig.~18 of][]{DR3-DPACP-176}. The overall spread of the eccentricity over the full 0 to 1 range in the observed data is well reproduced by our simulations (again $k=3$ and $k=5$ match best).

\begin{figure*}[t!]
    \centering
    \includegraphics[width=\textwidth]{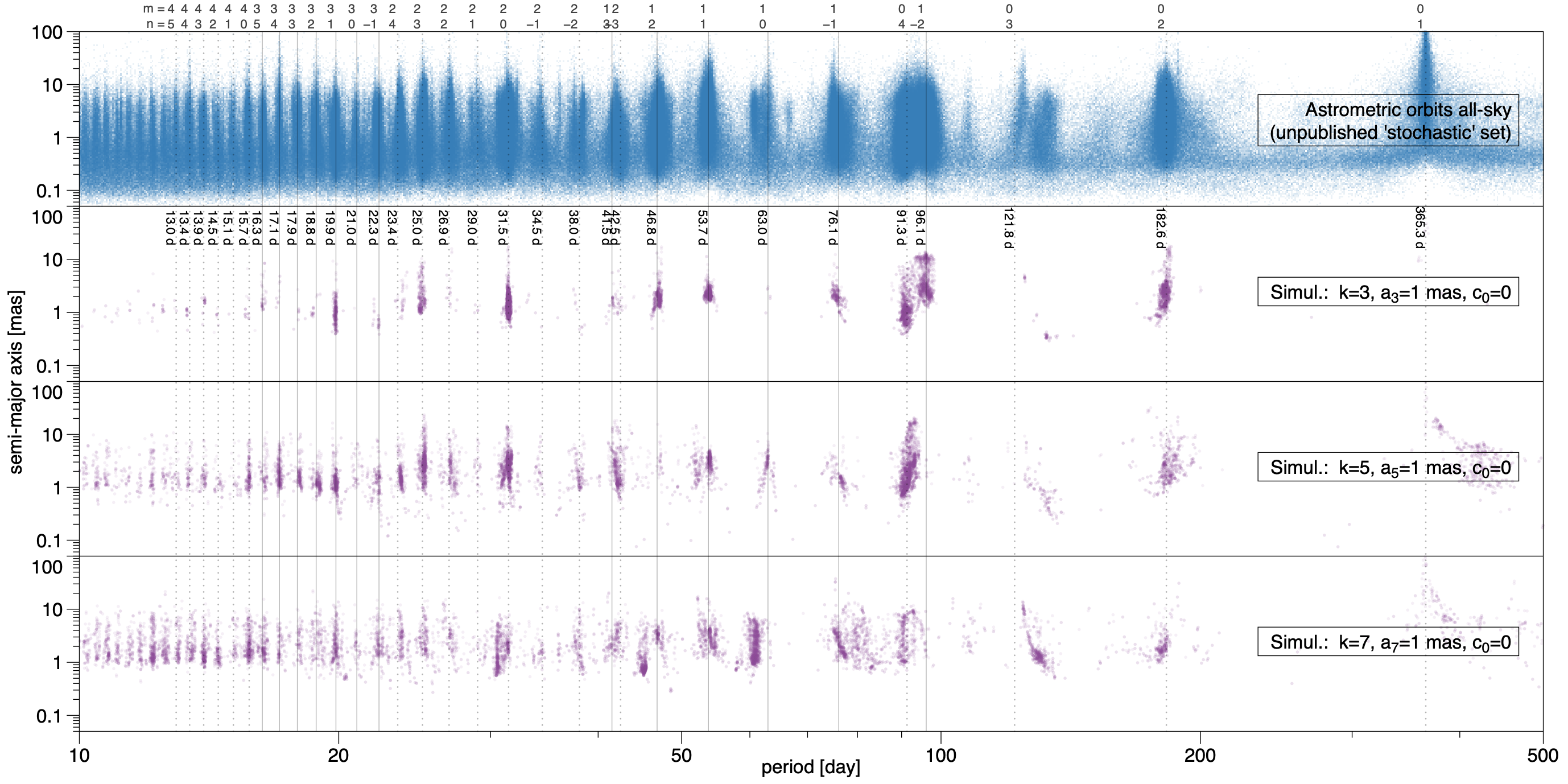}
    \caption{Same unpublished data as Fig.~\ref{fig:astrSimP}, now showing the period vs fitted semi-major axis, illustrating that the observed semi-major axes typically lie between 0.1 and 10~mas (top panel), and showing that the orbital solutions fitted to the noiseless sampled bias model with amplitude 1~mas induce semi-major axes of one to several milliarcseconds  (following panels).}
    \label{fig:astrSimPA0}
\end{figure*}

Additionally, we assess the fitted semi-major axes in Fig.~\ref{fig:astrSimPA0}, which for orbital fits to \gaia observations typically lie between 0.1 and 10~mas (top panel). The following panels (purple data) show that the 1~mas 
astrometric scan-angle bias signal is propagated into Keplerian orbital fits with semi-major axes of a few milliarcseconds, and sometimes even several dozen milliarcseconds.
When this scaling relation is applied to the available (residual) bias signals of Figs.~\ref{fig:alBiasExamples} and \ref{fig:alBiasExamplesQ0p23} (magenta lines) that broadly lie between 0.05 and 20~mas (for $q\,=\,0.23$ of a typical binary and $q\,=\,0.9$ of a near-equal mass binary for separations between 100 to 400~mas), it closely predicts the range of observed semi-major amplitudes. 

\begin{figure*}[t!]
    \centering
    \includegraphics[width=\textwidth]{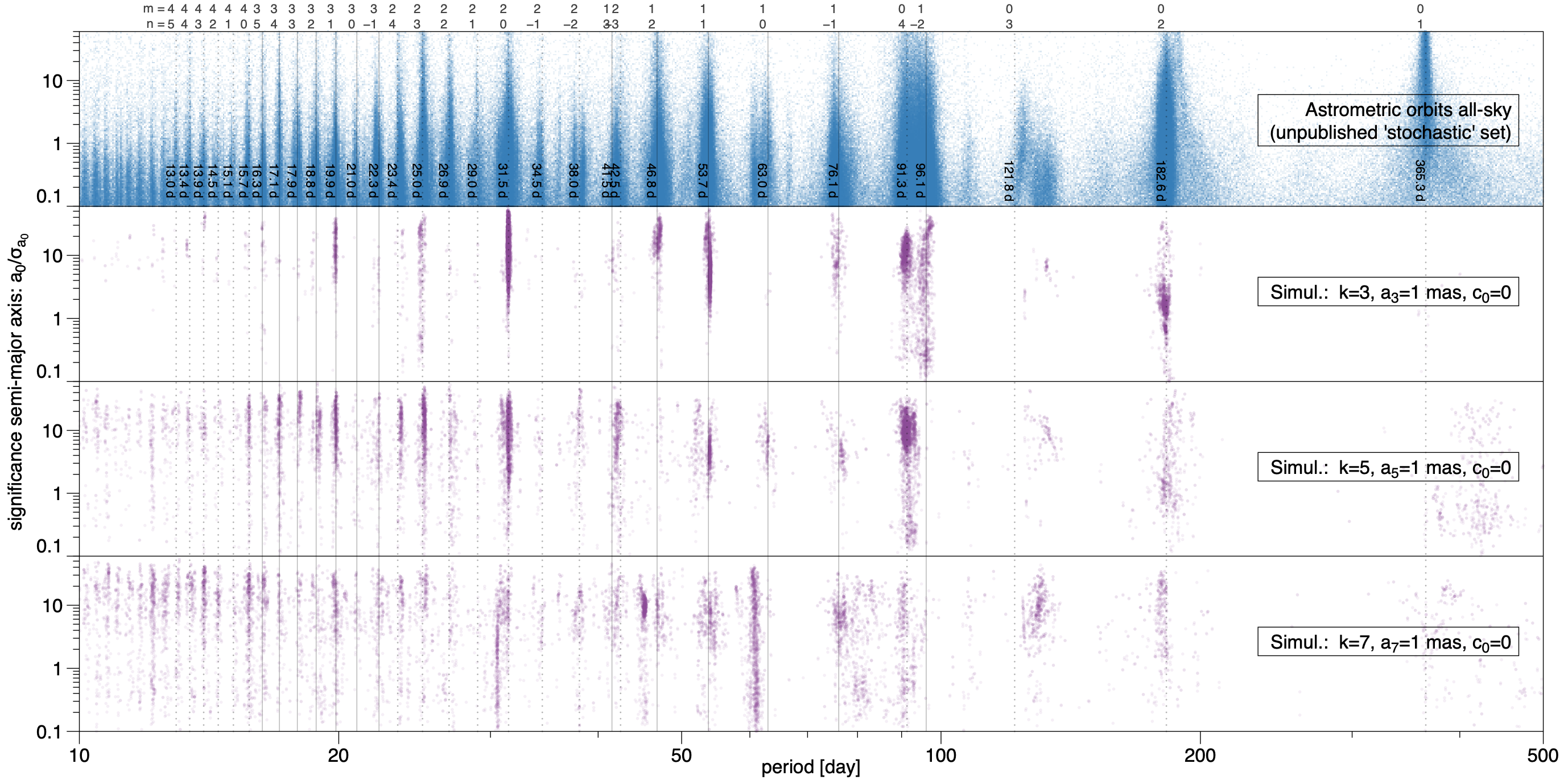}
    \caption{Same unpublished data as Fig.~\ref{fig:astrSimP}, now showing the period vs fitted semi-major axis significance, illustrating the highly significant nature of most of the observed orbits (top panel), and even more so for the orbital solutions fitted to the noiseless sampled bias model with amplitude 1~mas (following panels).}
    \label{fig:astrSimPA0signif}
\end{figure*}

Lastly, we similarly verified the significance of the solution by plotting the distribution of the significance of the semi-major axes in Fig.~\ref{fig:astrSimPA0signif}. The top panel shows that a large fraction of the orbits is rather significant. This is also predicted by the simulations that are shown in the following panels (purple data). 

Just as for the photometric data, all these analyses seem to suggest that \textbf{the vast majority of observed astrometric spurious orbital periods is due to close pairs with a fixed position angle on the sky}. Although our astrometric simulations appear to be consistent with the data, we would like to caution that the high degrees of freedom of the fitted orbital model and our simplified analyses of each harmonic separately might have led to perhaps (partially) biased predictions. More in-depth analyses of the impact of astrometric bias signals on astrometric (orbital) modelling is beyond the scope of this paper, but is foreseen to be studied for \gdr{4}.

Because the offset $c_0$ does matter in astrometry, we also ran a simulation (not shown) in which we set the offset to 1~mas, that is, $c_0=1$, and the scan-angle dependent part to zero, that is, $a_k=0$ for all $k$. This effectively traces out a circle on the sky and always results in orbits with a period of 1~yr with an extremely high orbital significance ($>10^6$) and eccentricity of about 0.033. This is twice that of the Earth's orbit (which is 0.0167).

\section{Detection of scan-angle-dependent signal} \label{sec:detectSaSignals}

With the knowledge that potential scan-angle-dependent signals might affect the time series of some sources, the obvious question arises how they might be identified.
Because photometric time series are available for a large sample of sources in \gdr{3,} we provide here several photometric diagnostics, 
and if they are not yet available in the \gdr{3} archive, we publish them in a special archive table (see Appendix~\ref{sec:auxTablesStats}) for all sources with available photometric time series. 
Because scan-angle-dependent signals also exist in astrometric and radial velocity time series, we are developing related diagnostics for these data that will be used in \gdr{4} processing. 

In this section, we first introduce two (partially complementary) correlation statistics: $r_\text{ipd}$ described in Sect.~\ref{ssec:ipdCorr}, and $r_\text{exf}$ described in Sect.~\ref{ssec:corExFlux}.
Next we discuss the use of the small-separation binary model (Eq.~\ref{eq:simu2}) described in Sect.~\ref{ssec:smallSepBinaryModelAsDetectionMethod}, and finally, we describe the (combined) use of available IPD and other parameters  in Sect.~\ref{ssec:ipdParamAsDetectionMethod}.

\begin{figure*}[t!]
    \centering
    \includegraphics[width=\textwidth]{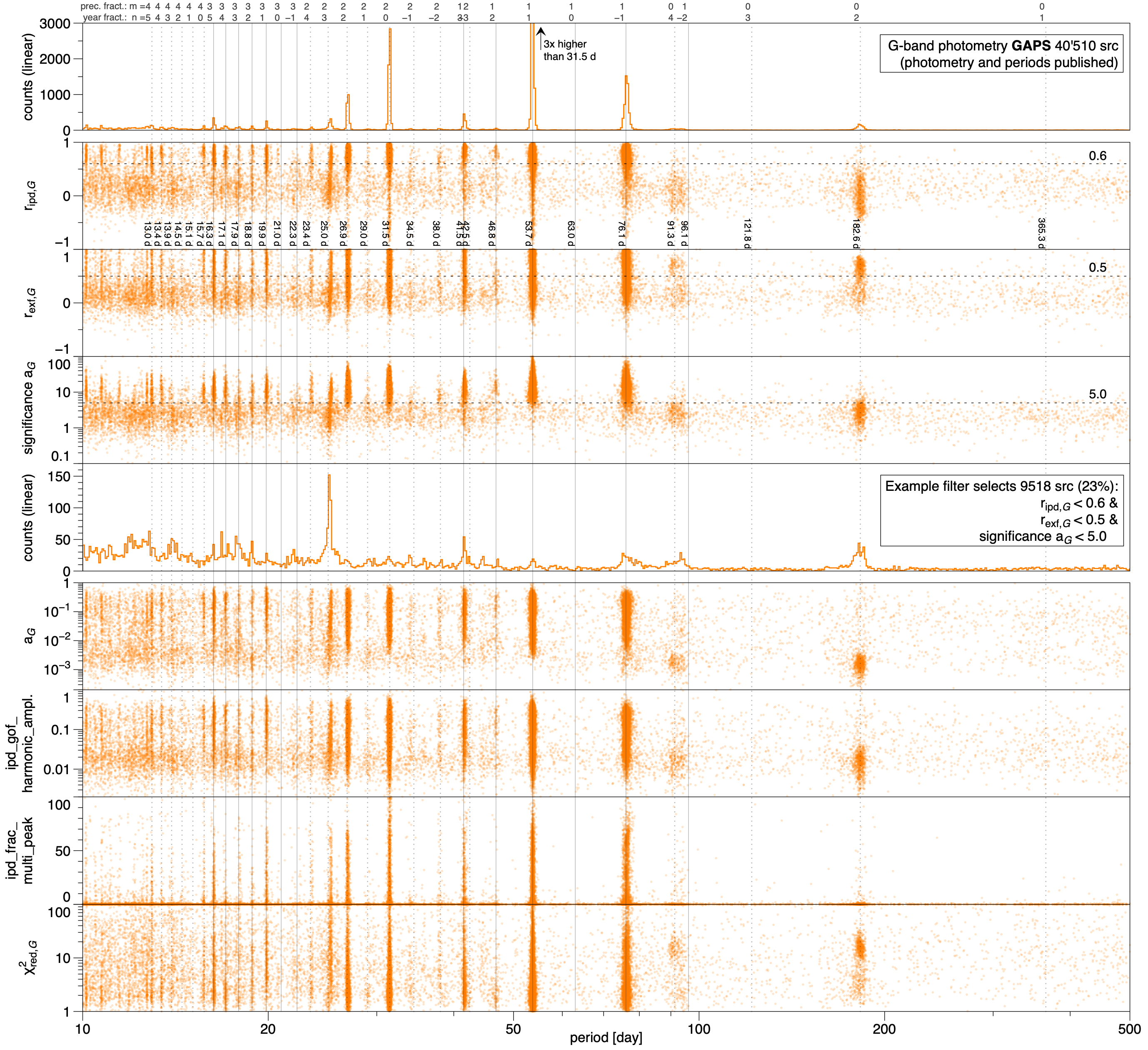}
    \caption{Distribution of sources published in GAPS. The top panel shows the same data as in the second panel of Fig.~\ref{fig:observedPhotPeriod}. The following panels illustrate the distributions of various statistics that can be used to diagnose possible scan-angle-dependent signals; see text for details. An example filter has been created to show the effect on the period distribution. The total source counts are only for the period range of 10 to 500~d.}
    \label{fig:GapsFilterExample}
\end{figure*}

We  demonstrate the distribution of these statistics for real \gaia data using the GAPS data set in Fig.~\ref{fig:GapsFilterExample}.

\subsection{Correlation with IPD model: $r_\text{ipd}$}\label{ssec:ipdCorr}

The IPD GoF model $M_\text{ipd}( \psi )$ of Eq.~\ref{eq:ipdScanAngleModel} gives us information about the phase (and amplitude) of a potential scan-angle-dependent signal. 
In order to quantify how much the \gmag , \gbp, \grp  is really affected by a scan-angle-dependent signal, we computed the IPD correlation: $r_\text{ipd}$. This is the Spearman correlation between a specific time series (for example \gmag in magnitude)
 and the IPD GoF model sampled at the scan angles of the time-series observations:
\begin{eqnarray}
r_\text{ipd}&=& f_\text{SpearmanCor}\left( \ { \{ \mathcal{S} \, G_i(\psi_i),\ M_\text{ipd}( \psi_i ) \ | \ i \in {1,...,N} \, } \ \}  \ \right) \\
&& \text{with} \ \mathcal{S} = \begin{cases}
     \ \ 1,& \text{if $G$ in flux }\\
     -1, & \text{if $G$ in magnitude, }
\end{cases} \nonumber
\label{eq:rIpd}
\end{eqnarray}
with $i$ being the observation index in the time series of a source with a total of $N$ observations, and $\psi_i$ the associated scan angle. The Spearman correlation is rank based, that is, it is insensitive to the specific magnitude differences of the values and only measures the level of correlated increase or decrease between the two input time series. This means that we do not need to normalise our values, and it makes the statistic also rather robust against (small numbers of) outliers. For example, if both time series fully coherently increase and decrease as a function of scan angle, then $r_\text{ipd}=1$, and --1 if the behaviour is inverted (that is, when one increases when the other decreases, or vice versa). This property means that $r_\text{ipd}$  will be  close to 1 if the signal has a scan-angle period of $\pi$ like the $M_\text{ipd}( \psi_i )$, and is phase-coherent with the \ipdGofHarmPhase (the amplitude does not have any effect), as can be seen in various examples provided in Sect.~\ref{ssec:afInstr}. 
We introduced the sign $\mathcal{S}$ to ensure that the correlation always gives the same value, independent of the units of the photometric input data.

Because the nominal scanning law means that scan angles do not vary smoothly between source observations separated by more than about four days (see Sect.~\ref{ssec:obsDistrSa}), varying time series that do not have a scan-angle-dependent signal will create an irregular signal when ordered by scan angle and thus will have an \rIpd close to zero. 
This is even more true for data of a constant source with randomly fluctuating noise.

The table published with this paper (Appendix~\ref{sec:auxTablesStats}) contains the IPD correlation computed for all available photometric time series with respect to \gmag, \gbp, and \grp, resulting in \rIpdG, \rIpdBp, and \rIpdRp. Observations during the ecliptic-pole scanning law (BJD -- 2455197.5 < 1693.14) and those rejected by the variability analyses ({\tt variability\_flag\_g\_reject}={\tt true} in the epoch photometry) were excluded from the computation. The resulting number of observations is listed in \numObsExclEpslGBpRp. Suggestions for minimum values for secure analyses are discussed in Appendix~\ref{sec:auxTablesStats}.
The clear correspondence of high \rIpdG with the location of the spurious period peaks in the GAPS data is illustrated in Fig.~\ref{fig:GapsFilterExample}.

A rather in-depth analyses of \rIpdG on the published \gdr{3} eclipsing binary candidates has been provided in \cite{2022arXiv221100929M}, though no sources were filtered based on this statistic.
Source filtering based on \rIpdG was applied to several \gdr{3} variability products as described in
\cite{DR3-DPACP-173},  \cite{,2022arXiv220706849C}, and  \cite{2022arXiv220605745L}.

For completeness, we would like to point out that the correlation can be computed by the user from public data. Although the photometry is readily available, obtaining the scan angle for each observation needs more work. To do this, the particular FoV of the observations first needs to be known, which can be extracted from the \texttt{transit\_id}\footnote{See \href{https://gea.esac.esa.int/archive/documentation/GDR3/Gaia_archive/chap_datamodel/sec_dm_photometry/ssec_dm_epoch_photometry.html\#epoch_photometry-transit_id}{\tt transit\_id} in  sect.~20.7.1 of \cite{2022gdr3.reptE....V}.} in the epoch photometry. 
With the FoV, the scan angle at the time of the observation can be looked up in the {\tt commanded\_scan\_law}\footnote{See \href{https://gea.esac.esa.int/archive/documentation/GDR3/Gaia_archive/chap_datamodel/sec_dm_auxiliary_tables/ssec_dm_commanded_scan_law.html}{\tt commanded\_scan\_law}, sect.~20.3.1  \cite{2022gdr3.reptE....V}}.
Since this is a rather complex procedure to implement, we decided to simply pre-compute
and provide these for all sources with public time series.
To be precise, we used the commanded scanning law for our computations. Although this will differ slightly from the actual law, this difference is negligible for the required precision on the scan angle.

\subsection{Correlation with the corrected excess flux factor: $r_\text{exf}$}\label{ssec:corExFlux}

To adequately filter spurious (periods of) solar-like variable sources, \cite{DR3-DPACP-173} developed a correlation statistic based on the Spearman correlation between the G-band magnitude and corrected excess flux $C^*$, 
\begin{equation}
r_\text{exf}=f_\text{SpearmanCor}\left( \ { \{ \, G(i),\ C^*(i) \ | \ i \in {1,...,N} \, } \ \}  \ \right)
\label{eq:exf}
,\end{equation}
using the corrected excess flux $C^{*}$ from \cite{2021A&A...649A...3R},
\begin{equation}
\label{eq:corExcessFactor}
C^{*}= C-f(\bpminrp ) 
,\end{equation}
where $C=(I_{BP} + I_{RP})/I_G$ is the  excess flux factor, $I_{BP}$, $I_{RP}$ , and $I_G$ are the cumulative fluxes in the \gbp, \grp , and \gmag bands,
and $f(\bpminrp )$ is the correction function defined in Table 2 of \cite{2021A&A...649A...3R}. 
Observations rejected by the variability analyses ({\tt variability\_flag\_g\_reject}={\tt true} in the epoch photometry) were omitted. Because $C^{*}$ requires that the transit flux from all three 
bands is available, the resulting number of observations \numAllBands used in the \rExf correlations can occasionally be quite low (suggestions for minimum values for secure analyses are discussed in Appendix~\ref{sec:auxTablesStats}). 
Sources without excess flux will have a G-band correlation \rExfG of about zero. 
In Sect.~\ref{ssec:xpInstr} and the associated Fig.~\ref{Fig:xpCstar}, we report some source examples with close to zero, close to 1, and close to --1 \rExf correlations for the different photometric bands. For completeness, the table published with this paper (Appendix~\ref{sec:auxTablesStats}) contains the excess flux correlation computed for all available photometric time series with respect to \gmag, \gbp, and \grp, resulting in \rExfG, \rExfBp, and \rExfRp. 

As discussed in \cite{DR3-DPACP-173}, \rExfG and \rIpdG are partially sensitive to the same scan-angle-dependent signals, which can also be seen from Fig.~\ref{fig:corrVsCorr}, although \rExfG is also sensitive to anomalies in sources with multiple window-gate configurations and to sources with strong emission lines whose intensity is correlated with \gmag. 
The clear correspondence of high \rExfG with the location of the spurious period peaks in the GAPS data is illustrated in Fig.~\ref{fig:GapsFilterExample}.

\subsection{Small-separation binary model fit diagnostics \label{ssec:smallSepBinaryModelAsDetectionMethod}}

As we showed in Sect.~\ref{sec:responseToSaSignal}, the simple small-separation binary model (Eq.~\ref{eq:simu2}) appears to be able to represent the bulk of the spurious signals observed in the \gaia data, as was illustrated in Figs.~\ref{fig:simCompareToPhotAllSky} and \ref{fig:simCompareToPhotGaps} in the $k=2$ panels.
We include the weighted fit of this model for all sources with photometric time series for all three bands \gmag, \gbp, and \grp in the table published with this paper (Appendix~\ref{sec:auxTablesStats}). A model fit like this will be implemented in \gdr{4} for all sources, just like the IPD parameters such as \ipdGofHarmAmpl that are currently available in \gdr{3}.

For the majority of sources (that do not have any scan-angle-dependent signal), the fit will be rather poor, which is reflected in a high \redChiSqG, as shown in Fig.~\ref{fig:rIpdGvsRedChi2} of Appendix~\ref{sec:auxTablesStats}. For a small subset, the \redChiSqG is rather good, for instance, lower than 9, indicating that this signal is well represented by the model. 
However, this will also be the case for constant stars (presumably producing uncorrelated noisy observations around their mean magnitude). Therefore, either an additional threshold should be set on the fitted amplitude, or it should be combined with other statistics, as we discuss in the next section. 


Figure~\ref{fig:magVsSamAmpl} shows the distribution of the fitted photometric model amplitude $a_G$ as function of median $G$ magnitude. The right panel basically traces the magnitude noise-floor because most of the sources in GAPS are (almost) constant. The variable stars (left panel) cause the fitted amplitude to be inflated over a wider range, but a clear cluster with amplitudes between 0.2-0.4 mag (horizontal white lines) still stands out. That this region is related to sources with high scan-angle-dependent signals is seen in the top left inset image, which shows the same area, but now colour-coded with the median \rIpdG between 0 (blue) and 1 (red), clearly highlighting the area above 0.2~mag for variable stars, and starting around 0.1~mag for the GAPS data. The second inset image shows the same colour-coding for \rExfG, clearly showing additionally regions with gate transitions at brighter magnitudes.
From this figure it is clear that any cuts to $a_G$ alone should be applied with great care. For example, a fixed cut at 0.01~mag, would remove a fraction of genuine variable sources that are fainter than G-band magnitude $\sim 13$ (as seen in the left panel), and basically all non-variable sources fainter than magnitude 19 (as seen in the right panel).

A better selection criterion seems to be the {significance} of $a_G$, which is computed from the (Eq.~\ref{eq:simu2}) fitted $c_2$ and $s_2$ and their correlation by means of eq.~2 and 3 of \cite{2022arXiv220605726H}.
Figures~\ref{fig:GapsFilterExample} and \ref{fig:signAmpl3Stats}  illustrate how well this statistic separates specific period peaks and how well it correlates with high values of \rIpdG and \rExfG.

For sources without public time series, no fitted photometric amplitude is available, but as shown in Fig.~\ref{fig:samAmplVsIpdAmpl} of Appendix~\ref{sec:auxTablesStats}, it is strongly correlated with \ipdGofHarmAmpl. The latter could therefore be used (for  \ipdGofHarmAmpl $\gtrsim 0.07$) as a lower-quality fall-back estimate of the photometric amplitude.

Finally, we note that we used the same selection for the observations as in Sect.~\ref{ssec:ipdCorr}, that is, we rejected outliers and excluded the period with the ecliptic poles scanning law (EPSL).

\subsection{Combining different indicators\label{ssec:ipdParamAsDetectionMethod}}

\begin{figure}
    \centering
    \includegraphics[width=0.45\textwidth]{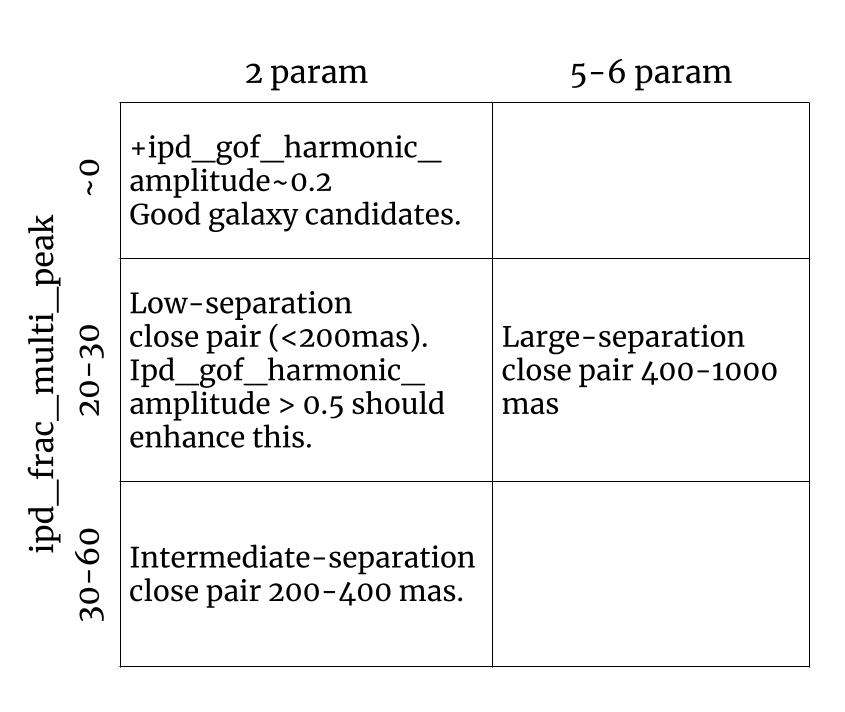}
    \caption{Parameter domain map for the filters based on the main source statistics.}
    \label{fig:parameterDomainTable}
\end{figure}

When the \gdr{3} catalogue is used as-is, we can make use of some of the indicators previously described in Sect.~\ref{sssec:ipdStats}. Statistics \ipdFracMultiPeak and \ipdGofHarmAmpl seem to be the most useful indicators. The type of AGIS solution (two-parameters versus five- or six-parameters) can also indicate that the source is problematic. In general, these features can be used to apply some filters such as those listed below (a summary of the main filters can be found in Fig. \ref{fig:parameterDomainTable}).
\begin{itemize}
    \item Sources with a moderate value in \ipdFracMultiPeak (for example 20 to 30) and a two-parameter astrometric solution even though enough transits are matched (that is, {\tt astrometric\_params\_solved}$=$3 and {\tt matched\_transits} of about 30 or more) are good candidates to be a low-separation close pair, probably below 200~mas. High values in \ipdGofHarmAmpl (for example 0.5 and above) should enhance this.
    \item Sources with the same conditions, but a quite higher \ipdFracMultiPeak value (for example 30 to 60) should also be good close pair candidates, but this time, with a higher separation, such as 200 to 400~mas.
    \item Sources with a five- or six-parameter astrometric solution and moderate \ipdFracMultiPeak (for example 20 to 30) should be close pairs with higher separations, such as 400 to 1000~mas.
    \item In addition to these conditions, we can also check for \ipdFracOddWin: moderate and high values (for example 20 and above) should select the weaker sources of the pairs, typically corresponding to the fainter components of the pairs, which are more strongly truncated in their windows. In contrast, low values should select the stronger sources, in which the IPD may have detected (and masked) the secondary peaks.
    \item When we select sources with a moderate \ipdGofHarmAmpl (for example 0.2 and above) and an \ipdFracMultiPeak value of zero or nearly zero, we should find good galaxy candidates with a relatively high ellipticity. Sources with two-parameter astrometric solutions and a high number of transits (for example 30 or more) should enhance this.
\end{itemize}

In addition to these built-in features, the new indicators \rExfG and \rIpdG provide even more reliable indications of scan-angle-dependent signals. Remarkably, the correlation with a corrected excess flux factor \rExfG for the G band exhibits values typically above 0.4 for close pairs, with values approaching 1.0 for very small separations. The correlation with IPD model \rIpdG seems to be more useful for extended sources (where \rExfG just shows moderate values), although it also exhibits high values in many cases of close pairs.

Combining a high \rIpdG (for example $>0.8$) with a low \redChiSqG (for example $<9$) potentially is a good way to select a sample of highly affected sources with a sinusoidal scan-angle-dependent signal, as shown in Fig.~\ref{fig:rIpdGvsRedChi2}. Many sources without particular variability easily produce a low \redChiSqG when the fitted amplitude is basically set to a very low value. This criterion should therefore only be used in combination with other indicators. A better single-parameter alternative probably is the significance of $a_G$, which we discussed in the previous section.

As must be clear from this section, there are a variety of ways to identify potential scan-angle-dependent signals. Although some are more powerful than others, it is difficult to provide a clear recipe of how to use them as this is highly dependent on (1) the sky location, (2) the use case at hand (see also the discussion in Sect.~\ref{ssec:shouldIReject}), and (3) the purity or contamination that is sought. To obtain an impression of how these indicators influence the period peaks, we refer to the visual overview in Fig.~\ref{fig:GapsFilterExample}, which shows the distribution of the main diagnostics discussed in this section for the public GAPS data for which we provide all parameters in Appendix~\ref{sec:auxTablesStats}. It also demonstrates a potential filter combining \rIpdG, \rExfG and the significance of \aG, which removes, or at least largely decreases, the spurious peaks.

\section{Discussion} \label{sec:discussion}

This section is meant to provide both a compact synthesis of this paper and answer some specific questions that may have arisen. It is to serve as a guide to available and planned developments.

\subsection{Scan-angle-dependent signals in a nutshell}\label{ssec:underlyingReasons}
In Sect.~\ref{sec:instrumentCal} and Appendix~\ref{sec:appendix-bcn-full-data}, we discuss that the \gaia instrument data reduction can introduce modelling errors (referred to as scan-angle-dependent signals) when on-sky source structure is present. This structure can be interpreted in a very broad way, covering: 
    (1) multiple close-pair ($\lesssim 1\arcsec$) sources of a point-like nature;
    (2) extended or non-symmetric sources (for example cores of galaxies or tidally distorted stars);
    (3) sources with a (much) brighter neighbour close by ($> 1\arcsec$);
    and (4) combinations of any of the above.
In principle, the structure can either be static, like a galaxy core or very long-period binaries ($P\gg 5$~y), or dynamic, like a short-term orbital binary. 

Subsequently, there are constraints on the data acquisition of sources by the \gaia instruments: 
    (1) observations consist of pixel values extracted in a limited window around each source;
    (2) observations are made with a large variety of on-sky scan angles due to the scanning law of the spacecraft (Sect.~\ref{sec:gaiaObserving});
    (3) pixels have an aspect ratio of roughly 1:3, the highest resolution of which is in the along-scan direction, that is, along the scan angle;
    and (4) for sources with $G\geq13$, all the data in the across-scan direction are collapsed into one-dimensional along-scan pixel counts, which enhances the effect of scan-angle-dependent signals.

The combined effect of structure, data acquisition, and the challenge of processing these data can cause scan-angle-dependent signals to appear in derived time-series values of \gdr{3}. In this paper, we abundantly demonstrate this to be the case in publicly available photometric G-band time series, but it equally exists in the (unpublished) astrometry and radial velocity time series (see Table~\ref{tab:ipdSignalsList} for examples).

\subsection{Spurious periods in a nutshell}\label{ssec:spuriousPerInShort}
Examples of observed distributions of pre-cleaned spurious period peaks are shown in Sect.~\ref{sec:spuriousPeriodsInGaiaData}.
To understand the emergence of spurious peaks caused by scan-angle-dependent signals, we need to first realise that the scan angle associated with each observation of the time series of a source is dictated by the scanning law, as explained in Sect.~\ref{sec:gaiaObserving}. Depending on the ecliptic position (mainly latitude) of a source, it causes different resonances in the data (Eq.~\ref{eq:periodPeaks}) that are related to harmonics of the $\sim63$~day spin-axis precession period and the one-year orbital period around the Sun. In Sect.~\ref{ssec:propSimSignals} we demonstrate using simulations that we can indeed qualitatively reproduce the structure of spurious period peaks in both photometry and astrometry by propagating scan-angle-dependent bias signals. For examples of public photometry that are phase folded with their spurious period, see Sect.~\ref{sec:simPeaksExamples}.

\subsection{Photometric amplitude of scan-angle-dependent signals\label{ssect:photAmplSam}}
In Sect.~\ref{sec:responseToSaSignal} we introduced two models for the scan-angle signal in the G-band photometry for a close binary. The model of Eq.~\ref{eq:simu} provides a prediction of the expected amplitudes based on various parameters such as the magnitude difference, while the model of Eq.~\ref{eq:simu2} is only accurate for small binary separations. The simplicity of the latter model allows it to be easily fitted to all time series and thus has been provided for the three photometric bands of all sources with published time series in Appendix~\ref{sec:auxTablesStats}. We can use the latter fitted model parameters for sources that show a strong indication of scan-angle-dependent signals to form an idea of the amplitude of the photometric effect. From the published data, we find \aG amplitudes of $0.2-0.4$~mag for sources with $G\gtrsim\,$16, as shown in Fig.~\ref{fig:magVsSamAmpl}. 
From internal tests on pre-filtered data (not shown), we find that amplitudes can range even wider: from $0.05-0.5$~mag down to $G\sim 13$; 
the scan-angle-dependent signal for brighter magnitudes is generally weaker (but not absent; see Fig.~\ref{fig:ep-sa-378810450446502400} as an example) due to the availability of two-dimensional window data.

For sources without public time series, the rather strong relation between \ipdGofHarmAmpl and \gmag photometric scan-angle model amplitude (\aG) for \ipdGofHarmAmpl $\gtrsim 0.07$ can also be used, as shown in Fig.~\ref{fig:samAmplVsIpdAmpl}. This allows estimating the latter \gmag scan-angle-dependent amplitude in magnitudes by the value of \ipdGofHarmAmpl (see Sect.~\ref{ssec:smallSepBinaryModelAsDetectionMethod} for more details).

\subsection{Work with scan-angle-dependent signals \label{sec:dealWithsa}}

As shown, multiple and extended sources can lead to scan-angle-dependent signals that can incorrectly be identified as astrophysical features of single point-like sources. The most typical example is the spurious detection of photometric variability, but we can also obtain spurious solutions of non-single stars, spurious extragalactic features, or other types of an incorrect astrophysical parameter determination. Despite the exhaustive validation made in DPAC, some of these spurious solutions may still exist, as  illustrated in this paper. Fortunately, the several quality, multiplicity, and scan-angle-dependence indicators discussed in Sect.~\ref{sec:detectSaSignals} (and  Sect.~\ref{sssec:ipdStats}) in many cases help to identify them.

\subsection{ Rejection of affected sources from a study\label{ssec:shouldIReject}}
First, 
 a possible rejection depends importantly on which parameters from the \gaia data are used. Especially when a period derived from \gaia data or other parameters is to be used that depends on the (phase-folded) shape of the time-series signal, we recommend to avoid sources that are affected in any significant way by using the statistics discussed in Sect.~\ref{sec:detectSaSignals}.

If this information is not needed, it needs to be determined whether the targets might exhibit some source structure (see definition Sect.~\ref{ssec:underlyingReasons}) that might produce a scan-angle-dependent signal. 
When the targets have no source structure, then we again recommend to avoid sources that appear to be affected by scan-angle-dependent signals (at the cost of some loss of completeness due to imperfect filtering statistics) because otherwise, they  might cause contaminate the sample.

When the targets have some source structure, then it becomes more delicate, as scan-angle-dependent signals might be expected to appear.
For example, when binaries with long orbital periods are to be identified, we can expect a strong scan-angle-dependent signal (and potentially a \gaia periodicity at one of the spurious peaks). If the purpose is to identify these binaries, then this signal might be used as valid selection criterion. However, it must then be verified that the source is in a non-dense environment and has no nearby bright stars (that is, the signal is not caused by background or nearby polluting stars or PSF spike pollution).

\subsection{Modelling the source environment around \gaia sources\label{ssec:sourceEnv}}

Following the discussion in Sect.~\ref{sssec:ipdModErrNonPointLike} related to the difficulty of extracting source information from a window that contains more than a single point-source, it might be wondered whether the source environment might be derived around all \gaia stars.

Starting from the fact that the AL-scan pixel size of \gaia is about 59~mas,
this means that Hubble-like resolution knowledge of the environment around each of the billion stars would be obtained. Although \gaia observations can certainly be affected by sources $\geq1$~arcsec away (meaning ground-based priors could certainly have their merit here), many of the complications related to close-binaries affect the 0.1--1.0 arcsec regime.

It is too computer intensive to do at the IPD level, but all available two- and one-dimensional \gaia observations of a source taken under a variety of scan angles can be used to try and reconstruct its environment. This is exactly the goal of two specific work-packages in \gaia:
(1) The CU4-EO surface brightness profile-fitting pipeline, which is dedicated to deriving the morphological parameters of galaxies observed by \gaia \citep[see][]{DR3-DPACP-153, DR3-DPACP-101}, the results of which are published in the DR3 \tabGalaxy archive table, and (2) the source environment analysis pipeline (SEAPipe). 
The aim of SEAPipe  is to combine the transit data for each source and to identify any additional sources in the local vicinity. 
The first operation in SEAPipe is image reconstruction, where a two-dimensional image is formed from the mostly one-dimensional transit data (\gmag$>\,$13~mag). The algorithm used to perform the image reconstructions is described in \cite{harrison11}. 
These images are then analysed and classified, based on whether the source is extended, whether additional sources are present, or whether the source is an isolated point source within the reconstructed image area (radius of $\sim 2^{\prime\prime}$).
The full SEAPipe analysis will be described in Harrison et. al. (in preparation) and is planned to be included in \gdr{4}.

As mentioned in Sect.~\ref{sssec:ipdModErrNonPointLike}, none of these methods will be used in the IPD centroid or flux estimation. However, they can nonetheless be used for a post-\gaia re-analysis of the window data accounting for (some) of the source structure and environment, together with publicly available large-scale  imaging survey data. This will allow improving the amount and accuracy of the information that can be extracted. Through the improved modelling, this might then also lead to the disappearance of scan-angle-dependent signals in the derived photometric and astrometric time series data. It seems likely that this might be beneficial for the subset of sources that were identified to be particularly non-point-like in the DPAC \gaia analyses.

\subsection{Fitting and resolving multiple peaks in IDU\label{ssec:iduClosePairs}}

As shown in Sect.~\ref{sssec:ipdStats}, IDU determines some quality and multiplicity indicators in the IPD, which are then combined at a source level by AGIS and later published in DR3. The \ipdFracMultiPeak relies on the detection of additional peaks in the window, as explained previously.
The next desirable step would be to fit the multiple peaks found in the window (instead of just masking them, as in DR3), determining one set of image parameters per detected peak. This detection and fit could take advantage of the several AF windows available per scan, aligning them using the instrumental calibration, and a combined processing might be performed. With this, the signal-to-noise ratio would be better and the PSF or LSF subsampling would be mitigated, enhancing the secondary peaks and allowing for a better fit.
With this approach, we would also achieve intra-scan consistency, that is, we would obtain consistent IPD results for each of the peaks throughout the several AF windows in each scan, avoiding swaps between the peaks. However, inter-scan consistency would also be required. This means that we need to  ensure that the peaks corresponding to the same astrophysical source from all scans are consistently matched, avoiding swaps between peaks and sources. This would require quite complex cross-matching and clustering algorithms. 
Peak detections from one-dimensional windows would require an adequate handling of their large uncertainty in the across-scan direction.

This approach has already been implemented and executed in IDU in preparation for DR4, with very good results. The first example shown in Sect.~\ref{sssec:demoSaSignalsPointSrc} (Fig.~\ref{fig:closePair130mas}) corresponds to a real, low-separation close pair that is correctly resolved with this approach. While the overall resolution for these cases has vastly improved in terms of new catalogue entries and astrometric solution, we are still evaluating the scan-angle dependence of their IPD GoF, and that of their epoch photometry when available.

\subsection{Co-modelling the scan-angle-dependent signal\label{ssec:coModelling}}
When it is known that time series might contain scan-angle-dependent signals, as in \gdr{3}, the most desirable option is to co-model the disturbing scan-angle signal in addition to the source signal that is searched for, for instance, following the concept of partial distance correlation presented in \cite{2022A&A...659A.189B}.
For photometry, the nuisance model could be parametrised as in Eq.~\ref{eq:simu}, or to first order as in Eq.~\ref{eq:simu2}. 
This is different from the \gmag model fit provided in Appendix \ref{sec:auxTablesStats}, which only fits Eq.~\ref{eq:simu2} and does not fit for any source model.

\section{Conclusions\label{sec:conclusions}}

We have presented an overview of the origin and background of scan-angle-dependent signals in the \gdr{3} data for the different instruments, which are generally caused by non-point-like object structure (mainly multiplicity or extendedness) or contamination from nearby (brighter) stars.
We qualitatively demonstrated that specific spurious periodic signals can be propagated into the photometric and astrometric time series when a dominating scan-angle-dependent signal is present at the right phase given the sky-position-dependent observation sampling; not all scan-angle-dependent signals therefore cause spurious periods. We would like to caution that the reverse is also true: not all features at the location of spurious periods are due to scan-angle-dependent signals, as these frequencies are specifically connected to regularities in the NSL that dictate the sampling and window function of the observations of each source.
For example, BH-1, the first nearby black hole companion discovered in \gdr{3} data \citep{2022arXiv220906833E,2022arXiv221005003C}, falls directly into a spurious period range with its 186~d orbital period and was therefore not mentioned in \cite{DR3-DPACP-100}.

This does not detract from our conclusion that the majority of spurious period peaks is caused by scan-angle-dependent signals originating from fixed-orientation optical pairs with a separation of $<0.5$\arcsec (including binaries with $P\gg 5$y) and (cores of) distant galaxies, as they are already well reproduced with our simple models. 

For the sample of sources with published epoch photometry in \gdr{3}, several statistics are published with this paper to help identify sources that might be affected, and thus reveal information of sub-arcsecond scale structure, especially in terms of binarity.

Although the vast majority of sources that are affected by these signals have been filtered out of the \gdr{3} 
 archive \texttt{nss\_two\_body\_orbit} and several \texttt{vari} tables, a certain fraction remains. Its existence should therefore be acknowledged (no sources were filtered from \texttt{gaia\_source}).
Improved modelling in the data processing will likely reduce the effect in future data releases.

\begin{acknowledgements}
We thank the anonymous referee and Floor van Leeuwen for their detailed feedback and suggestions that improved this paper.
This work has, in part, been carried out within the framework of the National Centre for Competence in Research PlanetS supported by SNSF.

This work presents results from the European Space Agency (ESA) space mission \gaia. \gaia\ data are being processed by the \gaia\ Data Processing and Analysis Consortium (DPAC). Funding for the DPAC is provided by national institutions, in particular the institutions participating in the \gaia\ MultiLateral Agreement (MLA). The \gaia\ mission website is \url{https://www.cosmos.esa.int/gaia}. The \gaia\ archive website is \url{https://archives.esac.esa.int/gaia}.
Acknowledgements are given in Appendix~\ref{ssec:gaiaAcknowledgements}.

This work made use of software from 
\href{https://www.oracle.com/java/}{Java},
\href{https://www.postgres-xl.org}{\textsc{Postgres-XL}},  \href{https://github.com/Tencent/TBase}{TBase} database management system, 
\href{https://www.visualdatatools.com/DataGraph/}{\textsc{DataGraph}},
\href{http://www.star.bris.ac.uk/~mbt/topcat/}{\textsc{TOPCAT}} \citep{2005ASPC..347...29T}, \href{http://www.gnuplot.info}{\textsc{gnuplot}}, and 
\href{https://www.mathworks.com/products/matlab.html}{\textsc{Matlab}}.

\end{acknowledgements}

%
%

\bibliographystyle{aa}
\bibliography{local}

\begin{appendix}



\section{\gaia archive table with photometric correlation coefficients and a  small-separation binary model fit for all sources with published time series \label{sec:auxTablesStats}} 

\begin{figure*}[t]
    \includegraphics[width=1.0\textwidth]{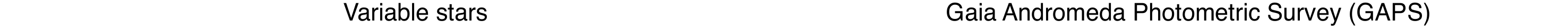}
    \includegraphics[width=0.5\textwidth]{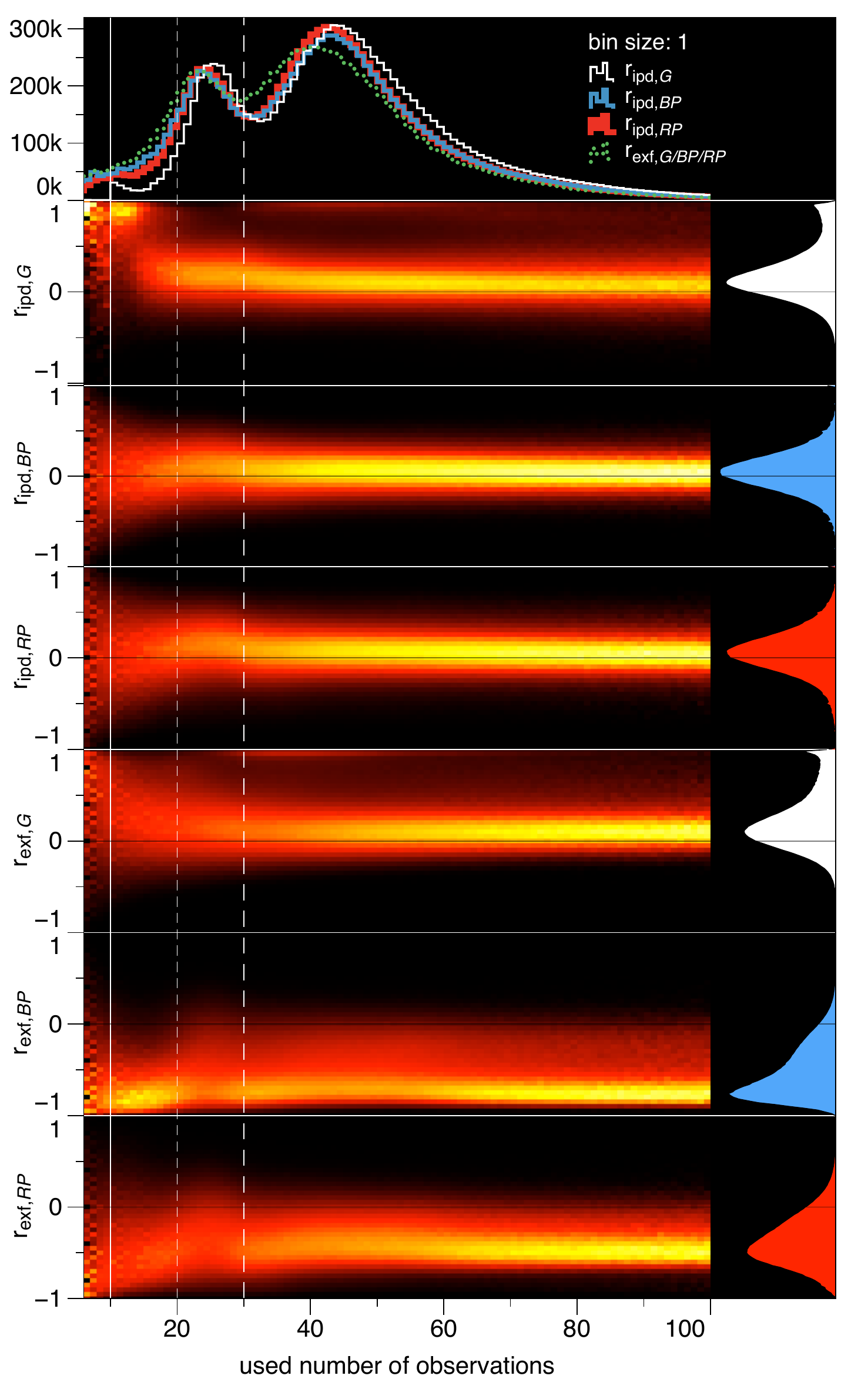}
    \includegraphics[width=0.5\textwidth]{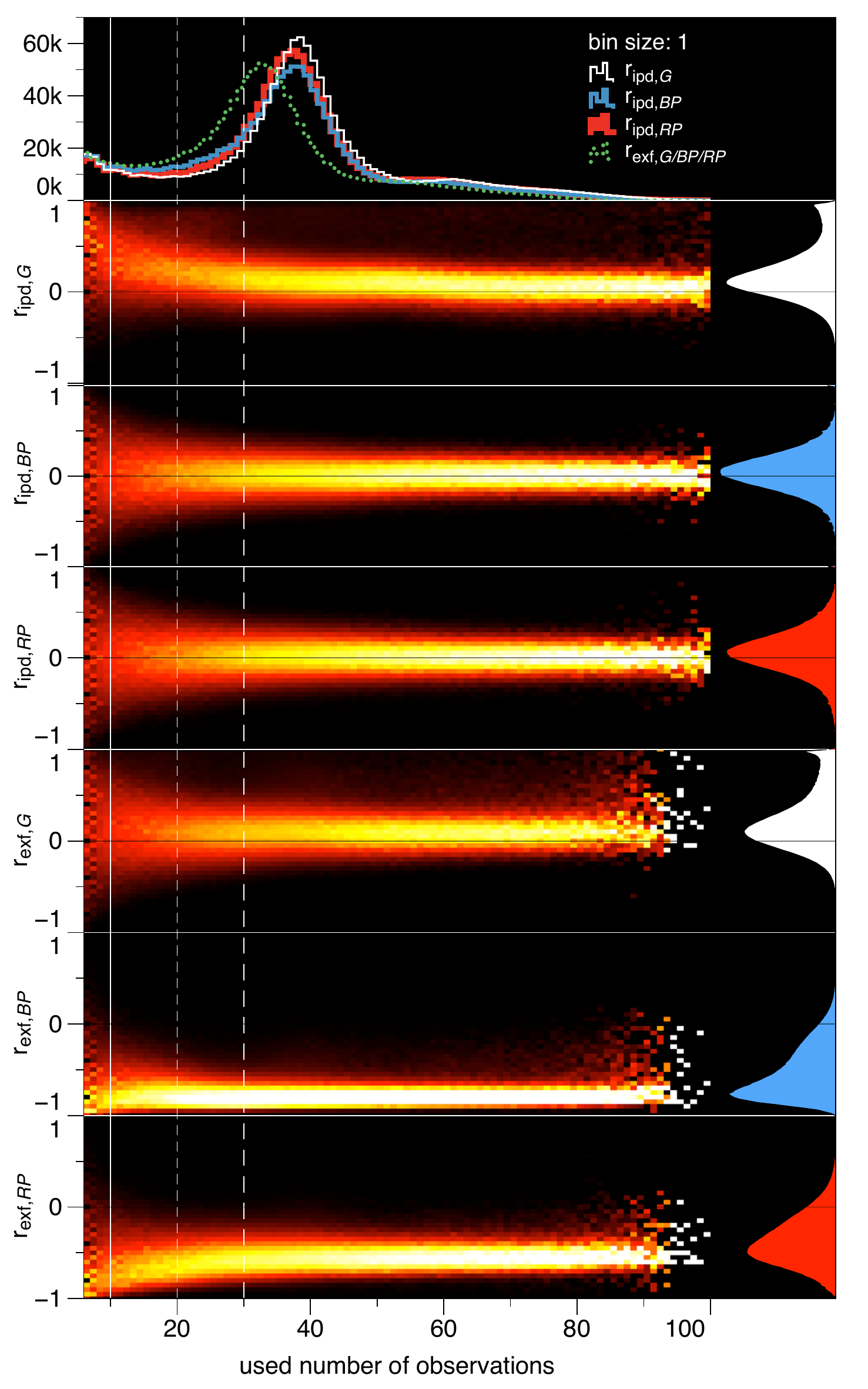}
\caption{\label{fig:corrVsNumber} Correlations based on the published photometric time series in \gdr{3}. Top panels: Histograms of the number of available observations for the different correlation coefficients. Second to fourth panel rows: ipd correlation coefficients (Sect.~\ref{ssec:ipdCorr}). Fifth to seventh panel rows: Corrected excess factor correlation coefficients (Sect.~\ref{ssec:corExFlux}). On the right side, a histogram illustrates the distribution of each correlation coefficient. 
The left panels contain the values for the 10.5~million variables, and the right panels show the values for the 1.3~million sources in GAPS. The heat maps are value normalised per number of observation bin to highlight the level of parameter spread. The histograms on the right sides represent the true count distribution for each parameter.
}
\end{figure*} 

\begin{figure*}[t]
\includegraphics[width=1.0\textwidth]{images/publishedTable/topLabelsVarAndGaps.pdf}
    \includegraphics[width=0.5\textwidth]{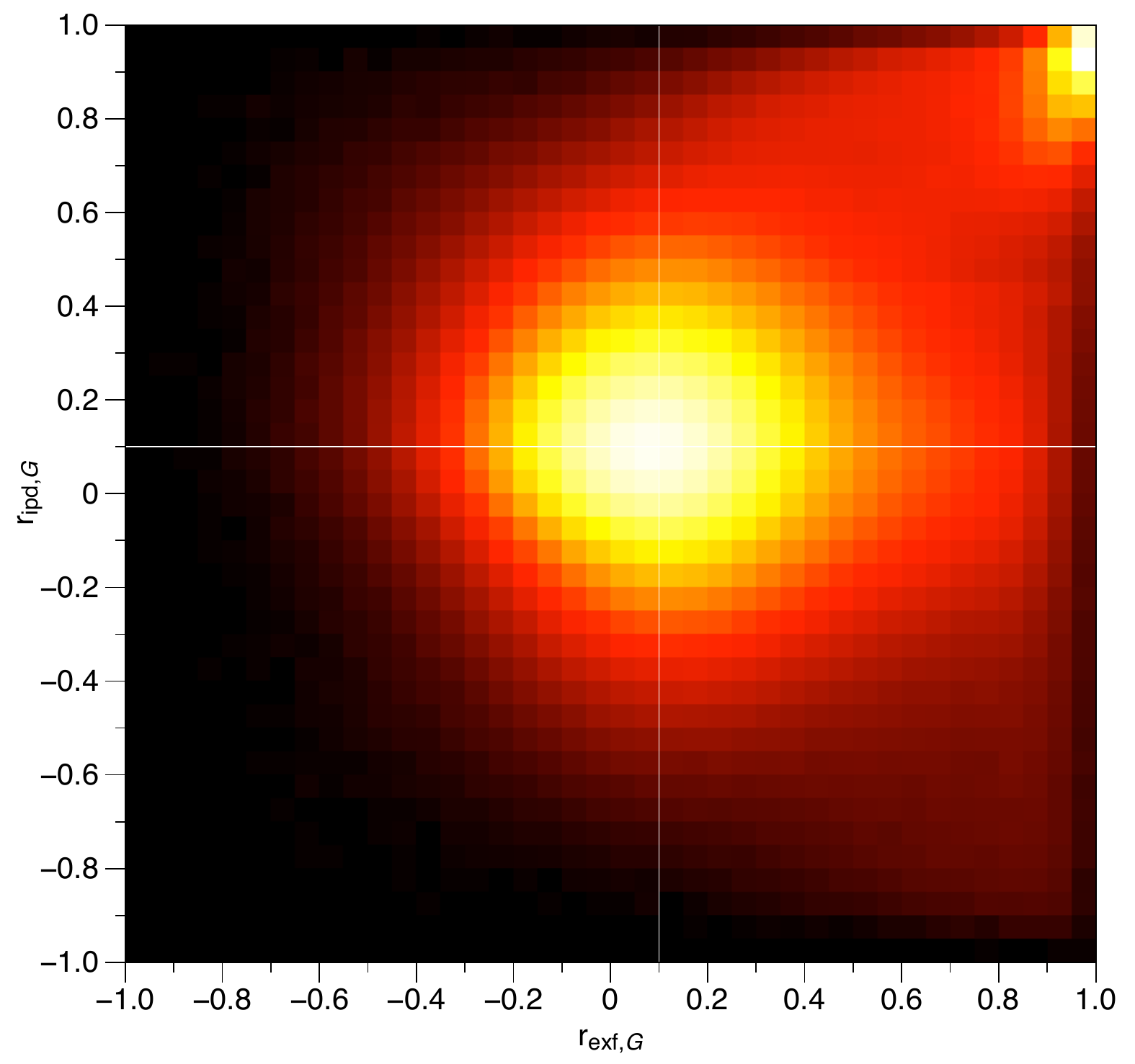}
    \includegraphics[width=0.5\textwidth]{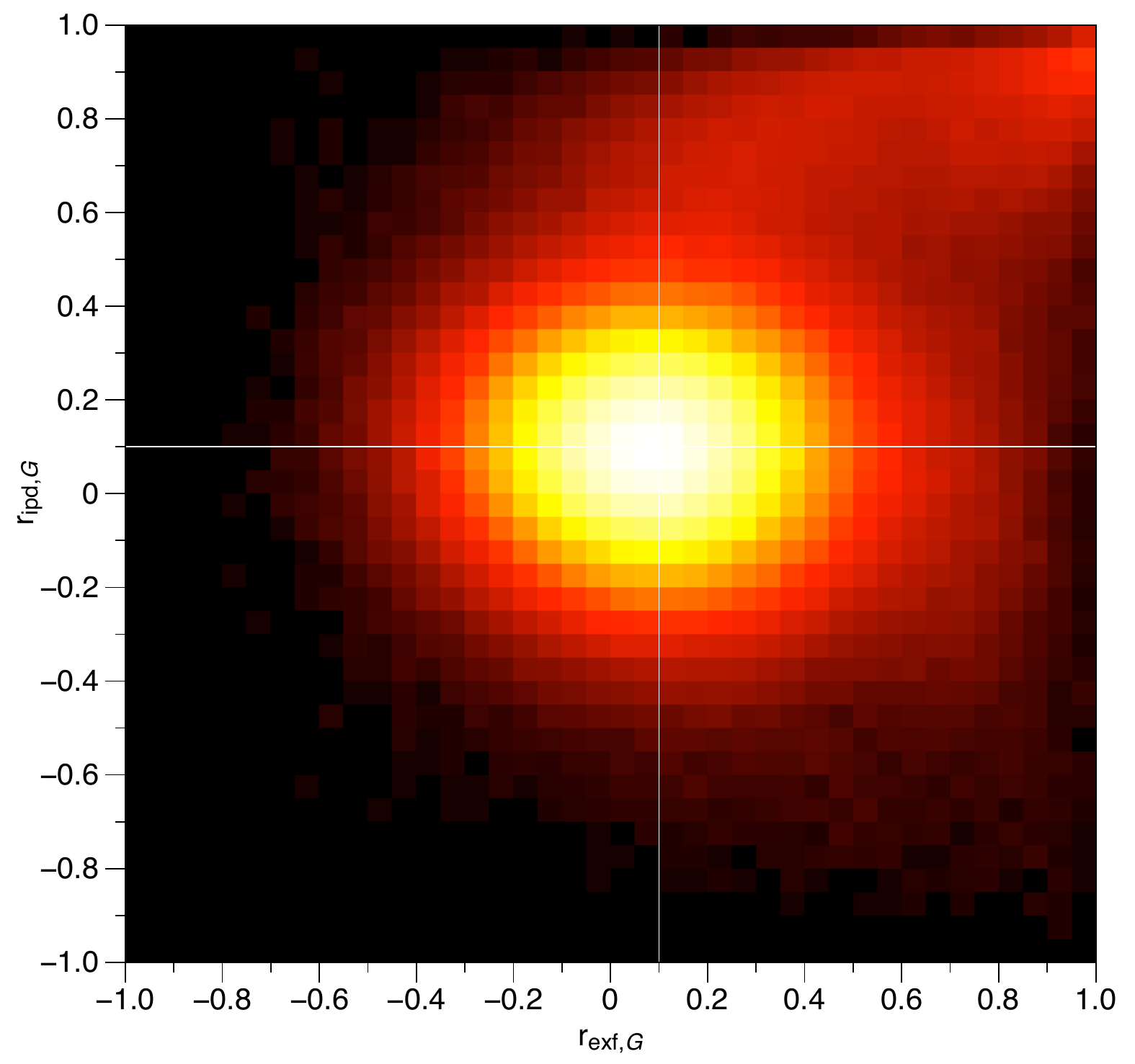}
\caption{ \label{fig:corrVsCorr} Density plots of the \rExfG vs \rIpdG correlation  with at least 20 observations for each statistic (\numAllBands $\geq 20$ and \numObsExclEpslG $\geq 20$). See Sect.~\ref{sec:detectSaSignals} for a discussion. 
}
\end{figure*} 

\begin{figure*}[t]
\includegraphics[width=1.0\textwidth]{images/publishedTable/topLabelsVarAndGaps.pdf}
    \includegraphics[width=0.5\textwidth]{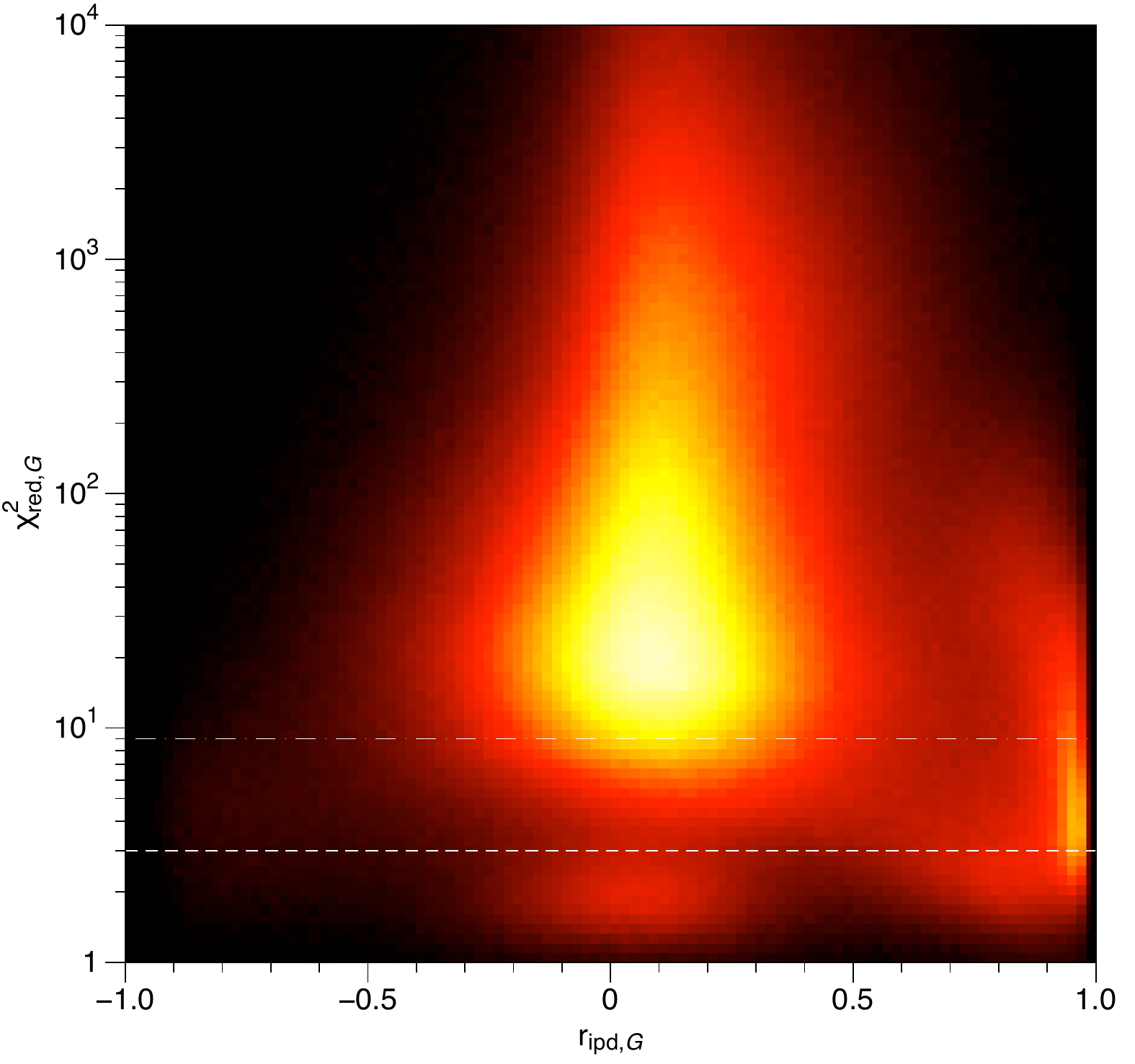}
    \includegraphics[width=0.5\textwidth]{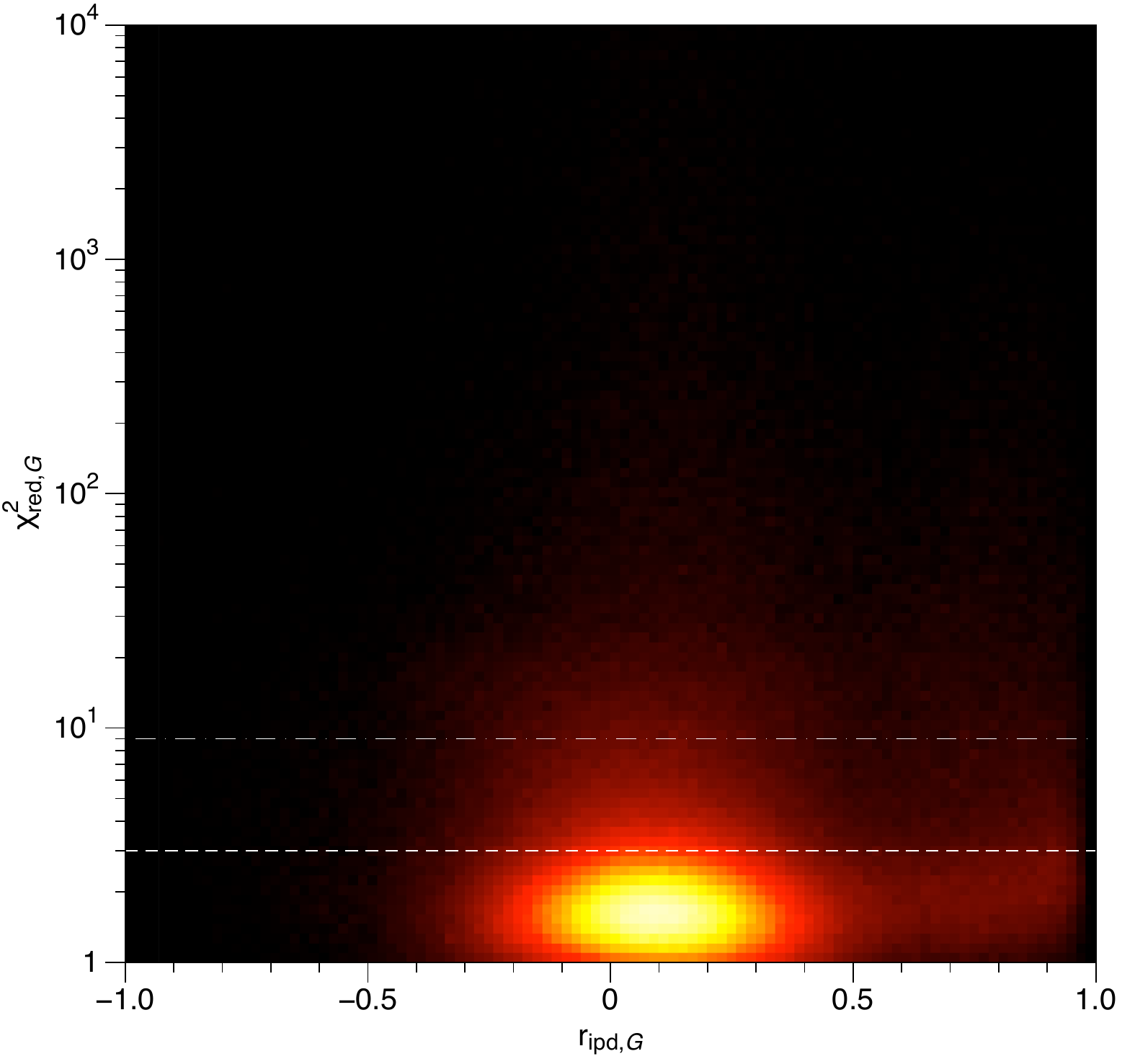}
\caption{ \label{fig:rIpdGvsRedChi2} Density plot of the relation between \rIpdG  and \redChiSqG of the small-separation binary model fit to the G-band photometry (Eq.~\ref{eq:simu2}). A low 
\redChiSqG together with a high \rIpdG suggests a strong scan-angle-dependent signal, while a low 
\redChiSqG in combination with a low correlation usually corresponds to non-variable (constant) stars with an insignificant amplitude, as seen (and expected) for the majority of the GAPS data set.
As a guide for low \redChiSqG values, we added horizontal lines at thresholds 3 (short-dashed) and 9 (longer-dashed; see Sect.~\ref{ssec:ipdParamAsDetectionMethod} for a more detailed discussion, and see also Fig.~\ref{fig:signAmpl3Stats} for the significance of the fitted amplitude). Plots for \numAllBands $\geq 20$ and \numObsExclEpslG $\geq 20$.
}
\end{figure*} 

\begin{figure*}[t]
    \includegraphics[width=1.0\textwidth]{images/publishedTable/topLabelsVarAndGaps.pdf}
    \includegraphics[width=0.5\textwidth]{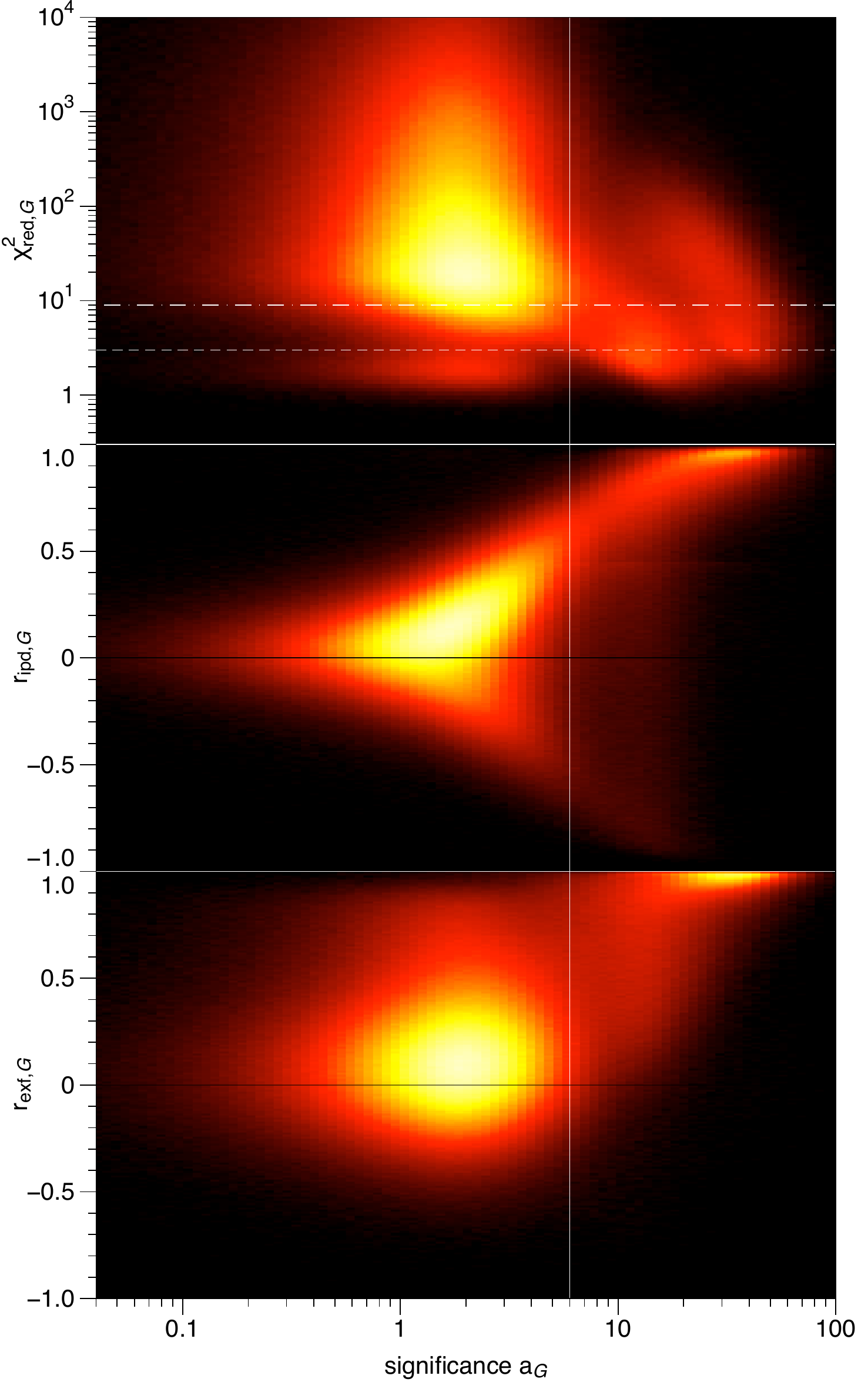}
    \includegraphics[width=0.5\textwidth]{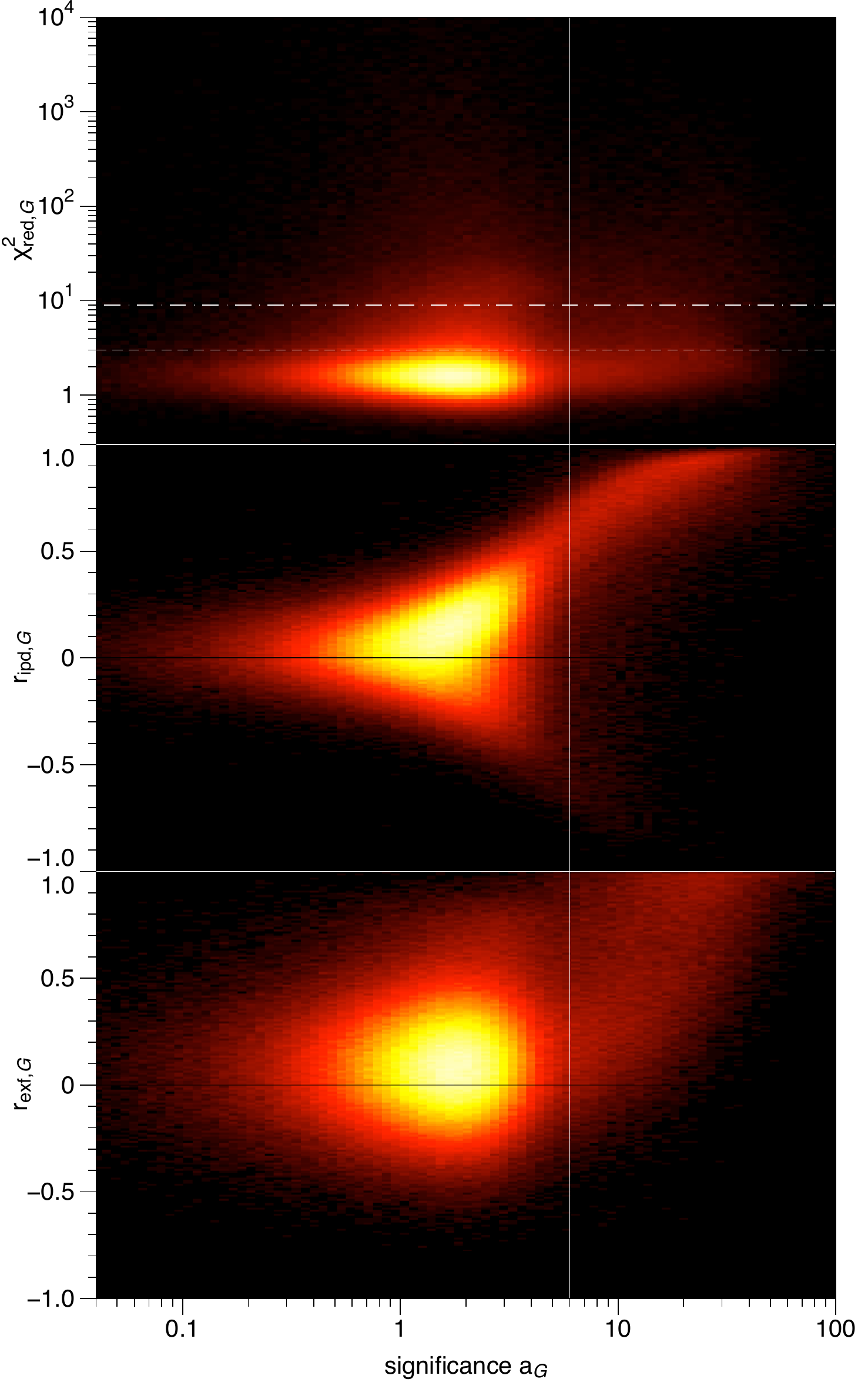}
\caption{\label{fig:signAmpl3Stats} Density plots of various parameters in relation to the significance of \aG, the amplitude of the small-separation binary model fit to the G-band photometry (Eq.~\ref{eq:simu2}). The white vertical line indicates a significance threshold of 6, above which the fitted amplitude is more likely due to a scan-angle-dependent signal. 
In the top panels, the same low \redChiSqG thresholds as in Fig.~\ref{fig:rIpdGvsRedChi2} are repeated at 3 and 9. The second and third panels 
show the strong relation between high-amplitude significance and high \rIpdG or \rExfG, respectively (see Sect.~\ref{ssec:ipdParamAsDetectionMethod} for further discussion).
Plots for \numAllBands $\geq 20$ and \numObsExclEpslG $\geq 20$.
}
\end{figure*}

\begin{figure*}[t]
    \includegraphics[width=1.0\textwidth]{images/publishedTable/topLabelsVarAndGaps.pdf}
    \includegraphics[width=0.5\textwidth]{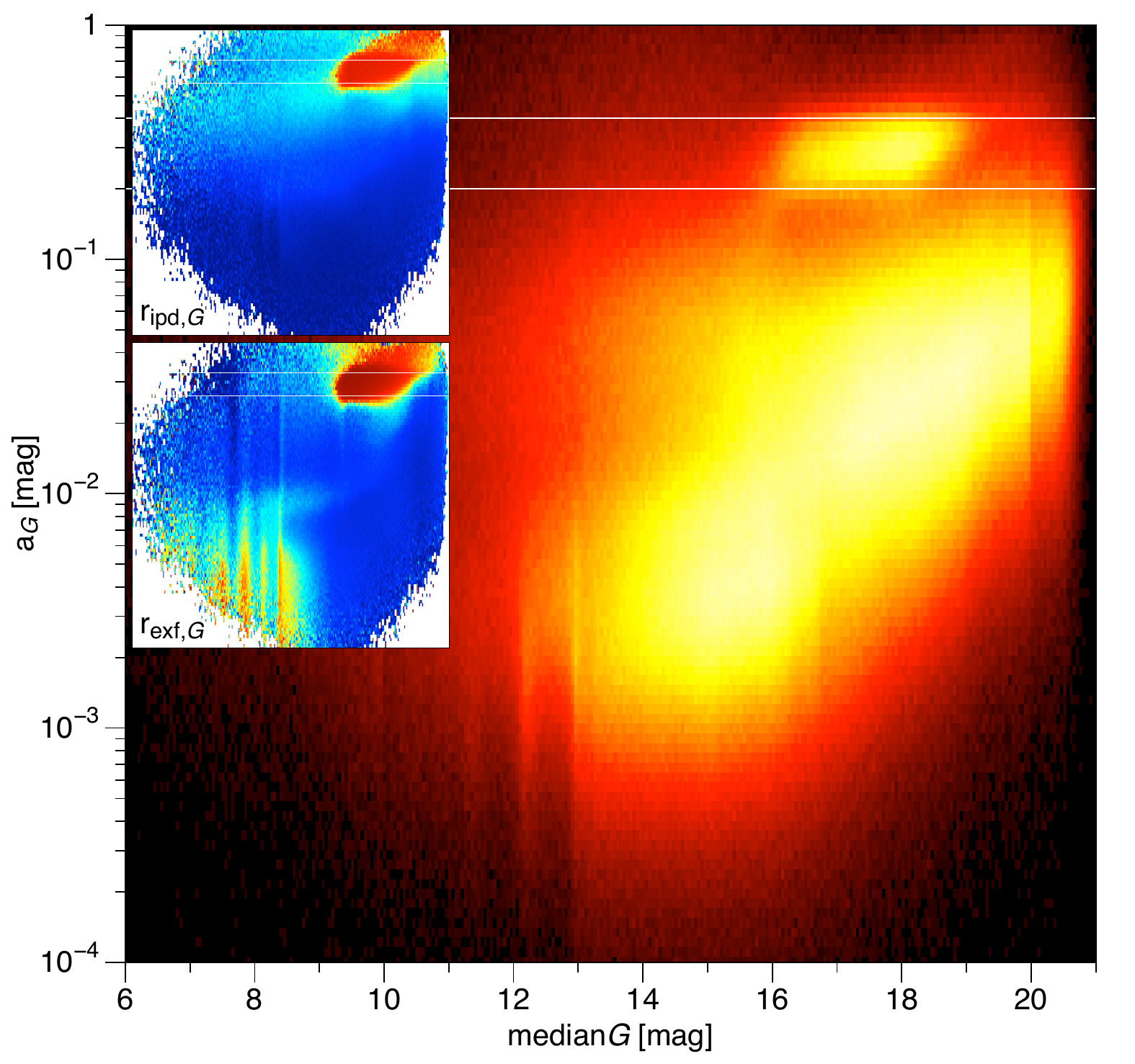}
    \includegraphics[width=0.5\textwidth]{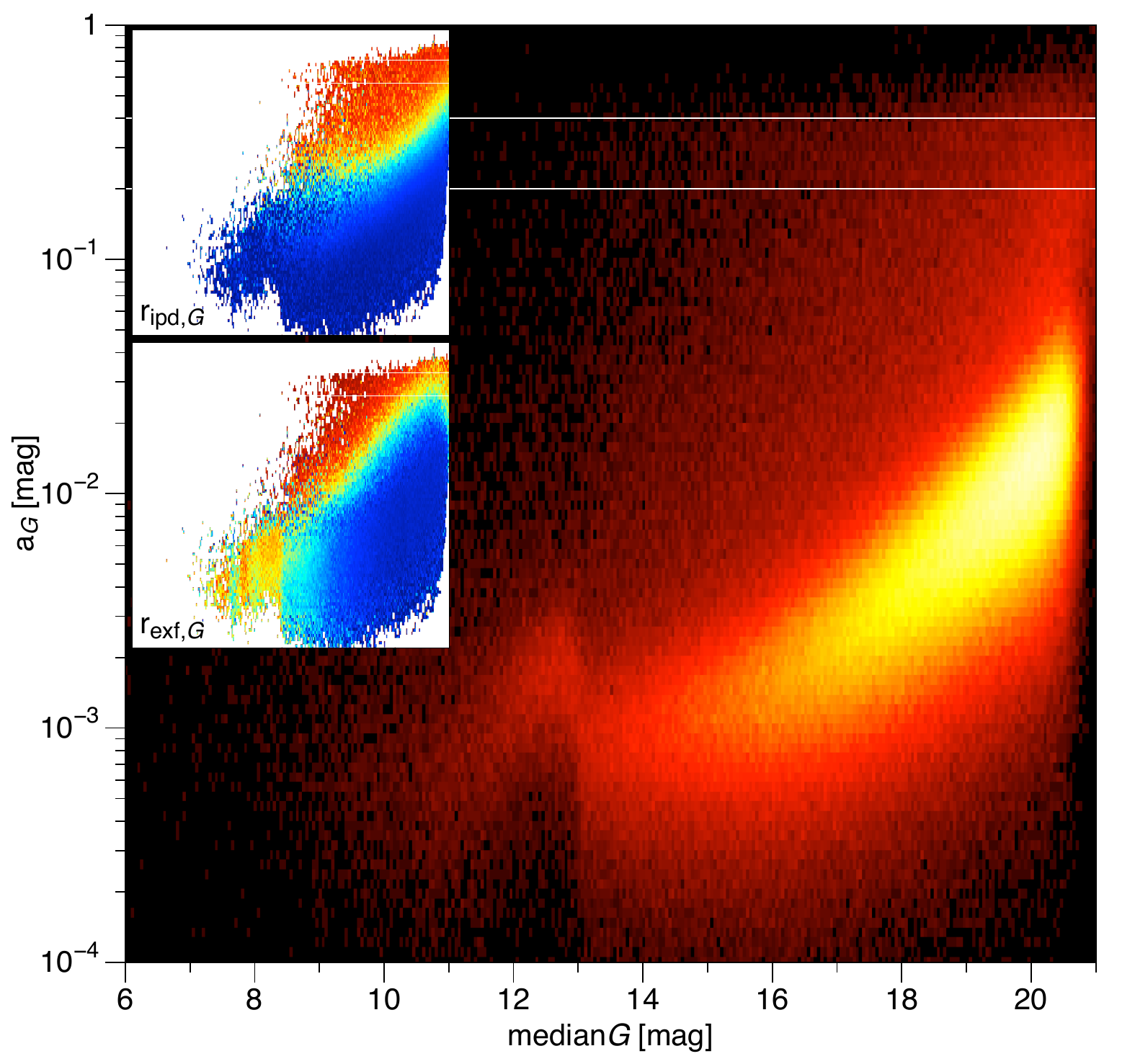}
\caption{\label{fig:magVsSamAmpl} Density of the median G-band magnitude dependence of \aG, the amplitude of the small-separation binary model fit to the G-band photometry (Eq.~\ref{eq:simu2}). 
The top left inset images shows the same data colour-coded with the median \rIpdG and \rExfG, respectively, both ranging between 0 (blue) and 1 (red). See Sect.~\ref{ssec:smallSepBinaryModelAsDetectionMethod} for discussion. Plots are restricted to sources with \numAllBands $\geq 20$ and \numObsExclEpslG $\geq 20$.
}
\end{figure*}

\begin{figure*}[t]
    \includegraphics[width=1.0\textwidth]{images/publishedTable/topLabelsVarAndGaps.pdf}
    \includegraphics[width=0.5\textwidth]{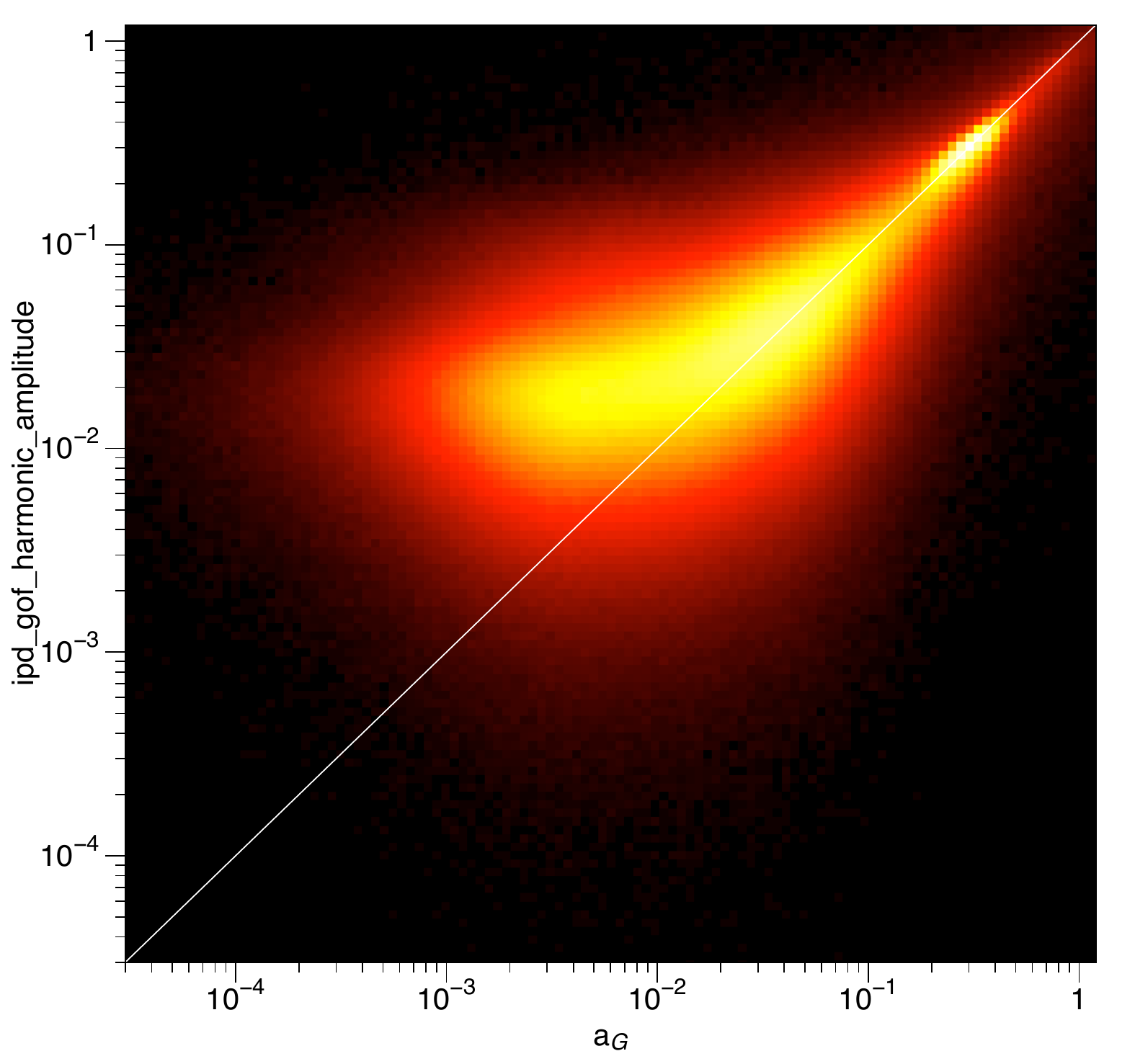}
    \includegraphics[width=0.5\textwidth]{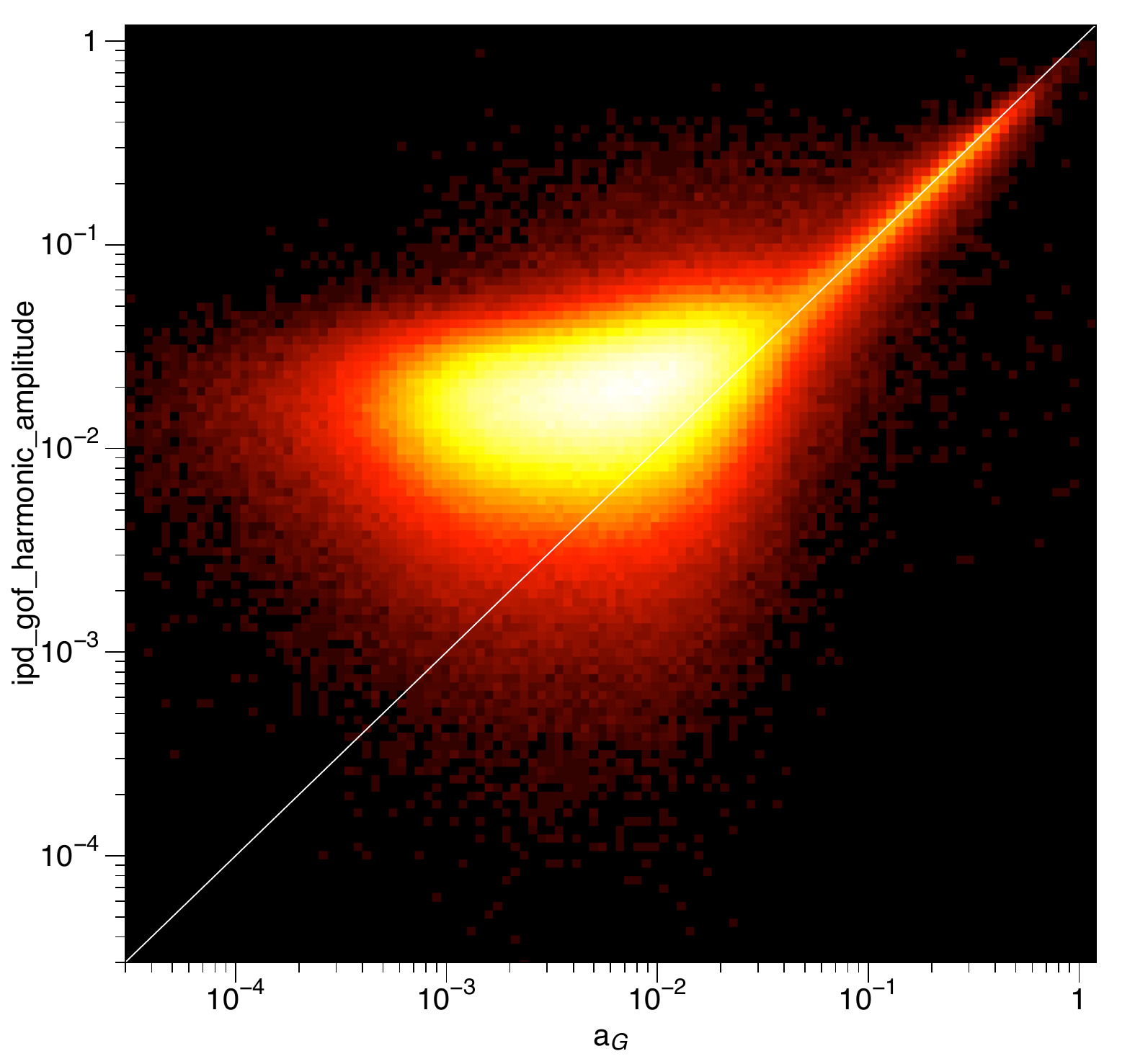}
\caption{\label{fig:samAmplVsIpdAmpl} Density plot of the relation between \ipdGofHarmAmpl and \aG (the amplitude of the fit to the G-band photometry (Eq.~\ref{eq:simu2}). For  \ipdGofHarmAmpl $\gtrsim 0.07,$ this relation is very strong (see Sects.~\ref{sssec:demoSaSignalsPointSrc}, \ref{ssec:smallSepBinaryModelAsDetectionMethod}, and \ref{ssect:photAmplSam} for a more detailed discussion). Plots for \numAllBands $\geq 20$ and \numObsExclEpslG $\geq 20$.
} 
\end{figure*}

As part of this paper, the table \texttt{vari\_spurious\_signals} 
is published in the \gdr{3} archive for all 11\,754\,237~sources with published photometric time series, that is, for sources in \texttt{gaia\_source} with \texttt{has\_epoch\_photometry=true}. Of these, 10\,509\,536 sources are variables \citep{DR3-DPACP-162} and 1\,257\,319 are part of GAPS \citep{DR3-DPACP-142}. They can be identified in \texttt{gaia\_source} by their \texttt{phot\_variable\_flag}=\texttt{VARIABLE} and \texttt{in\_andromeda\_survey}=\texttt{true} flags, respectively. We note that 12\,618 sources overlap between the two.

This table contains the parameters listed in Table~\ref{tab:mdbTableFields} (see also the \gaia archive documentation\footnote{\url{https://gea.esac.esa.int/archive/documentation/GDR3/Gaia_archive/chap_datamodel/sec_dm_performance_verification/ssec_dm_vari_spurious_signals.html}}): the $r_\text{ipd}$ and $r_\text{exf}$ correlations for the three photometric bands along with the available number of observations for each, as introduced in  Sect.~\ref{sec:detectSaSignals}. Additionally, for the G-band photometry, it contains a fit to the small-separation binary model introduced in Sect.~\ref{sec:responseToSaSignal}.
When fewer than six observations were available for the computation of a value, we found it to be generally unusable and set the value to be missing (null).
However, we caution that (much) more stringent cuts need to be applied on the number of observations for different parameter values. As illustrated in Fig.~\ref{fig:corrVsNumber}, the \rExf and  \rIpd are not really meaningful for fewer than 11 points, and some biases and features disappear only above 20 or 30. This is related to the fact that for the \gdr{3} 34 months of data, the scanning law should have provided sources all over the sky with at least 15--20 observations, and sources with fewer observations are generally found in crowded regions and/or are at the faint end, which makes them more susceptible to disturbances and/or non-detections. This increase in (scan-angle-dependent) disturbances is illustrated by the rise in G-band correlations towards fewer observations.  In the heat maps, we normalised the values for each number of observation bin to highlight the change in spread in the correlation distribution. True source-count distributions are provided by the histograms above and on the side of the heat maps.
For a more detailed discussion of the relation between the number of points and the significance of the Spearman correlation, see for example \cite{Bonett:2000tp}. The general increase in spread in all correlations towards fewer observations  illustrates this relation.

The parameters of the G-band photometry fit to the small-separation binary model of Eq.~\ref{eq:simu2} (fields labelled \texttt{scan\_angle\_model\_*}) are provided for all sources with six observations at least, but are only meaningful for sources with high G-band correlations, for instance \rExfG and/or  \rIpdG > 0.8 and with a sufficient number of observations, where we recommend \texttt{num\_obs\_excl\_epsl\_g\_fov} to be above 20 or 30. Parameters outside of this scope can be an arbitrary poor fit to the data, as the model is simply not applicable: these values therefore have to be used with appropriate care. Additionally, from Sect.~\ref{ssec:smallSepBinaryModelAsDetectionMethod}, the significance of the amplitude (\texttt{*ampl\_sig*}) and $\chi^2_\text{red}$ of the fit (\texttt{*red\_chi2*}) are provided. Finally, we also provide the F2, or Gaussianised $\chi^2$.

For this model fit, we used the same observations as for \rIpd, that is, we excluded the ecliptic pole scanning law observations due to their highly clustered scan angles, which easily bias the effective weight of the fit to only a small scan-angle range. For more information about the various diagnostic parameters in this table, see Sect.~\ref{sec:detectSaSignals}.

For completeness, we include the generalised least-squares (``\texttt{gls}'') period search result frequency, amplitude, signal detection efficiency \citep[SDE; see][]{2000ApJ...542..257A, 2002A&A...391..369K}, and Baluev false-alarm probability \citep{2009MNRAS.395.1541B} at the beginning of the table, and the frequency and estimated error from a subsequent non-linear harmonic modelling (\texttt{nhm}), as was described in Sect.~\ref{ssec:obsPerDistr}.

The additional Figs.~\ref{fig:corrVsCorr}, \ref{fig:rIpdGvsRedChi2}, \ref{fig:signAmpl3Stats}, \ref{fig:magVsSamAmpl}, and \ref{fig:samAmplVsIpdAmpl}  are presented here to support discussions in the body of this paper that are based on the data published with this paper.

\begin{sidewaystable*}
\caption{\label{tab:mdbTableFields} \texttt{gaia\_dr3.vari\_spurious\_signals} \gaia archive table fields. See the \href{https://gea.esac.esa.int/archive/documentation/GDR3/Gaia_archive/chap_datamodel/sec_dm_performance_verification/ssec_dm_vari_spurious_signals.html}{\gaia archive documentation} for more detailed descriptions.}
\begin{tabular}{
    lrlrlrrl }

\hline\hline
Field name & Symbol & Notes & Count (\ \ \ \ \ \ \ null) & Type & Min & Max & Units\\
\hline
\href{https://gea.esac.esa.int/archive/documentation/GDR3/Gaia_archive/chap_datamodel/sec_dm_performance_verification/ssec_dm_vari_spurious_signals.html#vari_spurious_signals-source_id}
{\tt source\_id                         } &               &                &  11\,754\,237 (\ \ \ \quad \quad 0) & long & & & \\
\href{https://gea.esac.esa.int/archive/documentation/lookup/gaiadr3/vari_spurious_signals/phot_variable_flag}
{\tt phot\_variable\_flag                         } & & \texttt{VARIABLE} or \texttt{NOT\_AVAILABLE}                            &  11\,754\,237 (\ \ \ \quad \quad 0) & String & & & \\
\href{https://gea.esac.esa.int/archive/documentation/lookup/gaiadr3/vari_spurious_signals/in_andromeda_survey}
{\tt in\_andromeda\_survey                         } &  & \texttt{true} for sources in GAPS               &  11\,754\,237 (\ \ \ \quad \quad 0) & boolean & & & \\
\href{https://gea.esac.esa.int/archive/documentation/lookup/gaiadr3/vari_spurious_signals/num_obs_common_all_bands}
{\tt num\_obs\_common\_all\_bands      } &  \numAllBands  &  Used in \texttt{*corr\_exf*} below                 &  11\,754\,237 (\ \ \ \quad \quad 0) & int & 0 & 252 & \\
\hline
\href{https://gea.esac.esa.int/archive/documentation/lookup/gaiadr3/vari_spurious_signals/num_obs_g_fov}
{\tt num\_obs\_g\_fov             } &         &   Used in \texttt{*freq*} below      &  11\,754\,237 (\ \ \ \quad \quad 0) & int & 0 & 265 & \\
\href{https://gea.esac.esa.int/archive/documentation/lookup/gaiadr3/vari_spurious_signals/gls_freq_g_fov}
{\tt gls\_freq\_g\_fov             } &         &  null if {\tt num\_obs\_g\_fov} < 5     &  11\,749\,374 (\  \ \ \ 4\,863) & double & 0.0008 & 25 & d$^{-1}$\\
\href{https://gea.esac.esa.int/archive/documentation/lookup/gaiadr3/vari_spurious_signals/gls_freq_ampl_g_fov}
{\tt gls\_freq\_ampl\_g\_fov             } &         &  null if {\tt num\_obs\_g\_fov} < 5     &  11\,749\,374 (\  \ \ \ 4\,863) & float & 0 & 1.0 & \\
\href{https://gea.esac.esa.int/archive/documentation/lookup/gaiadr3/vari_spurious_signals/gls_freq_sde_g_fov}
{\tt gls\_freq\_sde\_g\_fov             } &         &  null if {\tt num\_obs\_g\_fov} < 5     &  11\,749\,374 (\  \ \ \ 4\,863) & float & 0 & & \\
\href{https://gea.esac.esa.int/archive/documentation/lookup/gaiadr3/vari_spurious_signals/gls_freq_fap_g_fov}
{\tt gls\_freq\_fap\_g\_fov            } &         &  null if {\tt num\_obs\_g\_fov} < 5     &  11\,749\,374 (\  \ \ \ 4\,863) & double & 1E-300 & 1 & \\
\href{https://gea.esac.esa.int/archive/documentation/lookup/gaiadr3/vari_spurious_signals/nhm_fund_freq_g_fov}
{\tt nhm\_fund\_freq\_g\_fov             } &         &   null if {\tt num\_obs\_g\_fov} < 5  or harmonic model failed     &  11\,714\,493 (\  \ 39\,744) & double & 0 & 25 & d$^{-1}$\\
\href{https://gea.esac.esa.int/archive/documentation/lookup/gaiadr3/vari_spurious_signals/nhm_fund_freq_error_g_fov}
{\tt nhm\_fund\_freq\_error\_g\_fov             } &         &  null if {\tt num\_obs\_g\_fov} < 5 or harmonic model failed    &  11\,714\,493 (\  \ 39\,744) & float & 0 &  & d$^{-1}$ \\

\hline
\href{https://gea.esac.esa.int/archive/documentation/lookup/gaiadr3/vari_spurious_signals/spearman_corr_exf_g_fov}
{\tt spearman\_corr\_exf\_g\_fov             } & \rExfG        &  Eq.~\ref{eq:corExcessFactor}; null when  \numAllBands < 6                   &  11\,406\,589 (347\,648) & float & -1 & 1 & \\

\href{https://gea.esac.esa.int/archive/documentation/lookup/gaiadr3/vari_spurious_signals/num_obs_excl_epsl_g_fov}
{\tt num\_obs\_excl\_epsl\_g\_fov            } & \numObsExclEpslG              &   Used in \texttt{*\_g\_fov} below               &  11\,754\,237 (\ \ \ \quad \quad 0) & int & 0 & 177 & \\
\href{https://gea.esac.esa.int/archive/documentation/lookup/gaiadr3/vari_spurious_signals/spearman_corr_ipd_g_fov}
{\tt spearman\_corr\_ipd\_g\_fov             } & \rIpdG        & Eq.~\ref{eq:rIpd}; null when \numObsExclEpslG < 6 &  11\,731\,048 (\;  23\,189)  & float & -1 & 1 & \\
\href{https://gea.esac.esa.int/archive/documentation/lookup/gaiadr3/vari_spurious_signals/scan_angle_model_offset_g_fov}
{\tt scan\_angle\_model\_offset\_g\_fov      } & $c_\text{0}$              & Eq.~\ref{eq:simu2}; null when \numObsExclEpslG < 6 &  11\,731\,048 (\;  23\,189)  & float &  & & mag\\
\href{https://gea.esac.esa.int/archive/documentation/lookup/gaiadr3/vari_spurious_signals/scan_angle_model_ampl_g_fov}
{\tt scan\_angle\_model\_ampl\_g\_fov        } & \aG               & Eq.~\ref{eq:simu2}; null when \numObsExclEpslG < 6 &  11\,731\,048 (\;  23\,189)  & float & 0 & & mag \\
\href{https://gea.esac.esa.int/archive/documentation/lookup/gaiadr3/vari_spurious_signals/scan_angle_model_ampl_sig_g_fov}
{\tt scan\_angle\_model\_ampl\_sig\_g\_fov } &               & Eq.~\ref{eq:simu2}; null when \numObsExclEpslG < 6 &  11\,731\,048 (\;  23\,189)  & float & 0 & &  \\
\href{https://gea.esac.esa.int/archive/documentation/lookup/gaiadr3/vari_spurious_signals/scan_angle_model_phase_g_fov}
{\tt scan\_angle\_model\_phase\_g\_fov       } & \thetaG              & Eq.~\ref{eq:simu2}; null when \numObsExclEpslG < 6 &  11\,731\,048 (\;  23\,189)  & float & 0 & 180 & deg\\
\href{https://gea.esac.esa.int/archive/documentation/lookup/gaiadr3/vari_spurious_signals/scan_angle_model_red_chi2_g_fov}
{\tt scan\_angle\_model\_red\_chi2\_g\_fov   } &    \redChiSqG          & Eq.~\ref{eq:simu2}; null when \numObsExclEpslG < 6 &  11\,731\,048 (\;  23\,189)  & float & 0 & \\
\href{https://gea.esac.esa.int/archive/documentation/lookup/gaiadr3/vari_spurious_signals/scan_angle_model_f2_g_fov}
{\tt scan\_angle\_model\_f2\_g\_fov   } &    $f_\text{2,$G$}$          & Eq.~\ref{eq:simu2}; null when \numObsExclEpslG < 6 &  11\,731\,048 (\;  23\,189)  & float &  & \\
\hline
\href{https://gea.esac.esa.int/archive/documentation/lookup/gaiadr3/vari_spurious_signals/spearman_corr_exf_bp}
{\tt spearman\_corr\_exf\_bp             } & \rExfBp        &   Eq.~\ref{eq:corExcessFactor}; null when  \numAllBands < 6                   &  11\,406\,589 (347\,648) & float & -1 & 1 & \\
\href{https://gea.esac.esa.int/archive/documentation/lookup/gaiadr3/vari_spurious_signals/num_obs_excl_epsl_bp}
{\tt num\_obs\_excl\_epsl\_bp           } & \numObsExclEpslBp              &   Used in \texttt{*\_bp} below                                 &  11\,754\,237 (\ \ \ \quad \quad 0)& int & 0 & 167 & \\
\href{https://gea.esac.esa.int/archive/documentation/lookup/gaiadr3/vari_spurious_signals/spearman_corr_ipd_bp}
{\tt spearman\_corr\_ipd\_bp            } & \rIpdBp       & Eq.~\ref{eq:rIpd}; null when \numObsExclEpslBp < 6 &  11\,445\,687 (308\,550)  & float & -1 & 1 & \\
\href{https://gea.esac.esa.int/archive/documentation/lookup/gaiadr3/vari_spurious_signals/scan_angle_model_offset_bp}
{\tt scan\_angle\_model\_offset\_bp     } &               & set to null when \numObsExclEpslBp < 6 &  11\,445\,687 (308\,550) & float &  &  & mag\\
\href{https://gea.esac.esa.int/archive/documentation/lookup/gaiadr3/vari_spurious_signals/scan_angle_model_ampl_bp}
{\tt scan\_angle\_model\_ampl\_bp       } &               & set to null when \numObsExclEpslBp < 6 &  11\,445\,687 (308\,550) & float & 0  &  & mag\\
\href{https://gea.esac.esa.int/archive/documentation/lookup/gaiadr3/vari_spurious_signals/scan_angle_model_ampl_sig_bp}
{\tt scan\_angle\_model\_ampl\_sig\_bp } &               & set to null when \numObsExclEpslBp < 6 &  11\,445\,687 (308\,550)  & float & 0  & &  \\
\href{https://gea.esac.esa.int/archive/documentation/lookup/gaiadr3/vari_spurious_signals/scan_angle_model_phase_bp}
{\tt scan\_angle\_model\_phase\_bp      } &               & set to null when \numObsExclEpslBp < 6 &  11\,445\,687 (308\,550) & float & 0 & 180 & deg\\
\href{https://gea.esac.esa.int/archive/documentation/lookup/gaiadr3/vari_spurious_signals/scan_angle_model_red_chi2_bp}
{\tt scan\_angle\_model\_red\_chi2\_bp  } &     \redChiSqBp          & set to null when \numObsExclEpslBp < 6 &  11\,445\,687 (308\,550) &  float & 0 & & \\
\href{https://gea.esac.esa.int/archive/documentation/lookup/gaiadr3/vari_spurious_signals/scan_angle_model_f2_bp}
{\tt scan\_angle\_model\_f2\_bp   } &    $f_\text{2,$BP$}$          & set to null when \numObsExclEpslBp < 6 &  11\,445\,687 (308\,550)  & float &  & \\

\hline
\href{https://gea.esac.esa.int/archive/documentation/lookup/gaiadr3/vari_spurious_signals/spearman_corr_exf_rp}
{\tt spearman\_corr\_exf\_rp             } & \rExfRp        &   Eq.~\ref{eq:corExcessFactor}; null when  \numAllBands < 6                   &  11\,406\,589 (347\,648) & float & -1 & 1 & \\
\href{https://gea.esac.esa.int/archive/documentation/lookup/gaiadr3/vari_spurious_signals/num_obs_excl_epsl_rp}
{\tt num\_obs\_excl\_epsl\_rp           } & \numObsExclEpslRp              &     Used in \texttt{*\_rp} below                               &  11\,754\,237 (\ \ \ \quad \quad 0) & int & 0 & 167 &\\
\href{https://gea.esac.esa.int/archive/documentation/lookup/gaiadr3/vari_spurious_signals/spearman_corr_ipd_rp}
{\tt spearman\_corr\_ipd\_rp           } & \rIpdRp       & Eq.~\ref{eq:rIpd}; null when \numObsExclEpslRp < 6 &  11\,504\,400 (249\,837) & float  & -1 & 1 & \\
\href{https://gea.esac.esa.int/archive/documentation/lookup/gaiadr3/vari_spurious_signals/scan_angle_model_offset_rp}
{\tt scan\_angle\_model\_offset\_rp     } &               & set to null when \numObsExclEpslRp < 6 &  11\,504\,400 (249\,837) & float &  &  & mag \\
\href{https://gea.esac.esa.int/archive/documentation/lookup/gaiadr3/vari_spurious_signals/scan_angle_model_ampl_rp}
{\tt scan\_angle\_model\_ampl\_rp       } &               & set to null when \numObsExclEpslRp < 6 &  11\,504\,400 (249\,837) & float & 0 & & mag  \\
\href{https://gea.esac.esa.int/archive/documentation/lookup/gaiadr3/vari_spurious_signals/scan_angle_model_ampl_sig_rp}
{\tt scan\_angle\_model\_ampl\_sig\_rp } &               & set to null when \numObsExclEpslRp < 6 &  11\,504\,400 (249\,837)  & float & 0 & &  \\
\href{https://gea.esac.esa.int/archive/documentation/lookup/gaiadr3/vari_spurious_signals/scan_angle_model_phase_rp}
{\tt scan\_angle\_model\_phase\_rp      } &               & set to null when \numObsExclEpslRp < 6 &  11\,504\,400 (249\,837) & float & 0  & 180 & deg \\
\href{https://gea.esac.esa.int/archive/documentation/lookup/gaiadr3/vari_spurious_signals/scan_angle_model_red_chi2_rp}
{\tt scan\_angle\_model\_red\_chi2\_rp  } &     \redChiSqRp           & set to null when \numObsExclEpslRp < 6 &  11\,504\,400 (249\,837) & float & 0 &  & \\
\href{https://gea.esac.esa.int/archive/documentation/lookup/gaiadr3/vari_spurious_signals/scan_angle_model_f2_rp}
{\tt scan\_angle\_model\_f2\_rp   } &    $f_\text{2,$RP$}$          & set to null when \numObsExclEpslRp < 6 &  11\,504\,400 (249\,837)  & float &  & \\
\hline
\end{tabular}
\end{sidewaystable*}

\FloatBarrier

\section{Additional examples of scan-angle-dependent signals in IPD outputs\label{sec:appendix-bcn-full-data}}

As an extension of Sect.~\ref{ssec:afInstr}, the following figures show some representative examples covering different cases, such as angular separations between close pairs of sources, observations with 
one- and two-dimensional windows, bright nearby sources, and single sources identified (and characterised) as galaxies.

\subsection{Unresolved close pairs in DR3}

The first set of cases includes sources that are delivered in DR3 as single sources, but have some quality indicators or features that make them reasonable candidates to be unresolved close pairs. The following examples have been confirmed as close source pairs by IDU during the data processing for DR4.

\begin{figure}[h]
\centering
  \includegraphics[width=0.45\textwidth]{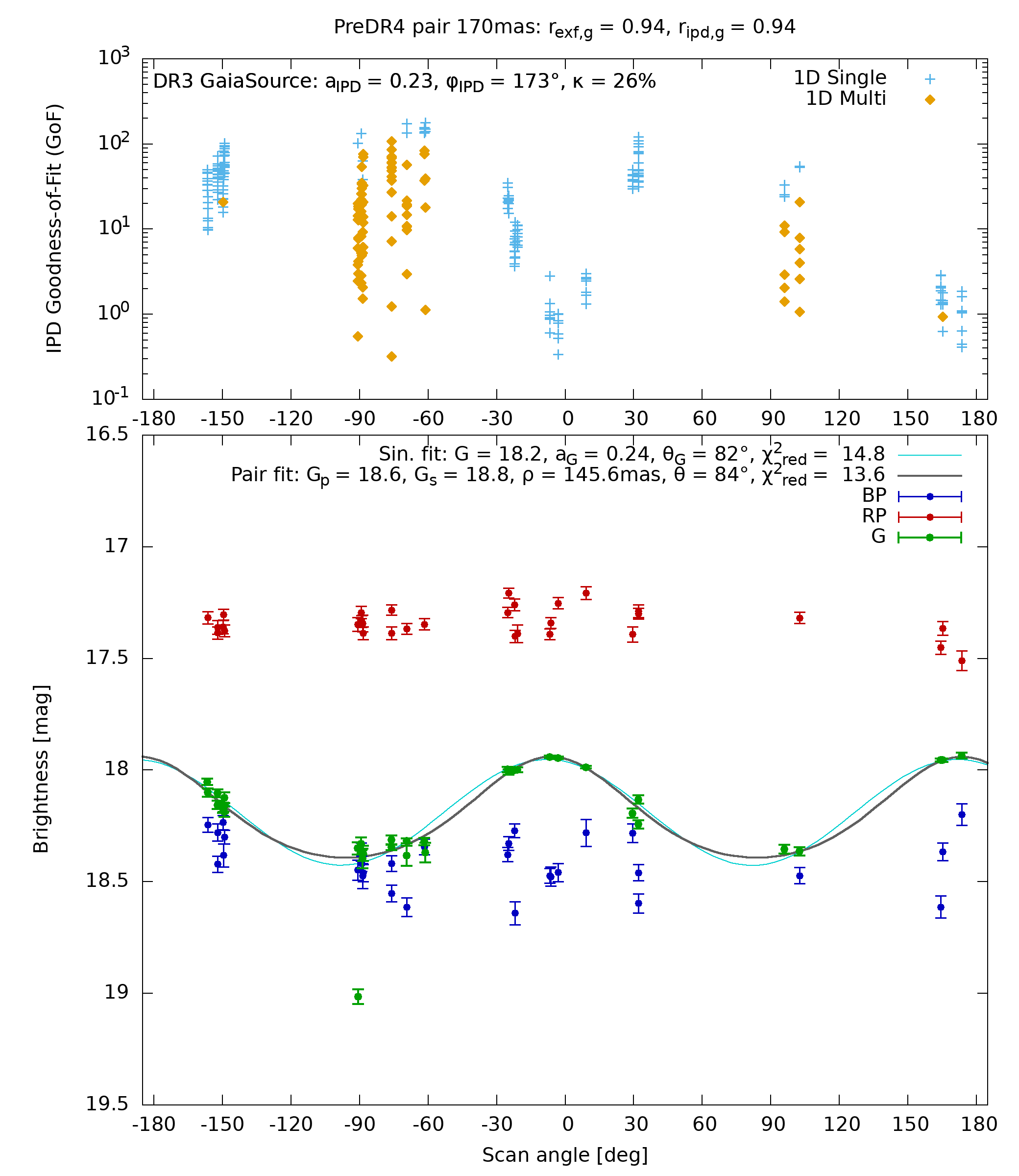}
\caption{\label{fig:ep-sa-388466602081536640} SourceID 388466602081536640: Scan-angle signatures.}
\end{figure}

The first example, already shown in Sect.~\ref{ssec:afInstr}, is taken from the outputs of query
{\tiny \begin{verbatim}
-- ADQL query on DR3, aiming at <200mas 1D: 
SELECT TOP 100 * FROM gaiadr3.gaia_source WHERE
in_andromeda_survey='TRUE'
AND phot_g_mean_mag>=14 AND phot_g_mean_mag<19 
AND ipd_frac_multi_peak<30 AND ipd_frac_multi_peak>20 
AND pmra IS NULL AND matched_transits>30 
AND visibility_periods_used>18 
AND astrometric_matched_transits > 30 ORDER BY random_index,
\end{verbatim}
}
\noindent which corresponds to a source with just a two-parameter solution from AGIS, a very high \rExfG and \rIpdG Spearman correlations (see Sect.~\ref{sec:detectSaSignals}), quite high IPD GoF harmonic amplitude, but a modest fraction of IPD multiple peaks. Fig.~\ref{fig:closePair130mas} shows that the epoch $G$ magnitude strongly varies with the scan angle, whereas the \gbp and \grp magnitudes both remain mostly constant. When we verified the IDU pre-DR4 data, we found that this DR3 source indeed is a close pair with a very small separation of just 130 mas, with one-dimensional windows and nearly the same magnitude in both sources. It is at the limit of the IDU on-ground detection capability of close source pairs for DR4. Figure~\ref{fig:ep-sa-388466602081536640} shows another example obtained from the same query, corresponding to a close pair with a slightly larger separation of about 170~mas.

\begin{figure}[h]
\centering
  \includegraphics[width=0.45\textwidth]{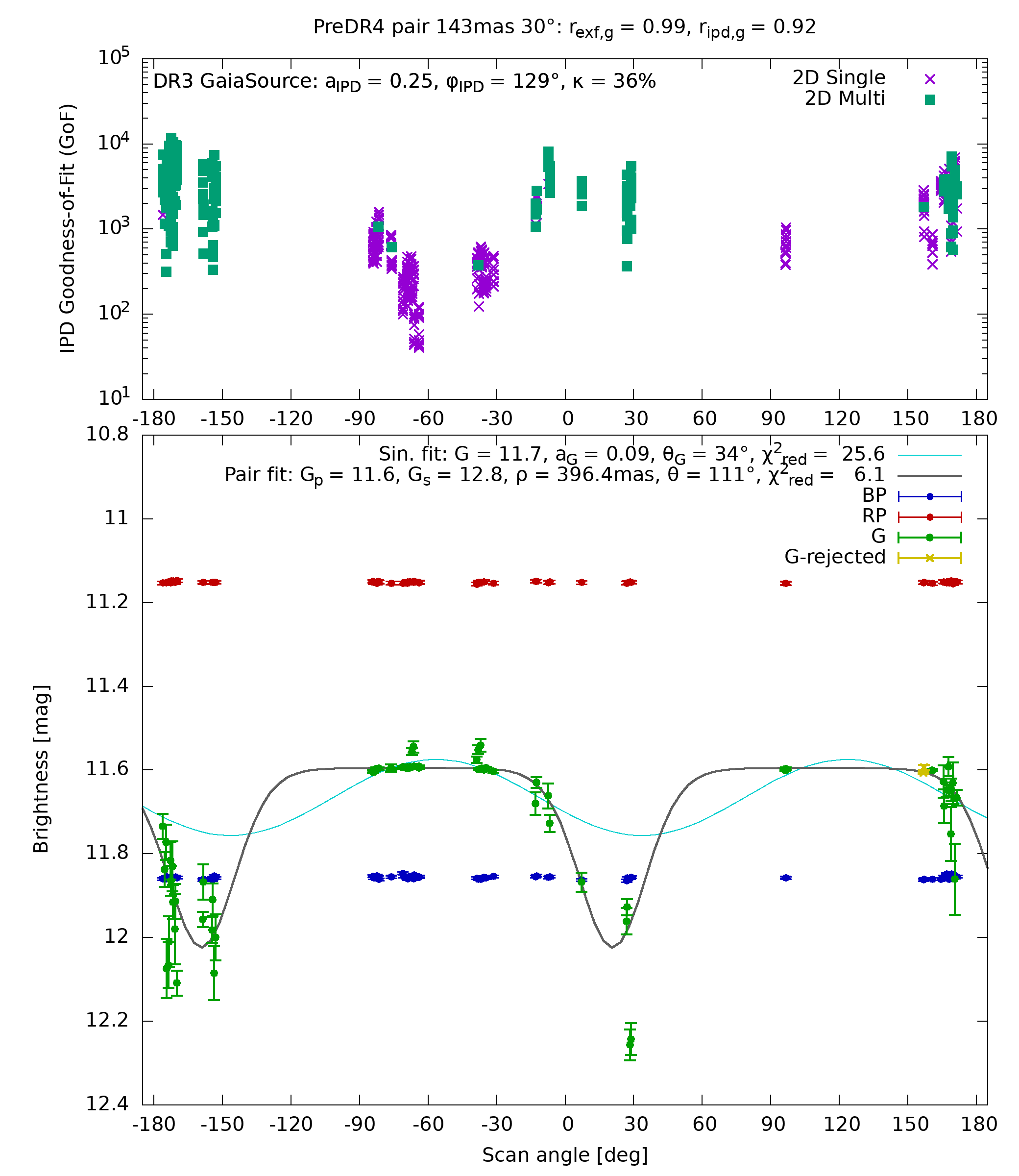}
  \includegraphics[width=0.45\textwidth]{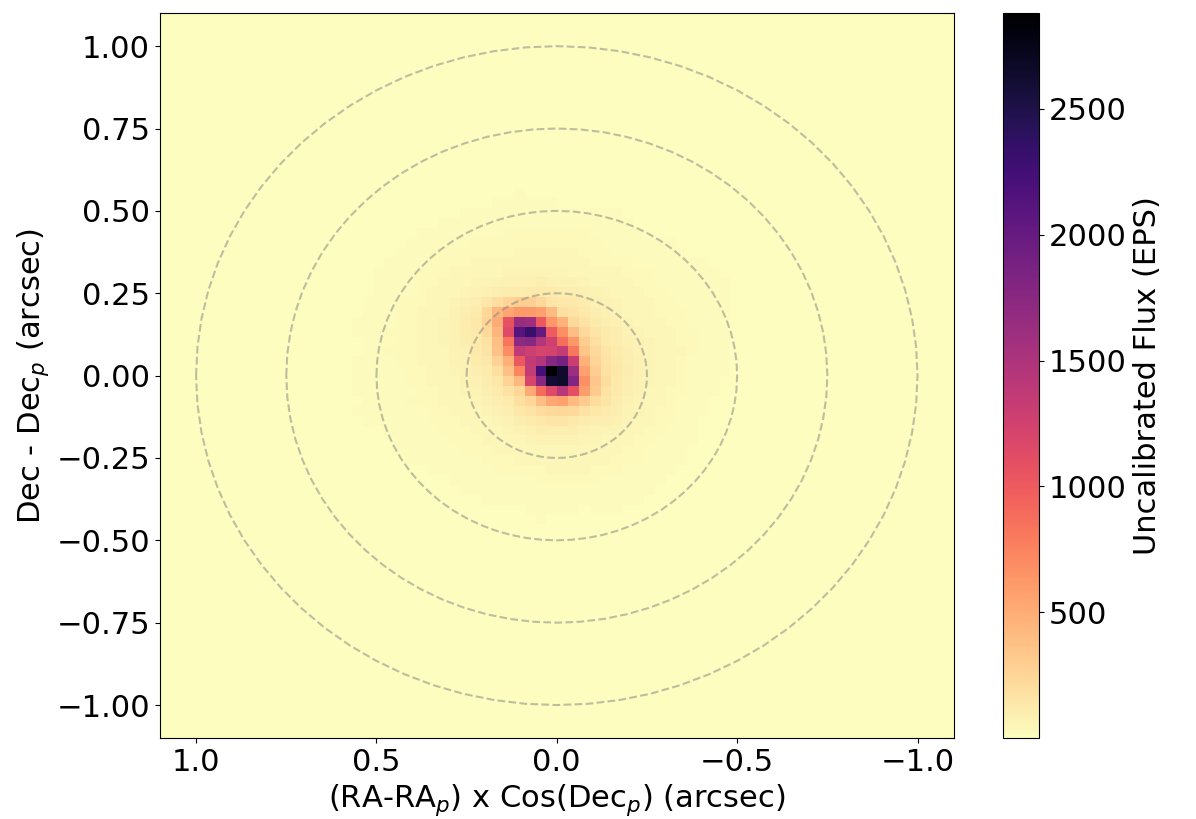}
\caption{\label{fig:ep-sa-378810450446502400} SourceID 378810450446502400: Scan-angle signatures and image reconstructed by SEAPipe.}
\end{figure}

The next example, taken from the outputs of query
{\tiny \begin{verbatim}
-- ADQL query on DR3, aiming at <400mas 2D:
SELECT TOP 100 * FROM gaiadr3.gaia_source WHERE
in_andromeda_survey='TRUE'
AND phot_g_mean_mag>=8 AND phot_g_mean_mag<12 
AND ipd_frac_multi_peak<60 AND ipd_frac_multi_peak>30 
AND pmra IS NULL AND matched_transits>30 
AND visibility_periods_used>18 
AND astrometric_matched_transits > 30 ORDER BY random_index,
\end{verbatim}
}
\noindent also corresponds to a two-parameter AGIS solution, but is now bright enough to have two-dimensional windows. Again, the \rExfG and \rIpdG  correlations are both very high, the IPD GoF harmonic amplitude is also quite high, and the IPD multi-peak fraction is slightly higher than in the previous case, with a secondary detected (and masked) peak in basically one-third of the transits. As shown in Fig.~\ref{fig:ep-sa-378810450446502400}, the BP and RP magnitudes are both nearly constant again, but the $G$ magnitude varies strongly (over 0.7 magnitudes) and is strongly correlated with the scan angle. It is worth noting that the fainter transits typically correspond to those with a secondary detected and masked peak, which means that the other peaks (in which no secondary source was detected) include unmodelled and unmitigated flux contamination. When we examined IDU pre-DR4 data, this DR3 source corresponded to a close pair with a separation of just 143~mas, a position angle of 30\degr, similar magnitudes, and a significant proper motion (with a very similar direction for both sources).

\begin{figure}[h]
\centering
  \includegraphics[width=0.45\textwidth]{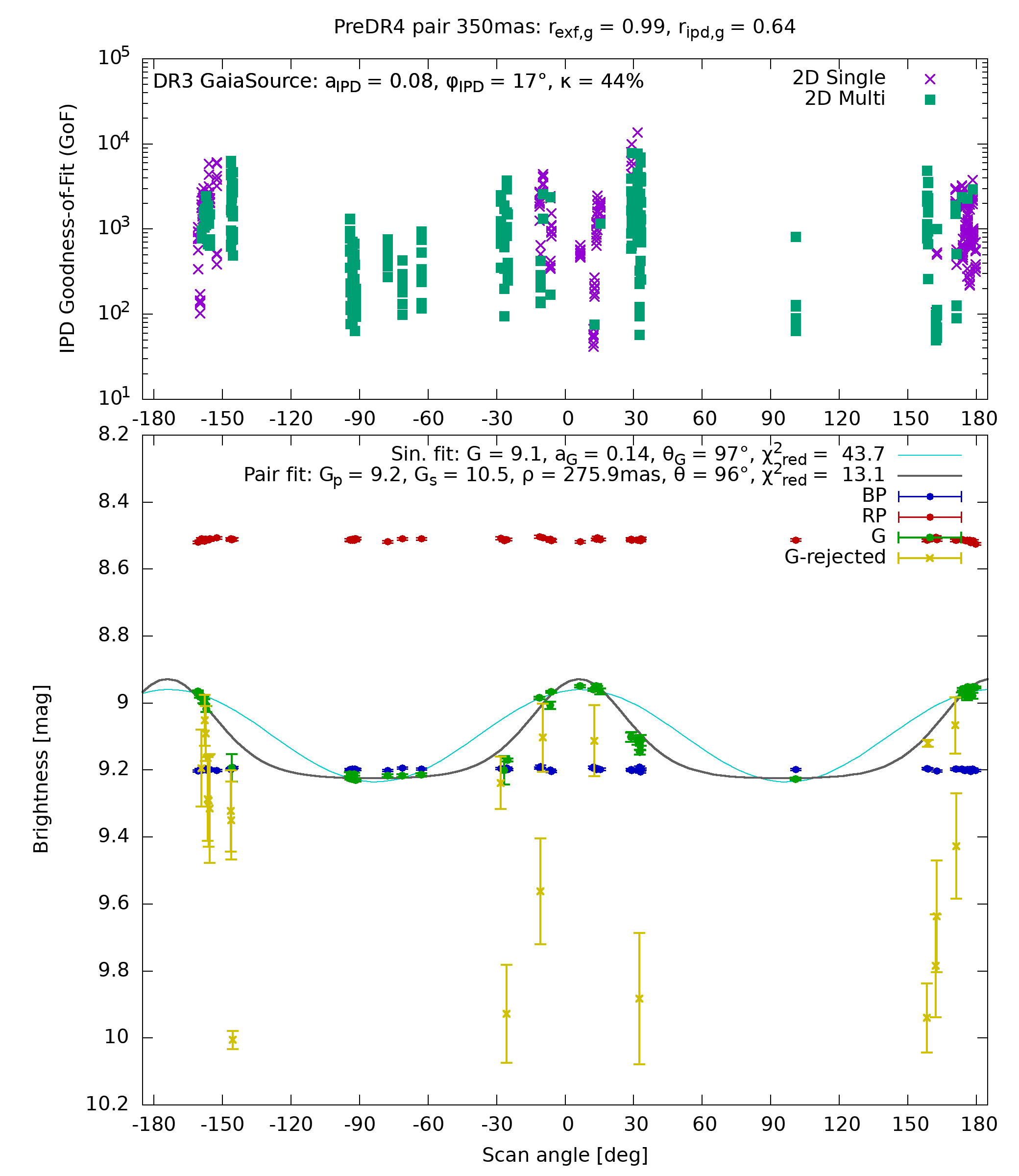}
  \includegraphics[width=0.45\textwidth]{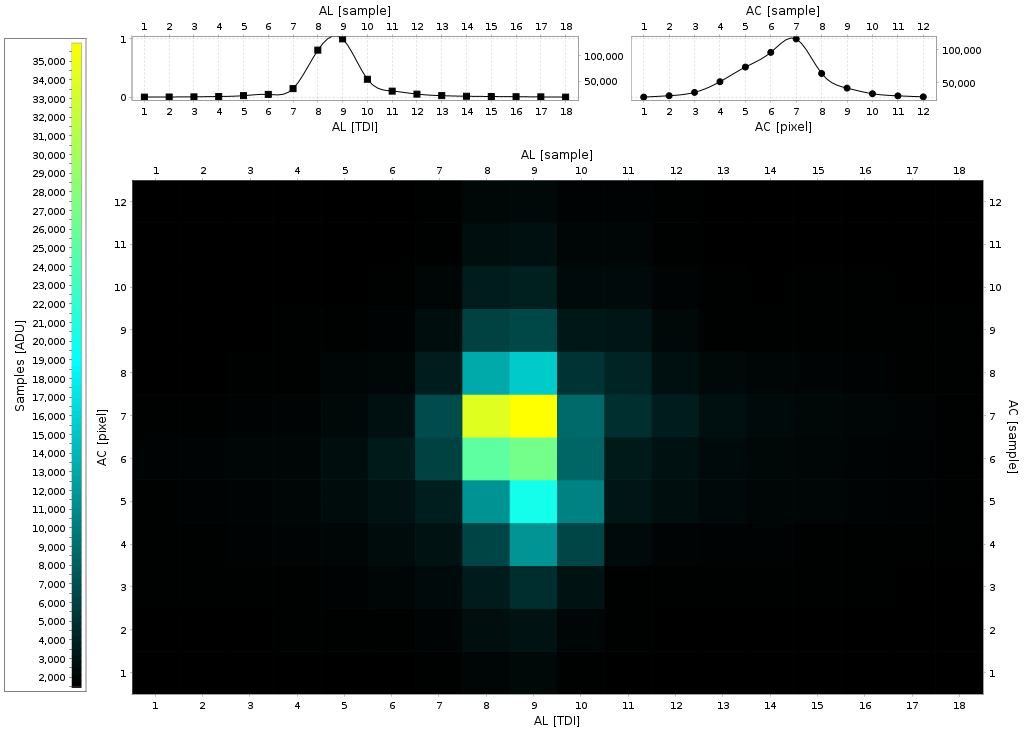}
  \includegraphics[width=0.45\textwidth]{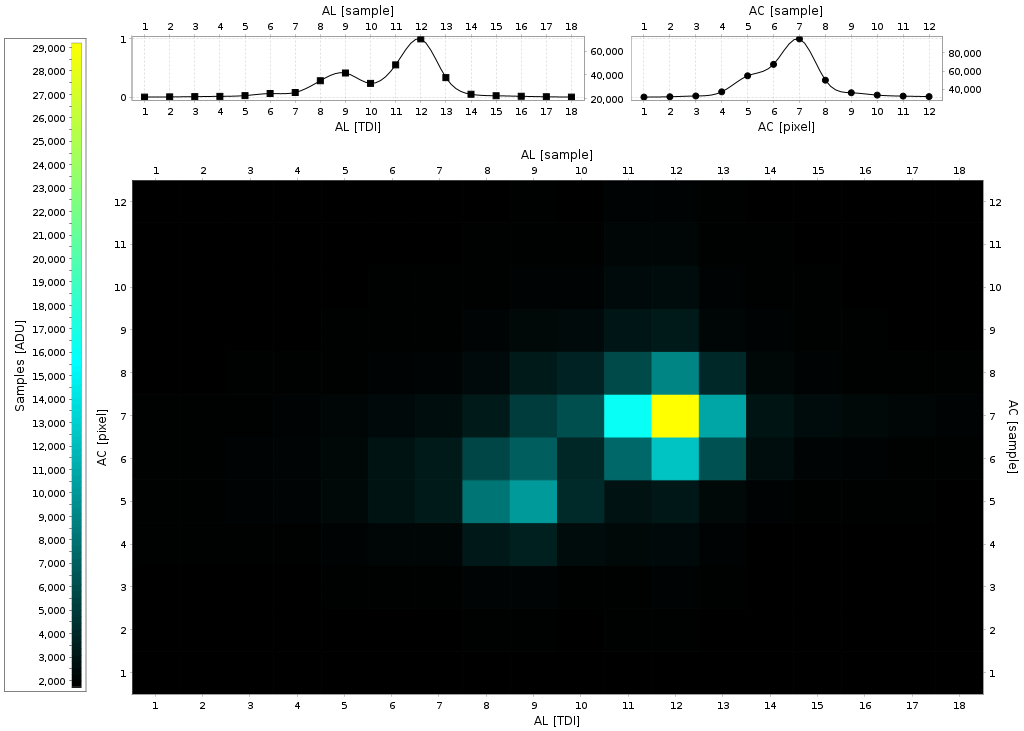}
\caption{\label{fig:ep-sa-385804856230839552} SourceID 385804856230839552: Scan-angle signatures (top panels) and raw windows for some of its transits (bottom panels).}
\end{figure}

Taking the same query, we can also find Fig.~\ref{fig:ep-sa-385804856230839552}, with a strong and correlated \gmag variation and very high \rExfG correlation. However, in this case, the \rIpdG correlation value is lower, as is the IPD GoF harmonic amplitude, although nearly half the transits have multiple peaks. In pre-DR4 IDU, it corresponds to a close pair separated by about 350 mas, two-dimensional windows, significant proper motion (with similar vectors), and a magnitude difference of about one between them. For illustrative purposes, we also show the raw windows
for some of its transits in the bottom panels of Fig.~\ref{fig:ep-sa-385804856230839552}. The two peaks can appear clearly separated or blended, depending on the orientation. \\

\begin{figure}[h]
\centering
  \includegraphics[width=0.45\textwidth]{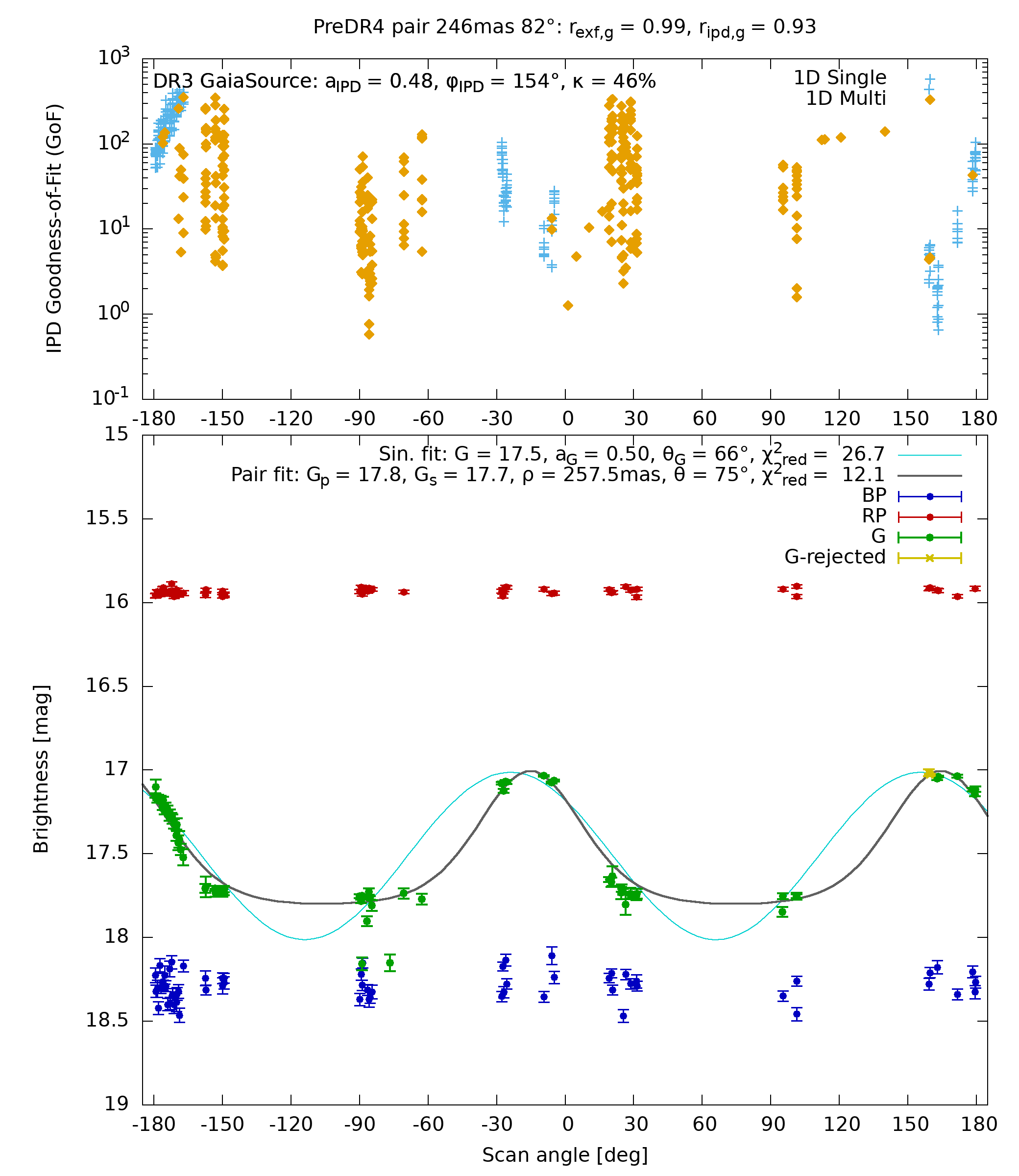}
  \includegraphics[width=0.45\textwidth]{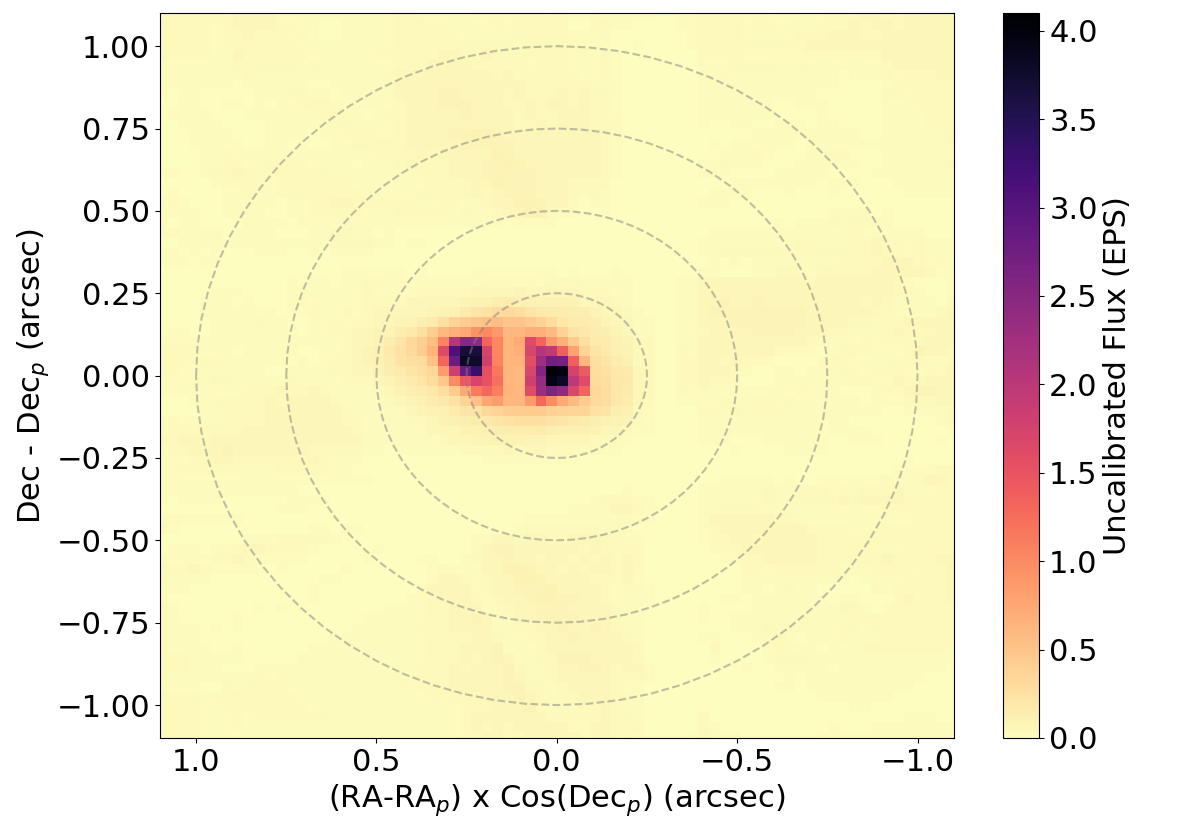}
  \includegraphics[width=0.45\textwidth]{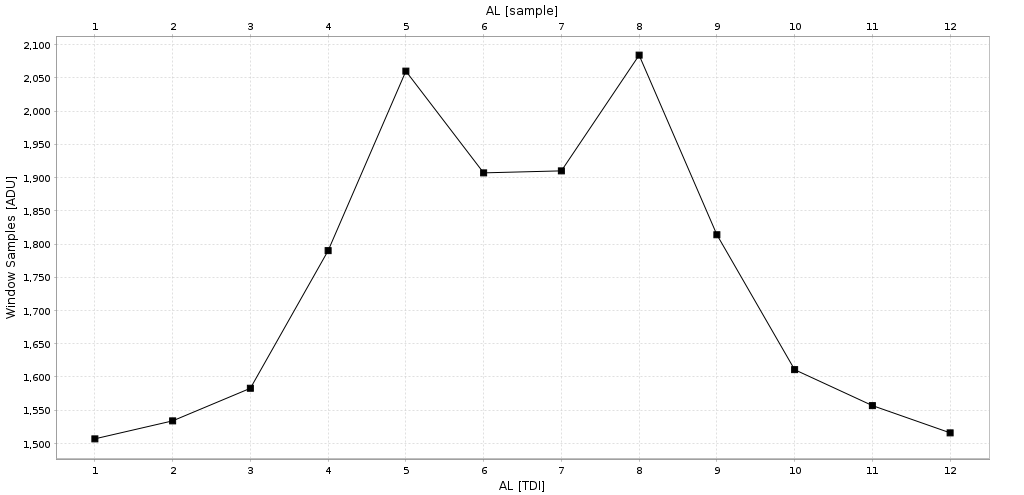}
  \includegraphics[width=0.45\textwidth]{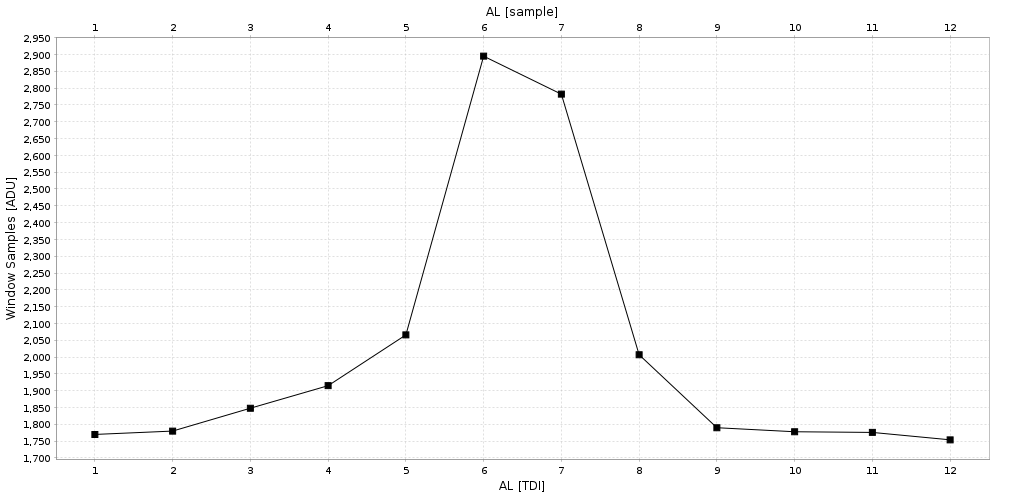}
\caption{\label{fig:ep-sa-382159975182923264} SourceID 382159975182923264: Scan-angle signatures, image reconstructed by SEAPipe, and two of the CCD observations taken at different scan angles.}
\end{figure}

Additional examples can be found from query
{\tiny \begin{verbatim}
-- ADQL query on DR3, aiming at 200-400mas 1D:
SELECT TOP 1000 * FROM gaiadr3.gaia_source WHERE
in_andromeda_survey='TRUE'
AND phot_g_mean_mag>=14 AND phot_g_mean_mag<19 
AND ipd_frac_multi_peak>30 AND pmra IS NULL 
AND matched_transits>30 AND visibility_periods_used>18 
AND astrometric_matched_transits > 30 ORDER BY random_index,
\end{verbatim}
}
\noindent which also yields a two-parameter AGIS solution, with one-dimensional windows as in the first example, but now with a higher fraction of IPD multiple peaks (nearly half the transits). The \rExfG and \rIpdG correlations are very high again, as is the IPD GoF harmonic amplitude. Here, Fig.~\ref{fig:ep-sa-382159975182923264} again shows the correlation with the scan angle for $G$, and also a number of variations in BP (although without a very clear correlation with the scan angle). Again, brighter transits correspond to those without detected secondary peaks. In IDU pre-DR4 data, this DR3 source corresponds to a close pair separated by 246~mas and feasible five-parameter astrometric solution. Because of the larger separation between the components, Eq.~\ref{eq:simu} provides a better fit and is able to reach quite good estimates for the separation and position angle.
For illustrative purposes, we show two of its CCD observations in the bottom panels of Fig.~\ref{fig:ep-sa-382159975182923264}, similarly as in Fig.~\ref{fig:ep-sa-385804856230839552}, but now in one-dimensional windows. It clearly reveals the two peaks in one of the scans, and just one blended peak in another scan.

\begin{figure}[h]
\centering
  \includegraphics[width=0.45\textwidth]{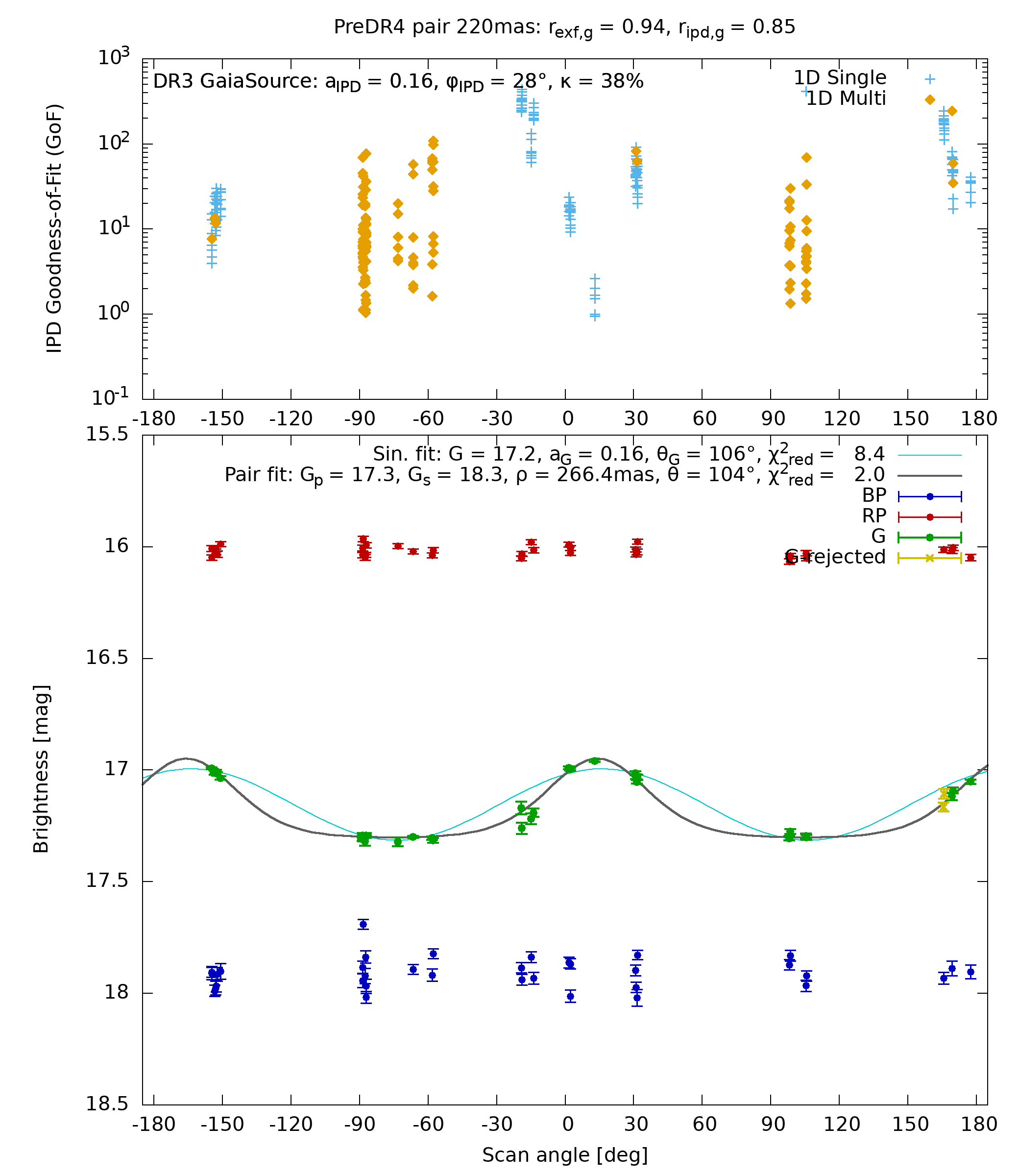}
\caption{\label{fig:ep-sa-376045247423005184} SourceID 376045247423005184: Scan-angle signatures.}
\end{figure}

\begin{figure}[h]
\centering
  \includegraphics[width=0.45\textwidth]{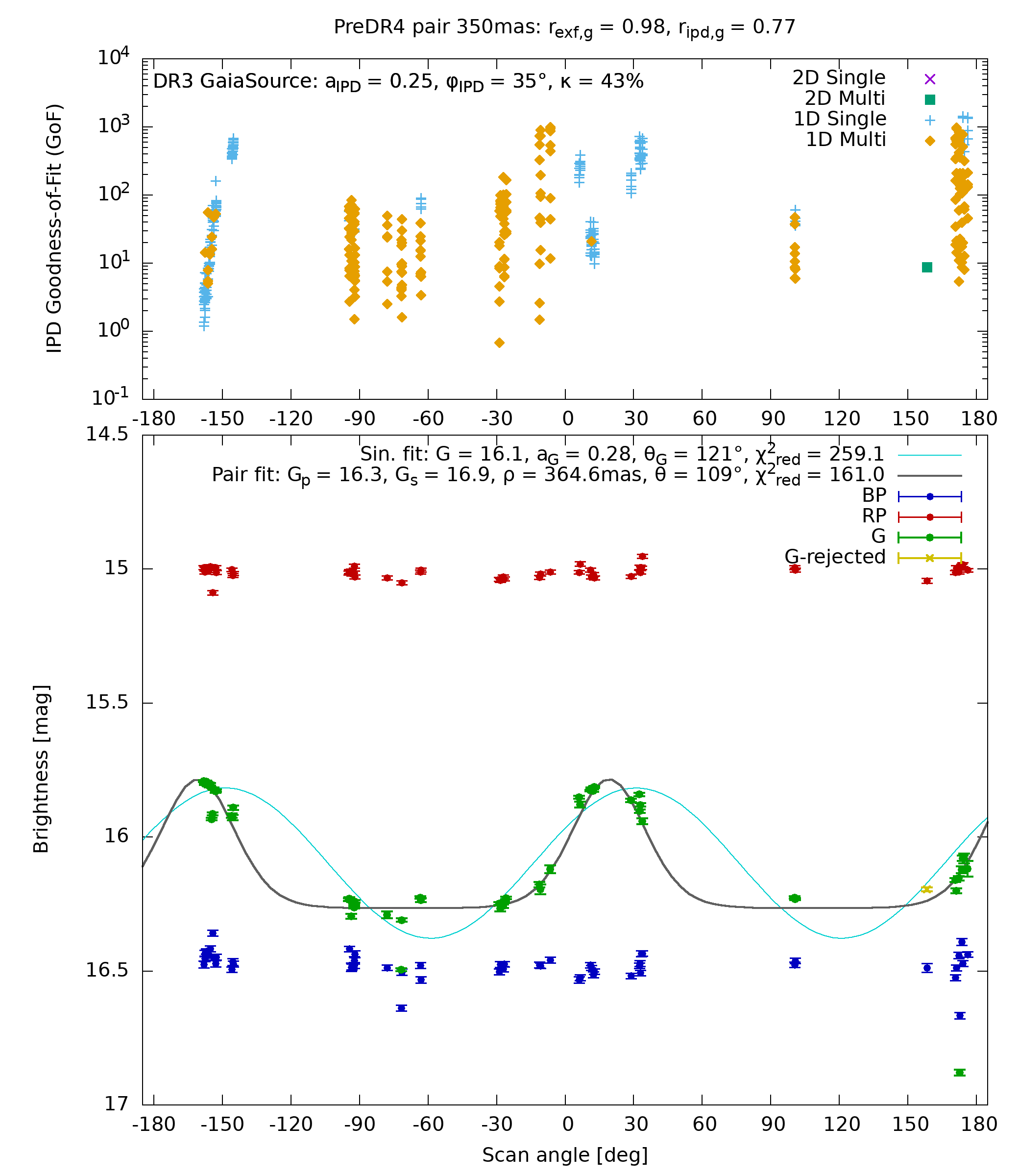}
\caption{\label{fig:ep-sa-385844060692409344} SourceID 385844060692409344: Scan-angle signatures.}
\end{figure}

Further examples from the same query are shown in Fig.~\ref{fig:ep-sa-376045247423005184} and Fig.~\ref{fig:ep-sa-385844060692409344}. \\

\begin{figure}[h]
\centering
  \includegraphics[width=0.45\textwidth]{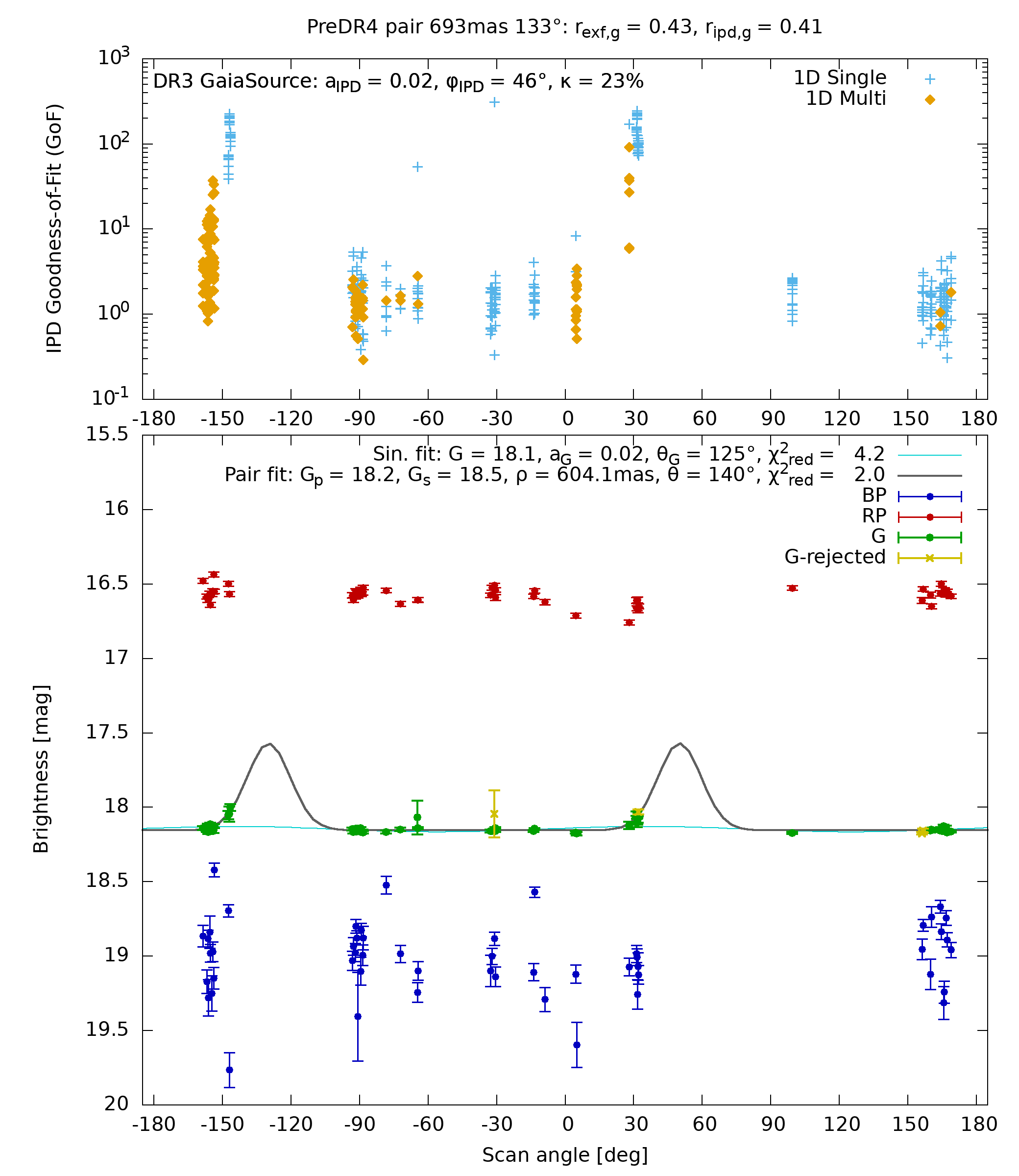}
  \includegraphics[width=0.45\textwidth]{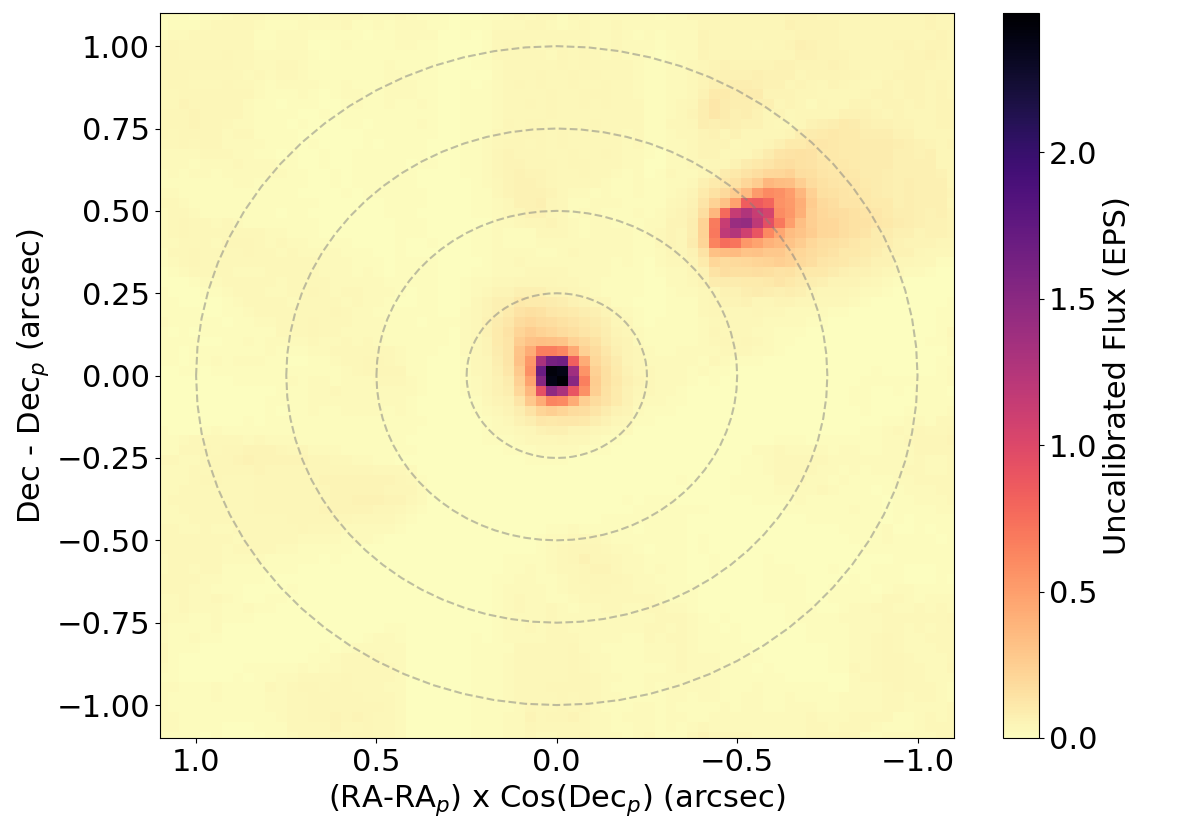}
\caption{\label{fig:ep-sa-385367010081173504} SourceID 385367010081173504: Scan-angle signatures and image reconstructed by SEAPipe.}
\end{figure}

\begin{figure}[h]
\centering
  \includegraphics[width=0.45\textwidth]{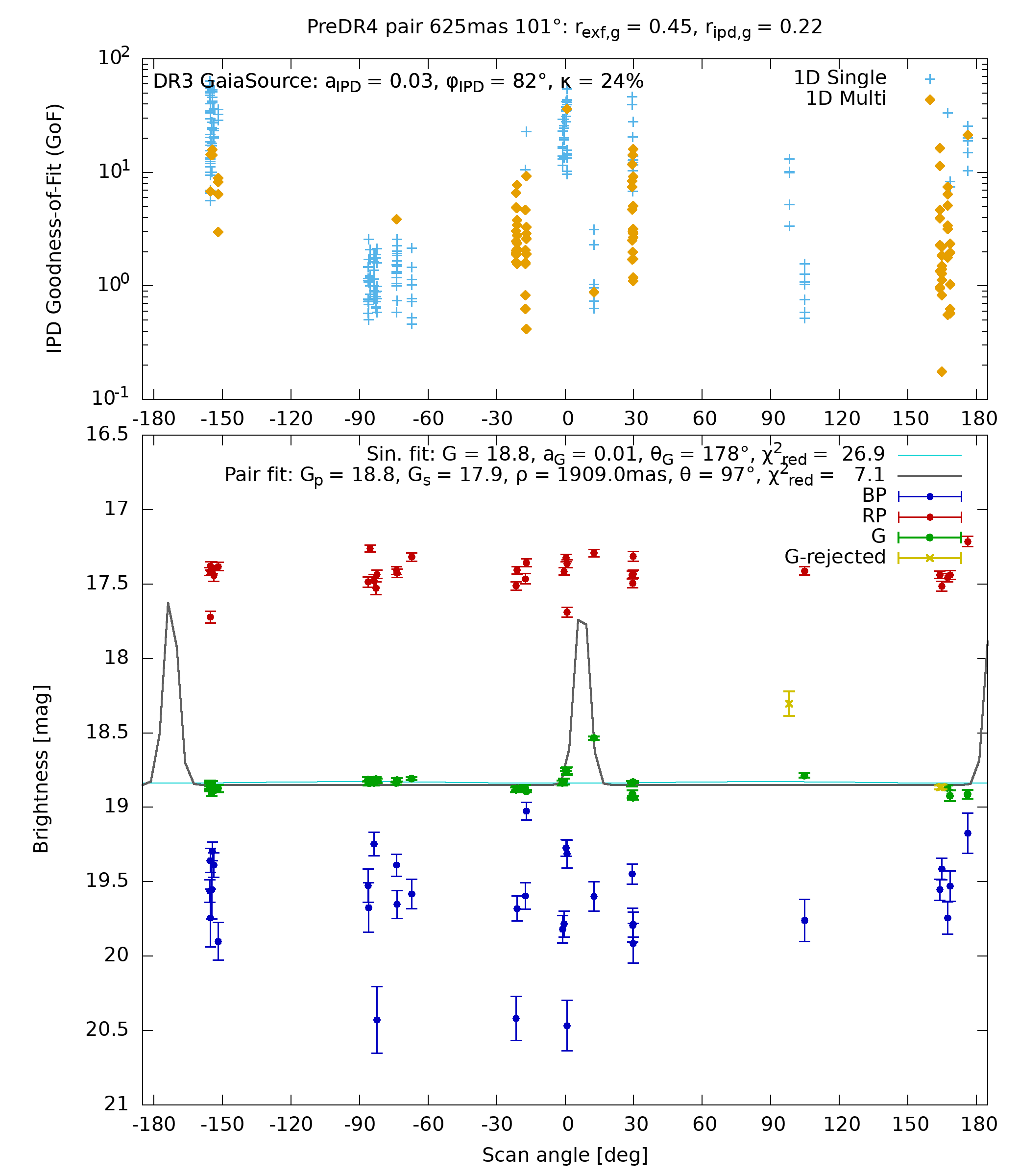}
\caption{\label{fig:ep-sa-387325652606842368} SourceID 387325652606842368: Scan-angle signatures.}
\end{figure}

Finally, with the query
{\tiny \begin{verbatim}
-- ADQL query on DR3, aiming at >400mas 1D:
SELECT TOP 1000 * FROM gaiadr3.gaia_source WHERE
in_andromeda_survey='TRUE'
AND phot_g_mean_mag>=14 AND phot_g_mean_mag<19 
AND ipd_frac_multi_peak<30 AND ipd_frac_multi_peak>20 
AND pmra IS NOT NULL AND matched_transits>30 
AND visibility_periods_used>18 
AND astrometric_matched_transits > 30 ORDER BY random_index,
\end{verbatim}
}
\noindent we find Fig.~\ref{fig:ep-sa-385367010081173504}, with rather modest $G$ variations, but it is stronger in BP and RP. The internal DR3 IDU data reveal a moderate scan-angle correlation in the epoch GoF values, although the DR3 harmonic amplitude is very small. The Spearman \rExfG and \rIpdG  correlations for $G$ are rather modest, but significant. Pre-DR4 IDU data reveal a 693~mas close pair here, which is also quite nicely revealed from Eq.~\ref{eq:simu} fitting results.
Yet another example from the same query is shown in Fig.~\ref{fig:ep-sa-387325652606842368}.

\subsection{Resolved close pairs in DR3}

\begin{figure}[h]
\centering
  \includegraphics[width=0.45\textwidth]{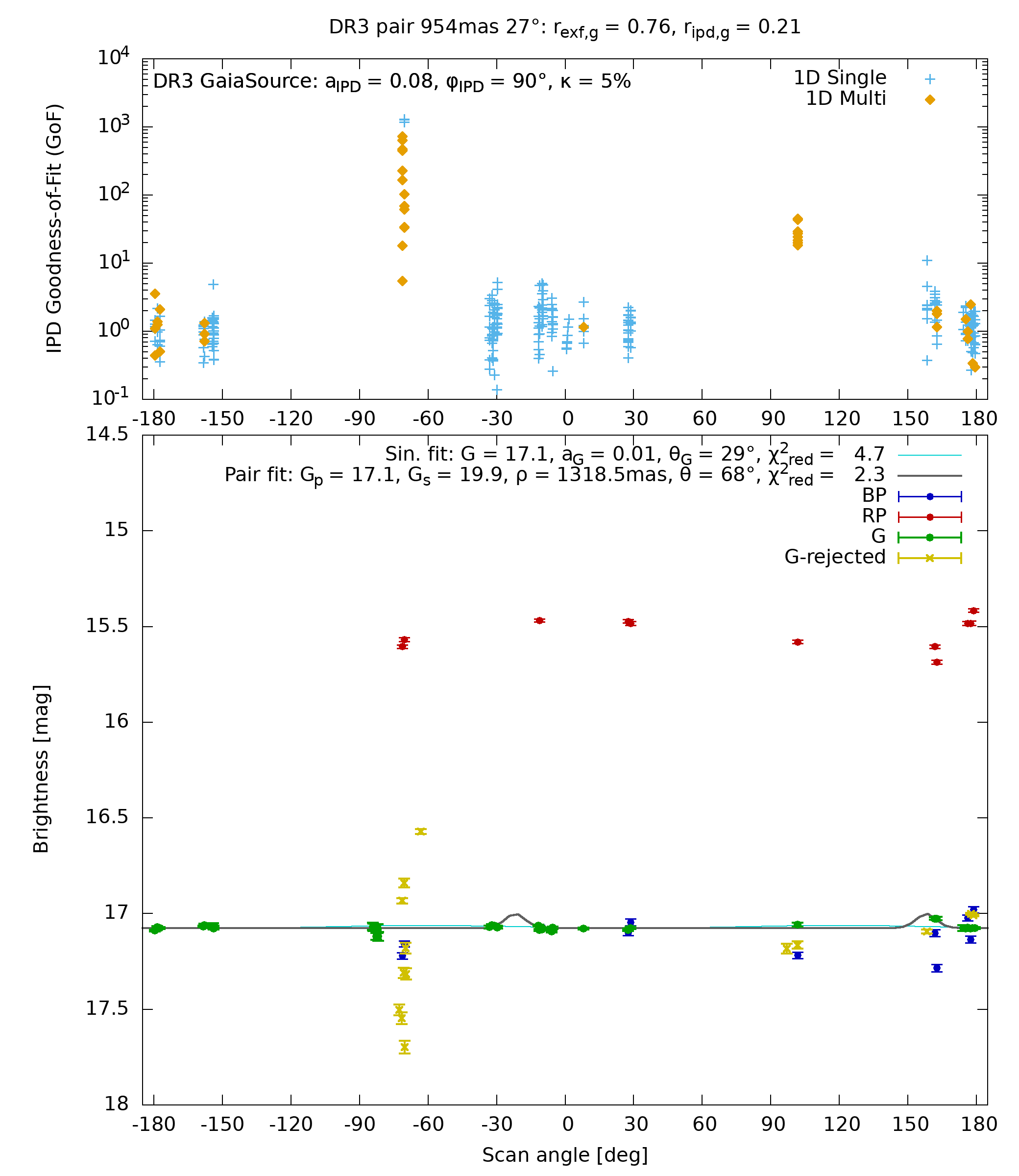}
  \includegraphics[width=0.45\textwidth]{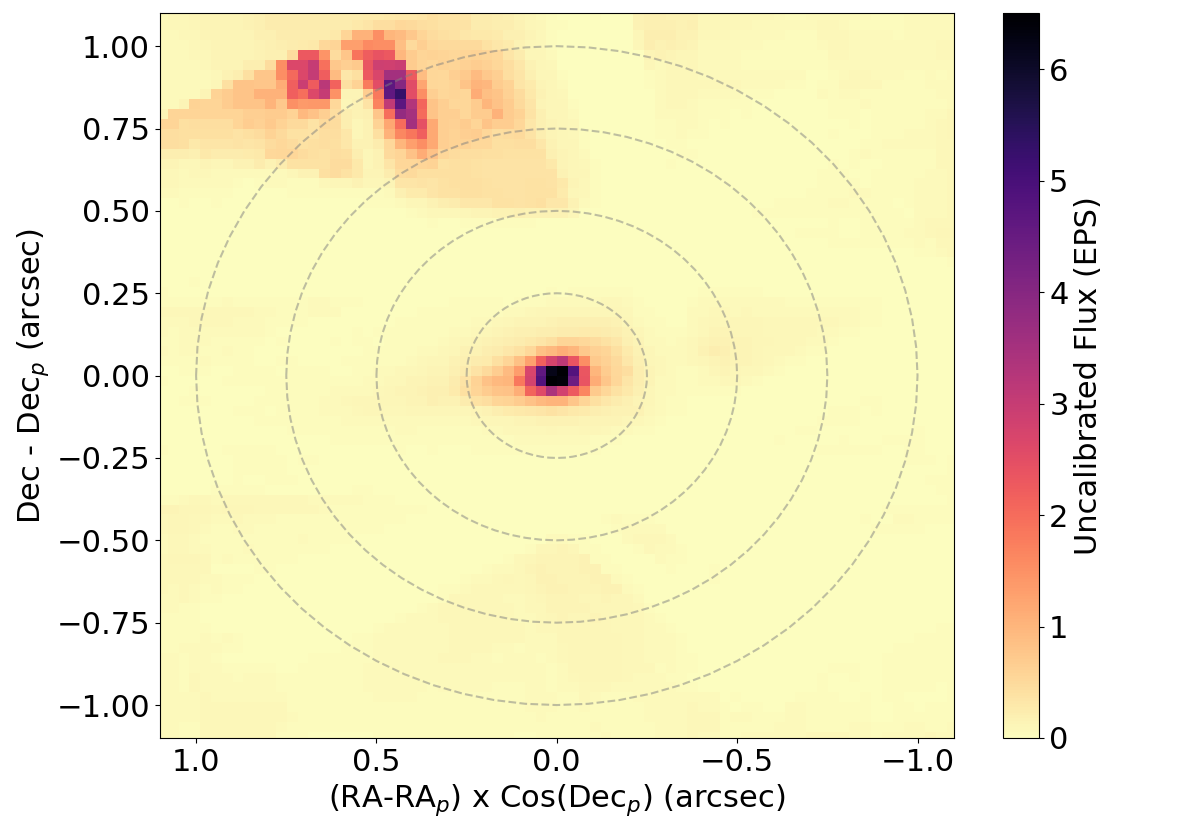}
\caption{\label{fig:ep-sa-379163256239241216} SourceID 379163256239241216: Scan-angle signatures and image reconstructed by SEAPipe.}
\end{figure}

The second set of examples corresponds to pairs that have been resolved as separate sources in DR3, obtained through the query
{\tiny \begin{verbatim}
-- ADQL query on DR3, aiming at sources with many transits 
-- and "intermediate" magnitudes: (leading to ~132k, 
-- after which a XM can be done)
SELECT source_id,ra,ra_error,dec,dec_error,parallax,
parallax_error,pmra,pmra_error,pmdec,pmdec_error,
astrometric_params_solved,matched_transits,
ipd_gof_harmonic_amplitude,ipd_gof_harmonic_phase,
ipd_frac_multi_peak,ipd_frac_odd_win,
phot_g_n_obs,phot_g_mean_mag,phot_variable_flag
FROM gaiadr3.gaia_source 
WHERE (in_andromeda_survey = 'true')
AND (matched_transits > 50)
AND (phot_g_mean_mag BETWEEN 12.0 AND 19.0),
\end{verbatim}
}
\noindent plus a simplistic nearest-neighbour cross match on the result.
The first example was already shown in Sect.~\ref{ssec:afInstr} in Fig.~\ref{fig:closePair371mas}.
The second example corresponds to a DR3 pair with a separation of 954~mas and nearly the same magnitudes, one of which is shown in Fig.~\ref{fig:ep-sa-379163256239241216}. In this case, we still see some scan-angle effect in $G$ (and some clear outliers that were rejected by variability processing), but BP and RP seem to be slightly more affected, which is otherwise expected due to the larger separation. Here there are very few transits with multiple detected peaks, and the IPD GoF harmonics have a small amplitude. Only the \rExfG correlation is able to indicate these effects clearly. The fit to the pair model is poor, probably due to the rejected transits.

\begin{figure}[h]
\centering
  \includegraphics[width=0.45\textwidth]{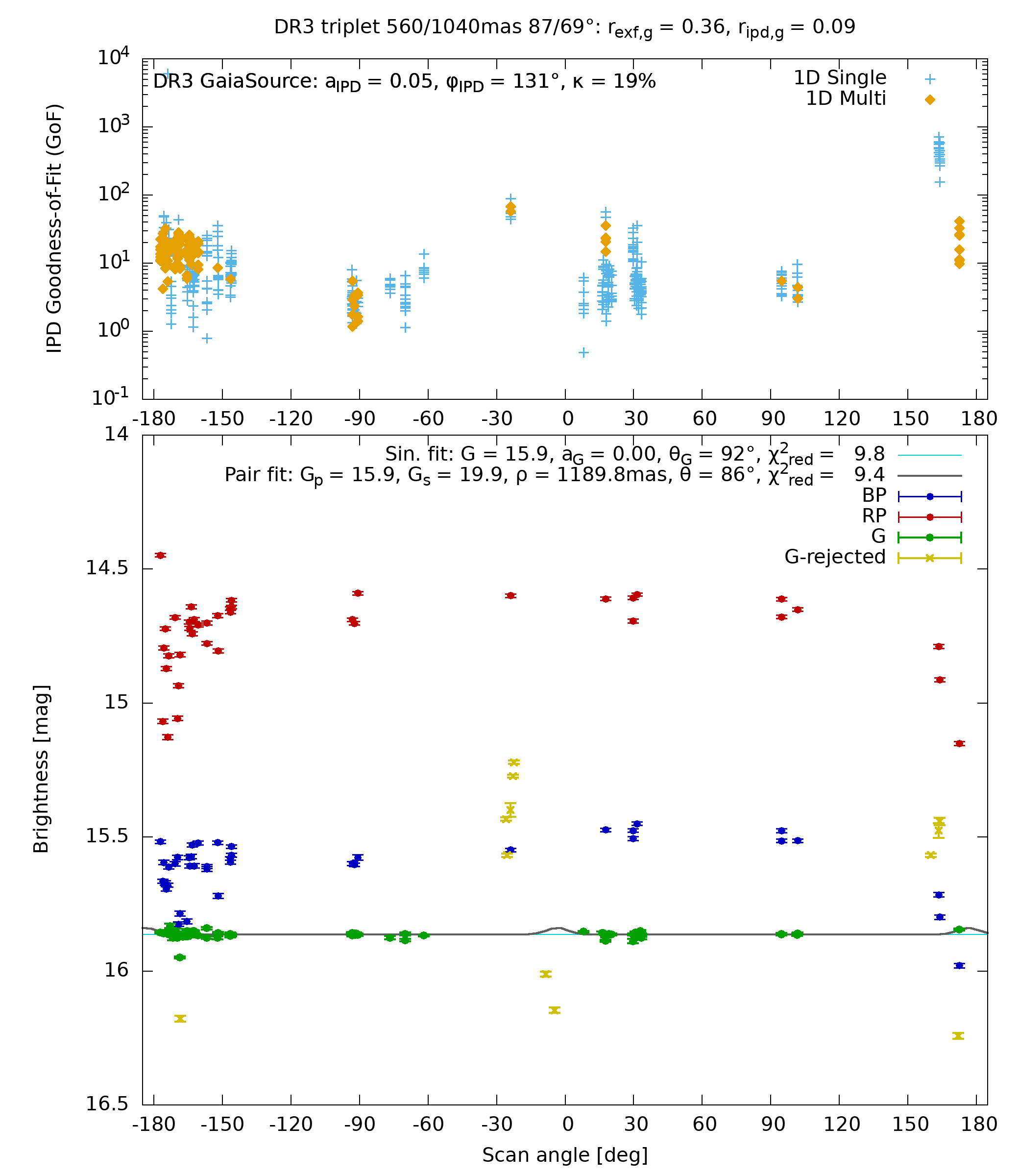}
  \includegraphics[width=0.45\textwidth]{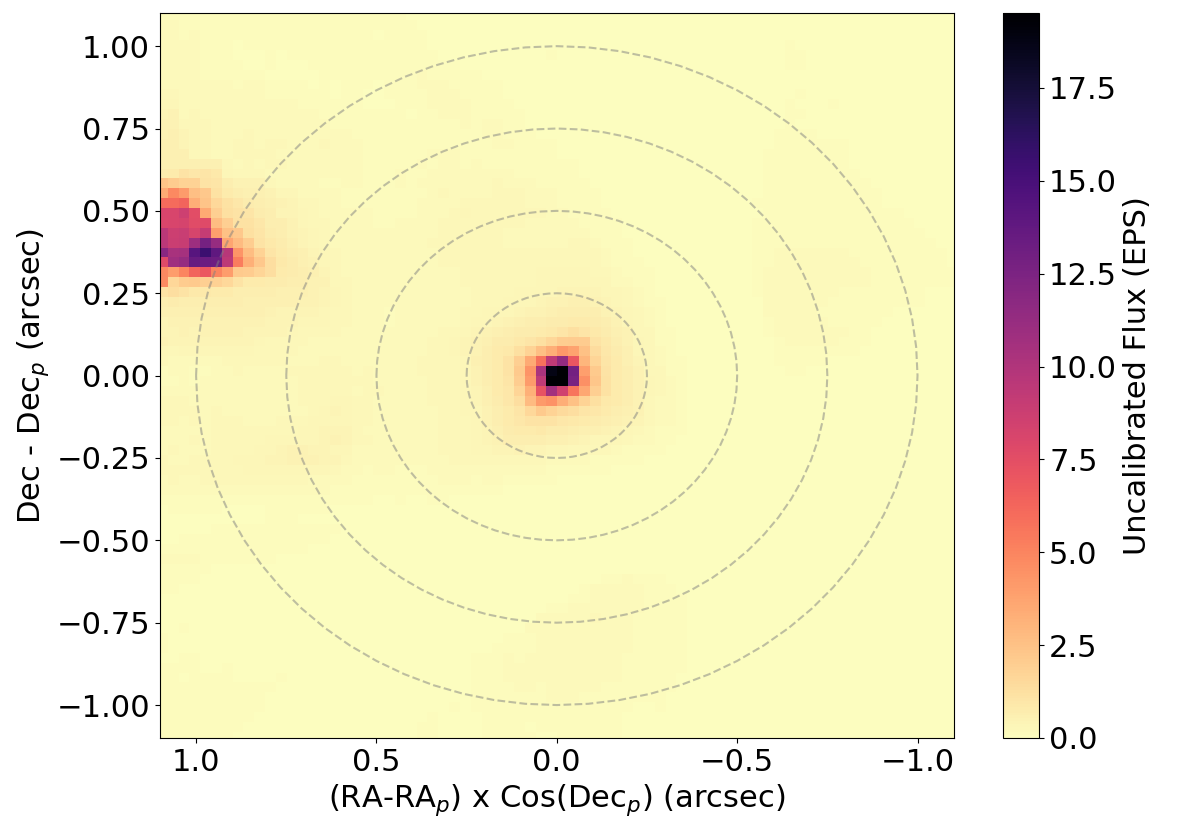}
\caption{\label{fig:ep-sa-388877407113709056} SourceID 388877407113709056: Scan-angle signatures and image reconstructed by SEAPipe.}
\end{figure}

Finally, Fig.~\ref{fig:ep-sa-388877407113709056} shows another example obtained with the same query. In this case, we have three sources, with separations of 560 and 1040~mas from the source shown in the figure. As otherwise expected, the effects on both $G$ and BP/RP are significant. Despite the rejected $G$ measurements, the pair model appears to determine sensible values.

\subsection{Wing of a bright star in DR3}

\begin{figure}[h]
\centering
  \includegraphics[width=0.45\textwidth]{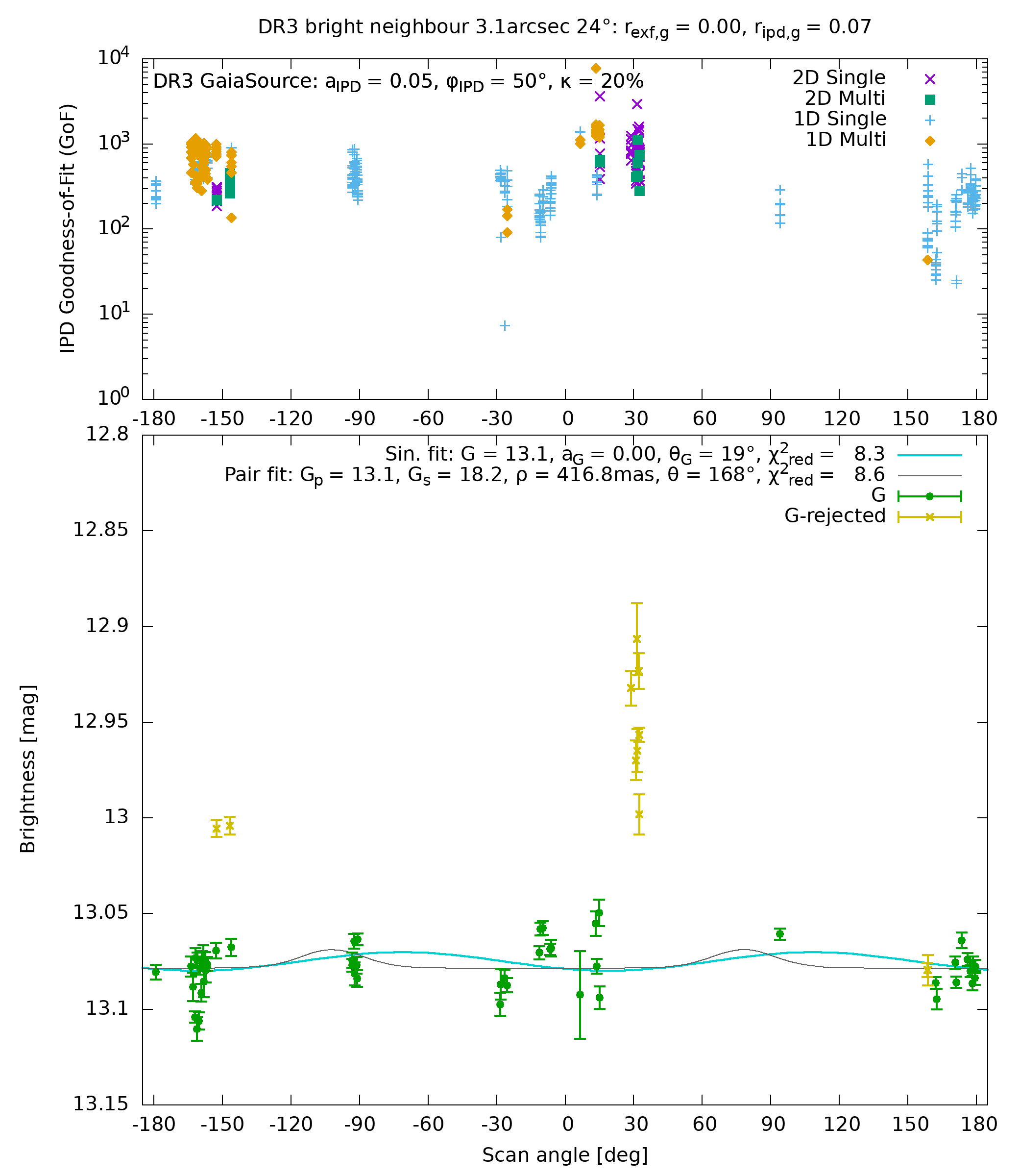}
  \includegraphics[width=0.45\textwidth]{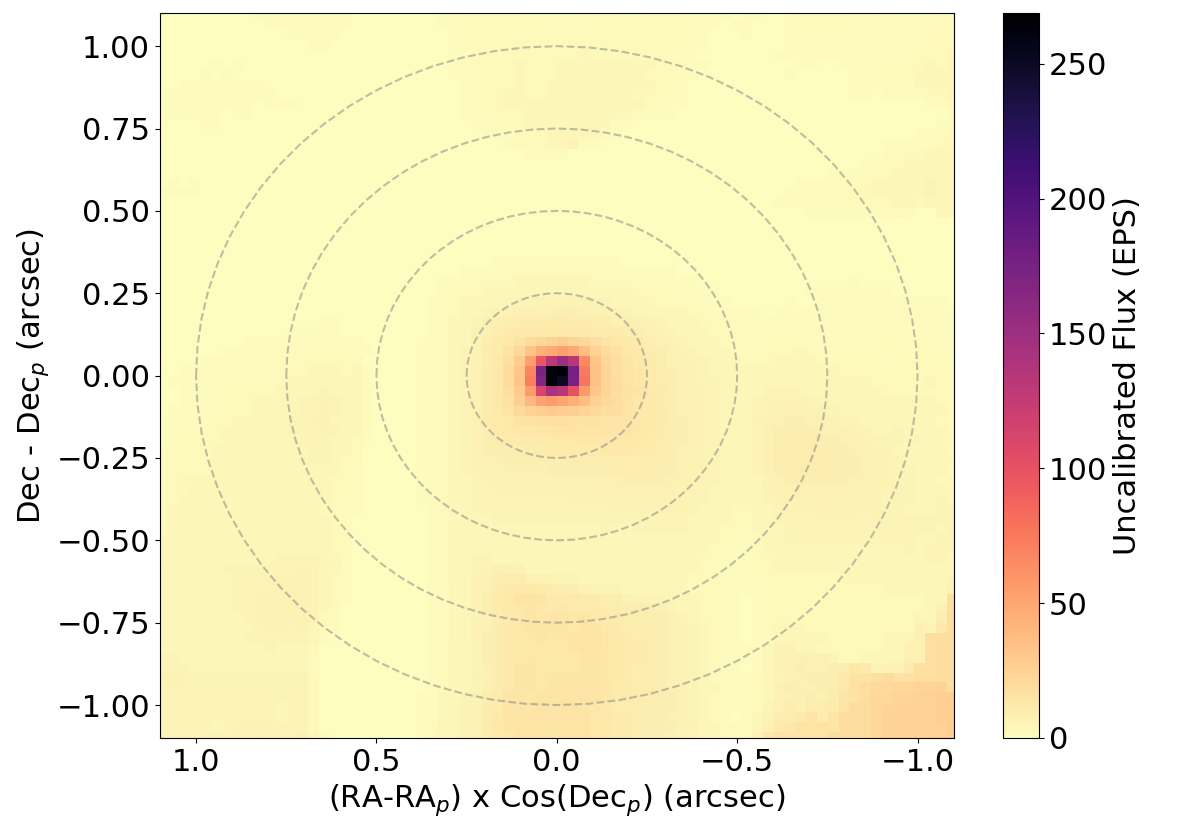}
\caption{Example of a source affected by the PSF wing of a nearby bright source.\label{fig:ep-sa-385771836519005184} SourceID 385771836519005184: Scan-angle signatures and image reconstructed by SEAPipe.}
\end{figure}

With the query
{\tiny \begin{verbatim}
-- ADQL query on DR3, aiming at bright stars,
-- then XM'ed with previous query:
SELECT source_id,ra,ra_error,dec,dec_error,parallax,
parallax_error,pmra,pmra_error,pmdec,pmdec_error,
astrometric_params_solved,matched_transits,
ipd_gof_harmonic_amplitude,ipd_gof_harmonic_phase,
ipd_frac_multi_peak,ipd_frac_odd_win,phot_g_n_obs,
phot_g_mean_mag,phot_variable_flag 
FROM gaiadr3.gaia_source 
WHERE (in_andromeda_survey = 'true')
AND (matched_transits > 50) 
AND (phot_g_mean_mag < 8.0)
\end{verbatim}
}
\noindent plus a simplistic cross match, we also obtained an example of a DR3 source that lies close to a bright source (magnitude 7.3). With this, we wished to verify whether the PSF wings of the nearby bright source can also cause scan-angle artefacts. Fig.~\ref{fig:ep-sa-385771836519005184} indeed shows a remarkable peak in the flux at a specific angle, with a smaller peak at 180 degrees of it. They were both rejected by variability processing. BP and RP photometry could not be determined here, most probably due to the contamination of the bright neighbour. Interestingly, in this case, the  \rExfG and \rIpdG  correlations are both nearly zero, as is the IPD GoF harmonics. Only the fraction of transits with multiple peaks provides an indication: Transits with multiple peaks are indeed at about the same angles at which we see the strong variations in $G$ photometry.

\subsection{Blind search for resolved or unresolved close pairs in DR3}

\begin{figure}[h]
\centering
  \includegraphics[width=0.45\textwidth]{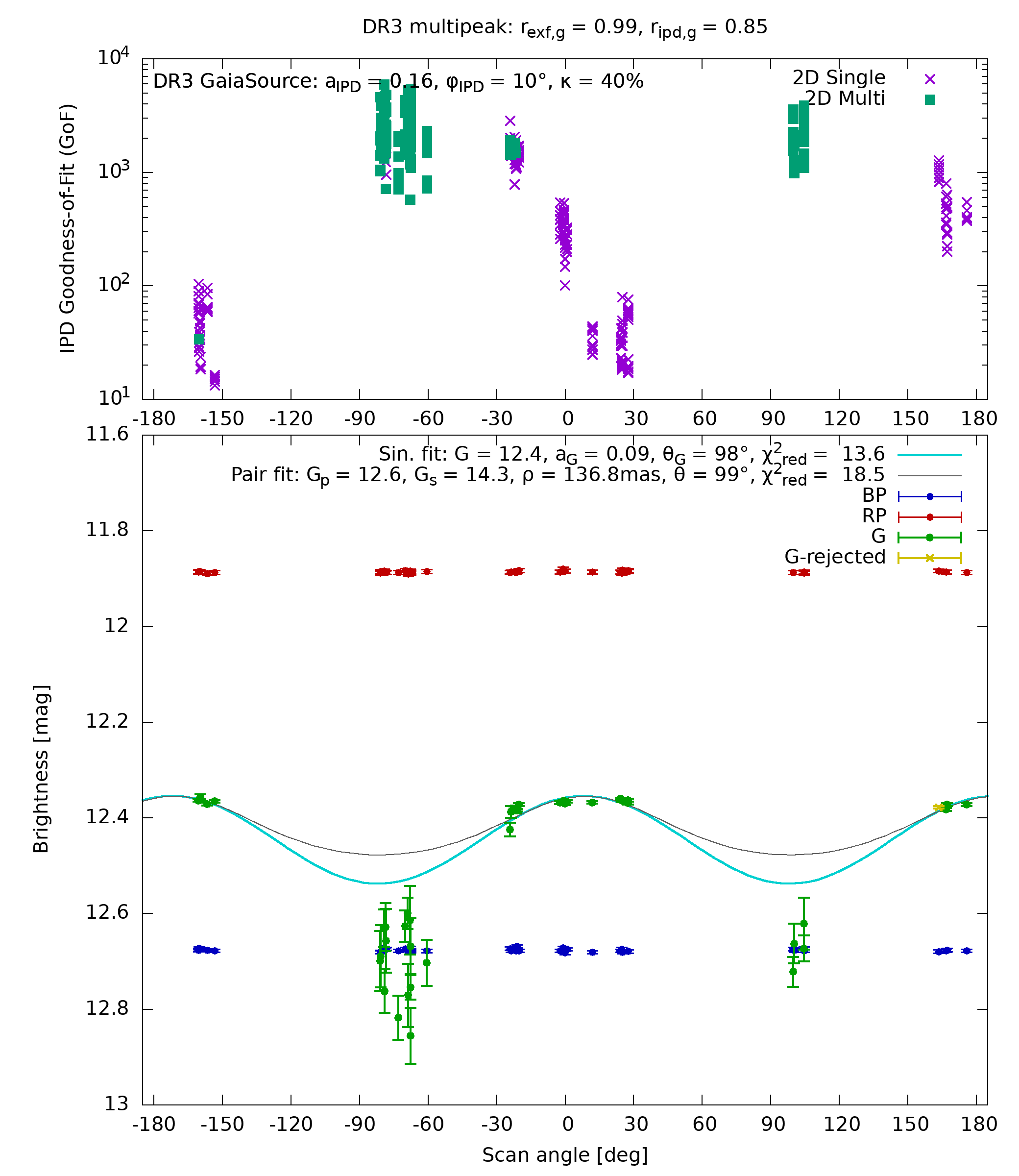}
\caption{\label{fig:ep-sa-367388551858425344} SourceID 367388551858425344: Scan-angle signatures.}
\end{figure}

\begin{figure}[h]
  \includegraphics[width=0.45\textwidth]{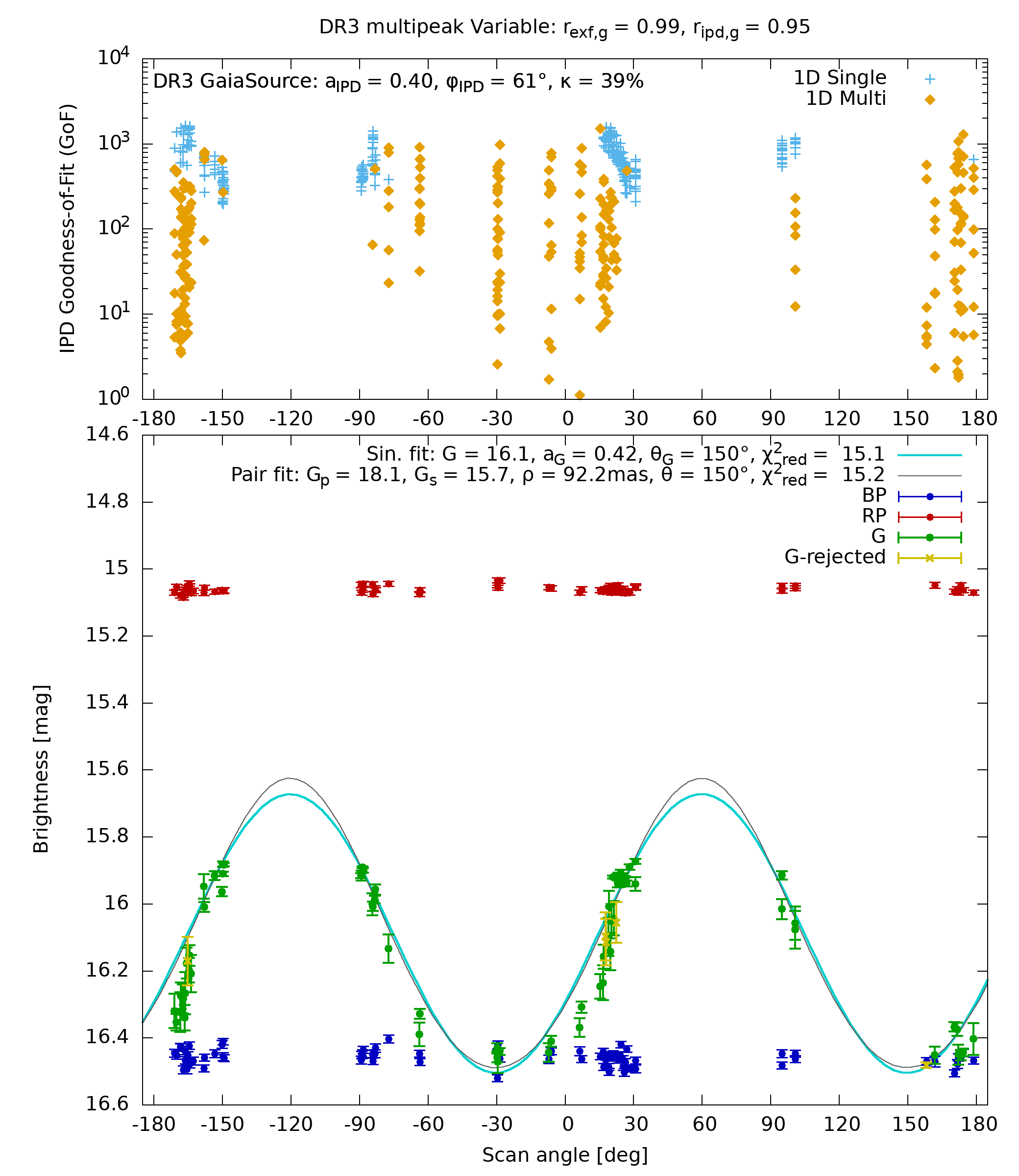}
\caption{\label{fig:ep-sa-383556286230747520} SourceID 383556286230747520: Scan-angle signatures.}
\end{figure}

\begin{figure}[h]
  \includegraphics[width=0.45\textwidth]{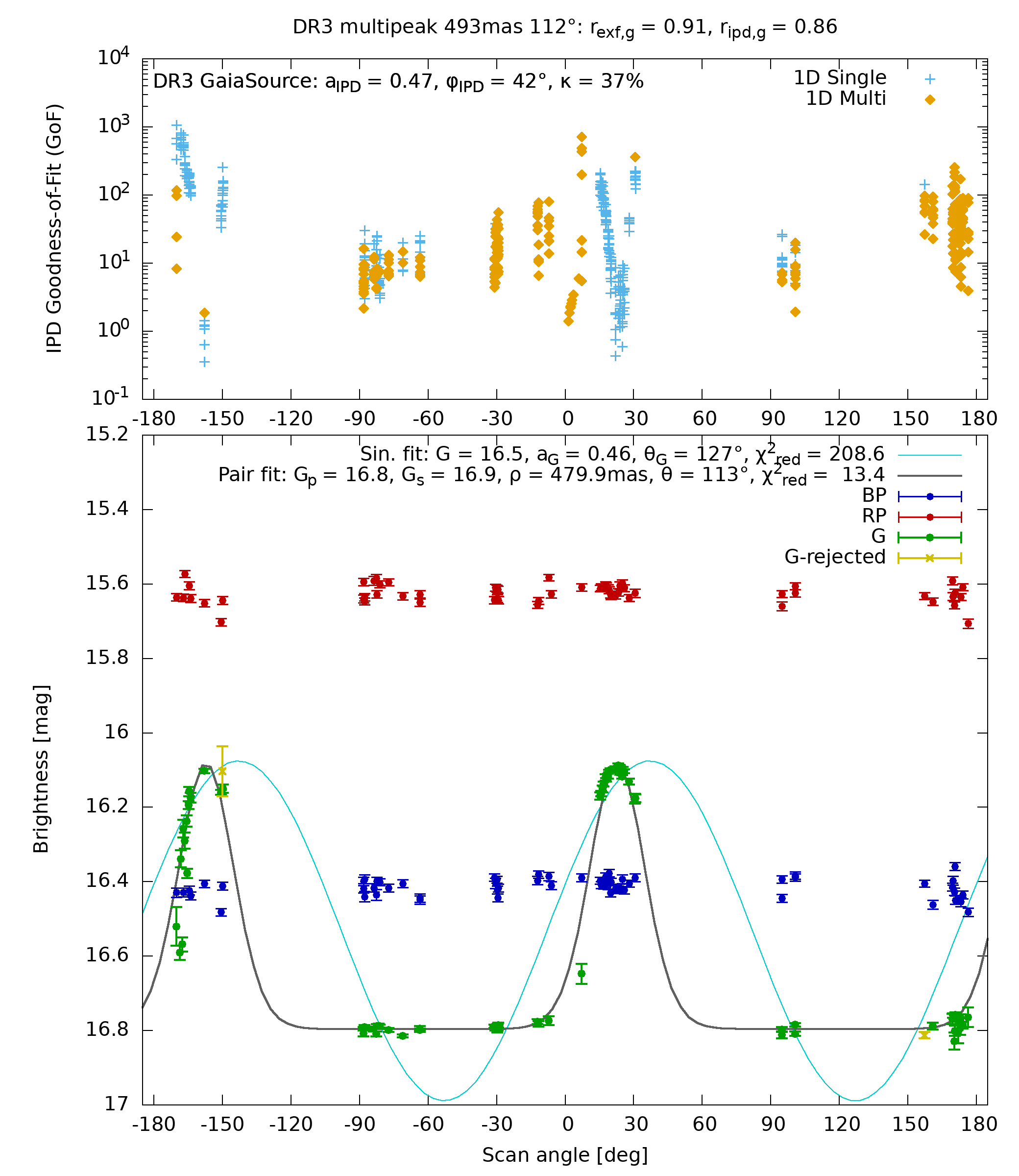}
  \includegraphics[width=0.45\textwidth]{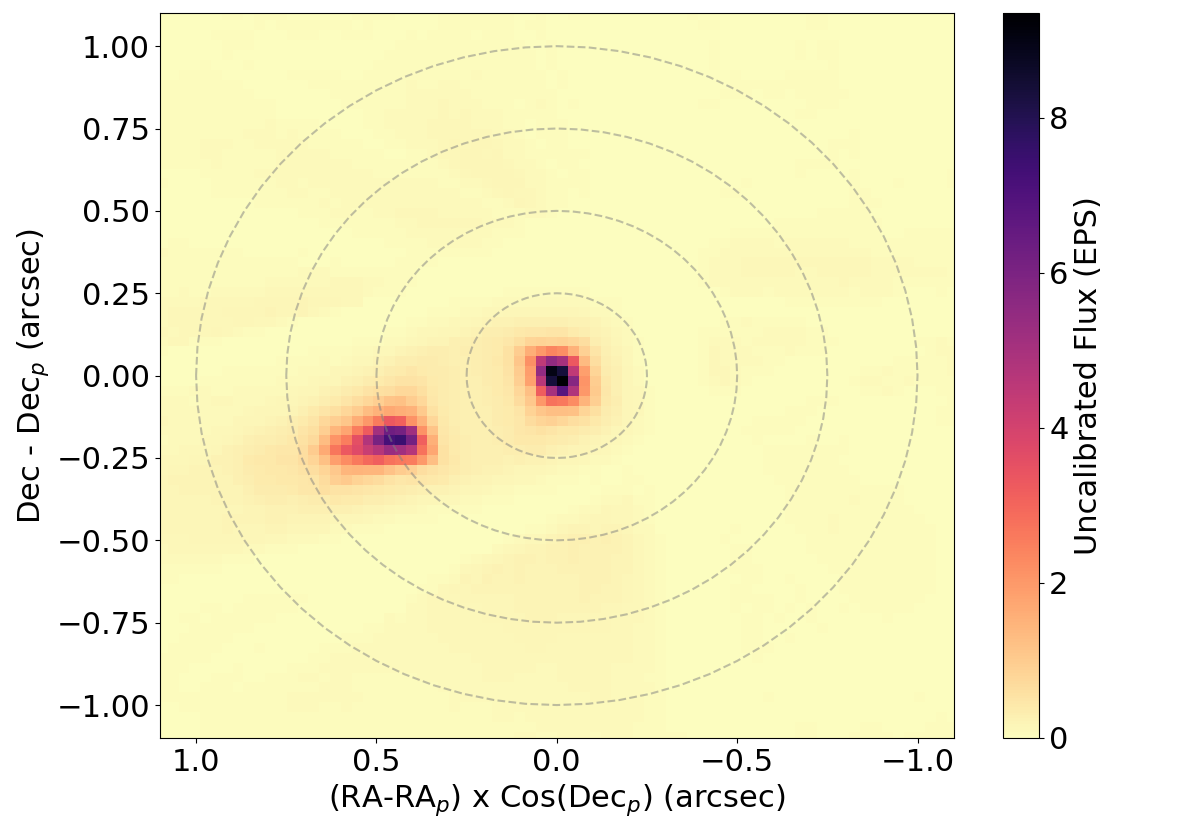}
\caption{\label{fig:ep-sa-380538569192874112} SourceID 380538569192874112: Scan-angle signatures and image reconstructed by SEAPipe.}
\end{figure}

With an adequate query on DR3 GaiaSource, making use of the several multiplicity indicators (\ipdFracMultiPeak, \ipdGofHarmAmpl),
{\tiny \begin{verbatim}
-- ADQL query on DR3, aiming at good candidates 
-- to be an unresolved pair:
SELECT source_id,ra,ra_error,dec,dec_error,
matched_transits,ipd_gof_harmonic_amplitude,
ipd_gof_harmonic_phase,ipd_frac_multi_peak,
ipd_frac_odd_win,phot_g_n_obs,phot_g_mean_mag,
phot_variable_flag FROM gaiadr3.gaia_source 
WHERE (in_andromeda_survey = 'true')
AND (ipd_frac_multi_peak BETWEEN 30 AND 50)
AND (ipd_gof_harmonic_amplitude >= 0.4)
AND (astrometric_params_solved = 3)
AND (phot_g_mean_mag BETWEEN 12 AND 19.0)
AND (matched_transits >= 50) 
AND (ipd_frac_odd_win < 20)
\end{verbatim}
}
\noindent  we can obtain good candidates of either resolved or unresolved close pairs. From the resulting list of sources, we can obtain the public epoch photometry, which can lead to figures such as Fig.~\ref{fig:ep-sa-367388551858425344}, Fig.~\ref{fig:ep-sa-383556286230747520}, or Fig.~\ref{fig:ep-sa-380538569192874112}. The second source is specially interesting because the Variability flag was set for this source. All sources show the usual effects: the brighter, scan-angle-dependent $G$ fluxes correspond to transits without a detected secondary peak. Thus, overall, it seems to demonstrate that the IDU IPD masking of secondary peaks behaved as expected. Regarding the IPD GoF harmonics, the third case shown here has one of the highest values, but in general, the Spearman  \rExfG  correlation seems to be a more reliable and useful indicator of contaminants.
The fit to the pair model achieved in Fig.~\ref{fig:ep-sa-380538569192874112} is excellent.

\subsection{Galaxies in DR3}

\begin{figure}[h]
  \includegraphics[width=0.45\textwidth]{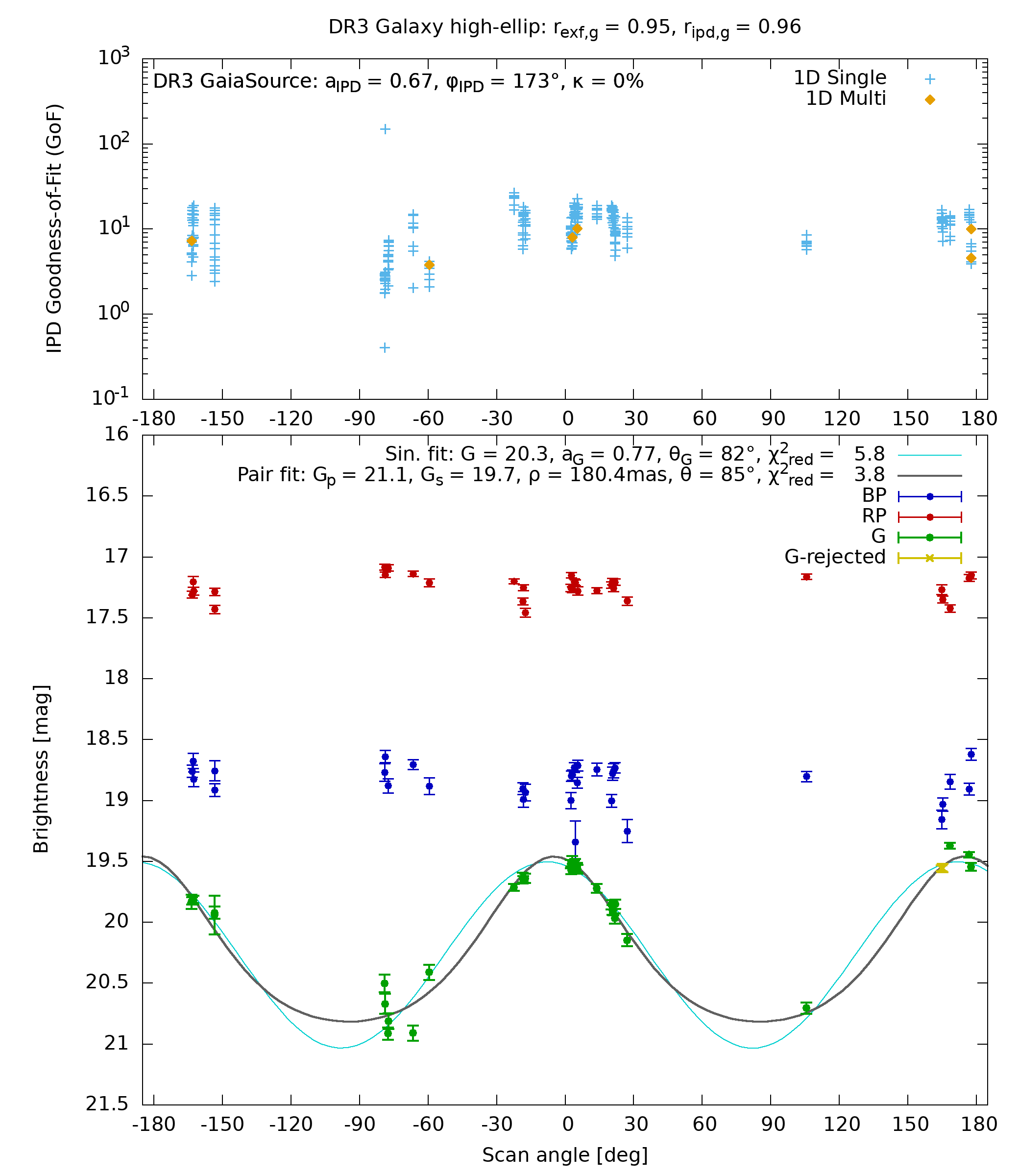}
  \includegraphics[width=0.45\textwidth]{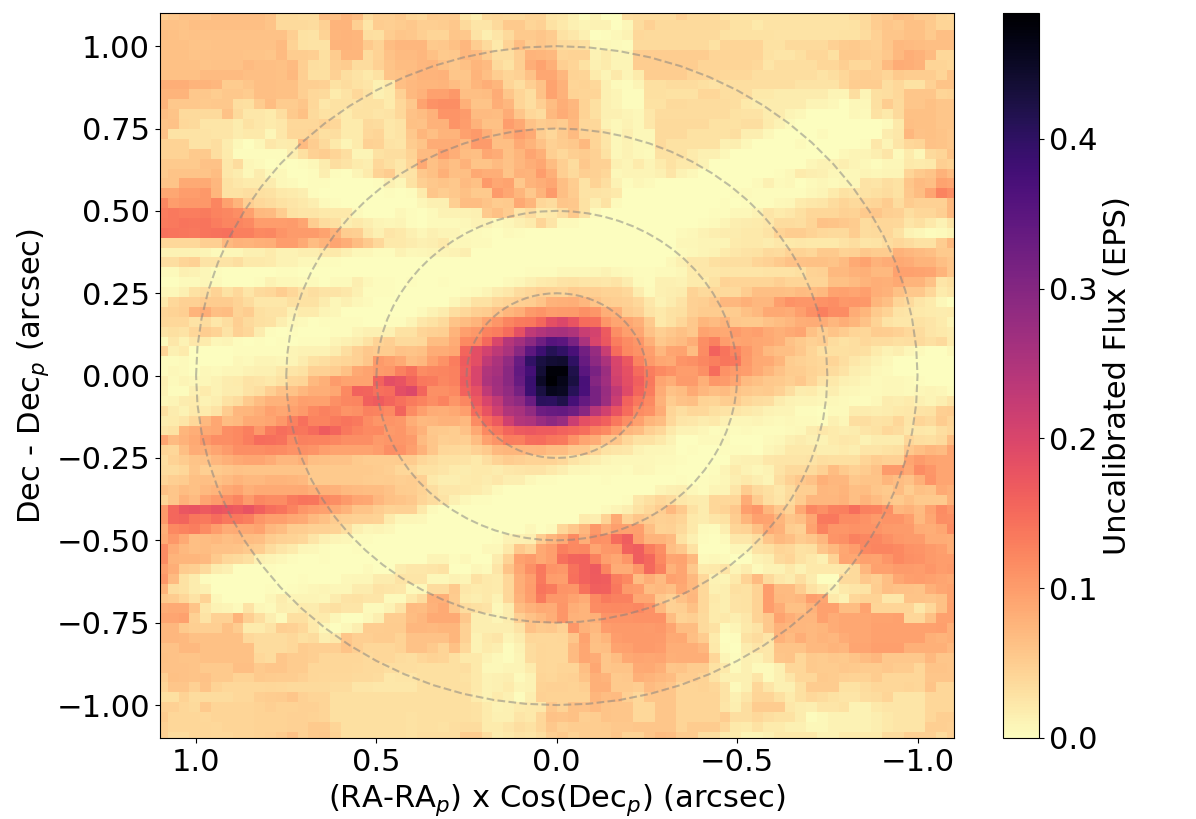}
\caption{\label{fig:ep-sa-364175332206026368} SourceID 364175332206026368: Scan-angle signatures and image reconstructed by SEAPipe.}
\end{figure}

\begin{figure}[h]
  \includegraphics[width=0.45\textwidth]{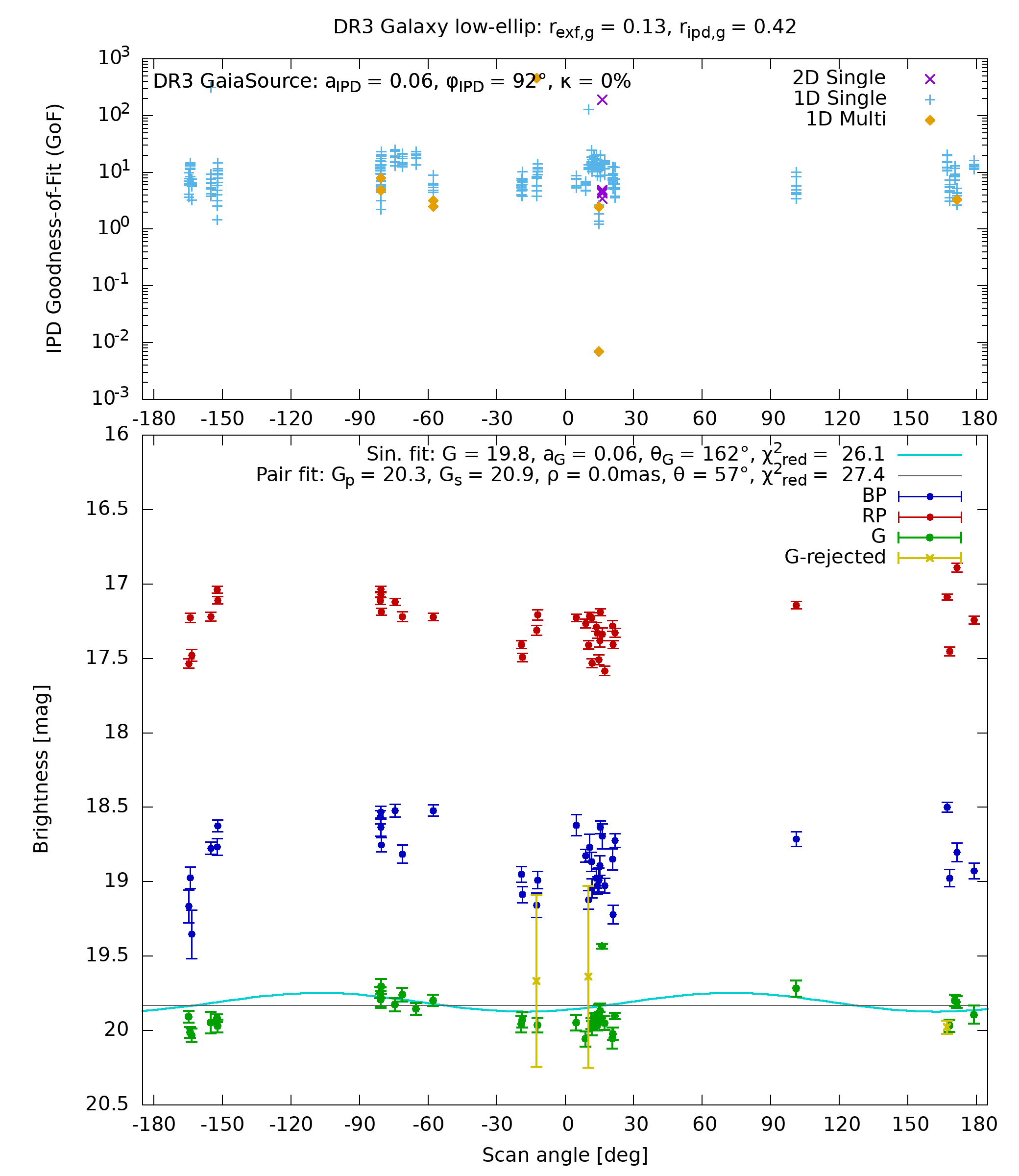}
  \includegraphics[width=0.45\textwidth]{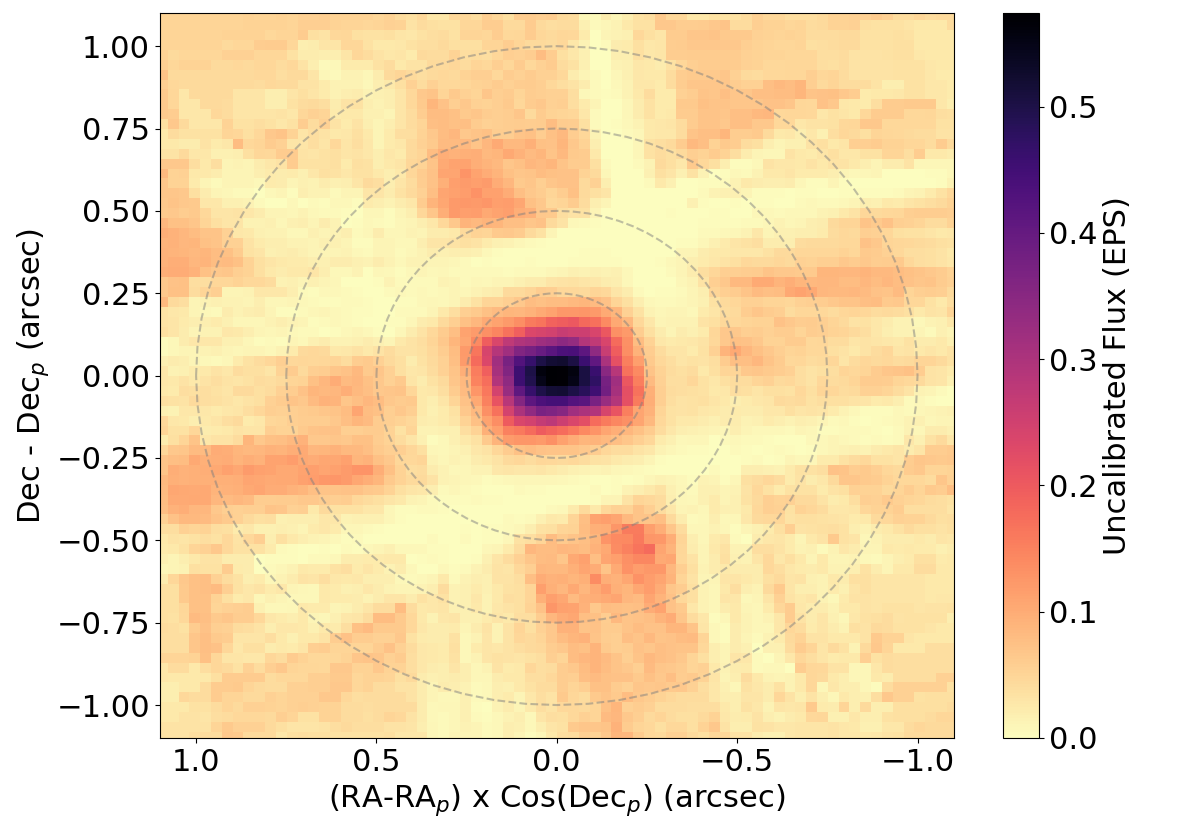}
\caption{\label{fig:ep-sa-373852271480563968} SourceID 373852271480563968: Scan-angle signatures and image reconstructed by SEAPipe.}
\end{figure}

Finally, with the query
{\tiny \begin{verbatim}
-- ADQL query on DR3, aiming at Galaxies with morph params:
SELECT source_id,ra,ra_error,dec,dec_error,
parallax,parallax_error,pmra,pmra_error,
pmdec,pmdec_error,astrometric_params_solved,
matched_transits,ipd_gof_harmonic_amplitude,
ipd_gof_harmonic_phase,ipd_frac_multi_peak,
ipd_frac_odd_win,phot_g_n_obs,phot_g_mean_mag,
phot_variable_flag FROM gaiadr3.gaia_source 
WHERE (in_andromeda_survey = 'true')
AND (matched_transits >= 30)
AND (phot_g_mean_mag <= 20.0)
AND (in_galaxy_candidates = 'true')
AND (astrometric_params_solved = 3)
AND (source_id IN (select source_id from 
gaiadr3.galaxy_candidates where 
gaiadr3.galaxy_candidates.radius_de_vaucouleurs IS NOT NULL)),
\end{verbatim}
}
\noindent  we also tested the case of galaxies for which we even have morphological values, such as the radius. The example shown in Fig.~\ref{fig:galaxymid} shows significant variations in all bands, but possibly a clearer scan-angle signature in \gmag (as otherwise expected). When the internal DR3 IPD epoch GoF values are inspected, the signature is clearer. This is revealed by the higher value in the \rIpdG Spearman correlation than in \rExfG. For completeness, Fig.~\ref{fig:ep-sa-364175332206026368} shows another example for a high-ellipticity galaxy, and Fig.~\ref{fig:ep-sa-373852271480563968} shows an example for a low-ellipticity galaxy.

\subsection{DR3 spectroscopic binaries}

\begin{figure}[h]
  \includegraphics[width=0.45\textwidth]{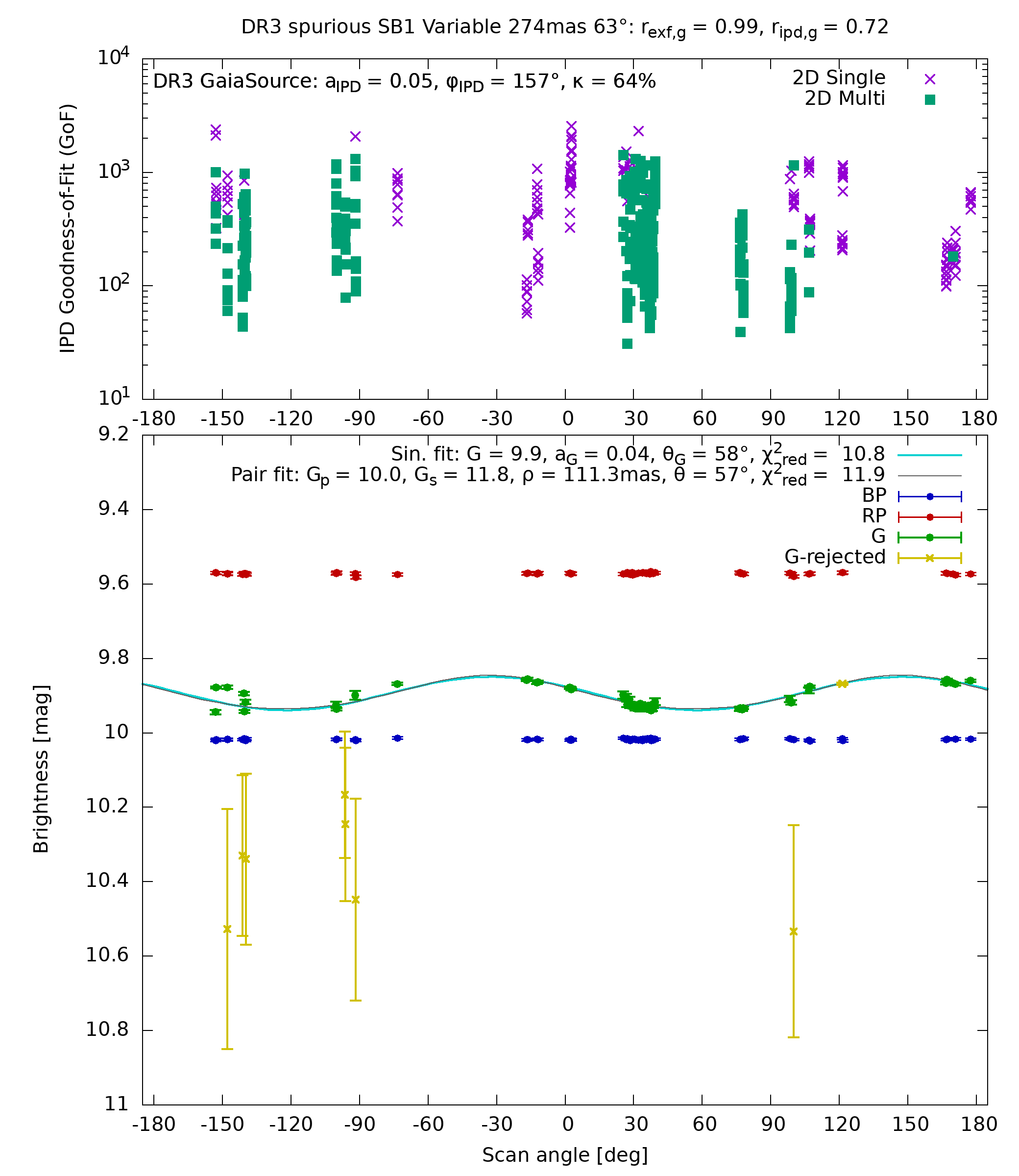}
  \includegraphics[width=0.45\textwidth]{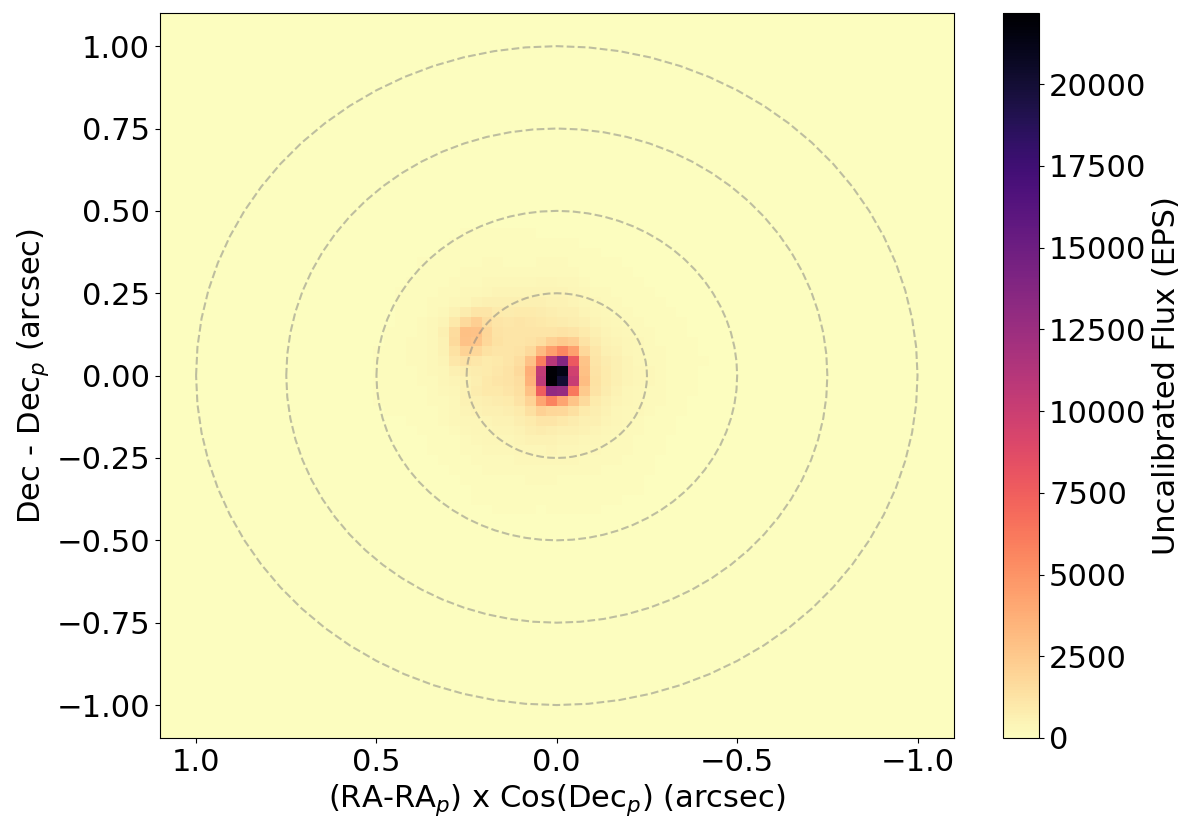}
\caption{\label{fig:ep-sa-415146526611154176} SourceID 415146526611154176: Scan-angle signatures and image reconstructed by SEAPipe.}
\end{figure}

\begin{figure}[h]
  \includegraphics[width=0.45\textwidth]{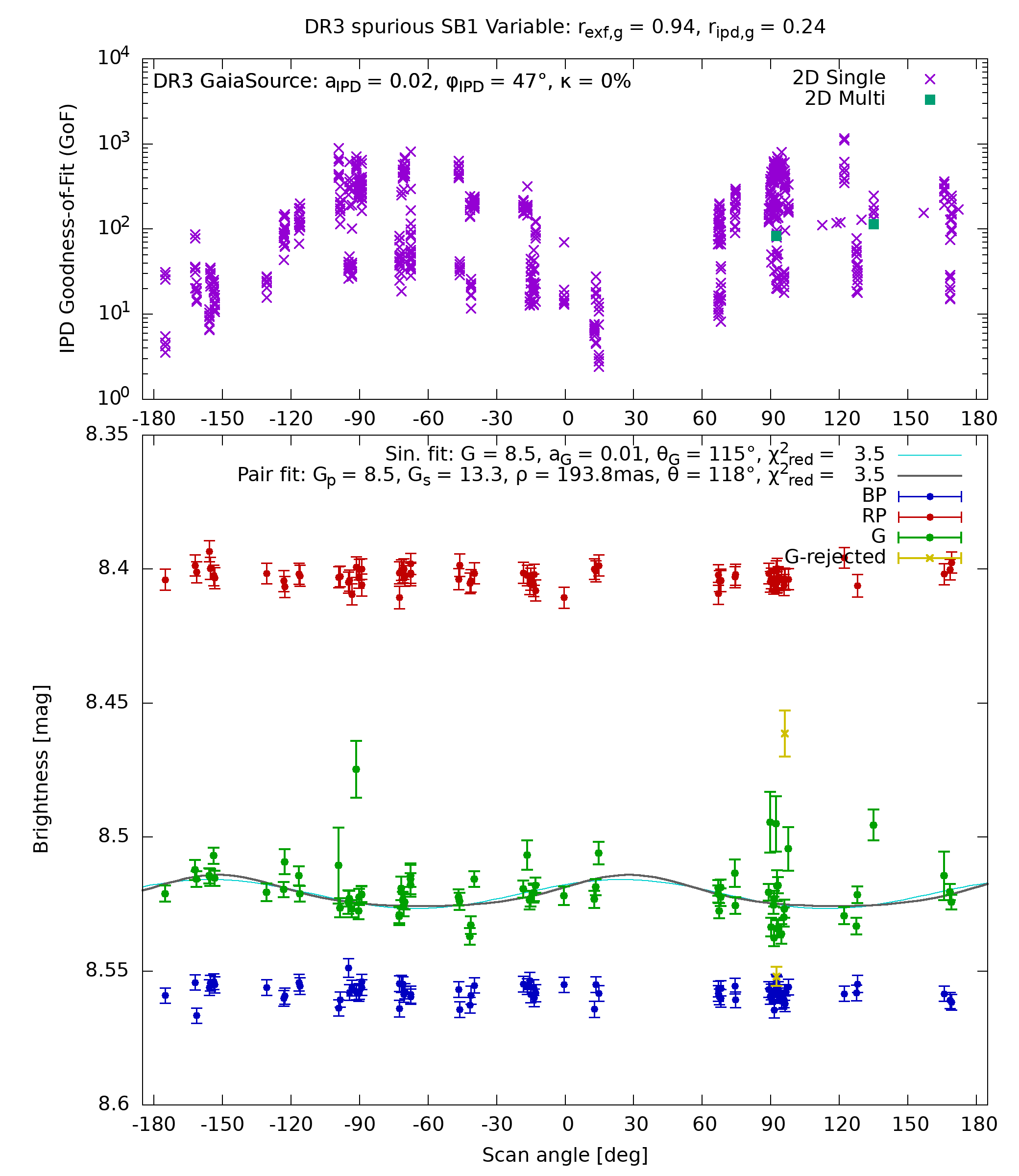}
  \includegraphics[width=0.45\textwidth]{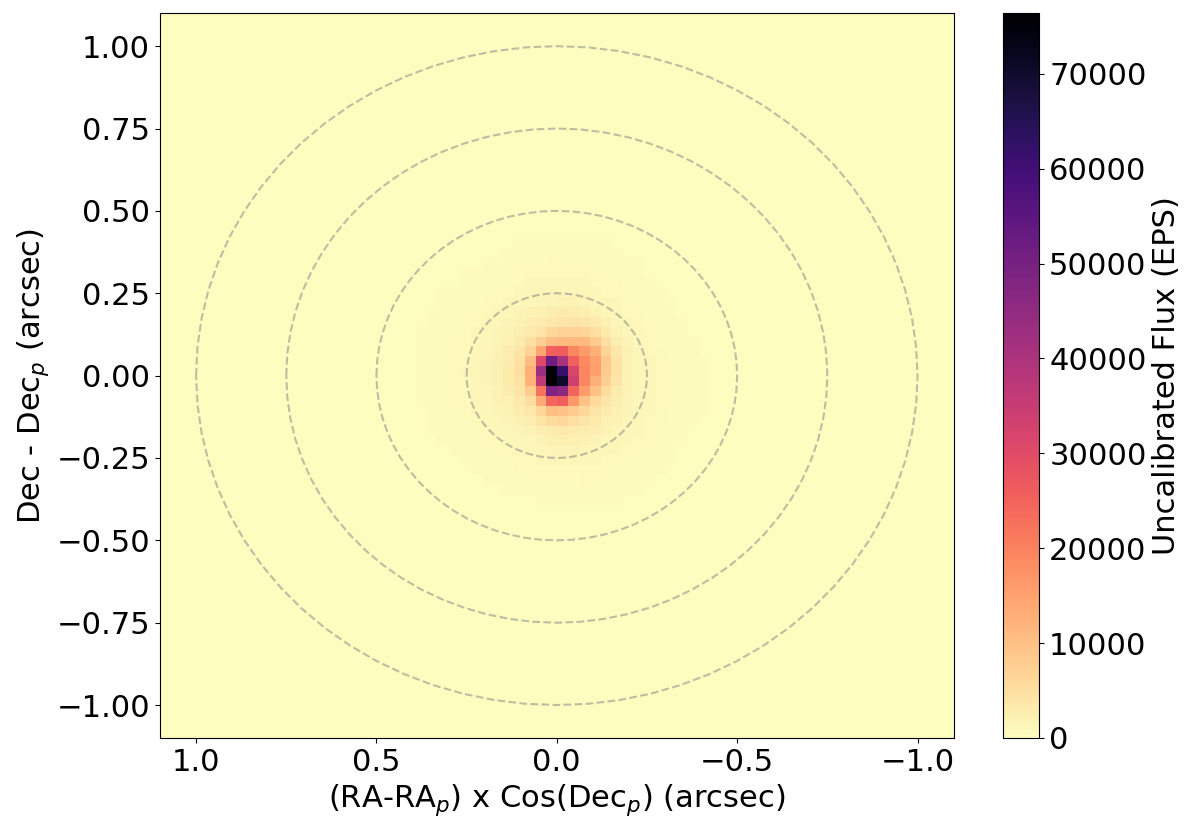}
\caption{\label{fig:ep-sa-5815369024263284352} SourceID 5815369024263284352: Scan-angle signatures and image reconstructed by SEAPipe.}
\end{figure}

\gaia DR3 contains some SB1 sources whose periods are very close to the precession period of the satellite (63 days), and thus they might be spurious. In Fig.~\ref{fig:ep-sa-415146526611154176} and~\ref{fig:ep-sa-5815369024263284352} we show two examples, which were furthermore identified as variable stars.

\FloatBarrier

\section{Sky distribution of observed and simulated spurious period peaks\label{sec:simPeaksExamples}}

\begin{figure}[h]
  \includegraphics[width=0.49\textwidth]{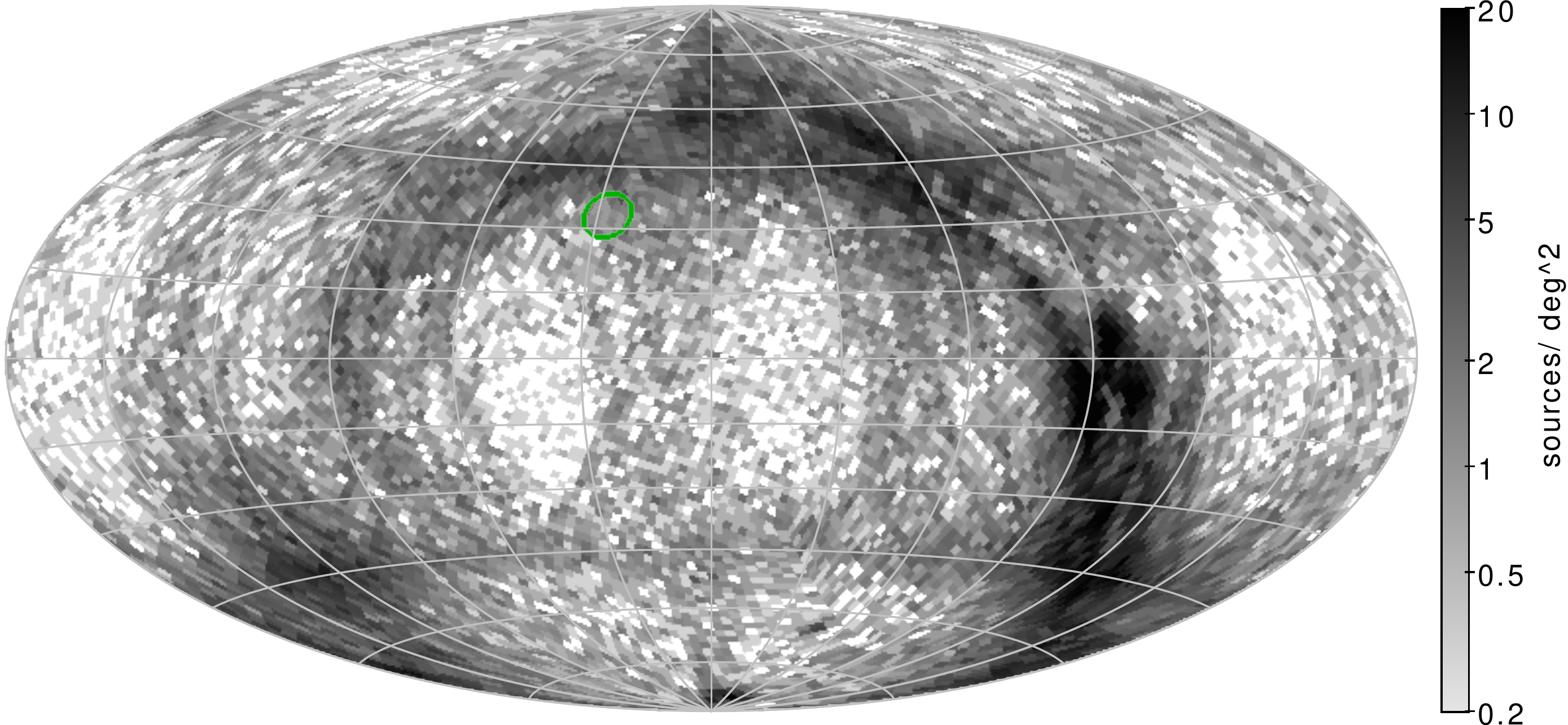}  
  \includegraphics[width=0.49\textwidth]{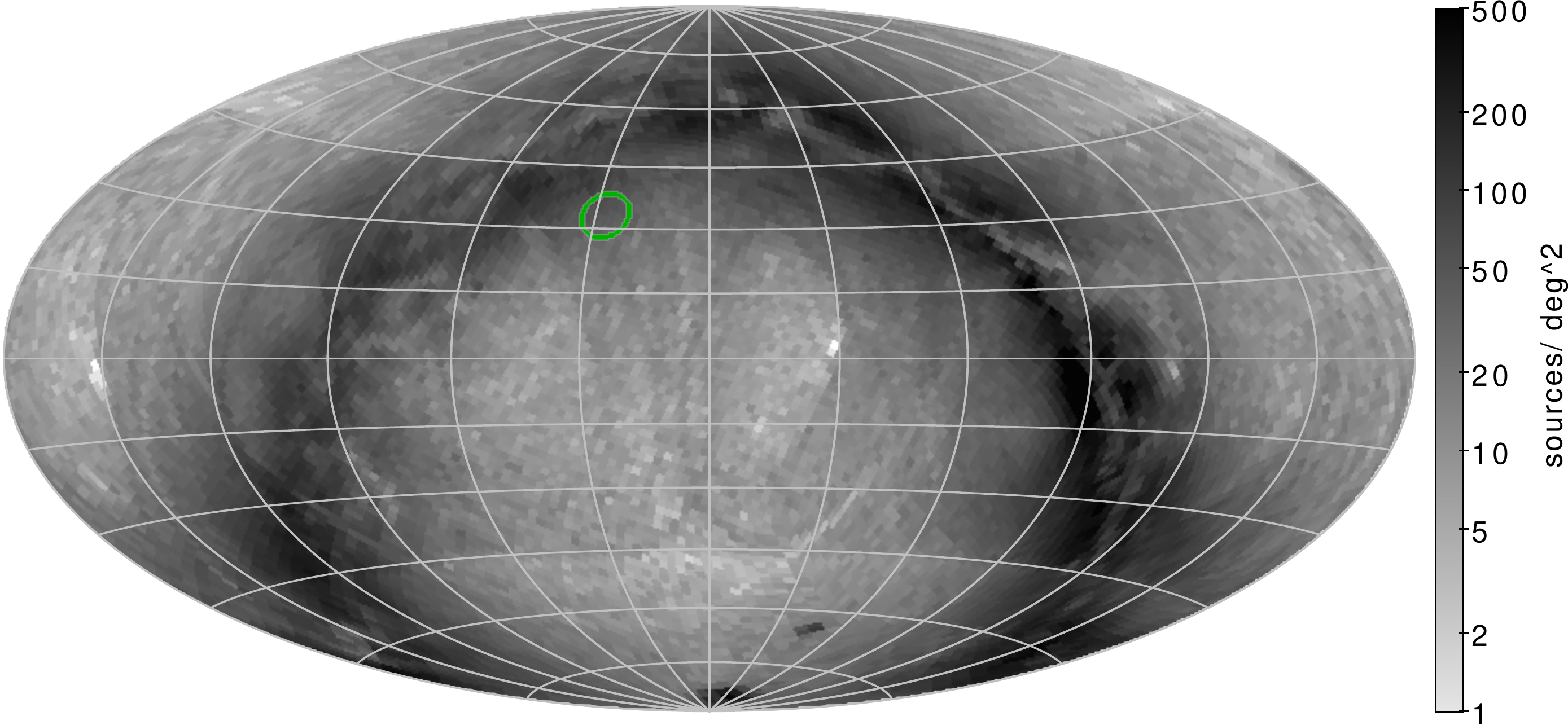} 
\caption{Ecliptic Aitoff projection with longitude zero at the centre and increasing to the left.
Top: Sky density of the (unpublished) all-sky photometric sample, subsets of which for different period ranges are shown in Figs.~\ref{fig:skyplotDistrPhotAllsky1} and \ref{fig:skyplotDistrPhotAllsky2}. Bottom:  Sky density of the (unpublished)  all-sky astrometric sample, subsets of which for different period ranges are shown in Figs.~\ref{fig:skyplotDistrAstroAllsky1} and \ref{fig:skyplotDistrAstroAllsky2}. Both samples are introduced in Sect.~\ref{ssec:obsPerDistr}. The green circle indicates the location of the GAPS catalogue (see Fig.~\ref{fig:simCompareToPhotGaps}).}
\label{fig:overallDensityCatalogs}
\end{figure}

\begin{figure*}[h]
\begin{tabular}{@{}lrr@{}}
\setlength{\tabcolsep}{0pt} 
\renewcommand{\arraystretch}{0} 
  \includegraphics[width=0.3\textwidth]{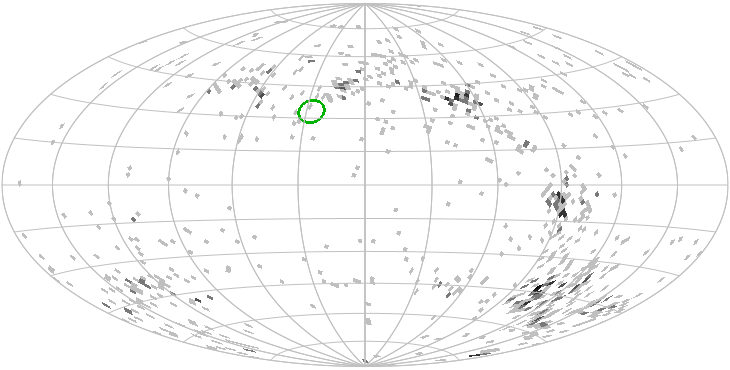}    
  & \includegraphics[width=0.3\textwidth]{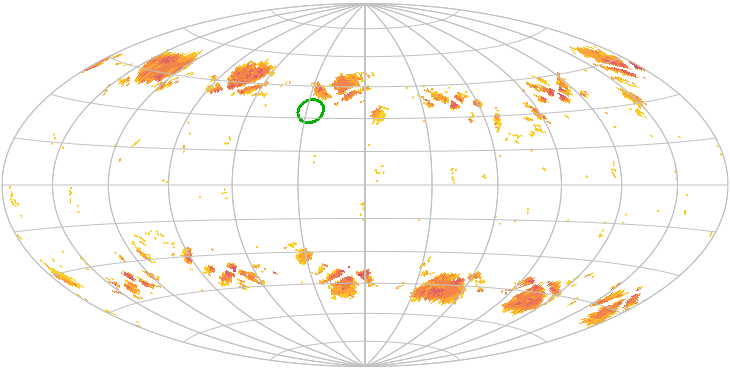}    
  & \includegraphics[width=0.3\textwidth]{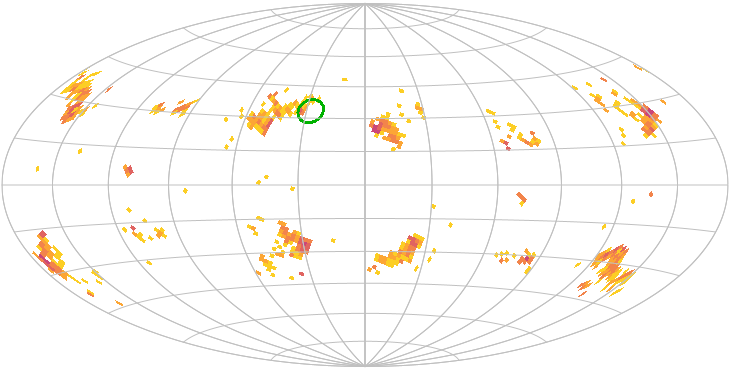}  \vspace{-3cm}\\
$13.95\pm 0.15$~d \quad \quad \quad \quad \ \ \  all-sky phot. & sim. $k=2$ & sim. $k=4$  \vspace{2.4cm} \\[6pt]
  \includegraphics[width=0.3\textwidth]{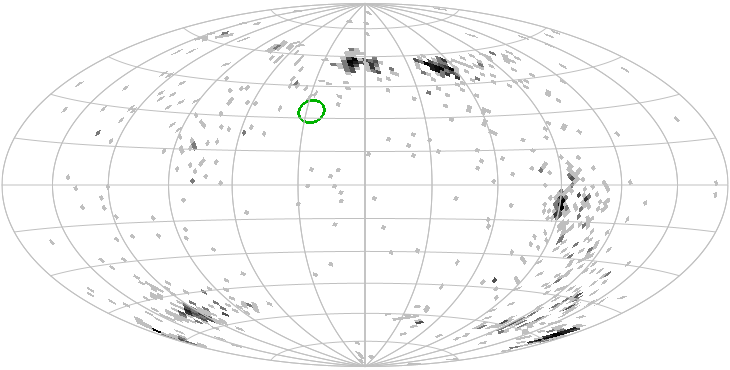}    
  & \includegraphics[width=0.3\textwidth]{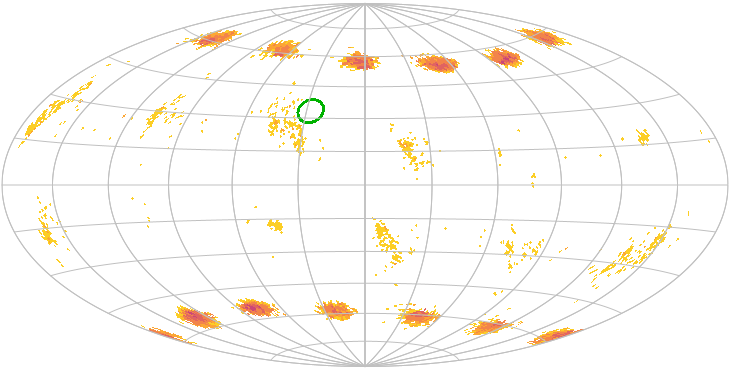}   
 & \includegraphics[width=0.3\textwidth]{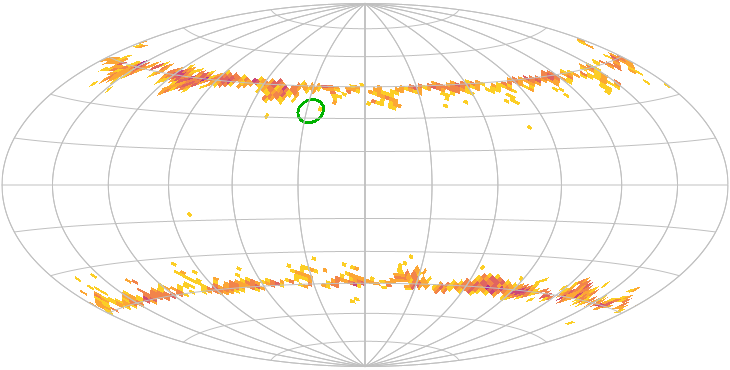} \vspace{-3cm}\\
$15.7  \pm  0.15$~d &  & \vspace{2.4cm}\\[6pt]
  \includegraphics[width=0.3\textwidth]{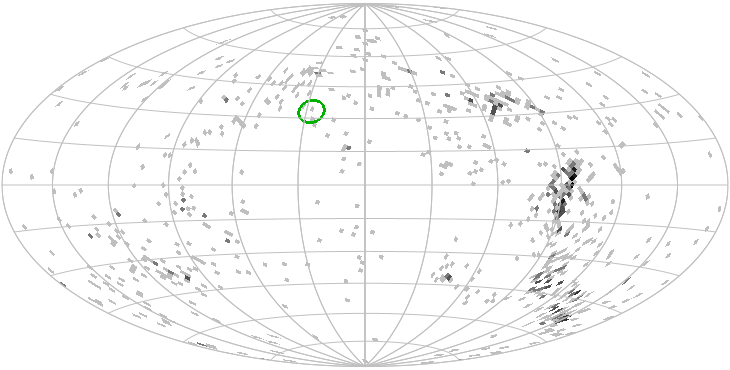}    
  & \includegraphics[width=0.3\textwidth]{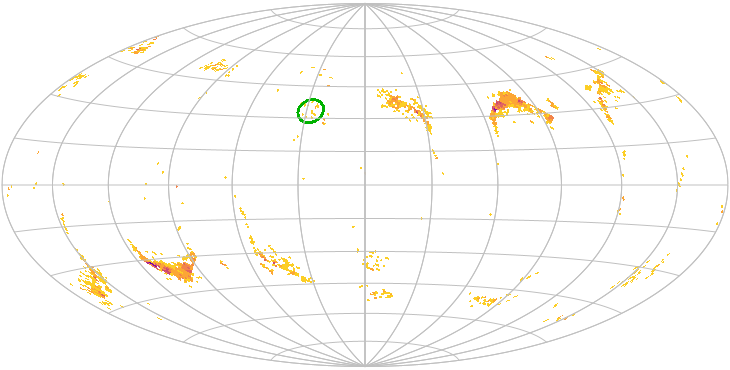}   
 & \includegraphics[width=0.3\textwidth]{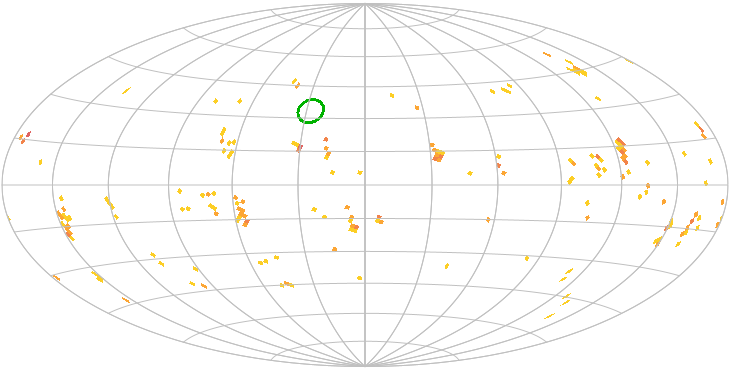}  \vspace{-3cm}\\
$16.35 \pm  0.2$~d &  &  \vspace{2.4cm}\\[6pt]

  \includegraphics[width=0.3\textwidth]{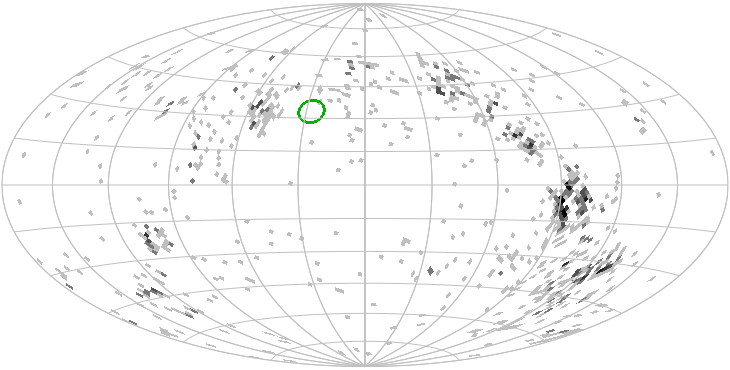}    
  &  \includegraphics[width=0.3\textwidth]{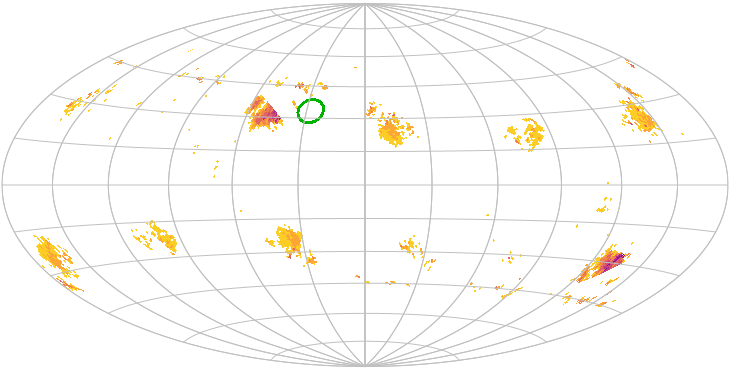}   
 & \includegraphics[width=0.3\textwidth]{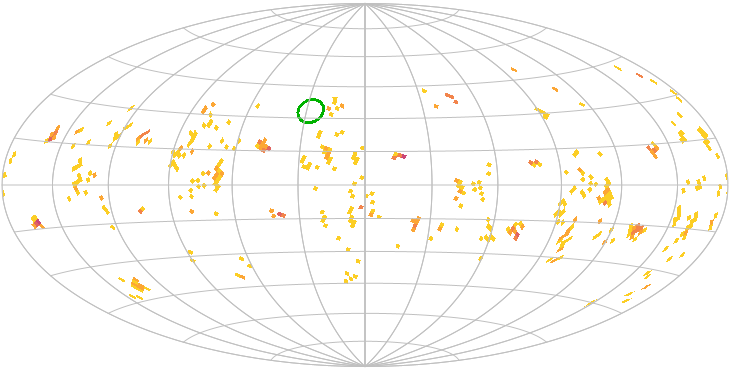}   \vspace{-3cm}\\
$18.8  \pm  0.2$~d &  &  \vspace{2.4cm}\\[6pt]
  \includegraphics[width=0.3\textwidth]{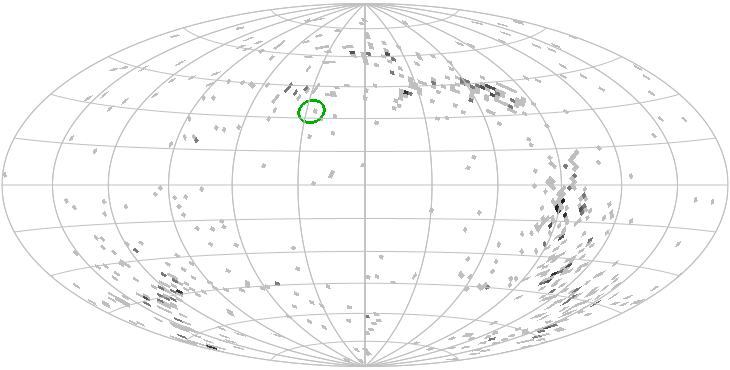}    
  &  \includegraphics[width=0.3\textwidth]{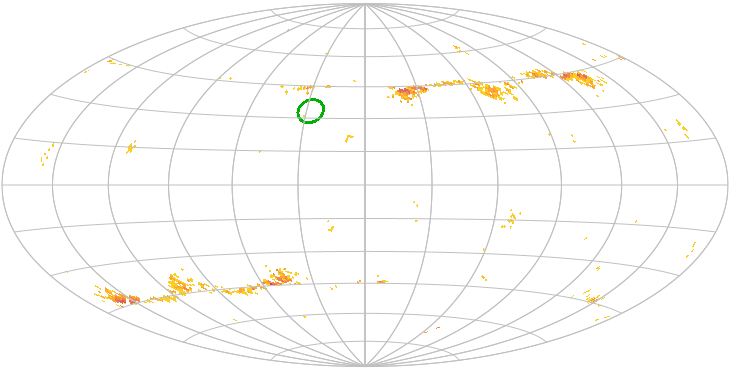}   
 & \includegraphics[width=0.3\textwidth]{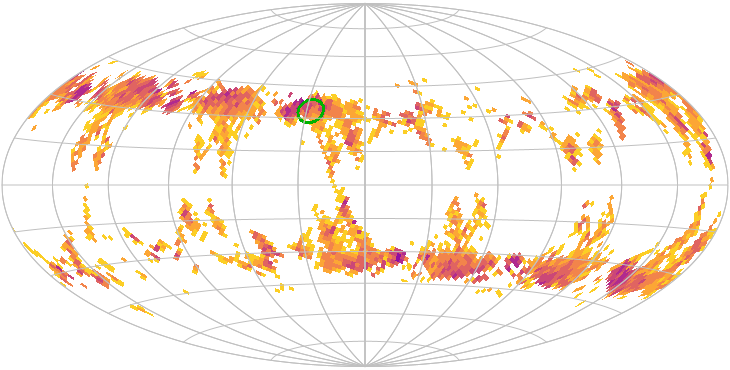}  \vspace{-3cm}\\
$19.9  \pm  0.2$~d &  &   \vspace{2.4cm}\\[6pt]
  \includegraphics[width=0.3\textwidth]{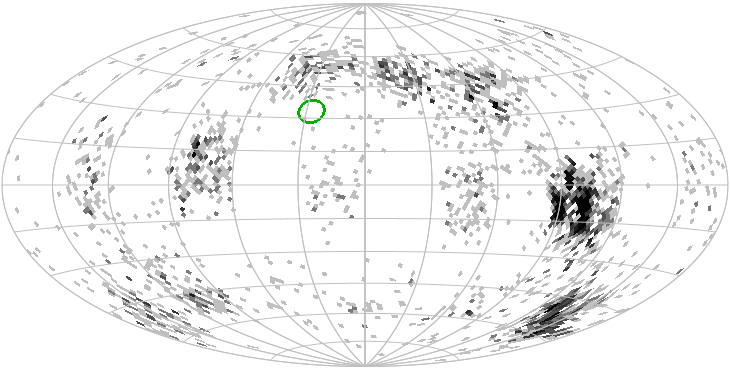}    
  &  \includegraphics[width=0.3\textwidth]{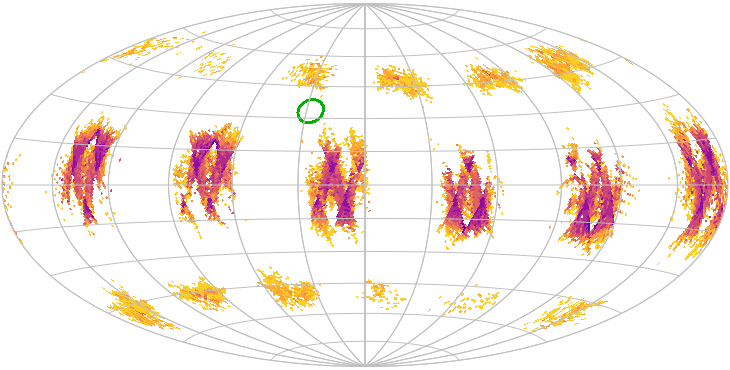}   
 & \includegraphics[width=0.3\textwidth]{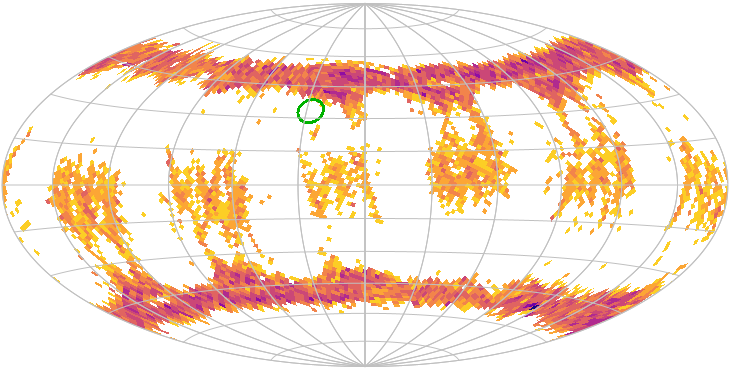}   \vspace{-3cm}\\
$25.1  \pm  0.3$~d &  & \vspace{2.4cm}\\[6pt]
   \includegraphics[width=0.3\textwidth]{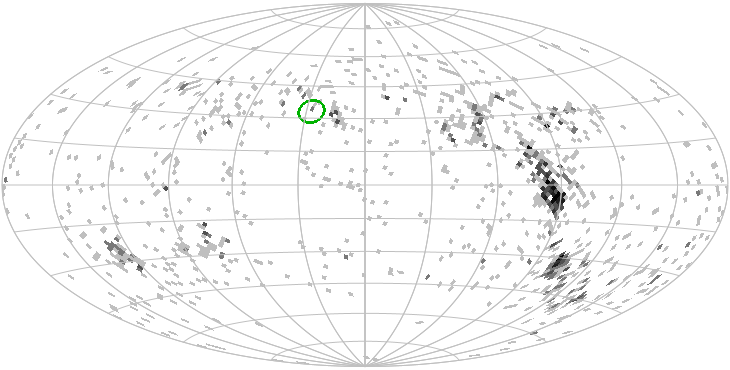}    
  & \includegraphics[width=0.3\textwidth]{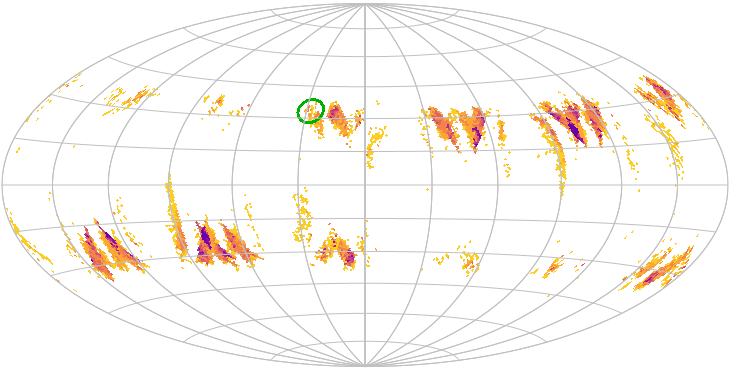}   
 & \includegraphics[width=0.3\textwidth]{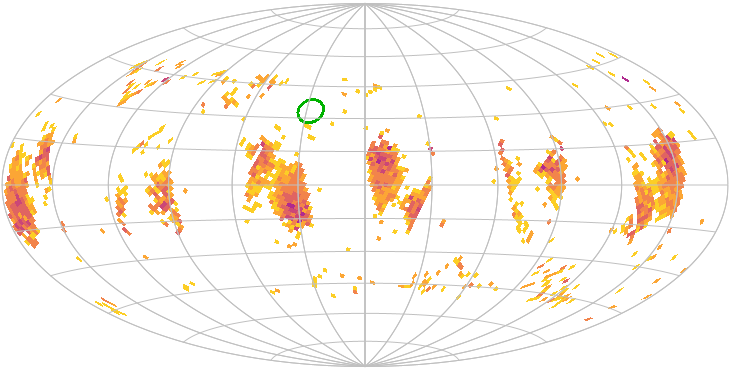}    \vspace{-3cm}\\
$26.9  \pm  0.4$~d &  &  \vspace{2.4cm}\\[6pt]
\multicolumn{3}{l}{\hspace{-0.32cm}\includegraphics[trim={0 3.3cm 0 0},clip,width=0.96\textwidth]{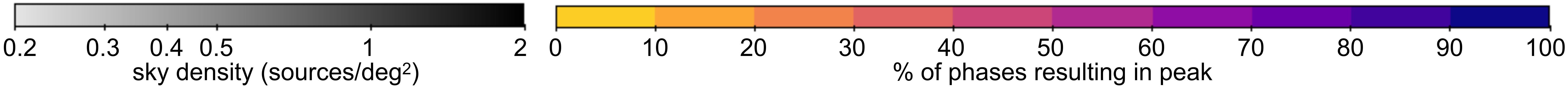} }  \vspace{-0.2cm}\\[6pt]
\end{tabular}
\caption{Ecliptic Aitoff projection  of the photometric data presented in Fig.~\ref{fig:simCompareToPhotAllSky}. Left: Source density of the photometric peaks for the all-sky sample (top panel of Fig.~\ref{fig:simCompareToPhotAllSky}). Right: Result of the  all-sky uniform simulations of our noiseless sampled bias model (panels 2 and 3 of Fig.~\ref{fig:simCompareToPhotAllSky}). They are
colour-coded with the percentage of phases (position angles) that results in this peak: A low value means that only specific phasing of the scan-angle signal will result in a particular peak being observed at the given location. The green circle indicates the location of the GAPS catalogue (see Fig.~\ref{fig:simCompareToPhotGaps}.) In all sky plots, longitude zero is at the centre and increasing to the left. }
\label{fig:skyplotDistrPhotAllsky1}
\end{figure*}

\begin{figure*}[h]
\begin{tabular}{@{}lrr@{}}
\setlength{\tabcolsep}{0pt} 
\renewcommand{\arraystretch}{0} 
  \includegraphics[width=0.3\textwidth]{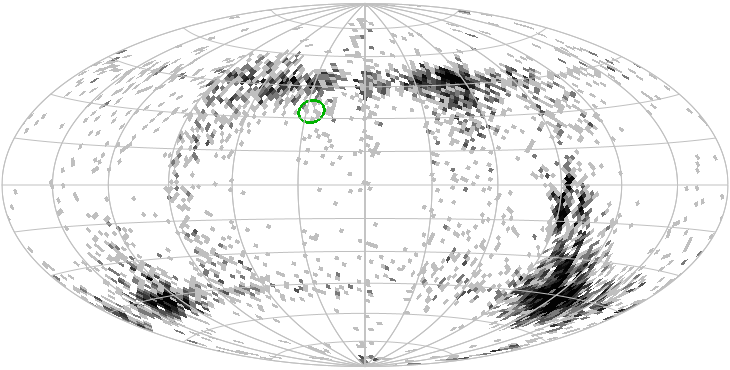}    
  & \includegraphics[width=0.3\textwidth]{images/simSpuriousPeriodSkyMaps/phot_ph5_k2_13p9d.png}    
  & \includegraphics[width=0.3\textwidth]{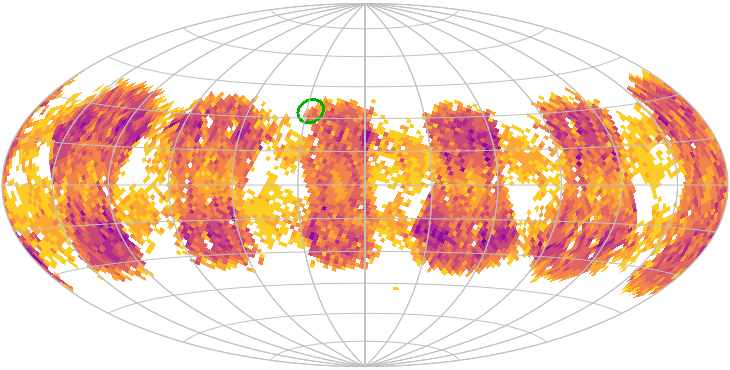}  \vspace{-3cm}\\
$31.5  \pm  0.6$~d \quad \quad \quad \quad \quad \ \ \ \  all-sky phot. & sim. $k=2$ & sim. $k=4$   \vspace{2.4cm} \\[6pt]
  \includegraphics[width=0.3\textwidth]{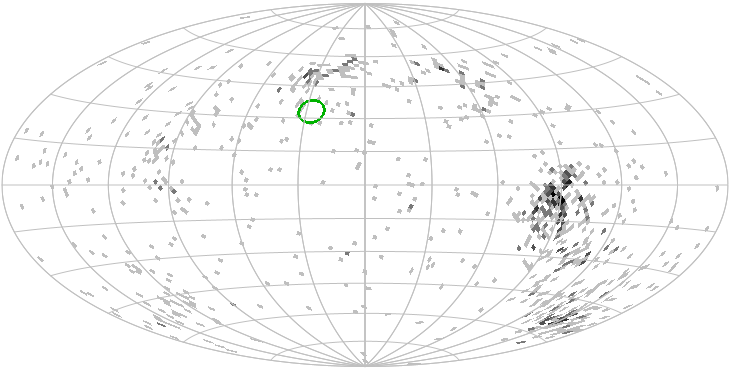}    
  & \includegraphics[width=0.3\textwidth]{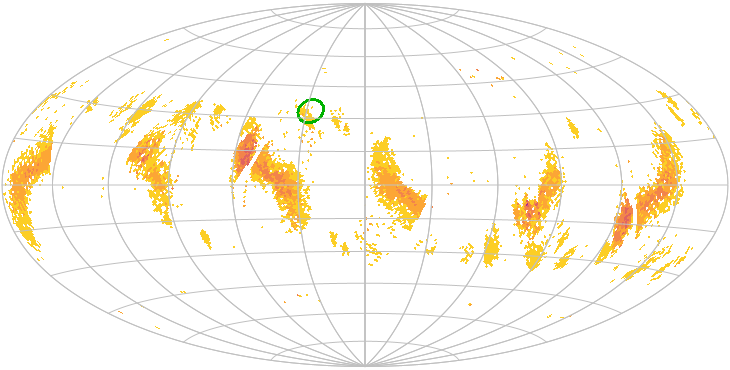}   
 & \includegraphics[width=0.3\textwidth]{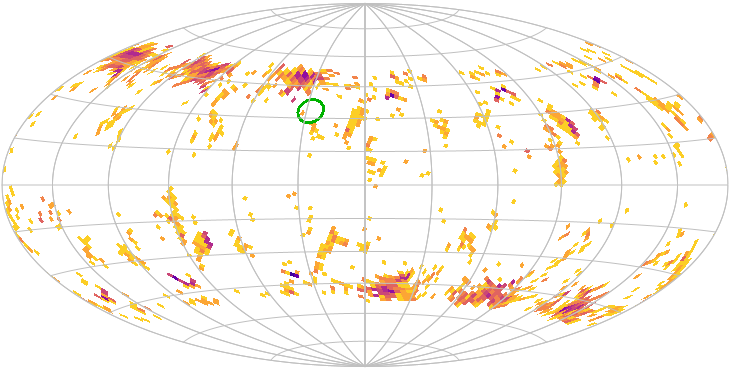} \vspace{-3cm}\\
$41.9  \pm  1.0$~d &  &  \vspace{2.4cm}\\[6pt]
  \includegraphics[width=0.3\textwidth]{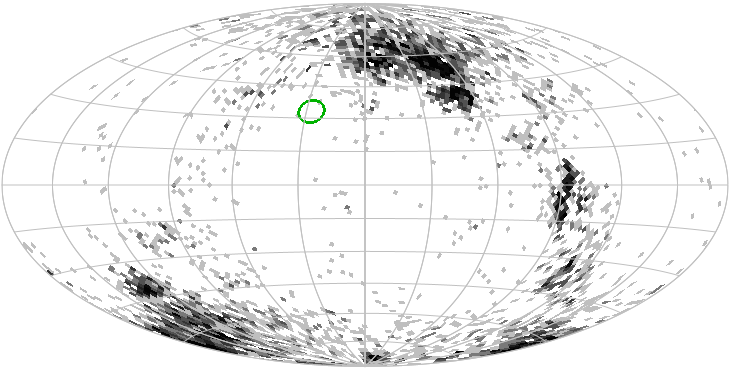}    
  & \includegraphics[width=0.3\textwidth]{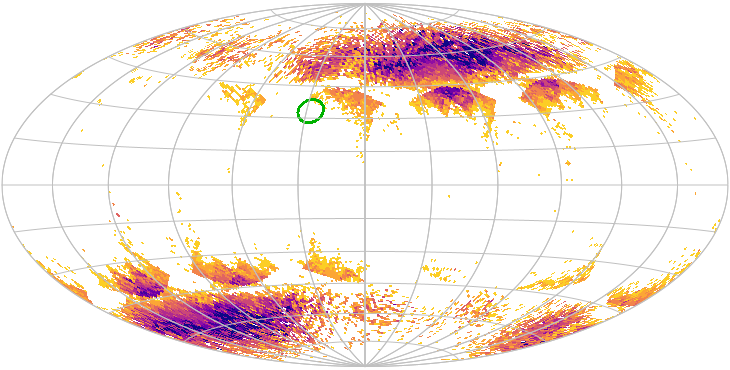}   
 & \includegraphics[width=0.3\textwidth]{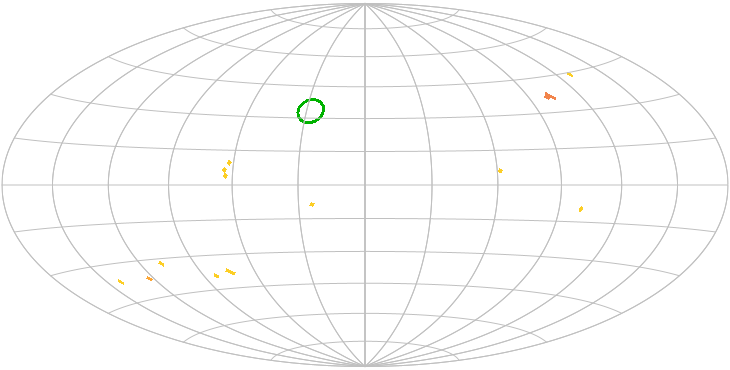}  \vspace{-3cm}\\
$46.8  \pm  1.1$~d &  &  \vspace{2.4cm}\\[6pt]
  \includegraphics[width=0.3\textwidth]{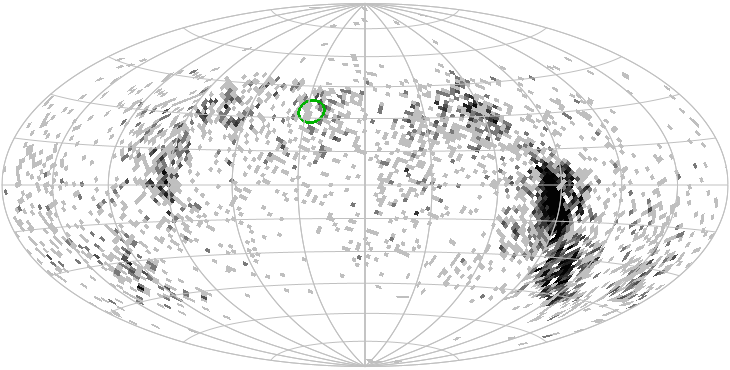}    
  &  \includegraphics[width=0.3\textwidth]{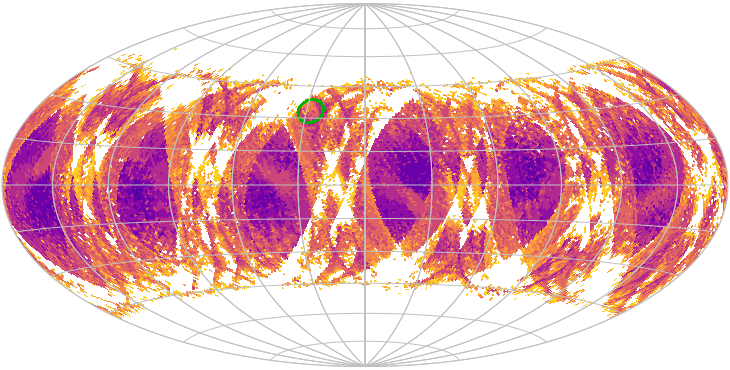}   
 & \includegraphics[width=0.3\textwidth]{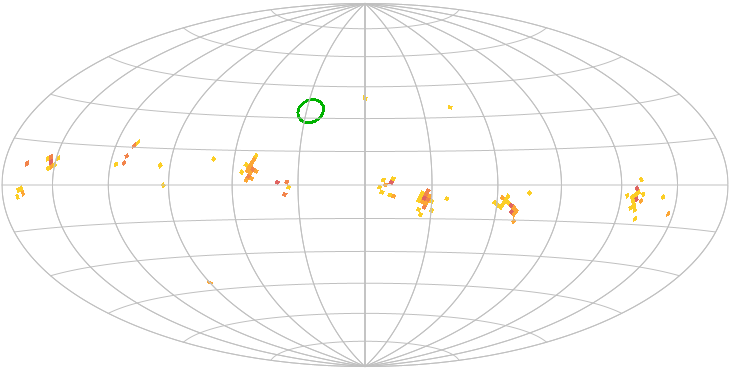}  \vspace{-3cm}\\
$53.7  \pm  1.1$~d &  &  \vspace{2.4cm}\\[6pt]
  \includegraphics[width=0.3\textwidth]{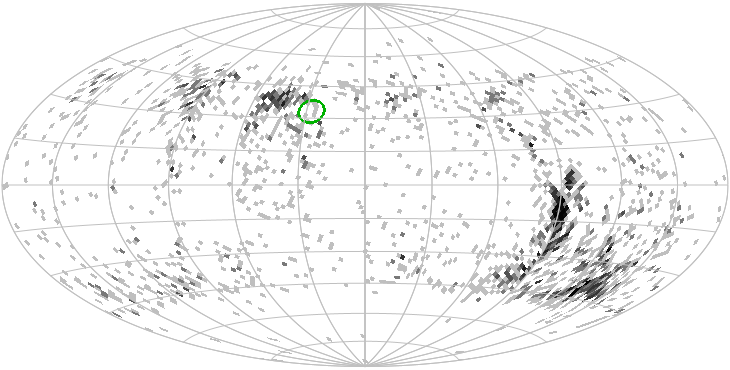}    
  &  \includegraphics[width=0.3\textwidth]{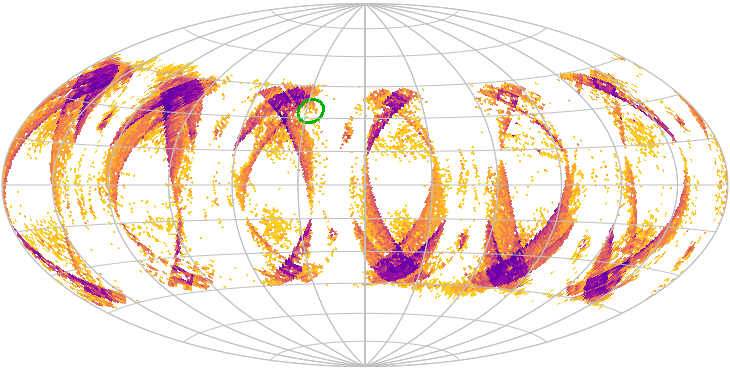}   
 & \includegraphics[width=0.3\textwidth]{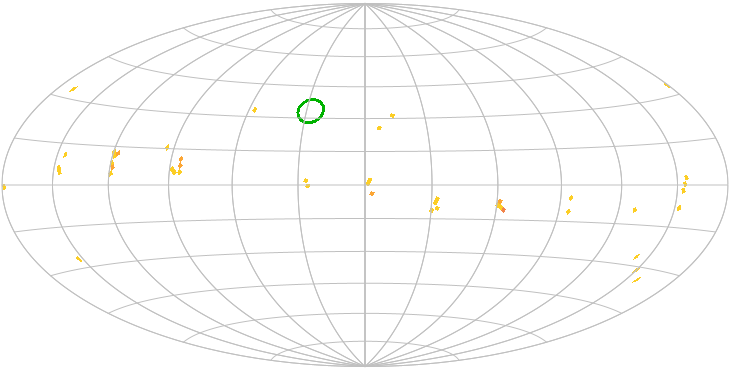}   \vspace{-3cm}\\
$76.1  \pm  1.7$~d &  & \vspace{2.4cm}\\[6pt]
  \includegraphics[width=0.3\textwidth]{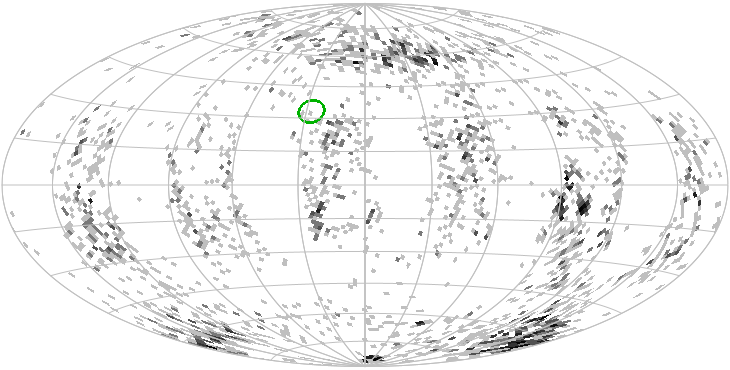}    
  &  \includegraphics[width=0.3\textwidth]{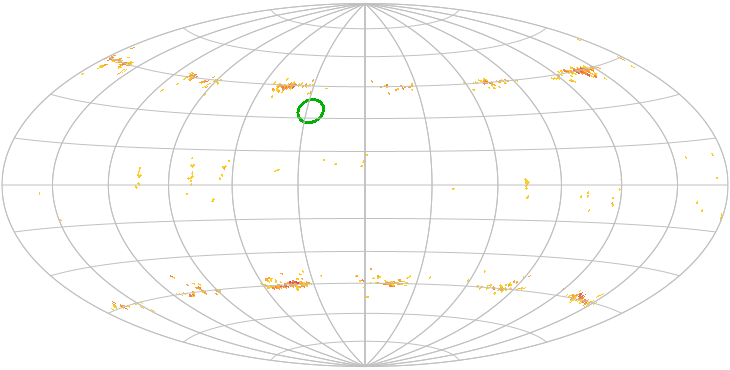}   
 & \includegraphics[width=0.3\textwidth]{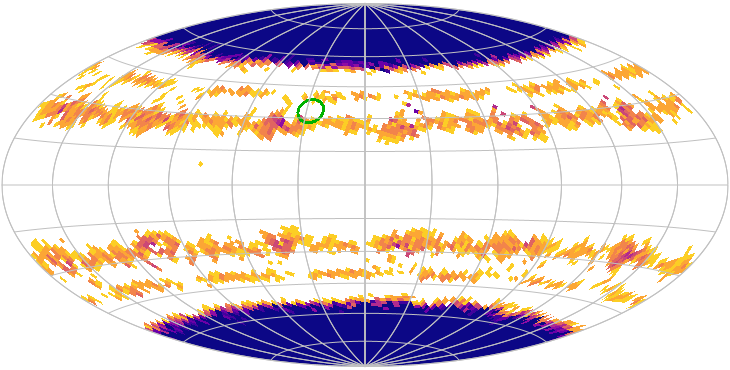}  \vspace{-3cm}\\
$91.3  \pm  2.4$~d &  &   \vspace{2.4cm}\\[6pt]
  \includegraphics[width=0.3\textwidth]{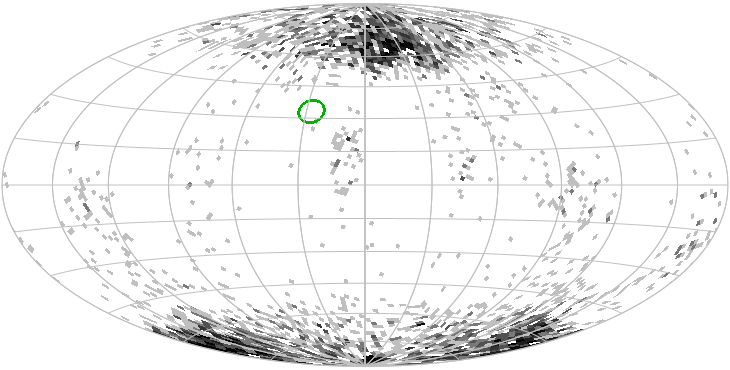}    
  &  \includegraphics[width=0.3\textwidth]{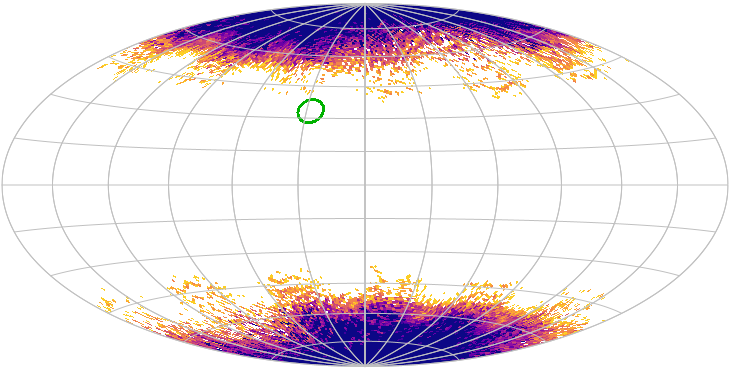}   
 & \includegraphics[width=0.3\textwidth]{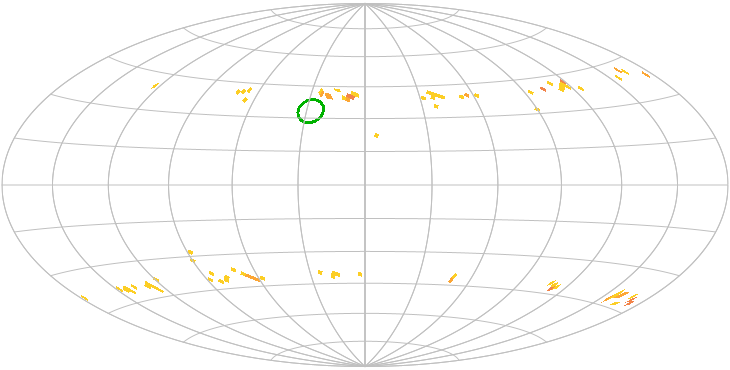}   \vspace{-3cm}\\
$96.1  \pm  2.4$~d &  & \vspace{2.4cm}\\[6pt]
   \includegraphics[width=0.3\textwidth]{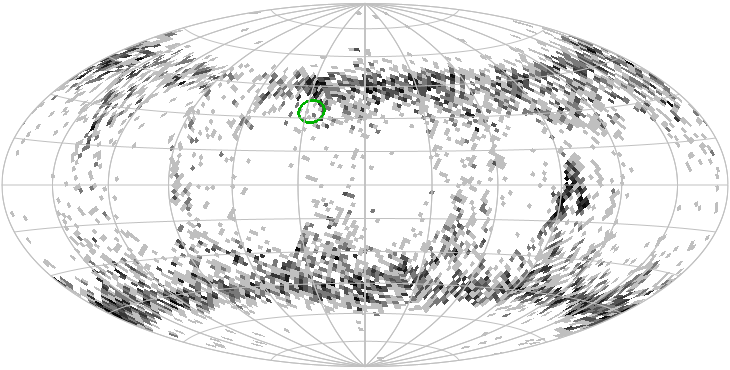}    
  & \includegraphics[width=0.3\textwidth]{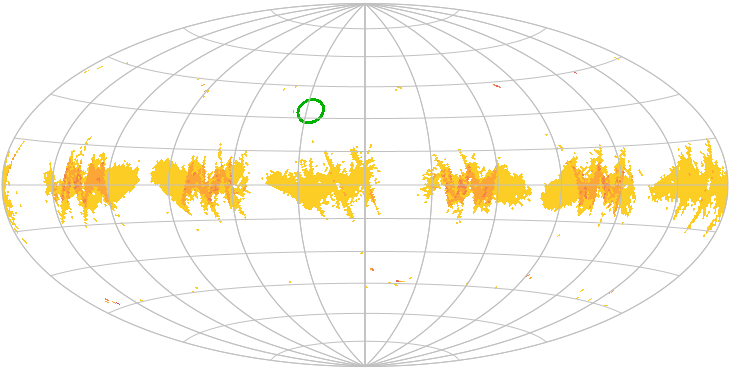}   
 & \includegraphics[width=0.3\textwidth]{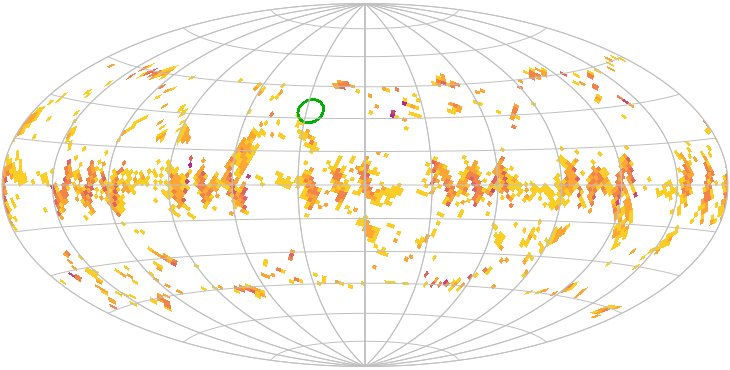}    \vspace{-3cm}\\
$182.6 \pm  10$~d &  &  \vspace{2.4cm}\\[6pt]
\multicolumn{3}{l}{\hspace{-0.32cm}\includegraphics[trim={0 3.3cm 0 0},clip,width=0.96\textwidth]{images/simSpuriousPeriodSkyMaps/FINAL_colourbar_phot6k.pdf} }  \vspace{-0.2cm}\\[6pt]
\end{tabular}
\caption{ Fig.~\ref{fig:skyplotDistrPhotAllsky1} ctd. for longer periods of photometric data.}
\label{fig:skyplotDistrPhotAllsky2}
\end{figure*}

In this section, we show the sky distribution of the most prominent photometric and astrometric spurious period peaks identified in Figs.~\ref{fig:simCompareToPhotAllSky}, \ref{fig:simCompareToPhotGaps}, and \ref{fig:astrSimP} of Sect.~\ref{ssec:propSimSignals}. It is important to realise that the photometric and astrometric unpublished data are sampled non-uniformly over the sky, as shown in Fig.~\ref{fig:overallDensityCatalogs}. The Galactic source density is very strongly imprinted on the sky sampling, meaning that this pattern will thus naturally be present in any of the observed period subsets shown in the left panels of Figs.~\ref{fig:skyplotDistrPhotAllsky1}, \ref{fig:skyplotDistrPhotAllsky2}, \ref{fig:skyplotDistrAstroAllsky1}, and \ref{fig:skyplotDistrAstroAllsky2}.

\begin{figure*}[h]
\begin{small}
\setlength{\tabcolsep}{0pt} 
\begin{tabular}{L{0.12\textwidth}R{0.3\textwidth}R{0.29\textwidth}R{0.29\textwidth}}
\renewcommand{\arraystretch}{0} 
  &  5808594933118768384  & 494246561341595648 & 5800597119919472768 \\
 \rotatebox{0}{\normalsize $13.95 \pm 0.15$~d}
    & \includegraphics[height=0.15\textwidth]{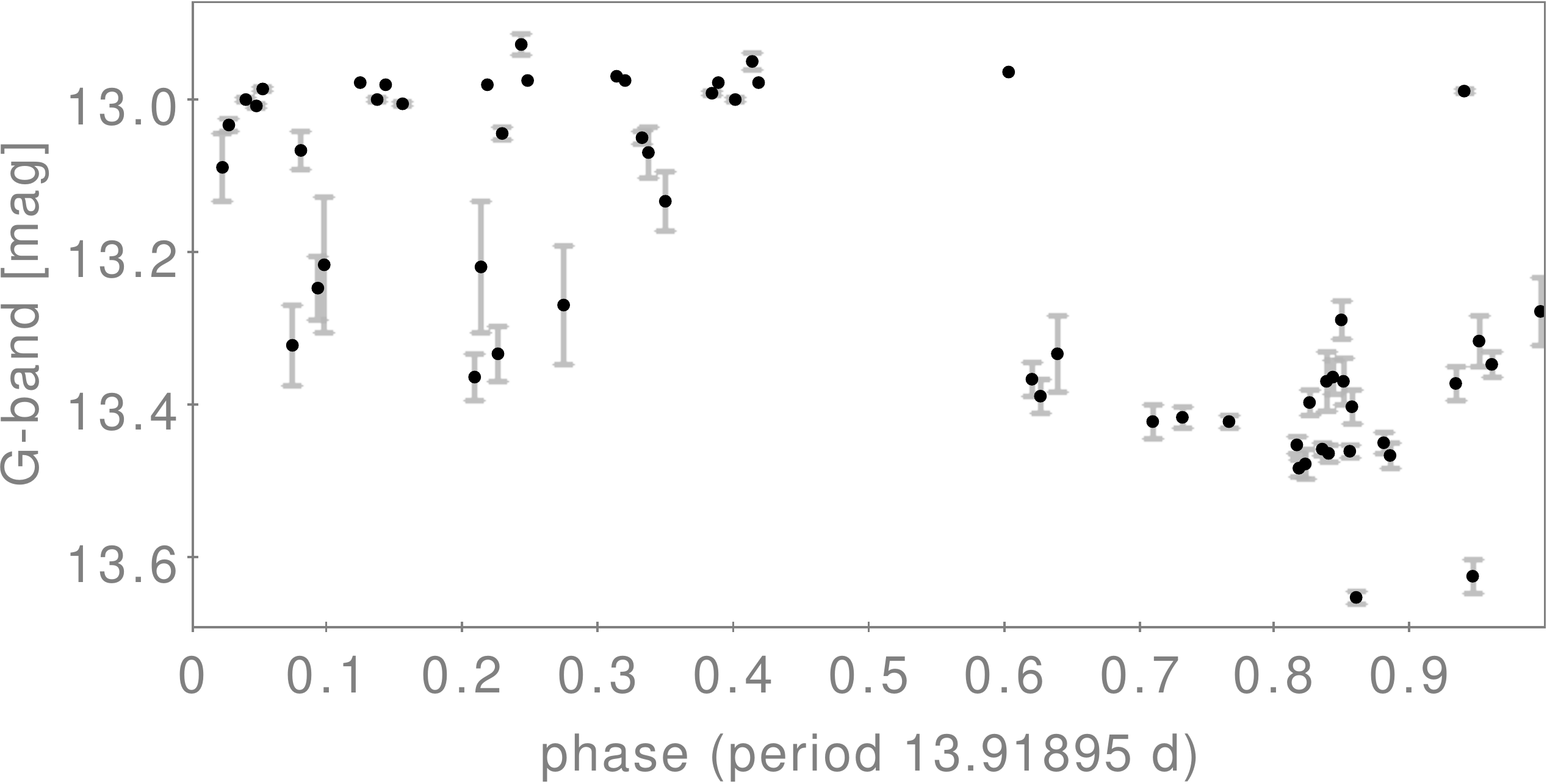}    
  & \includegraphics[trim=1.0cm 0 0 0, clip, height=0.15\textwidth]{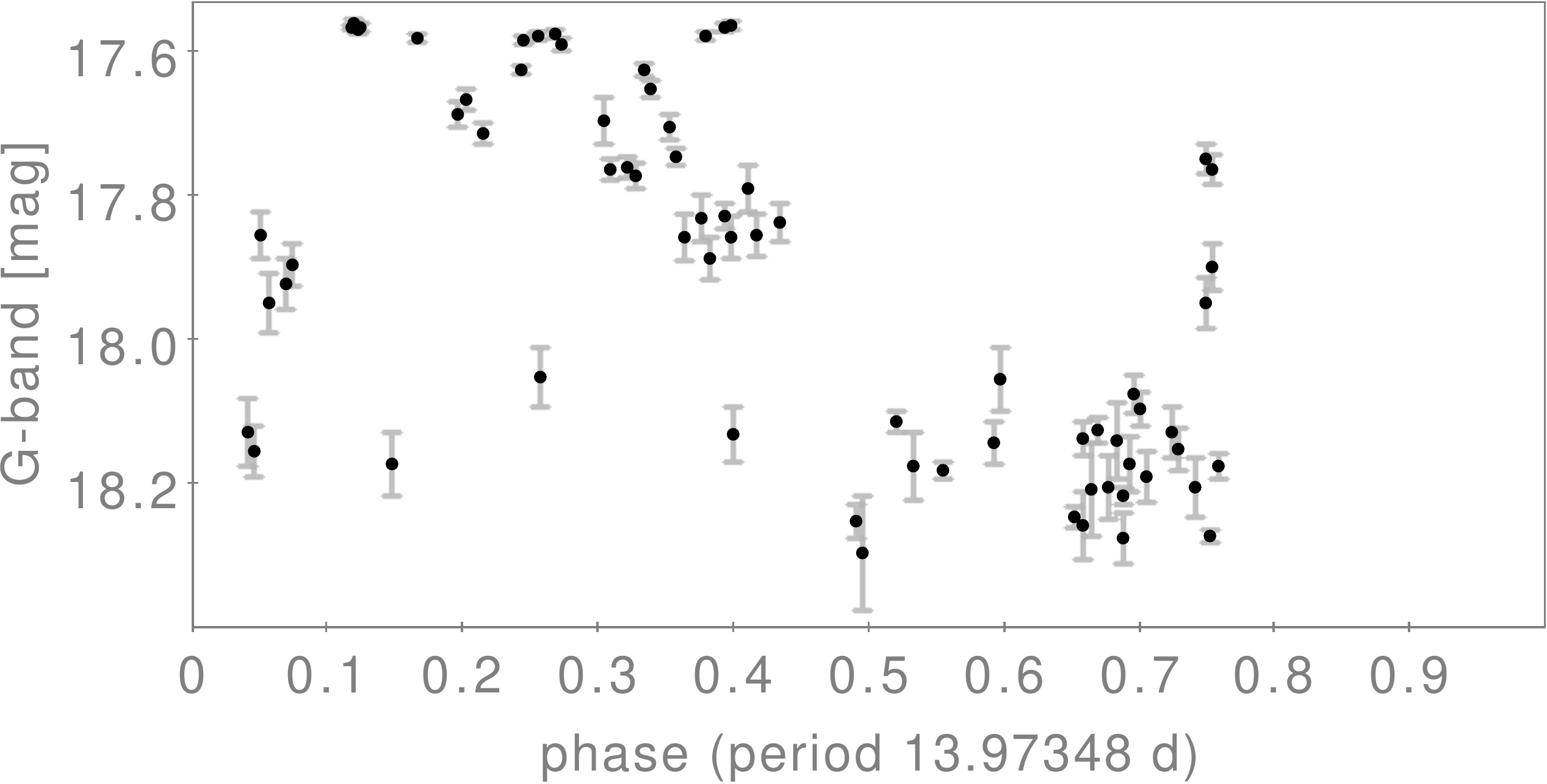}   
  & \includegraphics[trim=1.0cm 0 0 0, clip, height=0.15\textwidth]{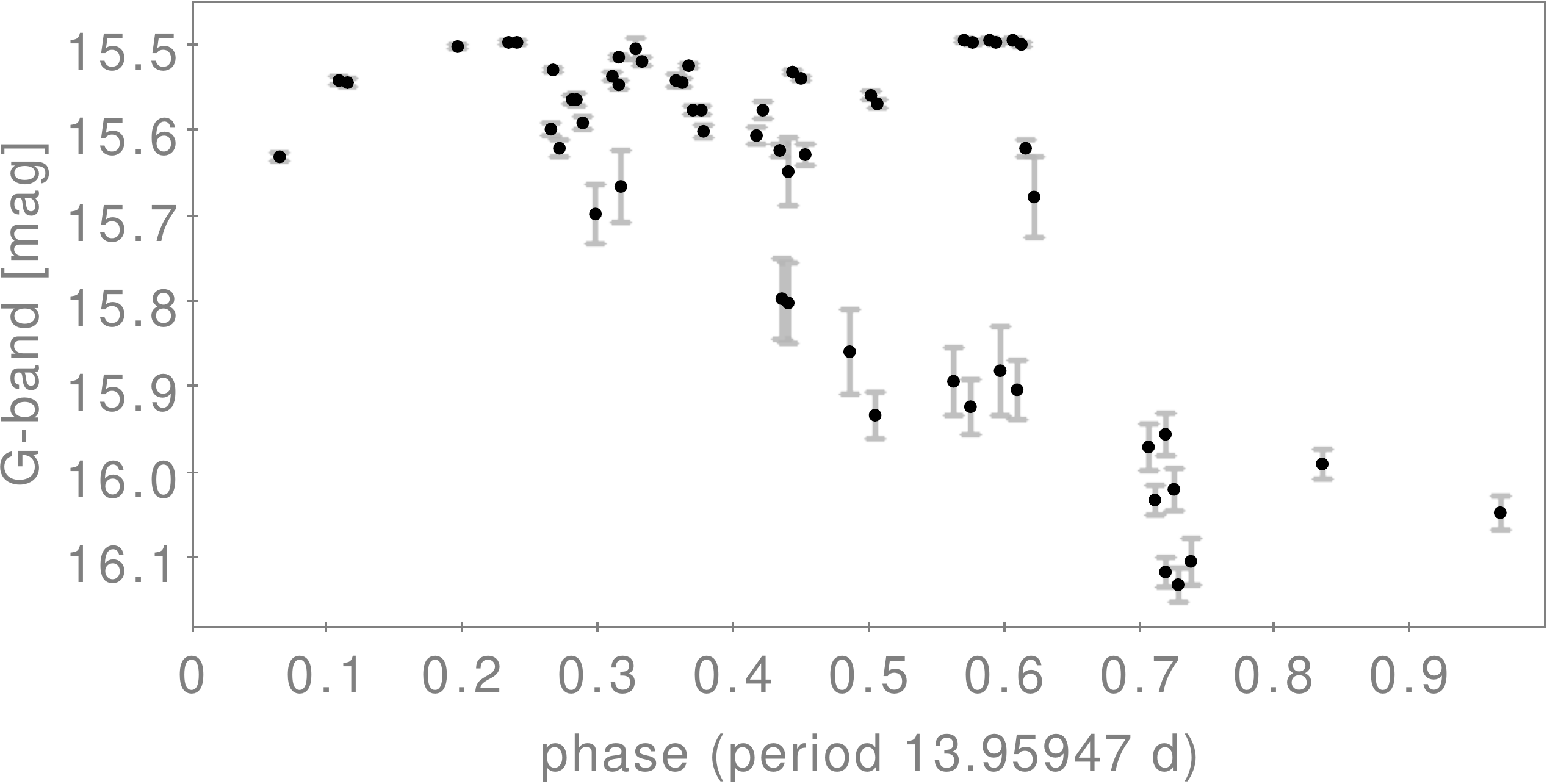}   \\
\vspace{0.1cm}  & 2000990628590812672 & 1098895626386959360 & 5593903914718590592\\
\rotatebox{0}{\normalsize $15.7  \pm  0.15$~d} 
  & \includegraphics[height=0.15\textwidth]{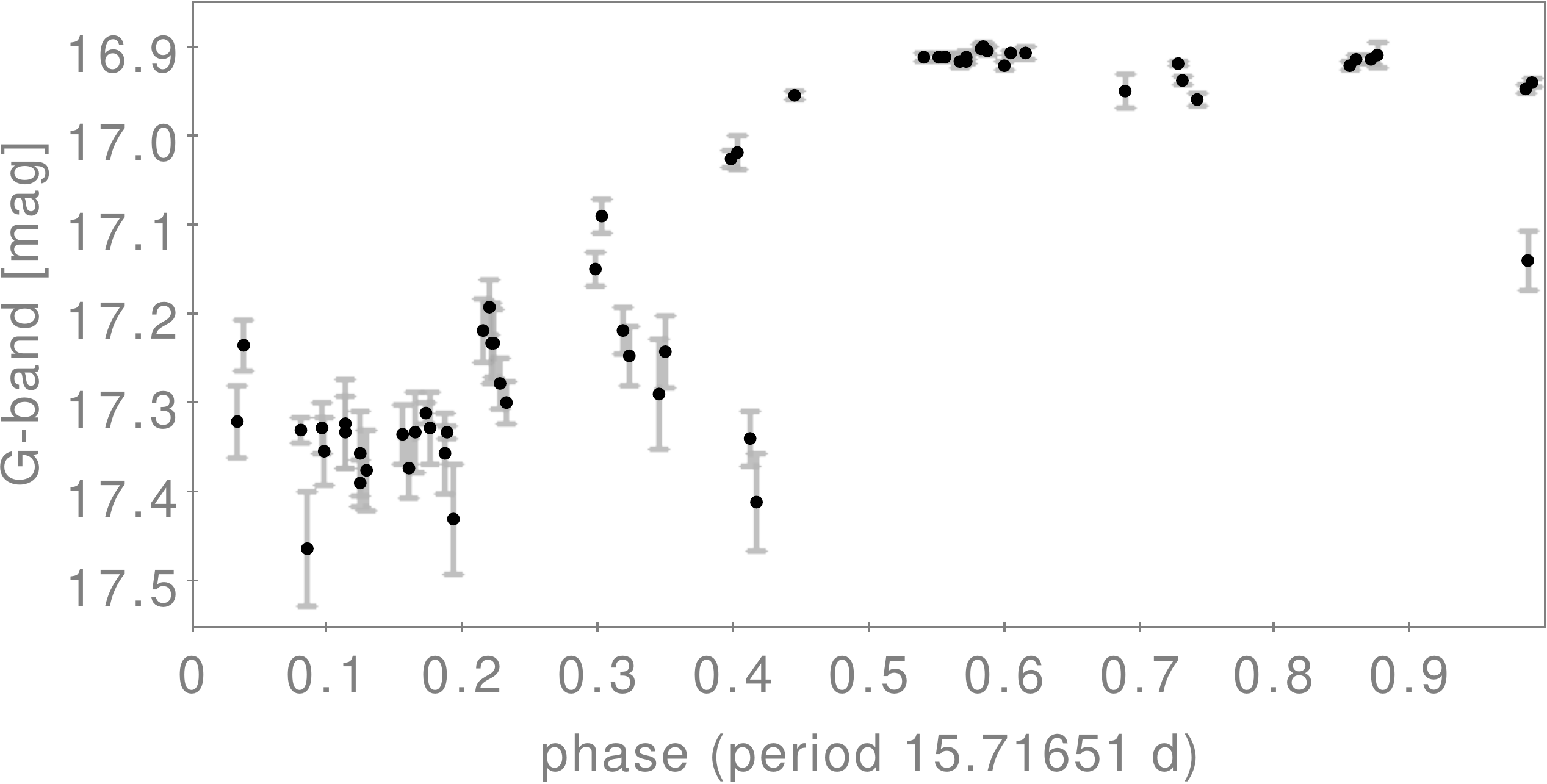}    
  & \includegraphics[trim=1.0cm 0 0 0, clip, height=0.15\textwidth]{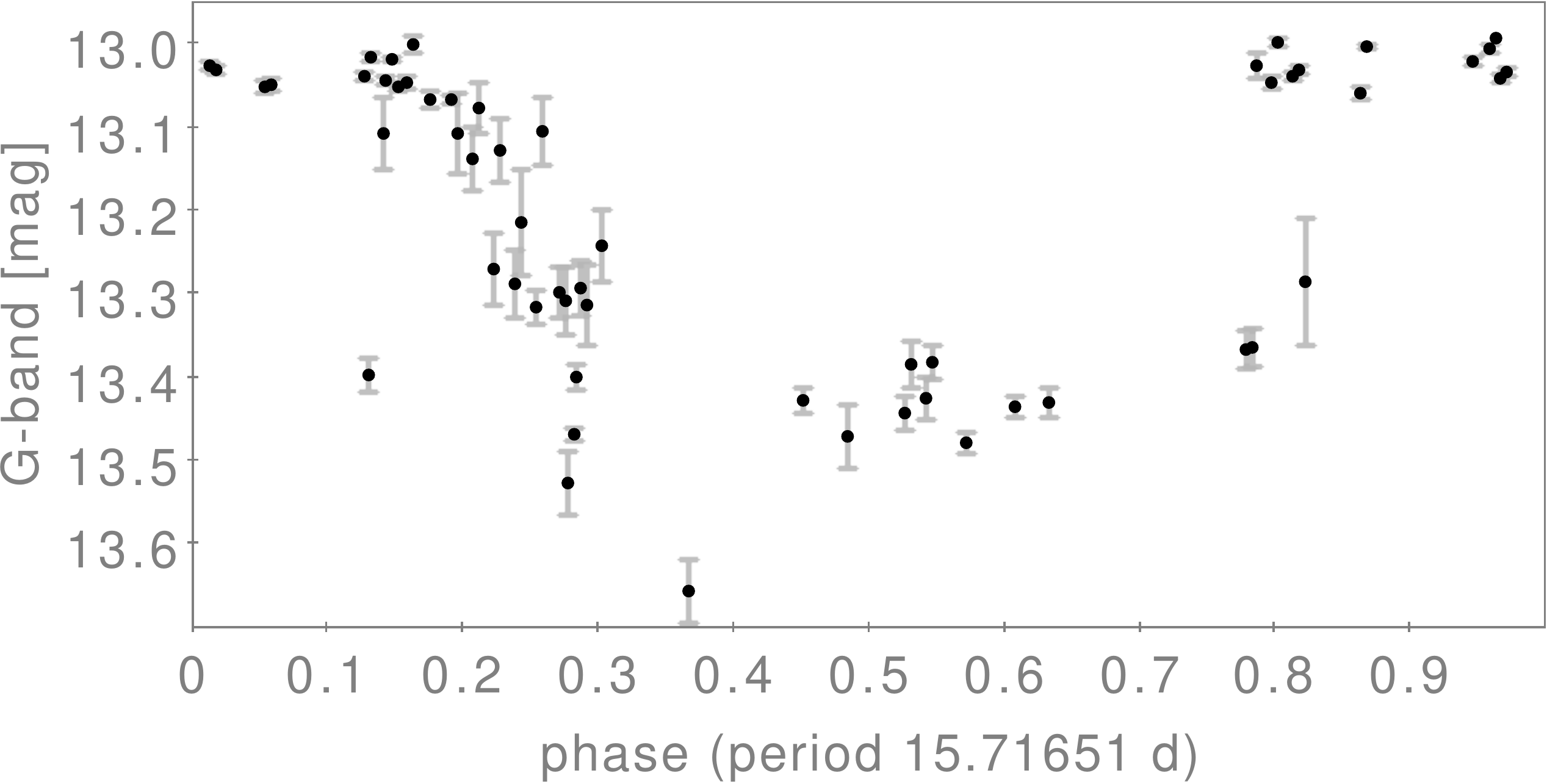}   
  & \includegraphics[trim=1.0cm 0 0 0, clip, height=0.15\textwidth]{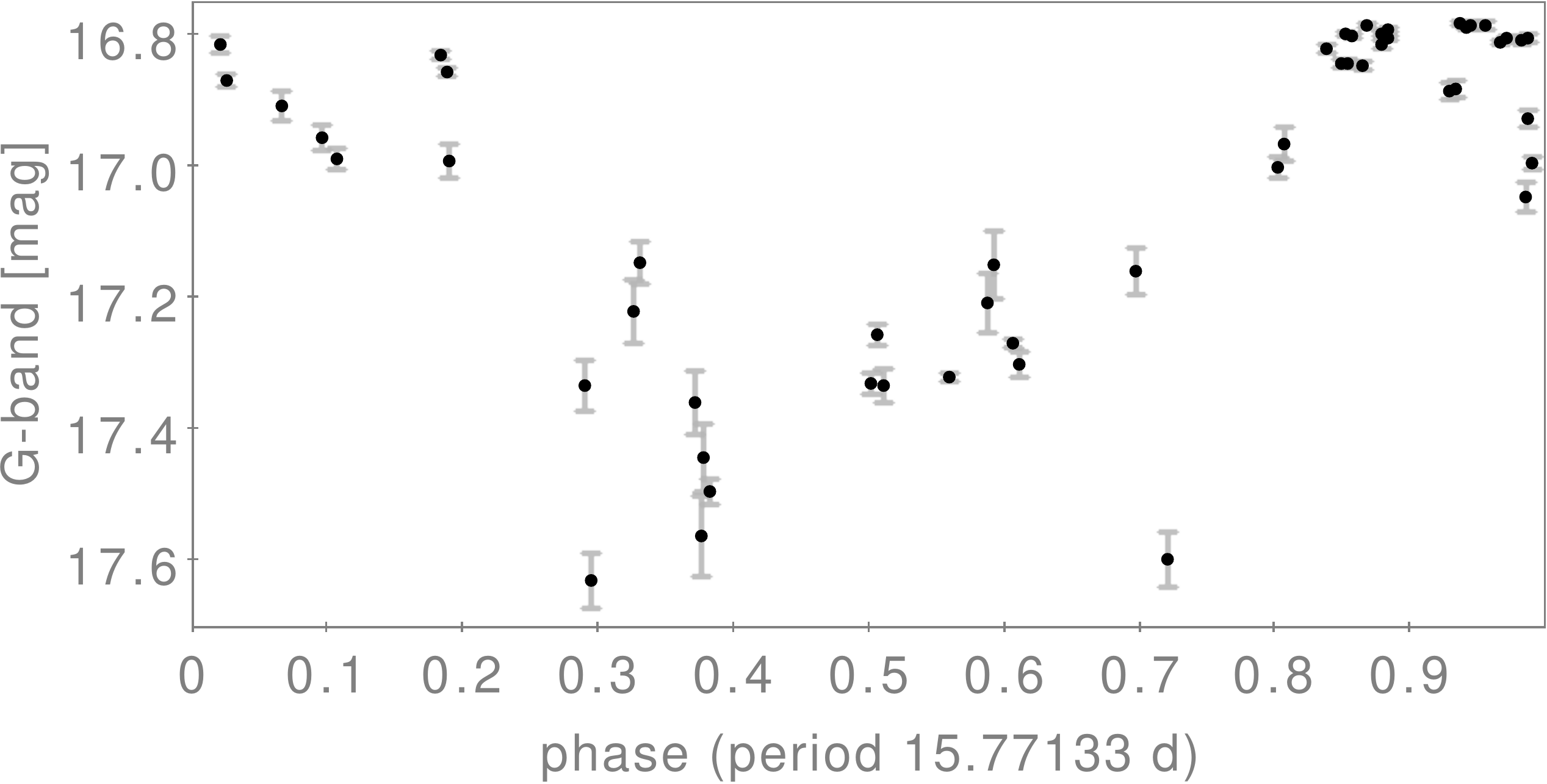}  \\
\vspace{0.1cm}    & 3012937395145831808 & 3047159827703139712 & 3012937223347148800\\
\rotatebox{0}{\normalsize $16.35 \pm  0.2$~d}
  & \includegraphics[height=0.15\textwidth]{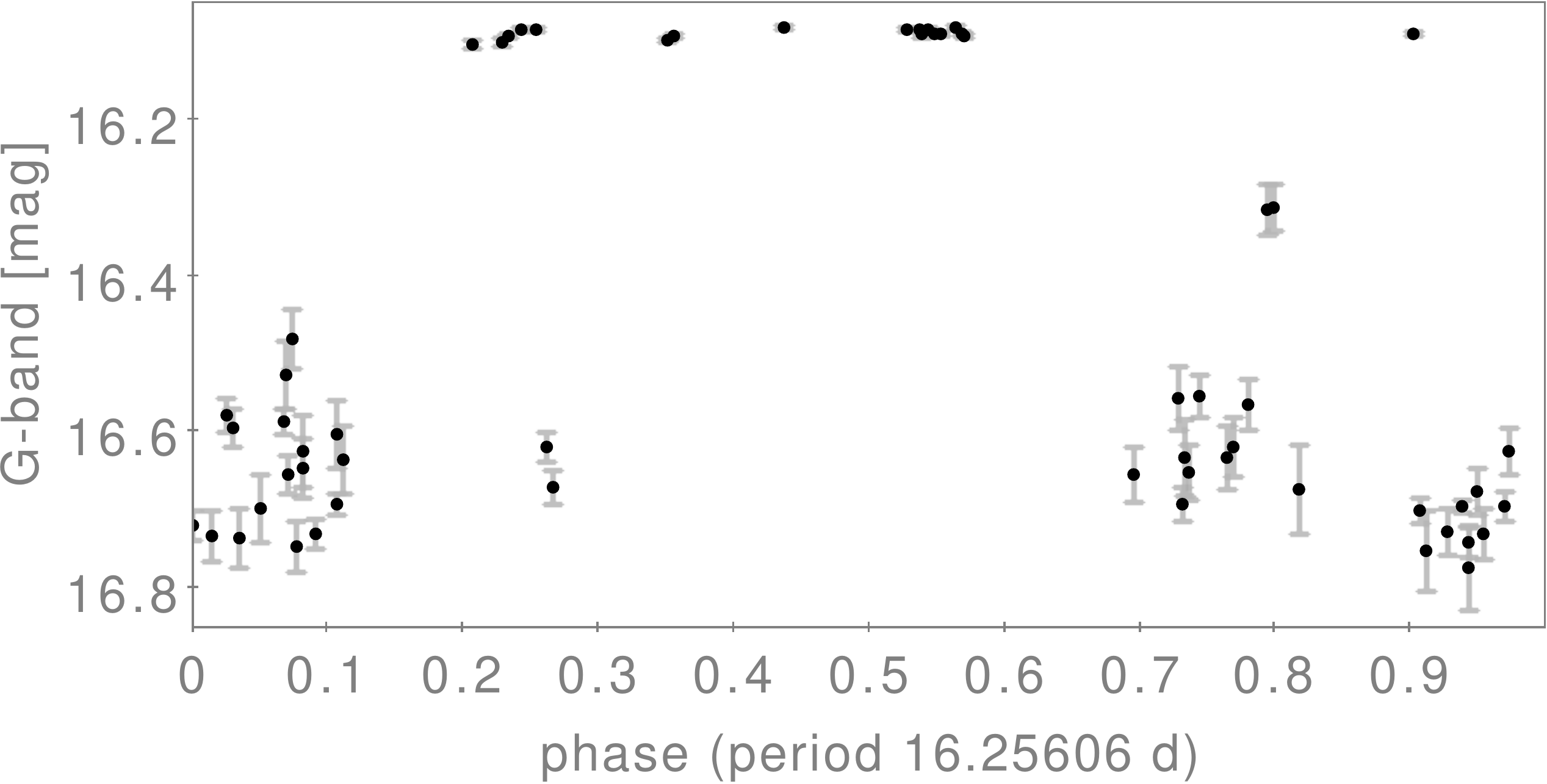}    
  & \includegraphics[trim=1.0cm 0 0 0, clip, height=0.15\textwidth]{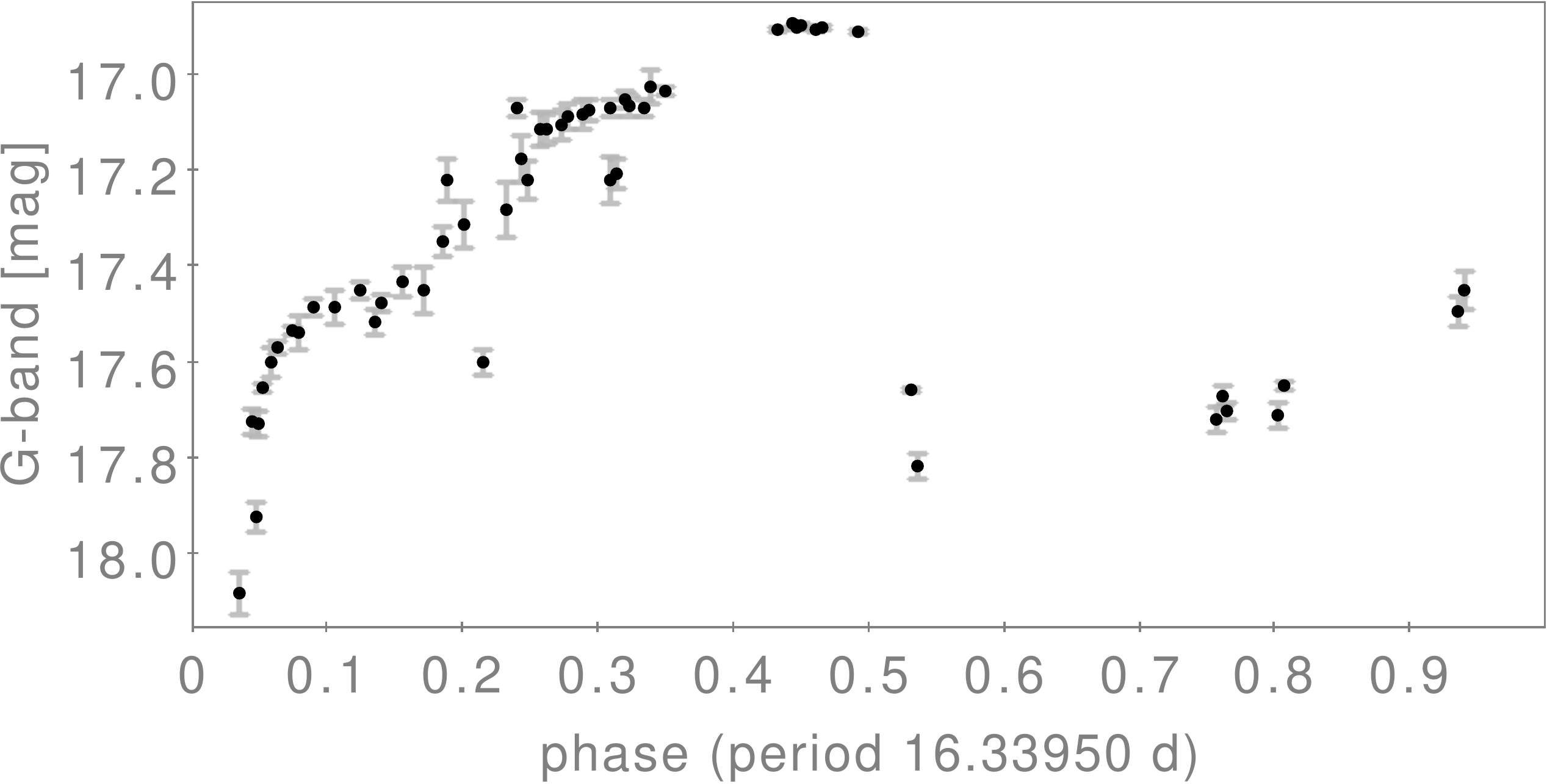}   
  & \includegraphics[trim=1.0cm 0 0 0, clip, height=0.15\textwidth]{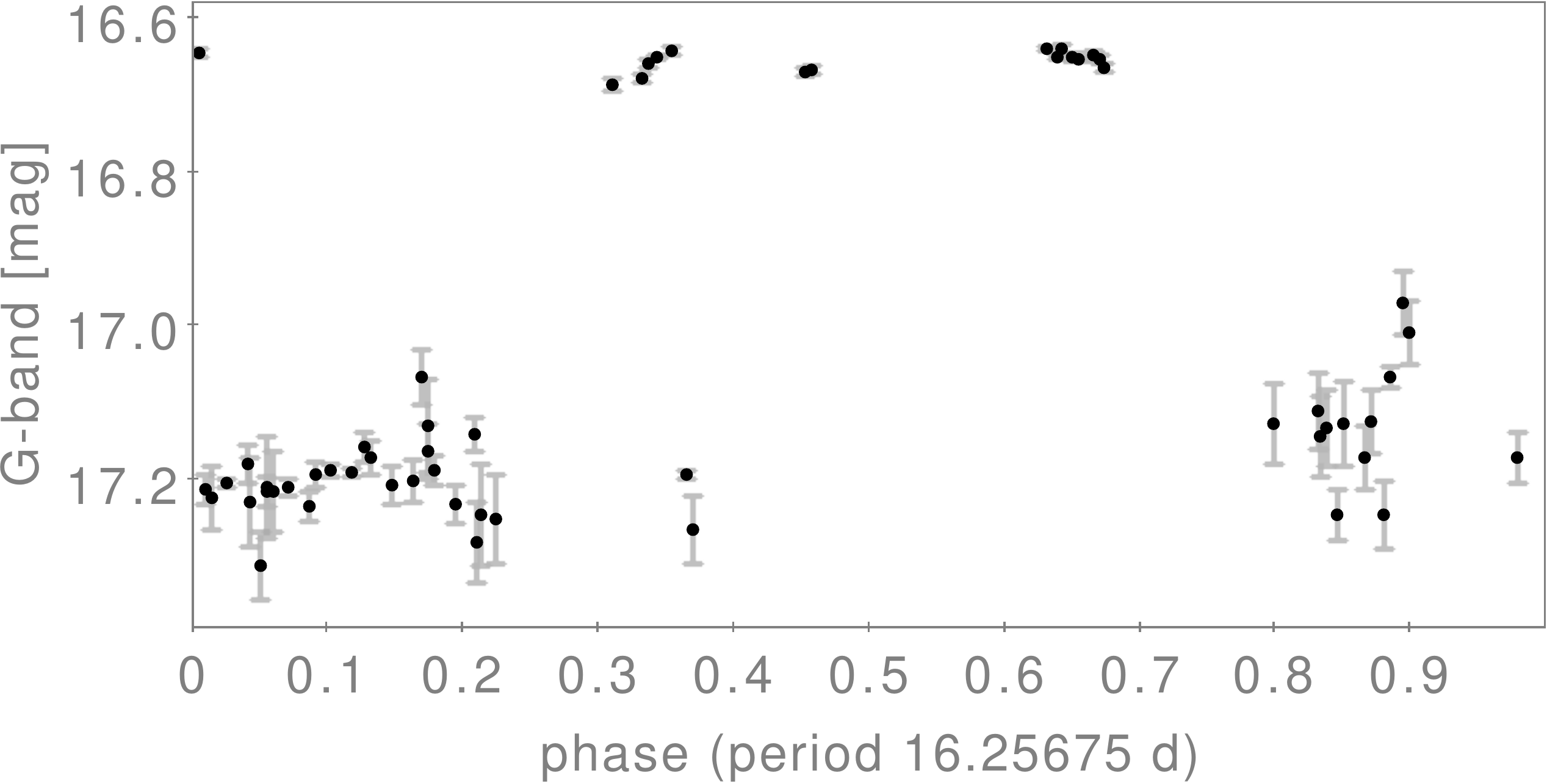}\\
\vspace{0.1cm} & 5893274748891268864 & 6086491132614741632 & 5867210963160082944 \\
\rotatebox{0}{\normalsize $18.8  \pm  0.2$~d}
& \includegraphics[height=0.15\textwidth]{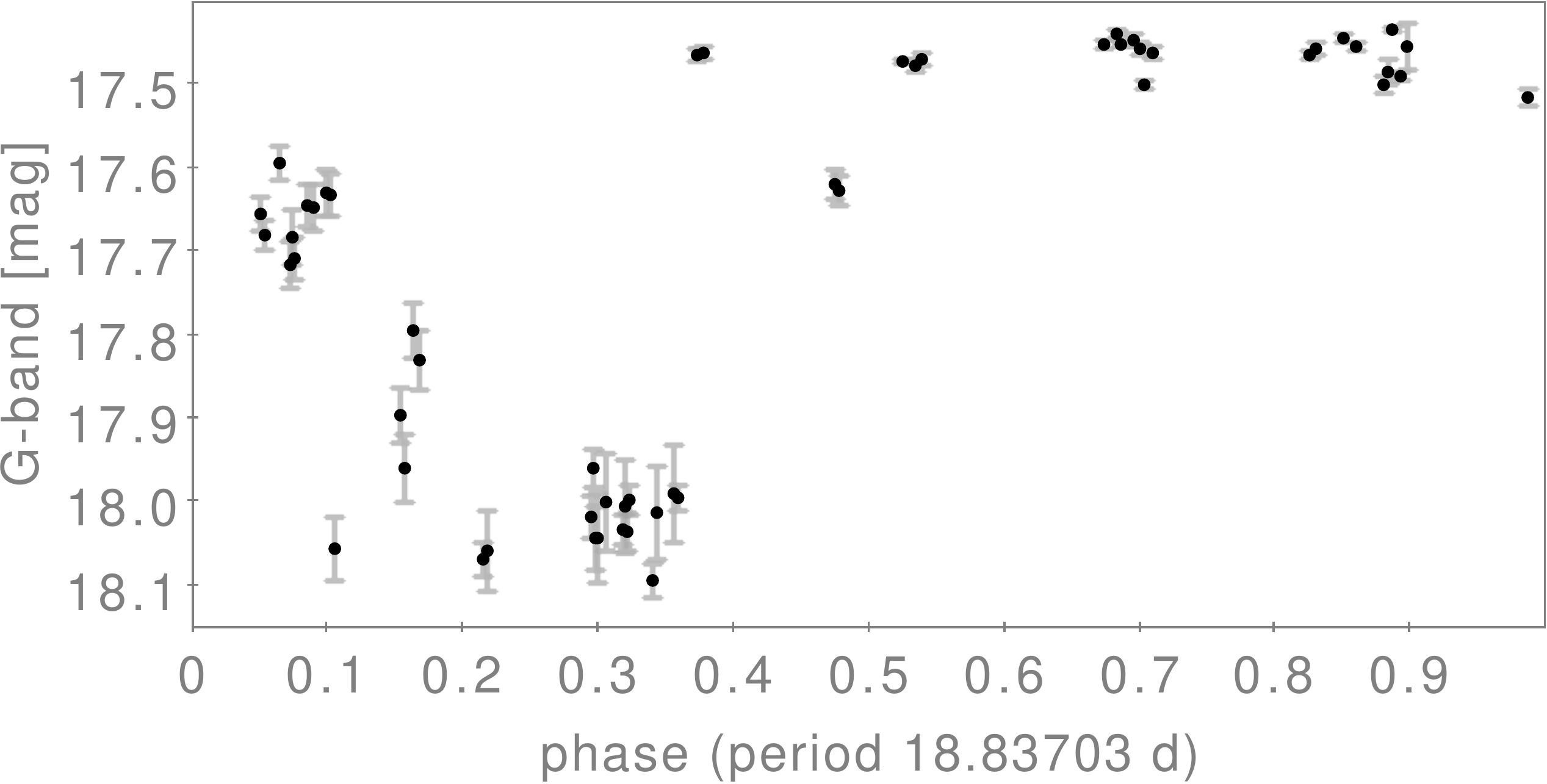}    
  & \includegraphics[trim=1.0cm 0 0 0, clip, height=0.15\textwidth]{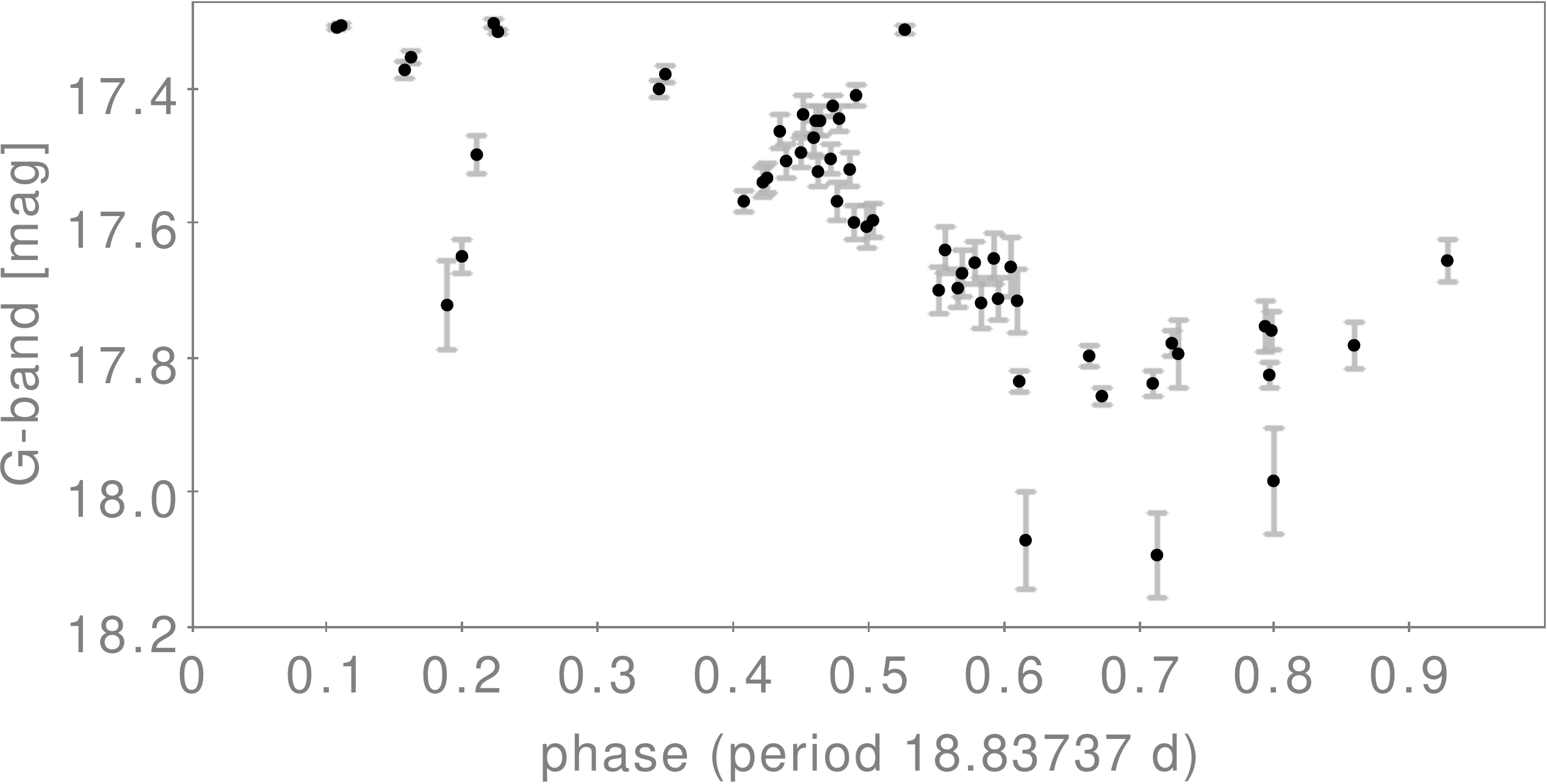}   
  & \includegraphics[trim=1.0cm 0 0 0, clip, height=0.15\textwidth]{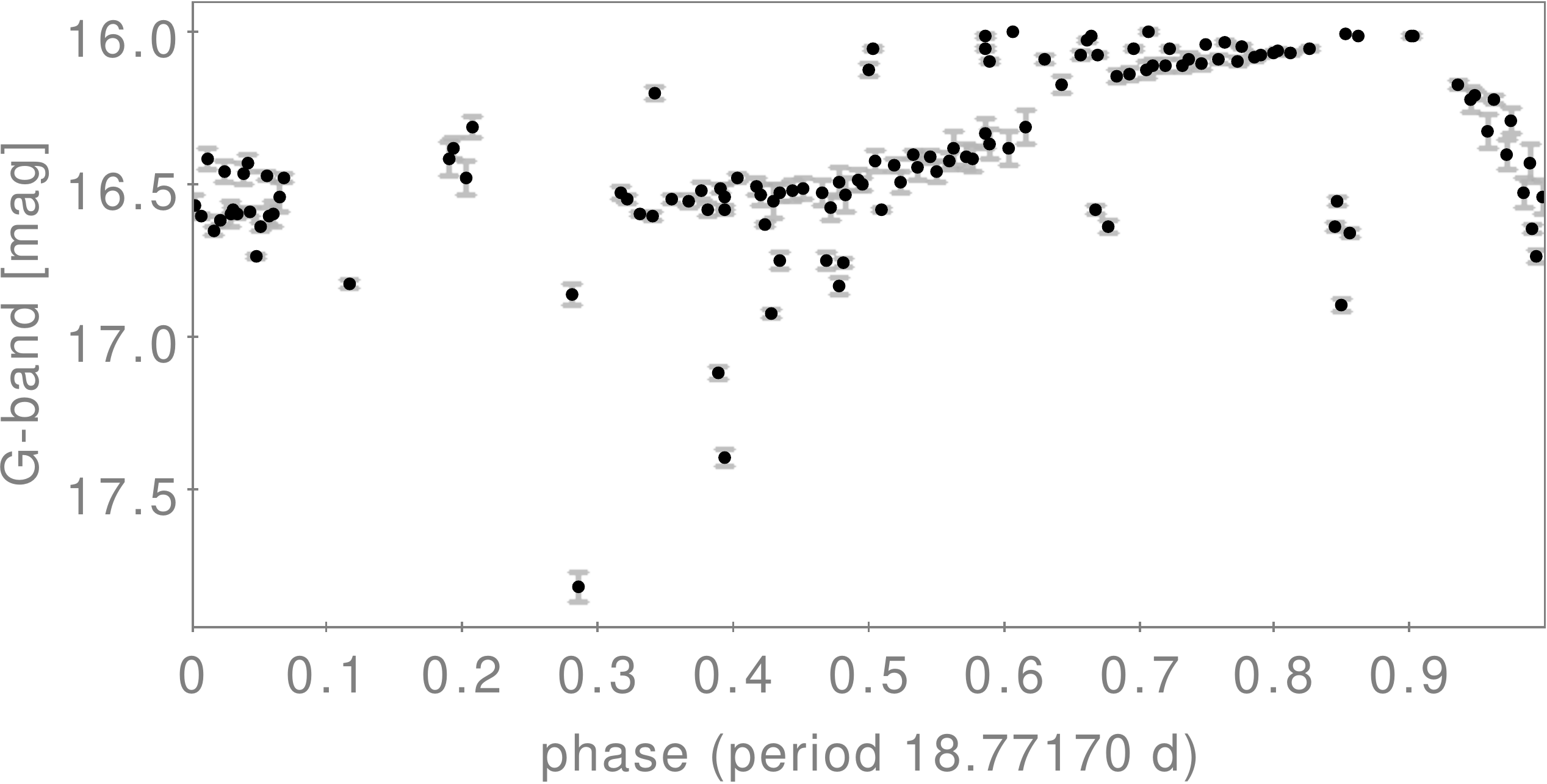}   \\
\vspace{0.1cm}    & 382037139119435008 & 4512271926570691328 & 2929859739665870976  \\
\rotatebox{0}{\normalsize $19.9  \pm  0.2$~d}
  & \includegraphics[height=0.15\textwidth]{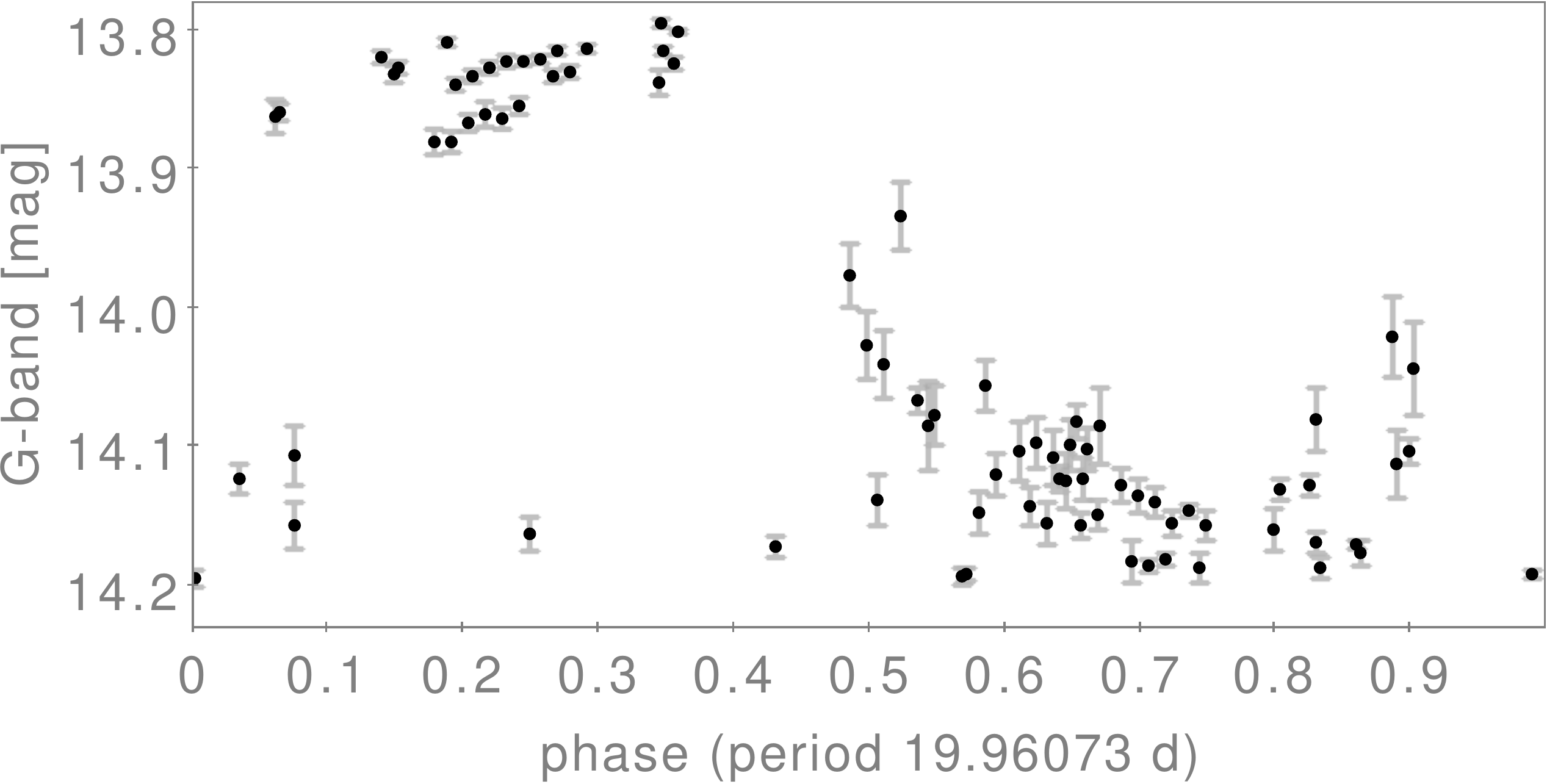}    
  & \includegraphics[trim=1.0cm 0 0 0, clip, height=0.15\textwidth]{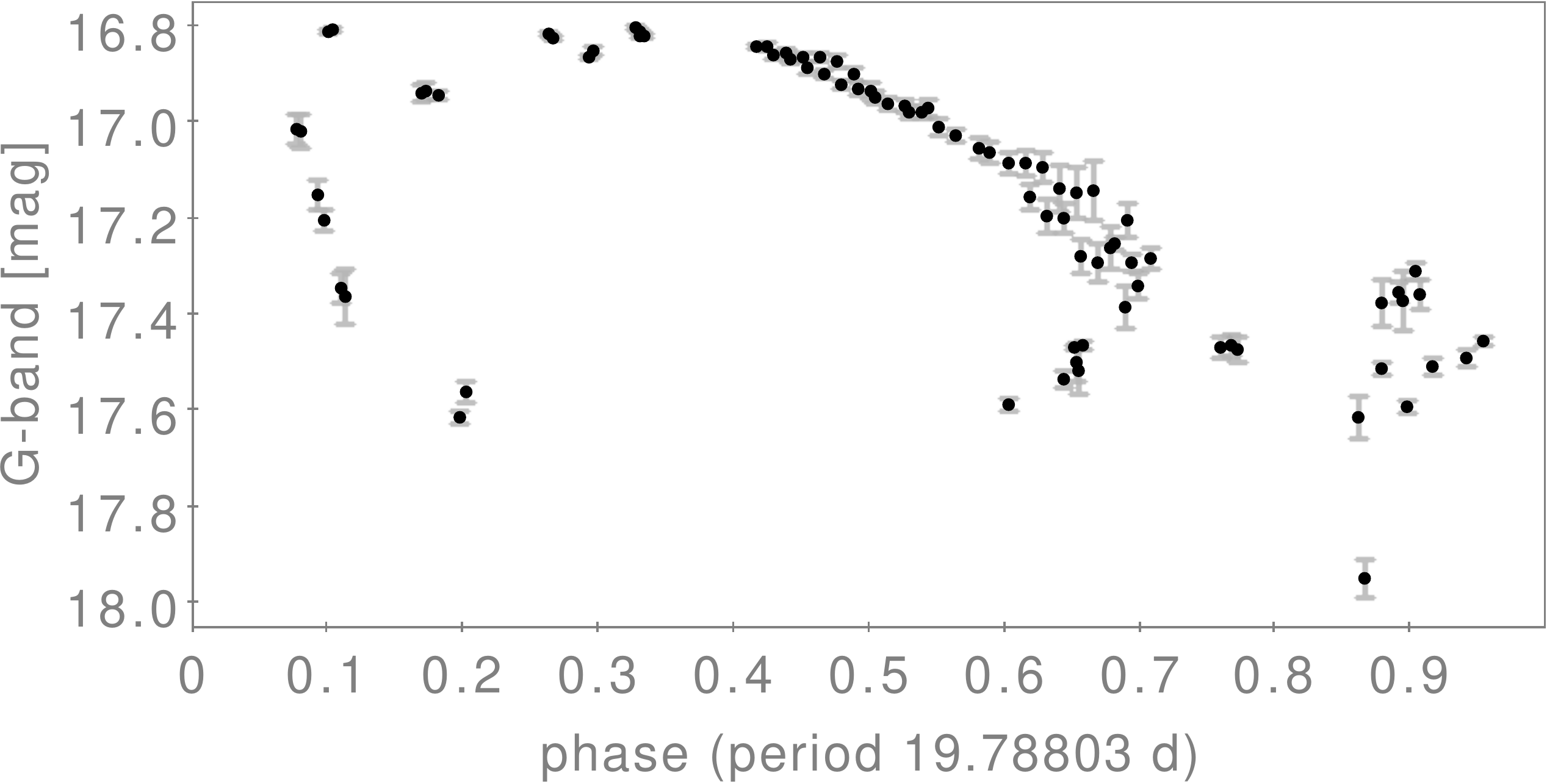}   
  & \includegraphics[trim=1.0cm 0 0 0, clip, height=0.15\textwidth]{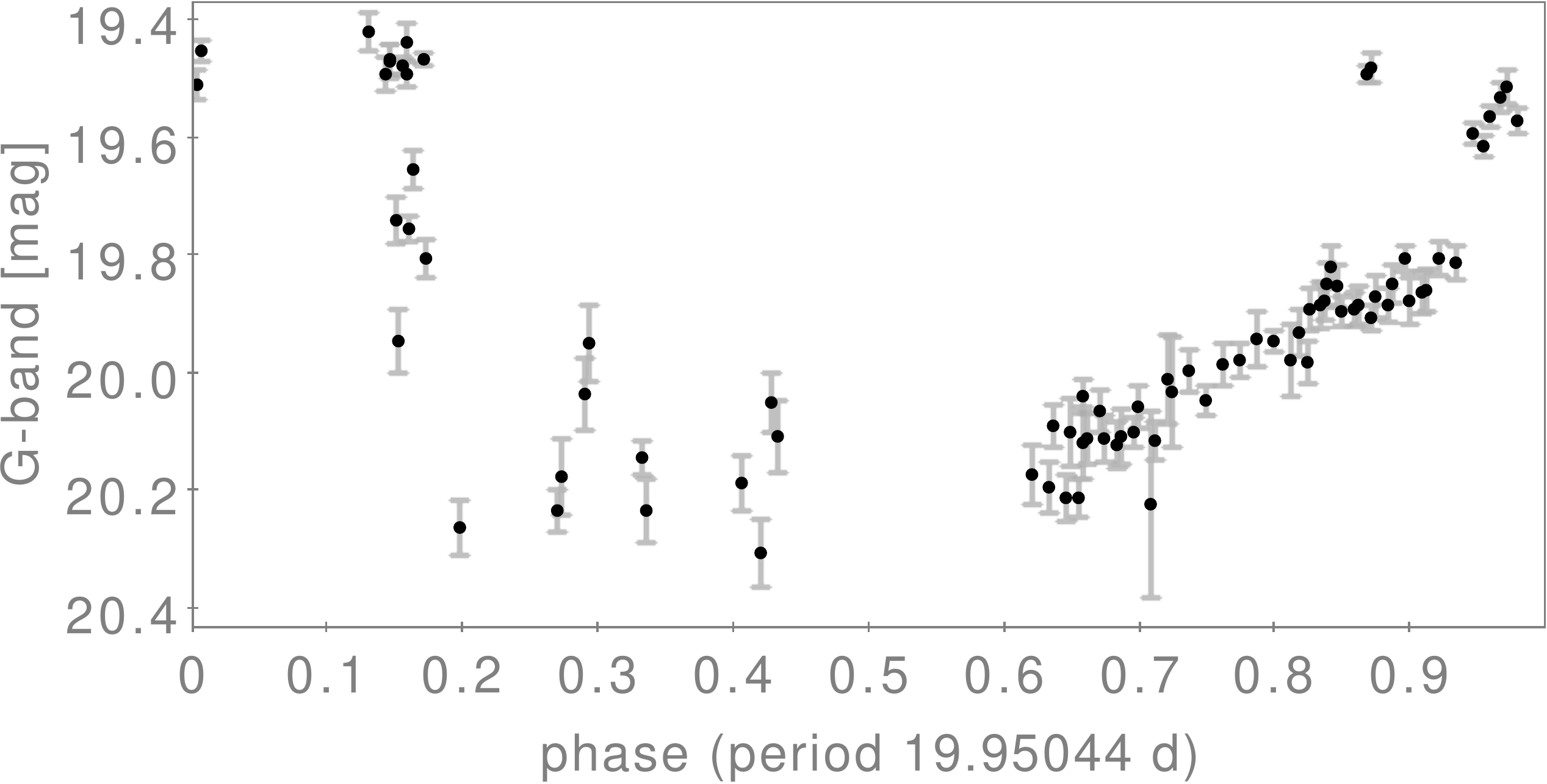} \\
\vspace{0.1cm} & 6814479533711680896 & 5629197413336858752 &  6076139338033445248 \\
\rotatebox{0}{\normalsize $25.1  \pm  0.3$~d}
  & \includegraphics[height=0.15\textwidth]{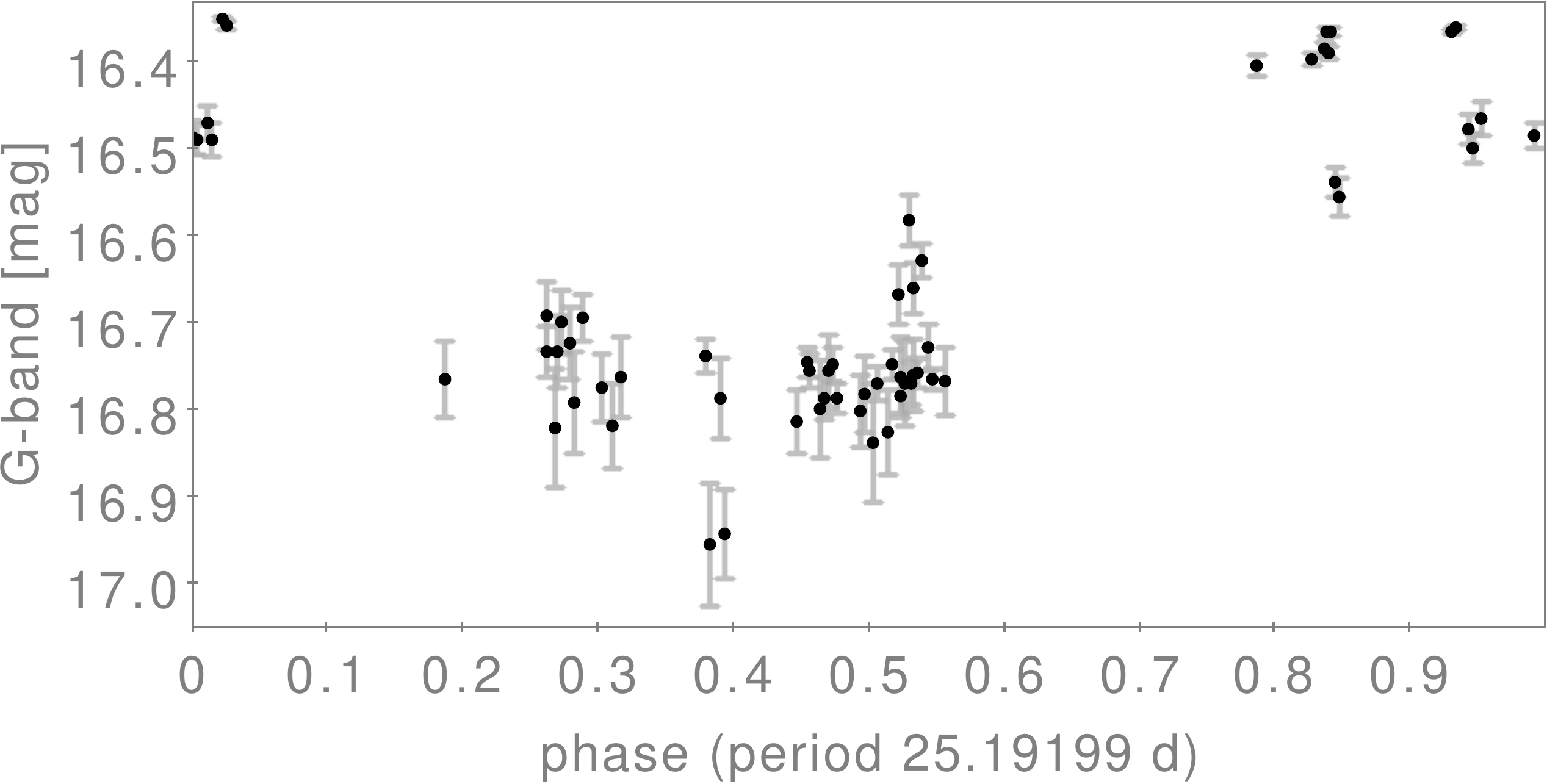}    
  & \includegraphics[trim=1.0cm 0 0 0, clip, height=0.15\textwidth]{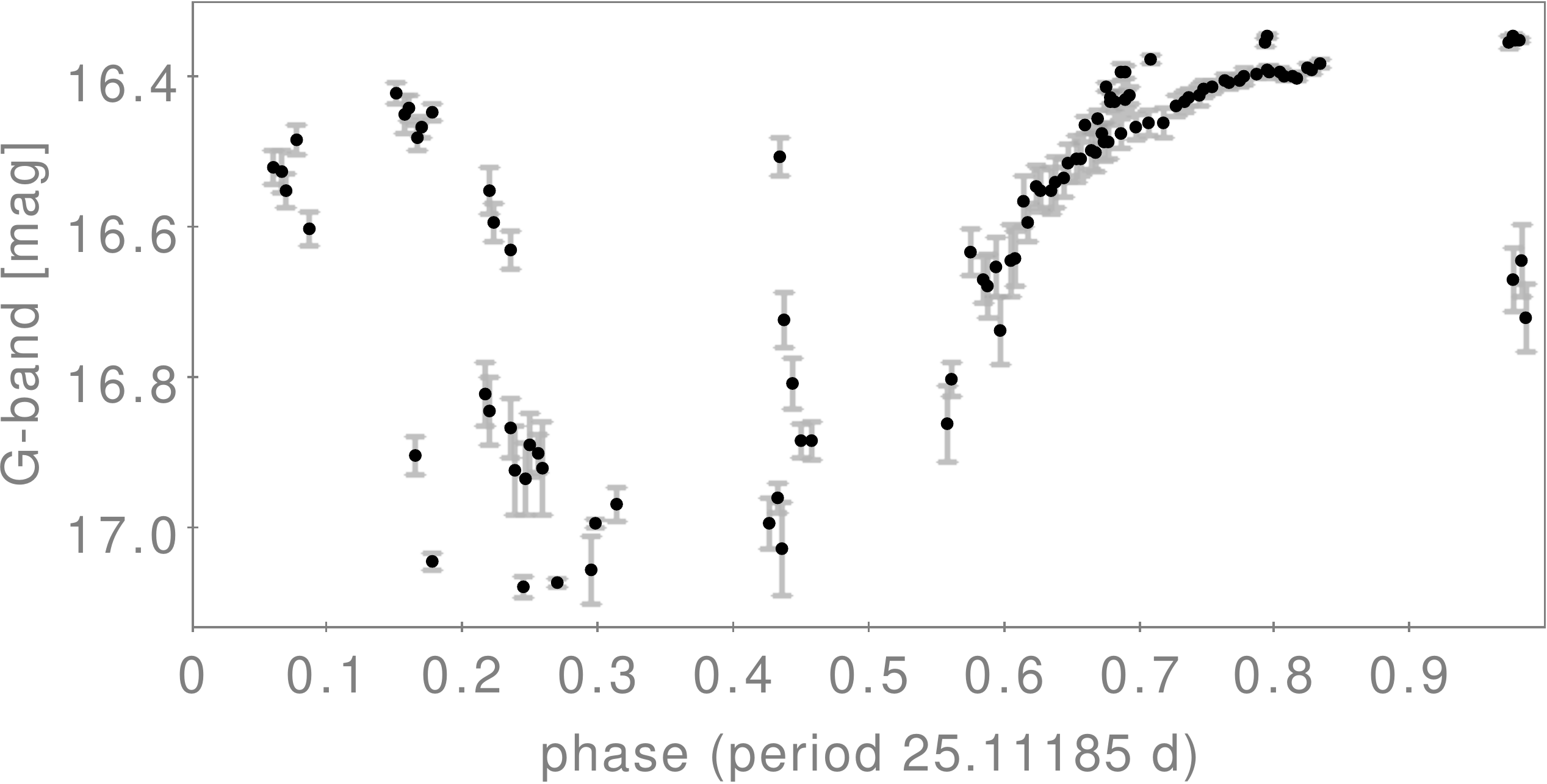}   
  & \includegraphics[trim=1.0cm 0 0 0, clip, height=0.15\textwidth]{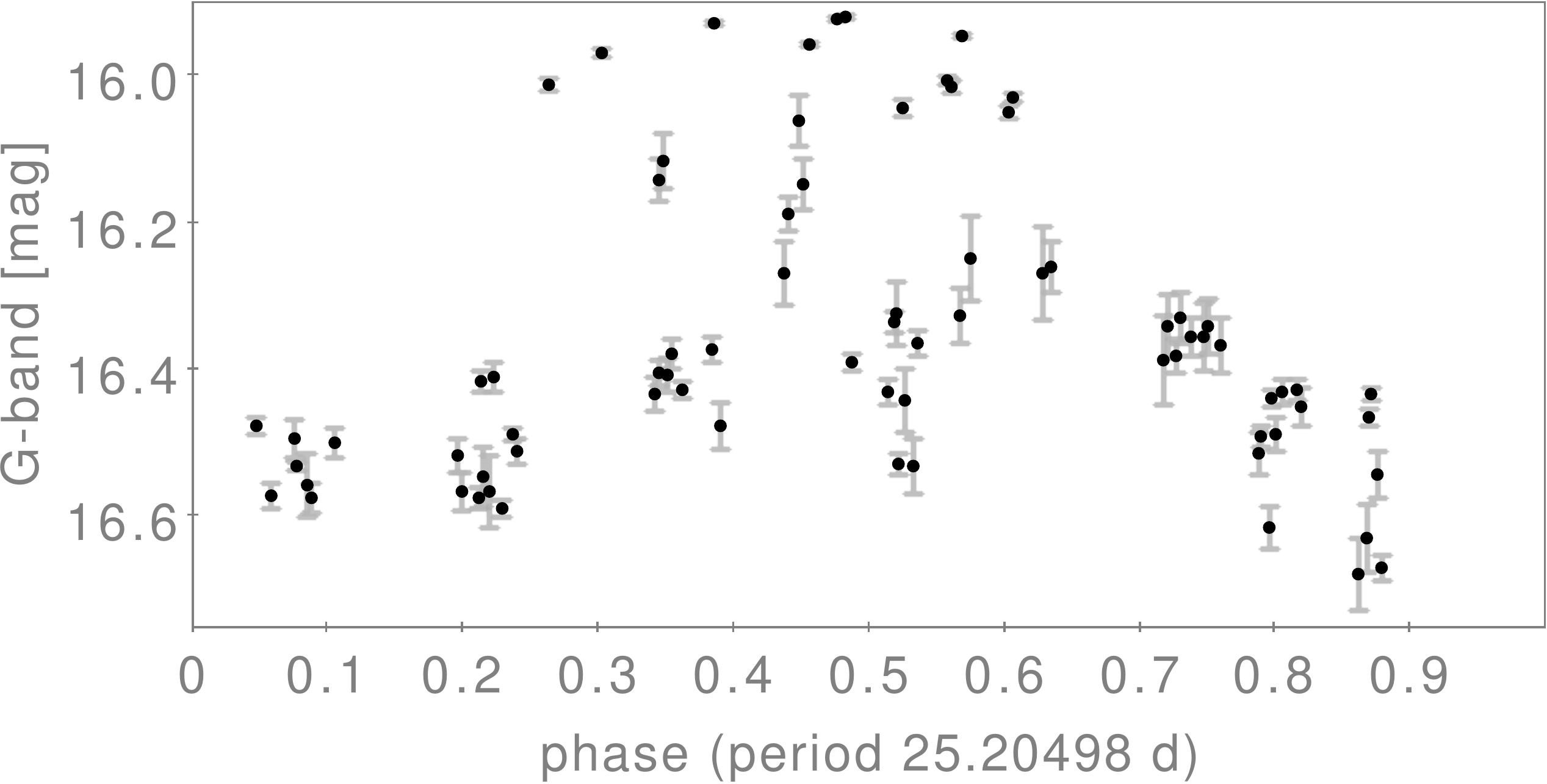}    \\
\vspace{0.1cm}  & 367388551858425344*& 380538569192874112*& 385844060692409344*\\
 \rotatebox{0}{\normalsize $26.9  \pm  0.4$~d}
  & \includegraphics[height=0.15\textwidth]{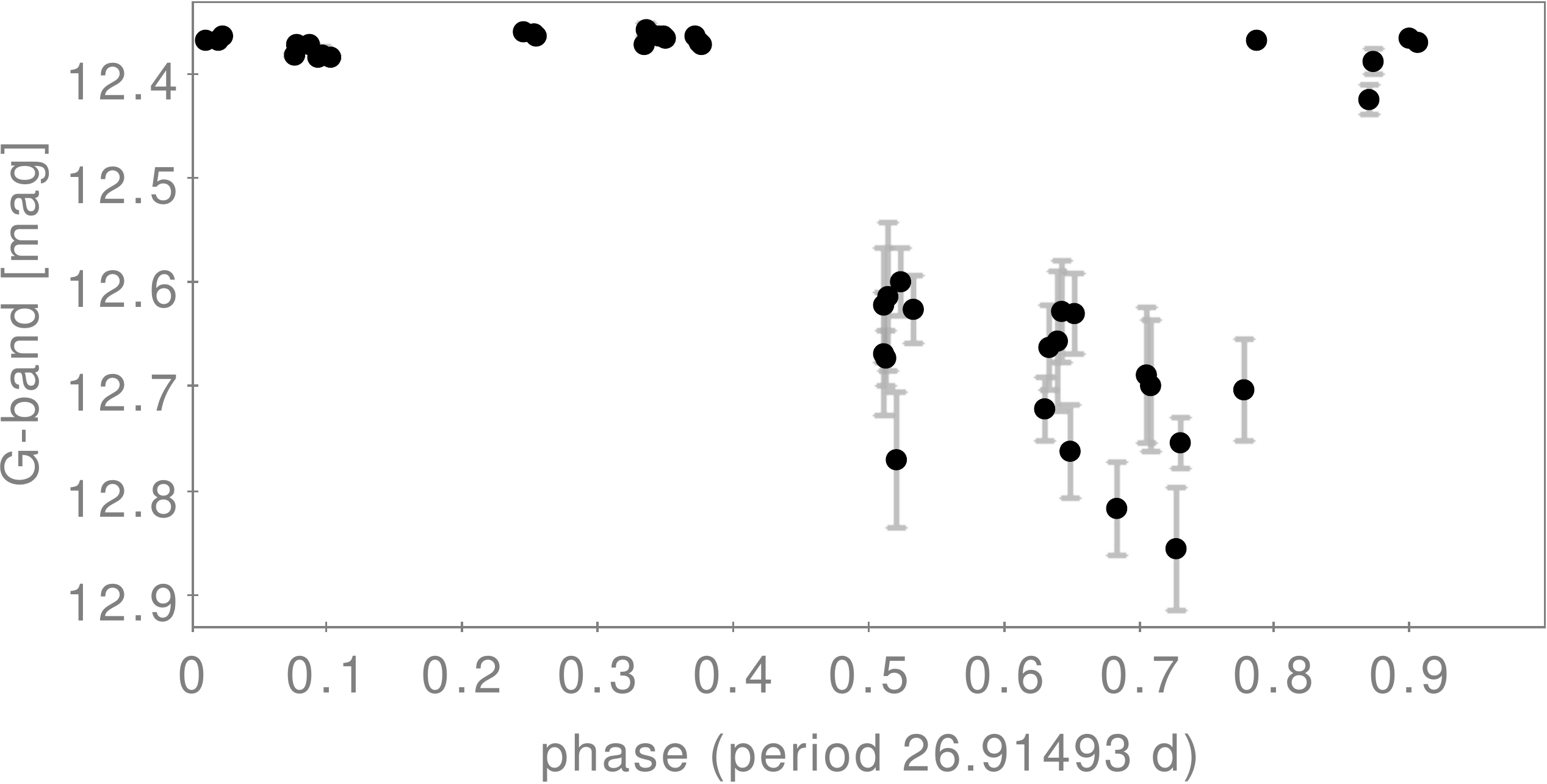}    
  & \includegraphics[trim=1.0cm 0 0 0, clip, height=0.15\textwidth]{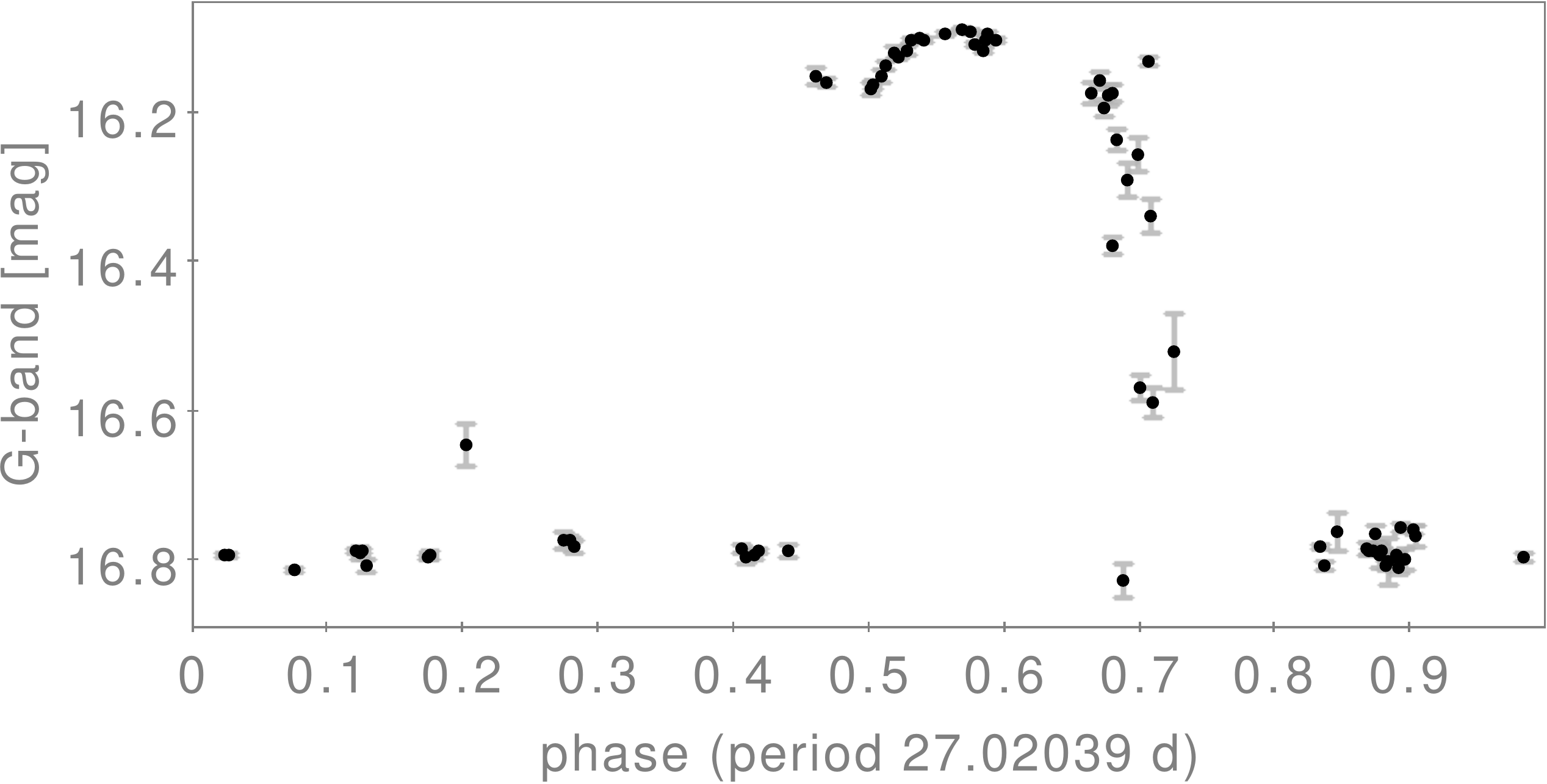}   
  & \includegraphics[trim=1.0cm 0 0 0, clip, height=0.15\textwidth]{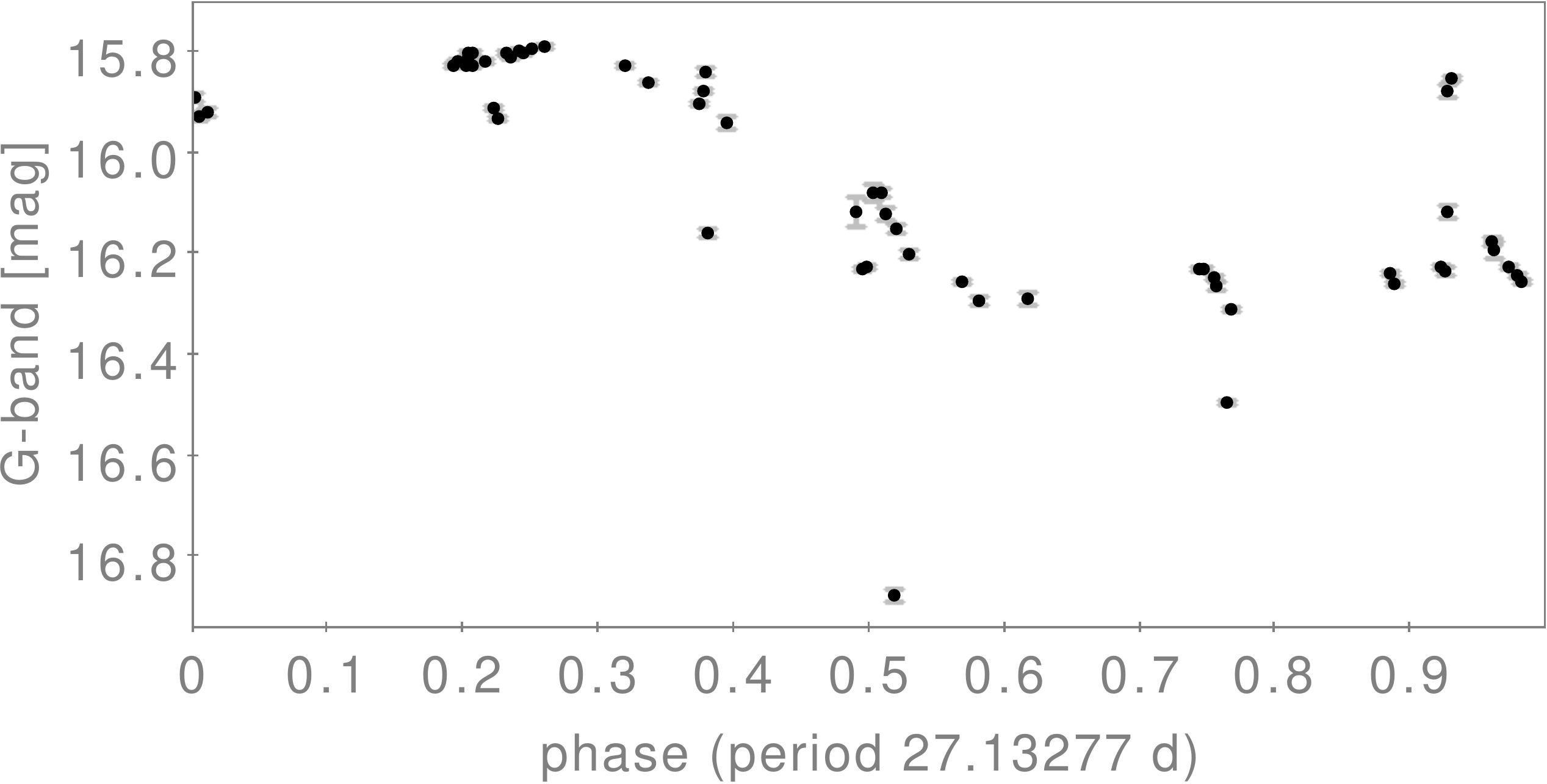} \\
\end{tabular}
\caption{Example public folded G-band light curves for different period peaks, derived by the multi-harmonic modelling following a generalised least-squares period search as described in Sect.~\ref{ssec:obsPerDistr}. The source id is provided in the top right corner. Additional information is available in Table~\ref{tab:ipdSignalsList} for sources with an asterisk. In general, the sources either have very high \rIpdG, \rExfG, or \aG significance (but always a low false-alarm probability).}
\label{fig:foldedPeriodGExamples1}
\end{small}
\end{figure*}

\begin{figure*}[h]
\begin{small}
\setlength{\tabcolsep}{0pt} 
\vspace{-0.4cm}
\begin{tabular}{L{0.12\textwidth}R{0.3\textwidth}R{0.29\textwidth}R{0.29\textwidth}}
\renewcommand{\arraystretch}{0} 
&  366951667785042688* & 382159975182923264* & 383556286230747520* \\
\rotatebox{0}{\normalsize $31.5  \pm  0.6$~d} 
  & \includegraphics[height=0.15\textwidth]{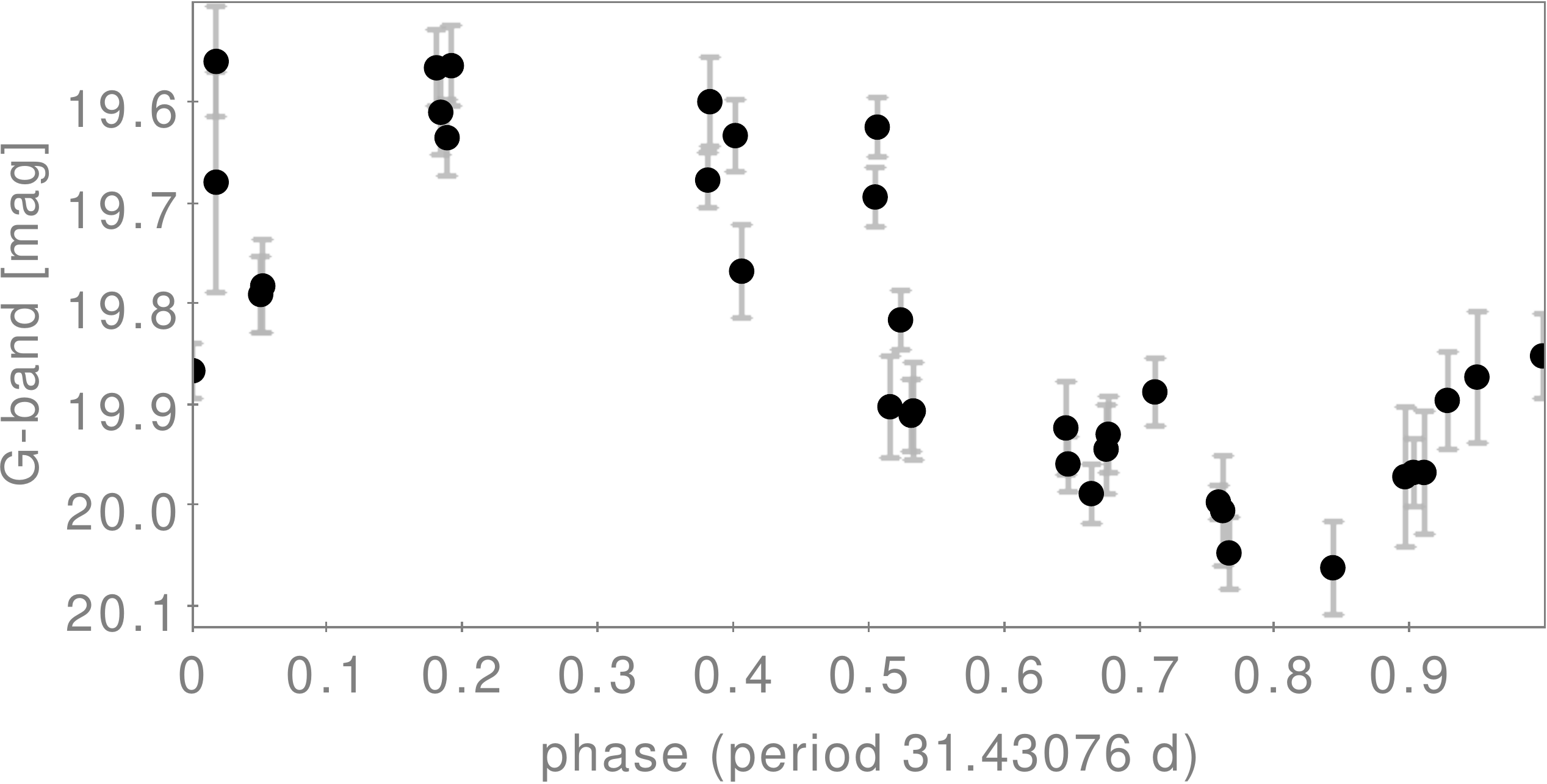}       
  & \includegraphics[trim=1.0cm 0 0 0, clip, height=0.15\textwidth]{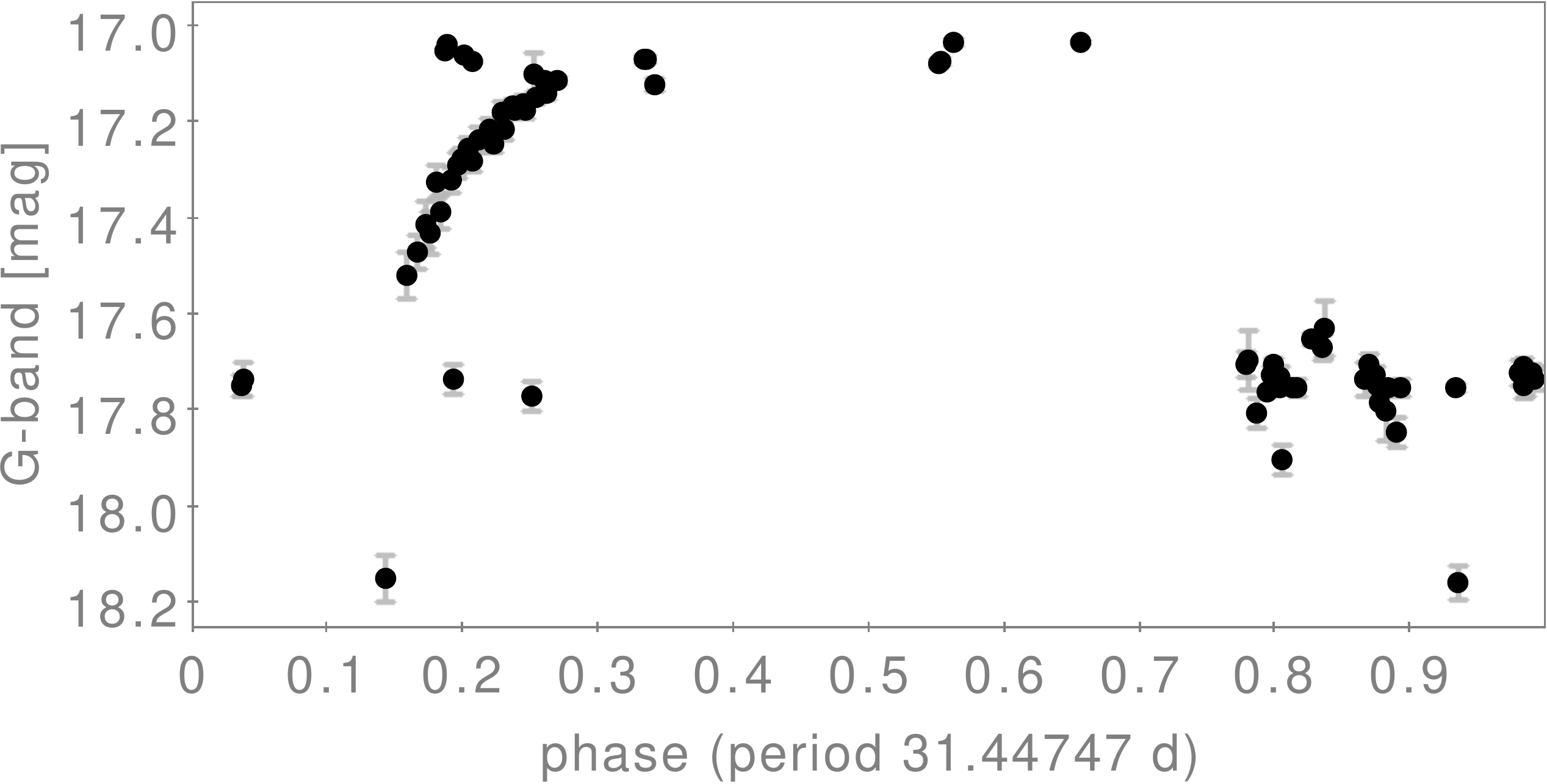}    
  & \includegraphics[trim=1.0cm 0 0 0, clip, height=0.15\textwidth]{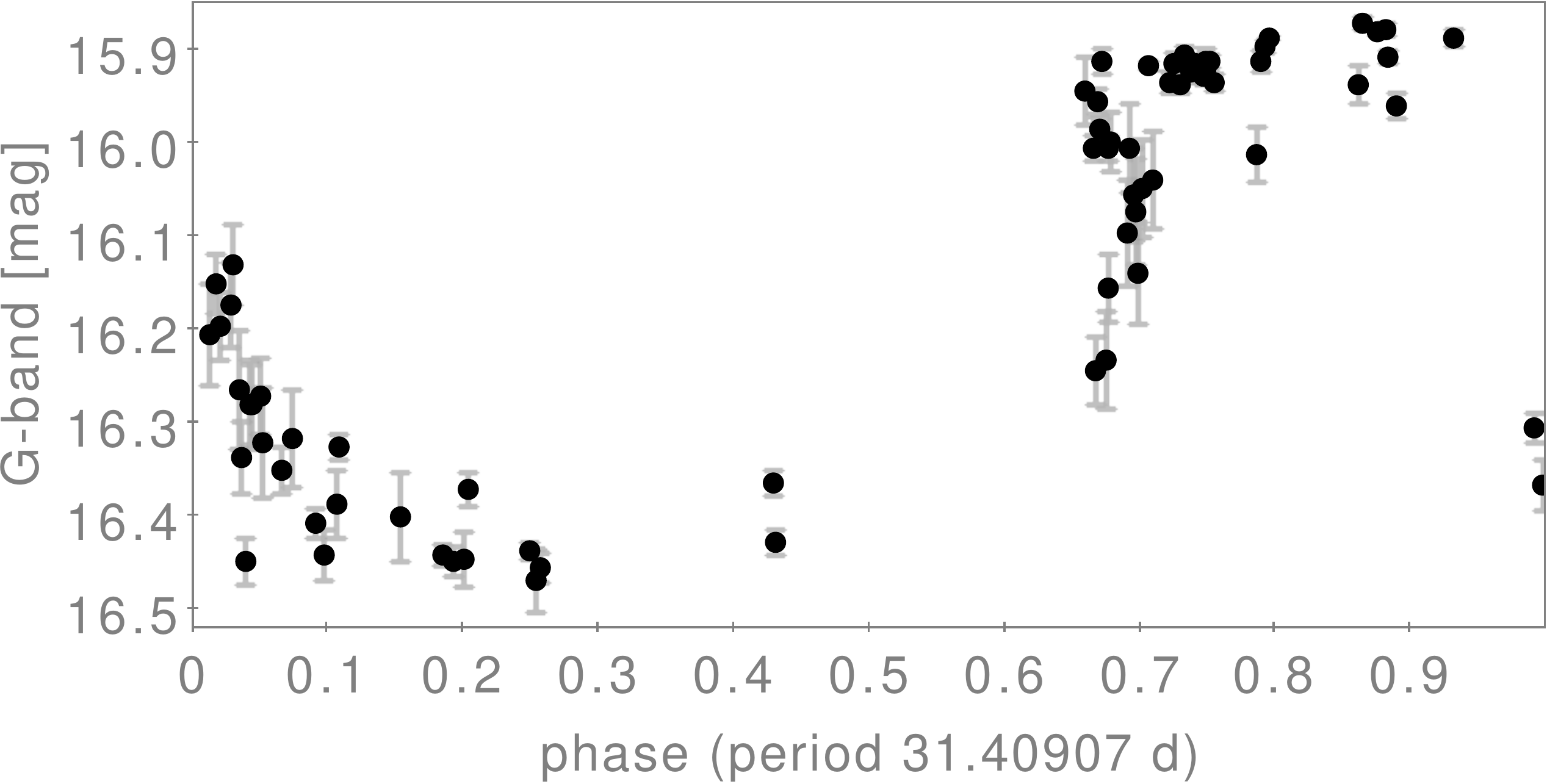}    
  \\
& 6072038743786686592   & 375565821697320320 &  4506964820148820864  \\
\rotatebox{0}{\normalsize $41.9  \pm  1.0$~d}  
  & \includegraphics[height=0.15\textwidth]{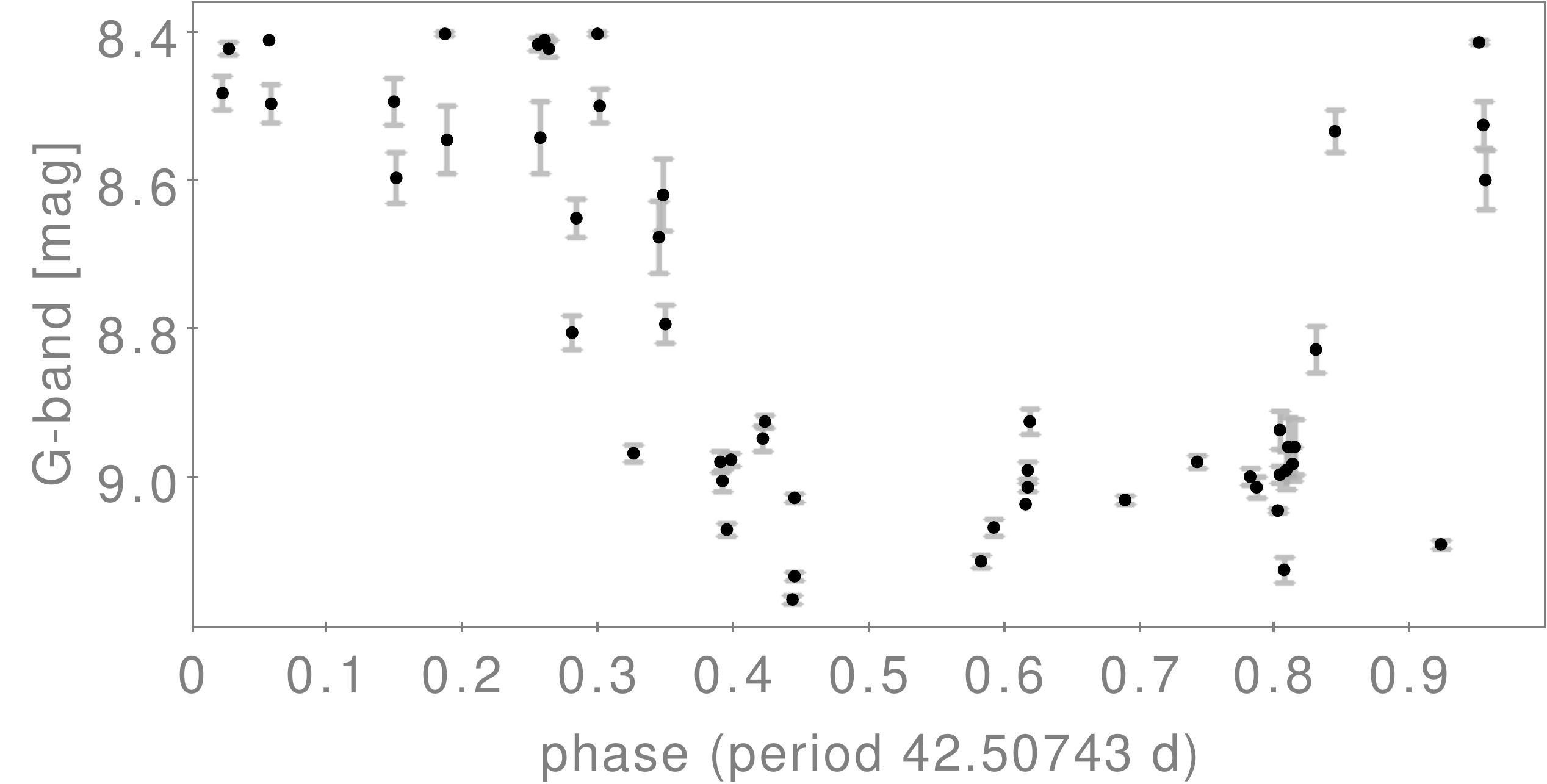}    
  & \includegraphics[trim=1.0cm 0 0 0, clip, height=0.15\textwidth]{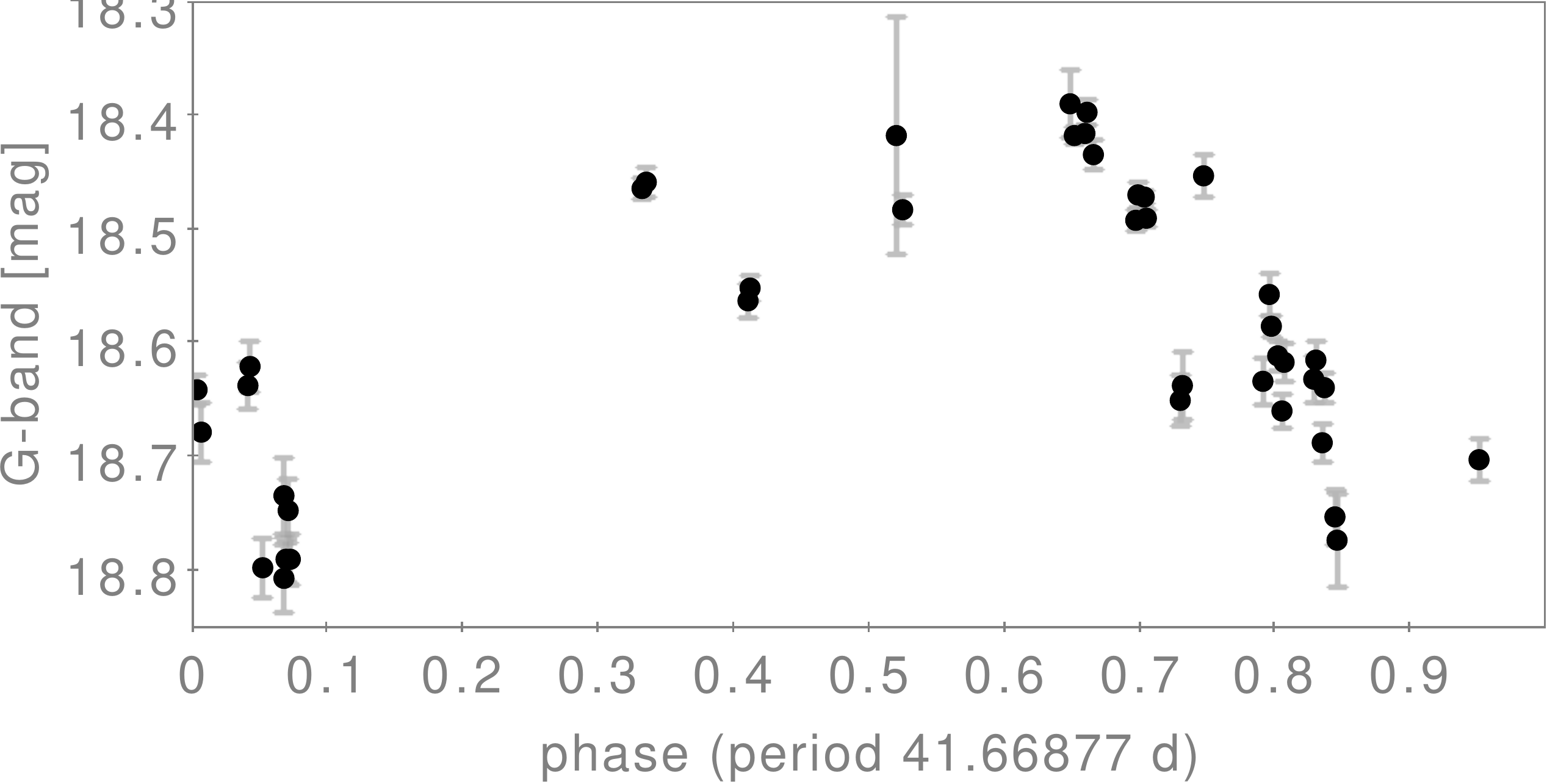}   
  & \includegraphics[trim=1.0cm 0 0 0, clip, height=0.15\textwidth]{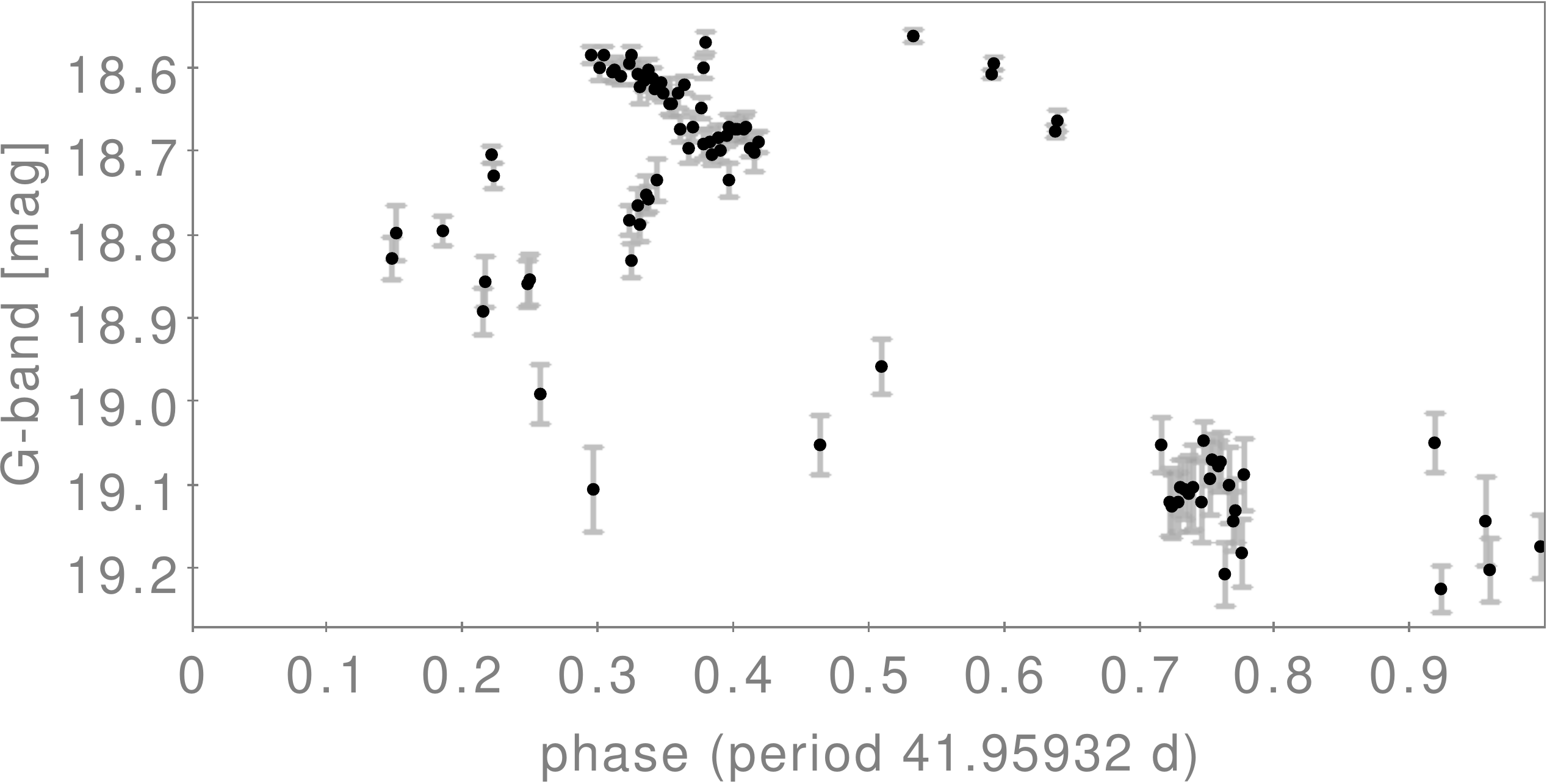}   \\
& 4370876960309439616 & 5533716682576596352 & 2927496884874462336 \\
\rotatebox{0}{\normalsize $46.8  \pm  1.1$~d} 
  & \includegraphics[height=0.15\textwidth]{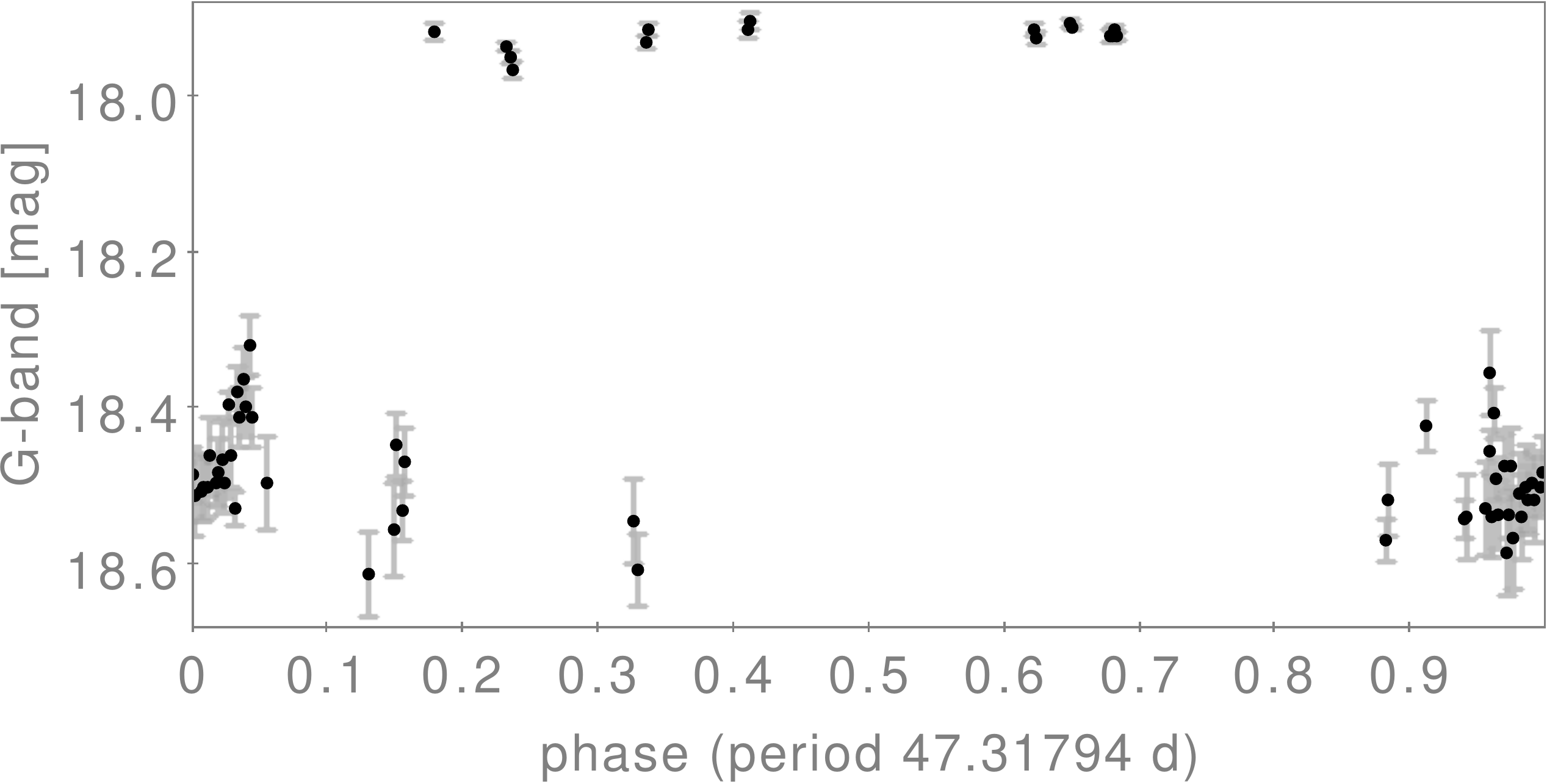}    
  & \includegraphics[trim=1.0cm 0 0 0, clip, height=0.15\textwidth]{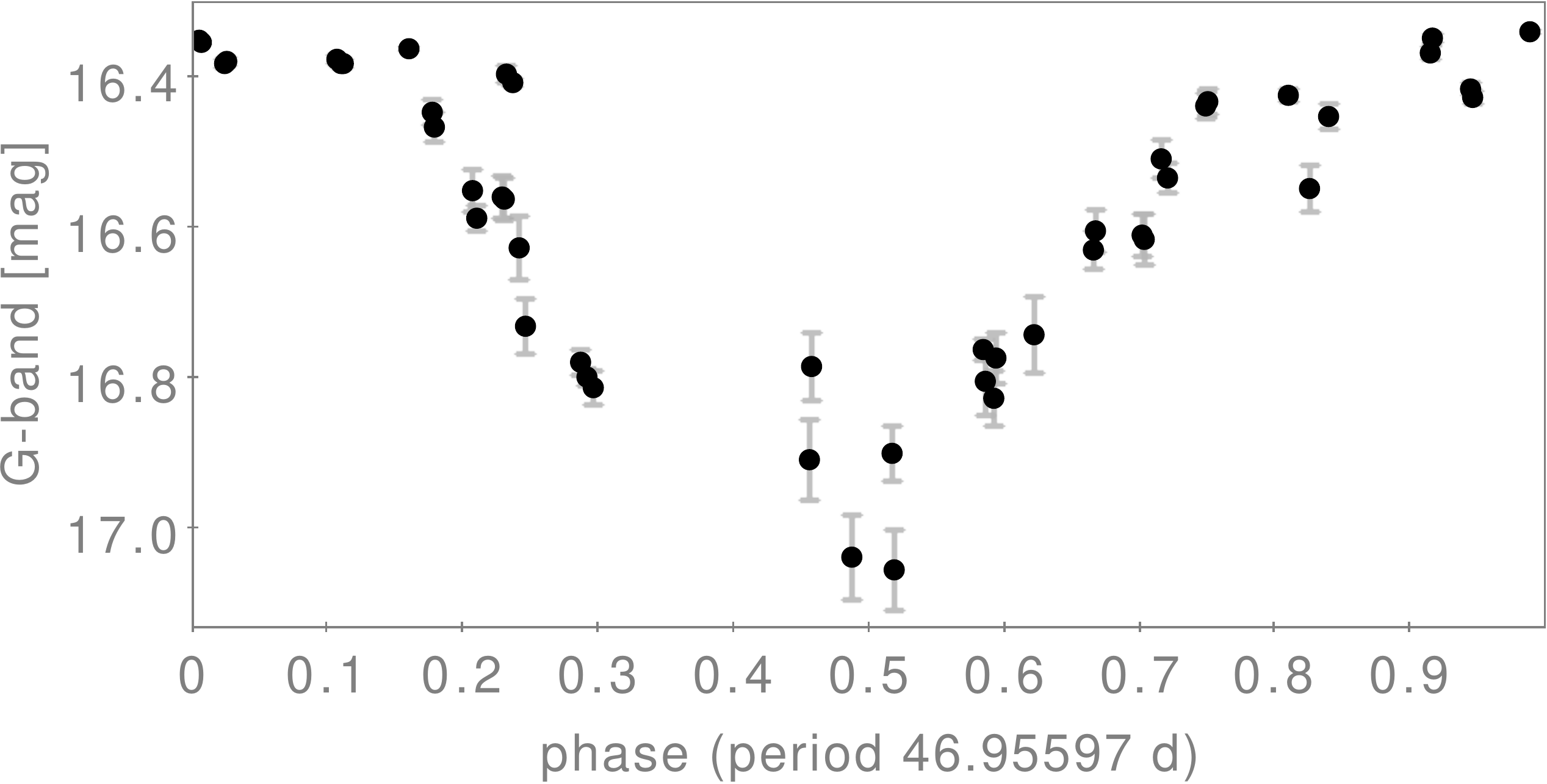}   
  & \includegraphics[trim=1.0cm 0 0 0, clip, height=0.15\textwidth]{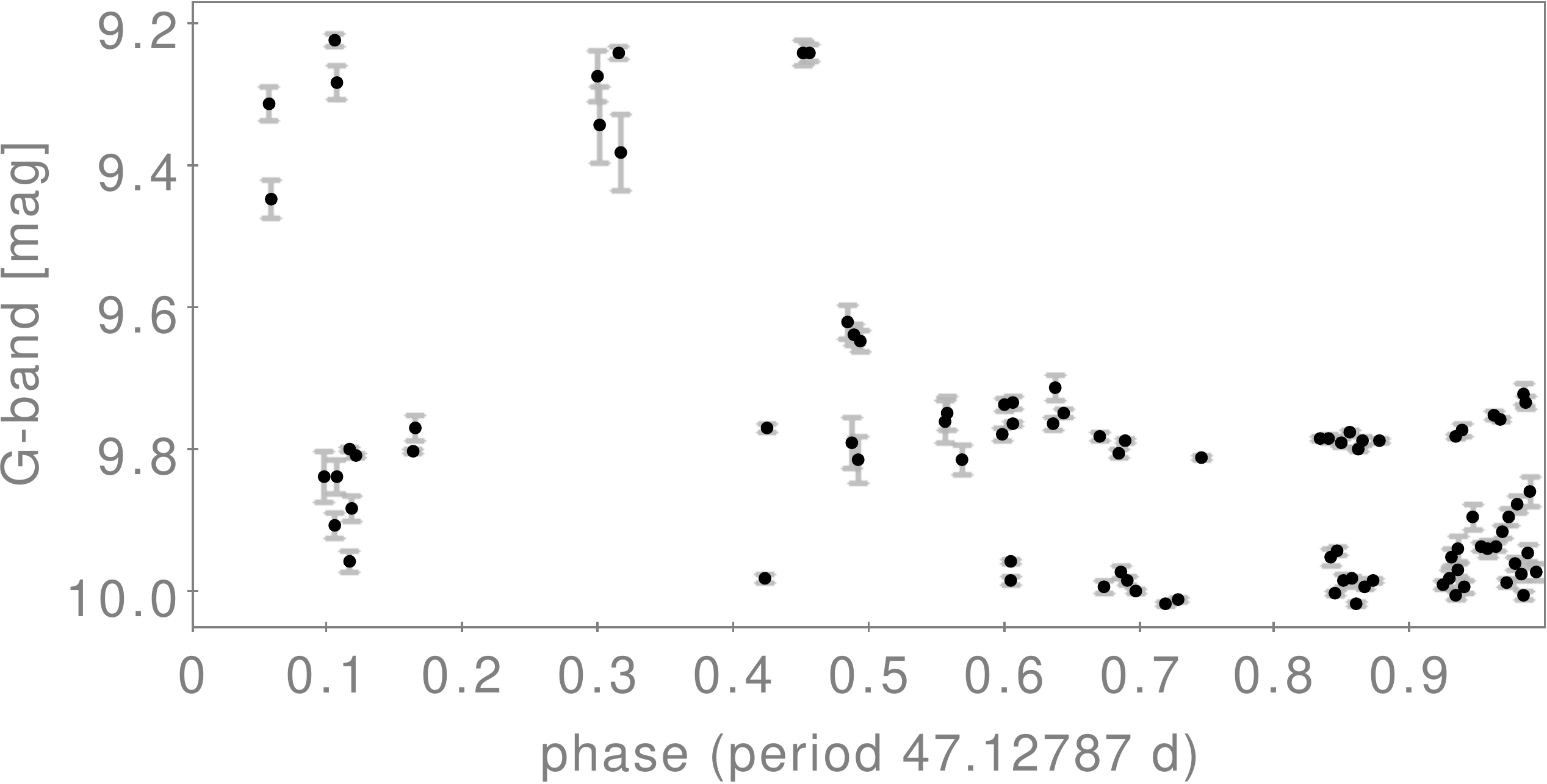} \\
 & 378810450446502400*& 389636619892245248*& 376045247423005184*\\
\rotatebox{0}{\normalsize $53.7  \pm  1.1$~d}
  & \includegraphics[height=0.15\textwidth]{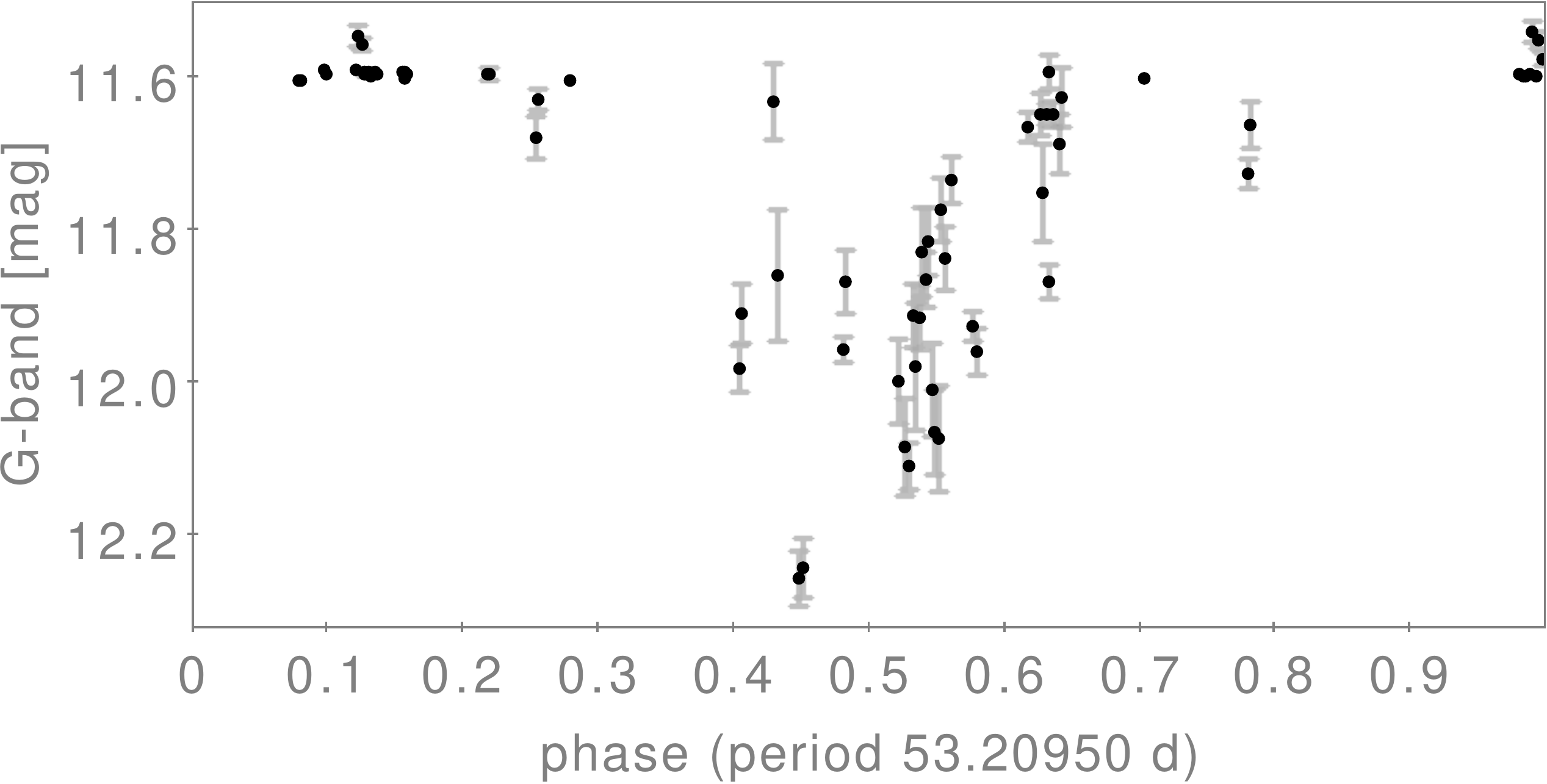}    
  & \includegraphics[trim=1.0cm 0 0 0, clip, height=0.15\textwidth]{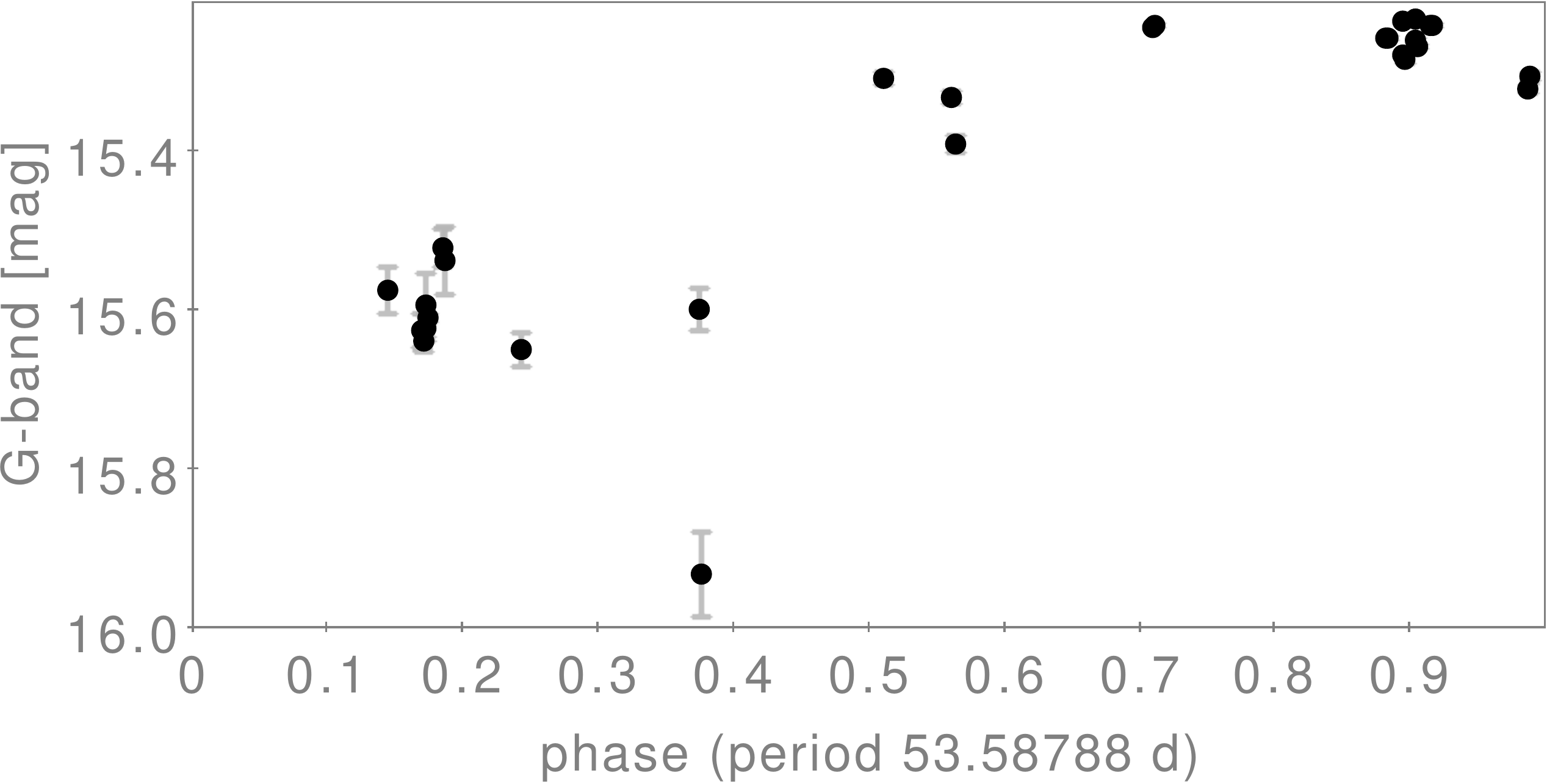}   
  & \includegraphics[trim=1.0cm 0 0 0, clip, height=0.15\textwidth]{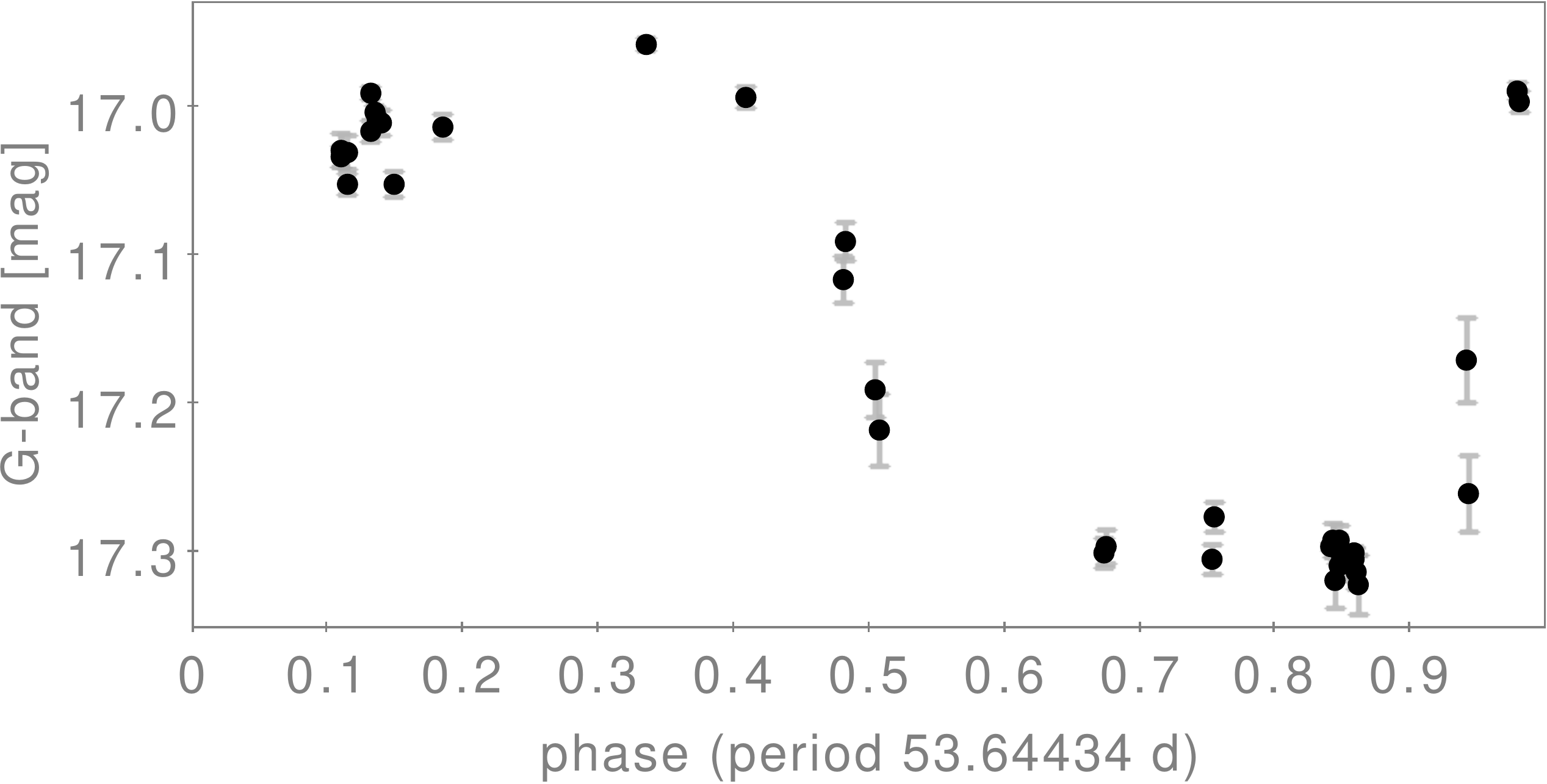}  \\
 & 388466602081536640*& 235109950053908096 & 2940536478605189504\\
 \rotatebox{0}{\normalsize $76.1  \pm  1.7$~d}
  & \includegraphics[height=0.15\textwidth]{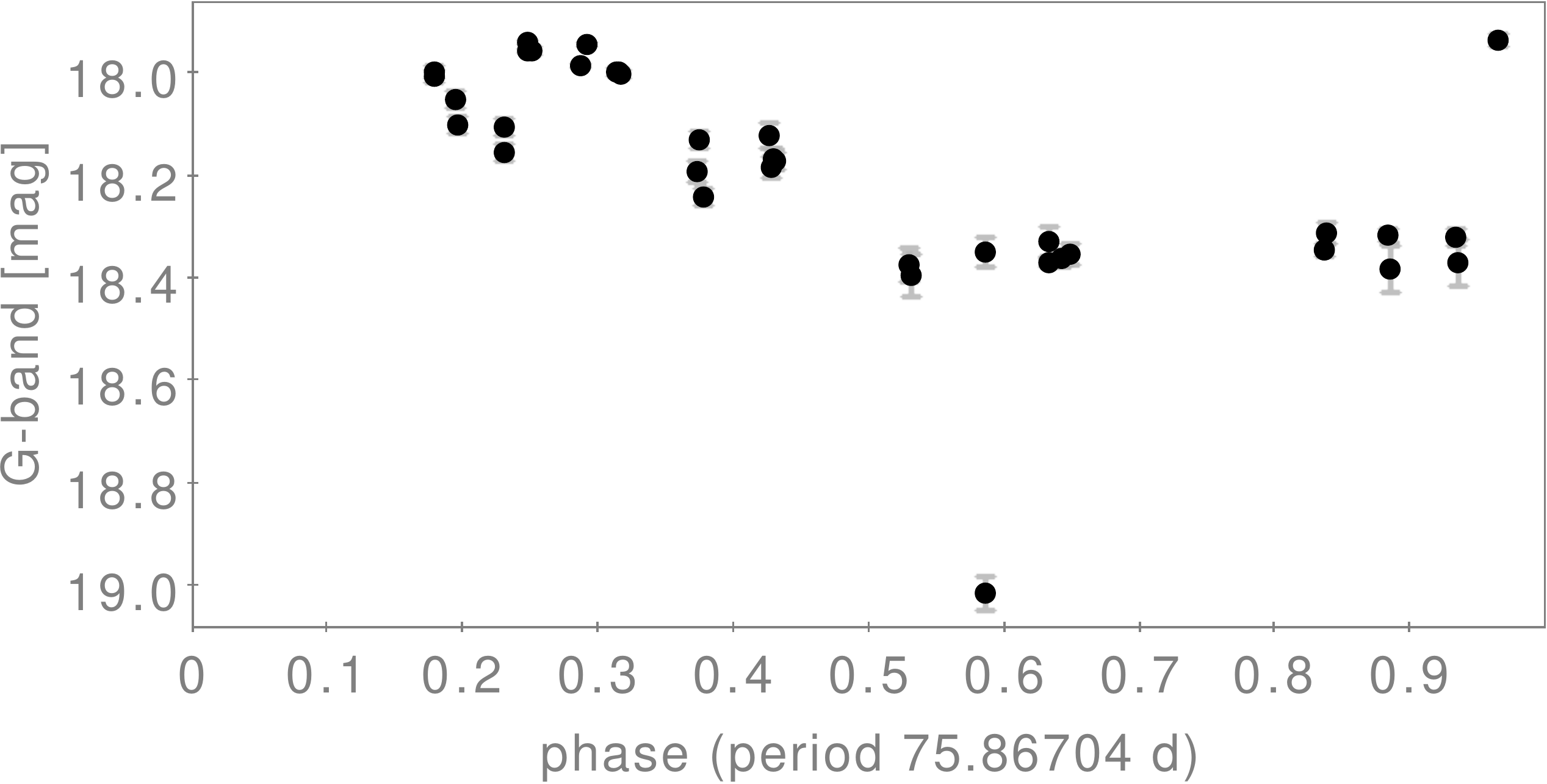}    
  & \includegraphics[trim=1.0cm 0 0 0, clip, height=0.15\textwidth]{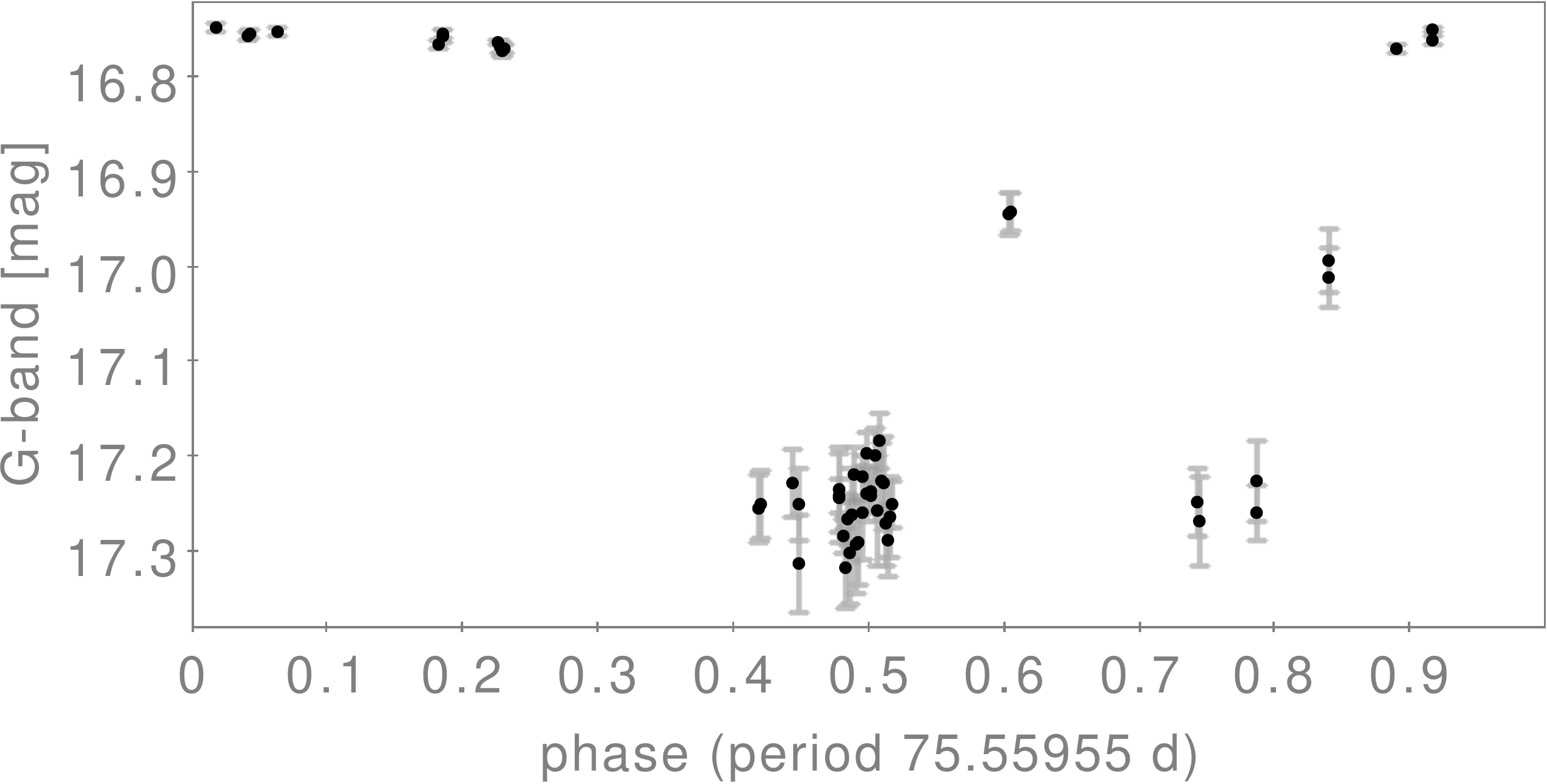}   
  & \includegraphics[trim=1.0cm 0 0 0, clip, height=0.15\textwidth]{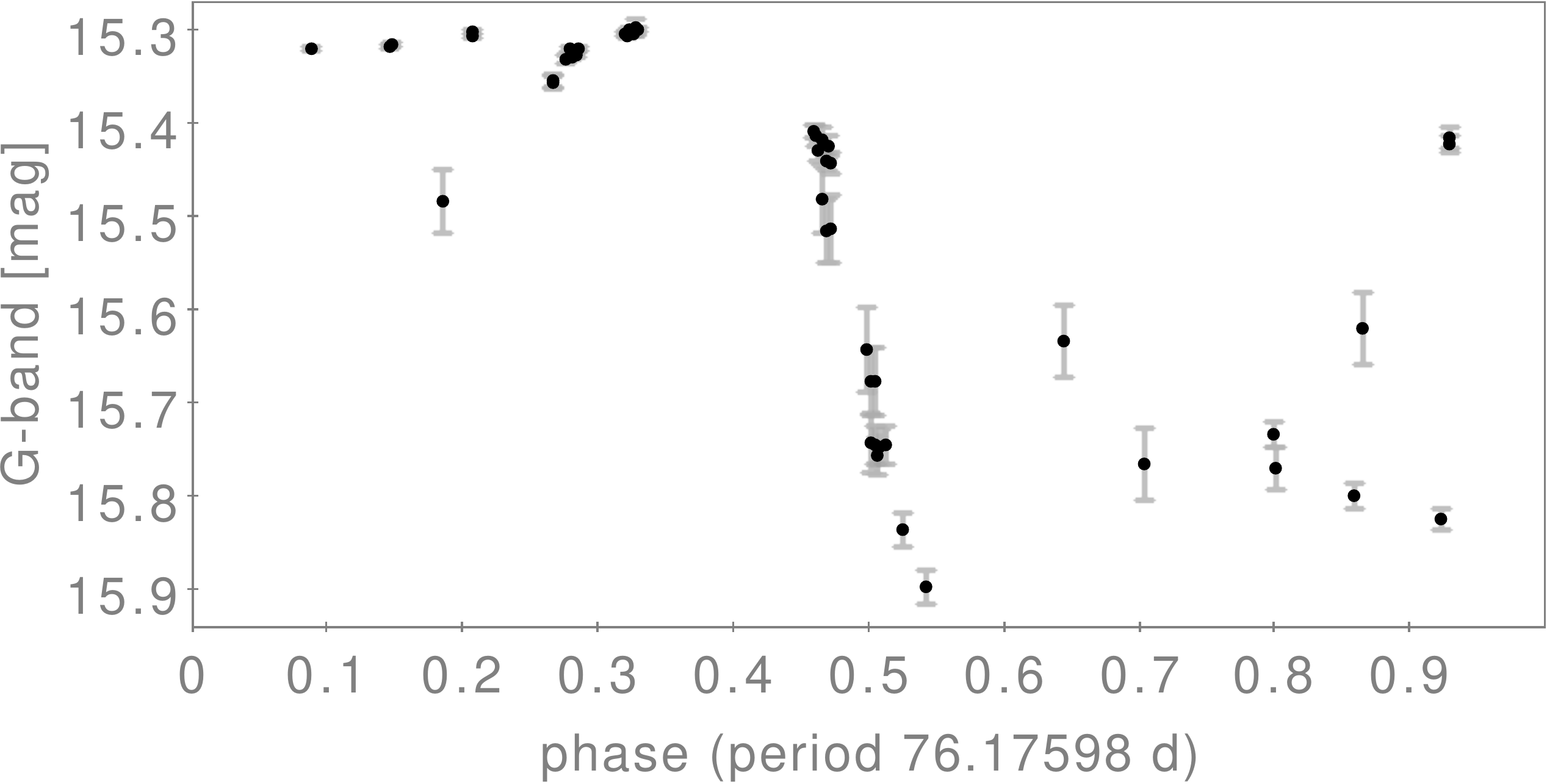} \\
 & 2090165114714003328 & 2230871022168218240 & 2005646617966572416 \\
 \rotatebox{0}{\normalsize $91.3  \pm  2.4$~d}
  & \includegraphics[height=0.15\textwidth]{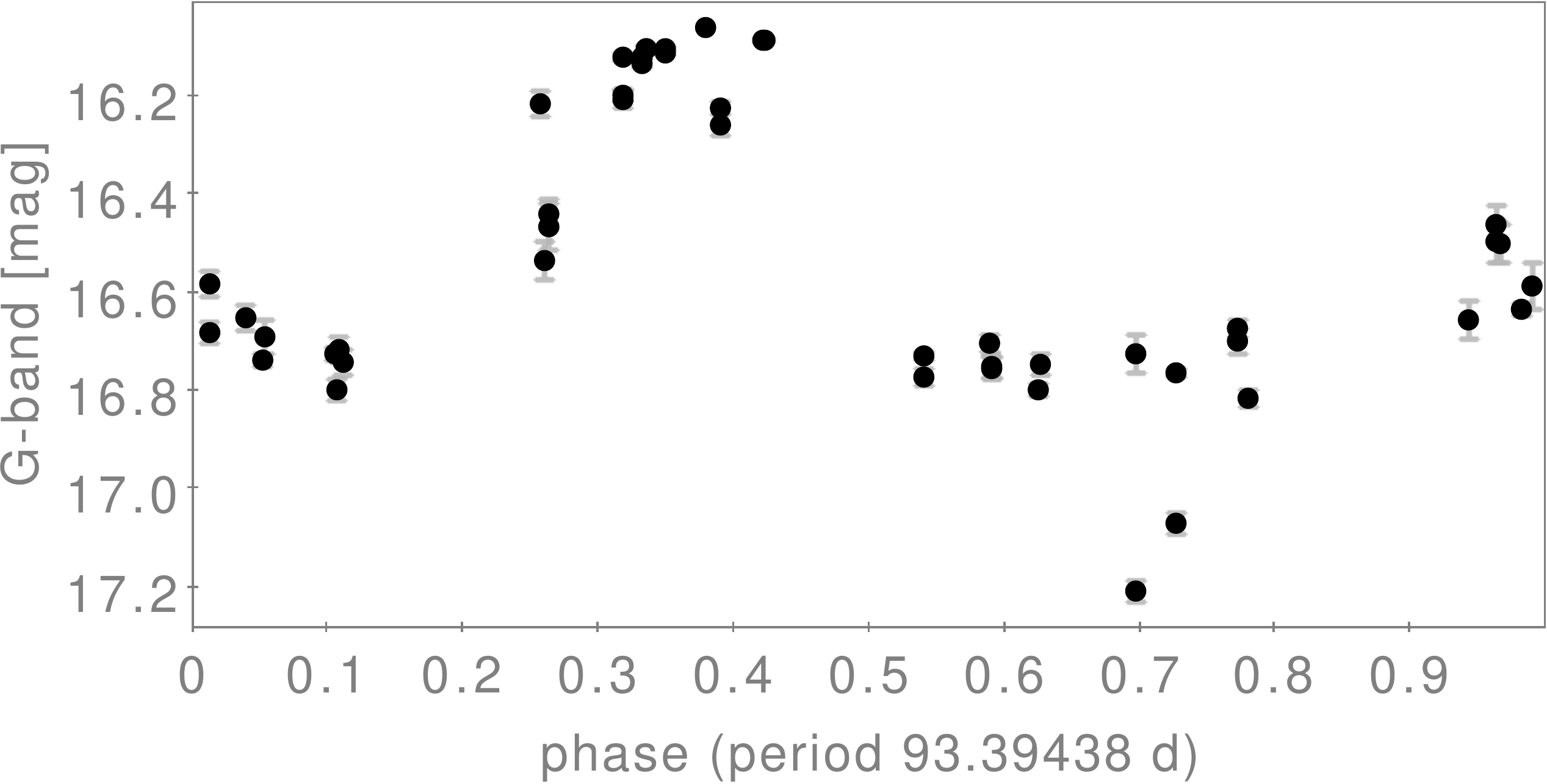}    
  & \includegraphics[trim=1.0cm 0 0 0, clip, height=0.15\textwidth]{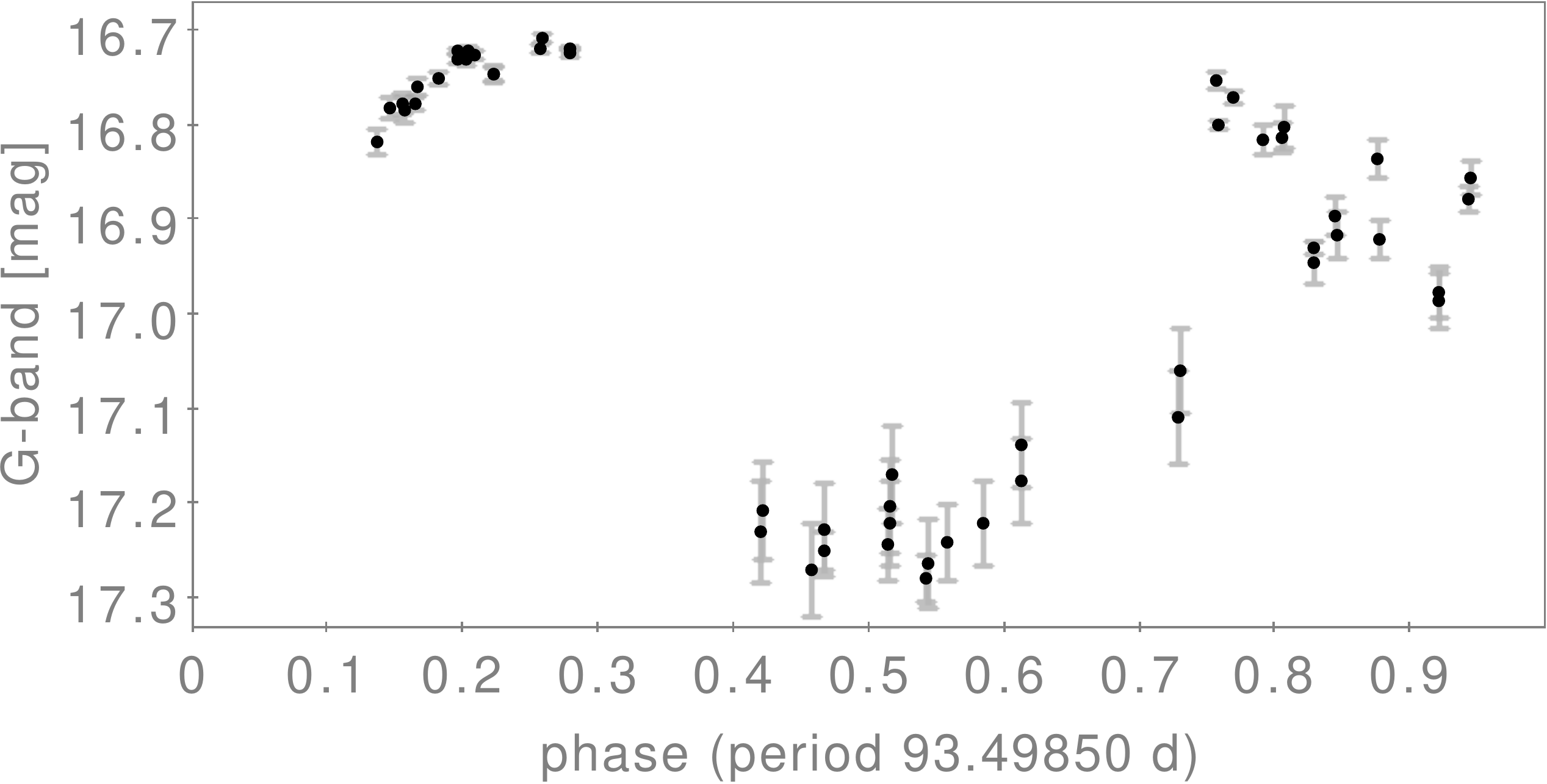}   
  & \includegraphics[trim=1.0cm 0 0 0, clip, height=0.15\textwidth]{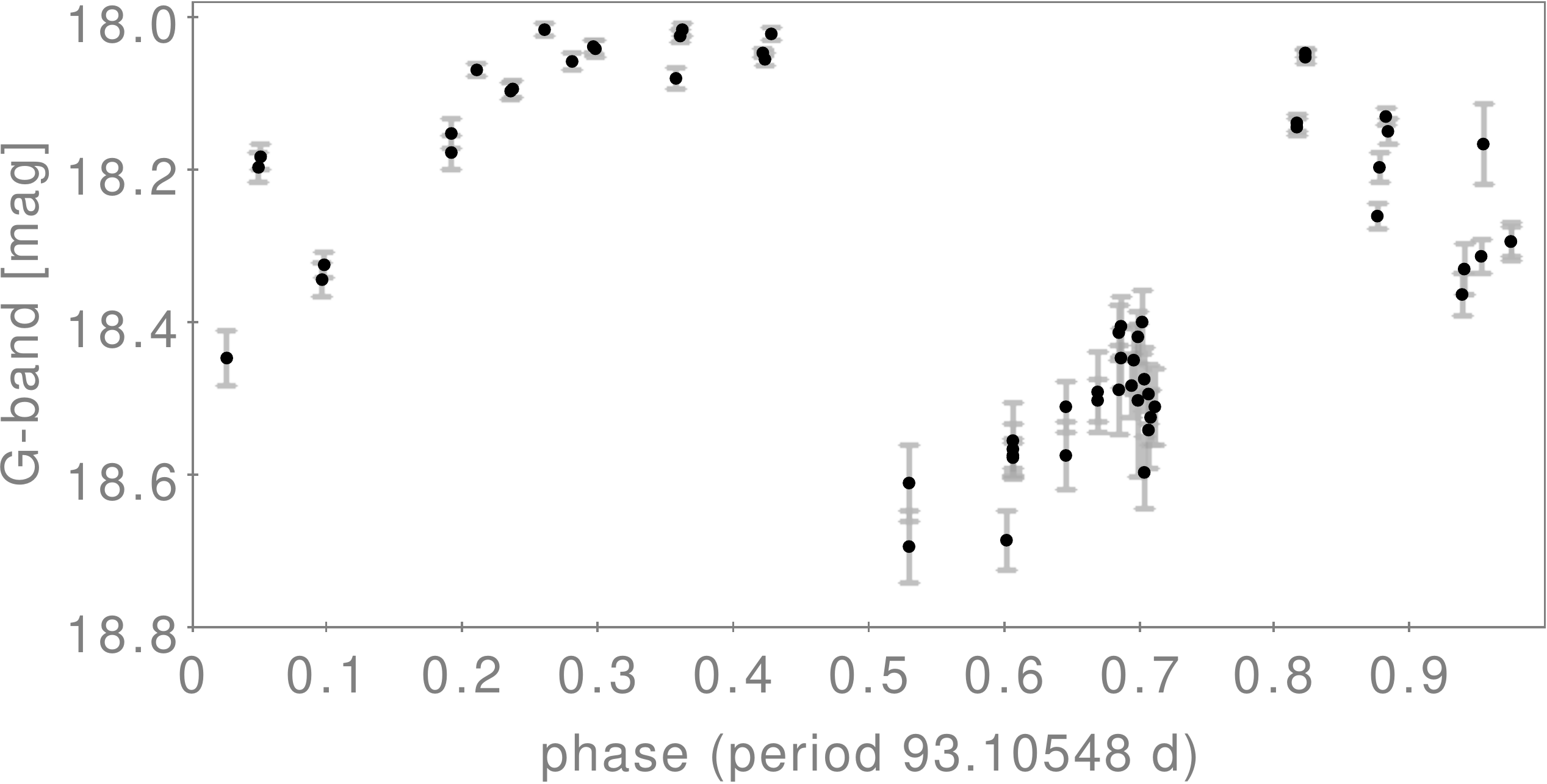} \\
 & 5309834605981982720 & 5312836307075200000 & 5297091884671648512 \\
 \rotatebox{0}{\normalsize $96.1  \pm  2.4$~d}
  & \includegraphics[height=0.15\textwidth]{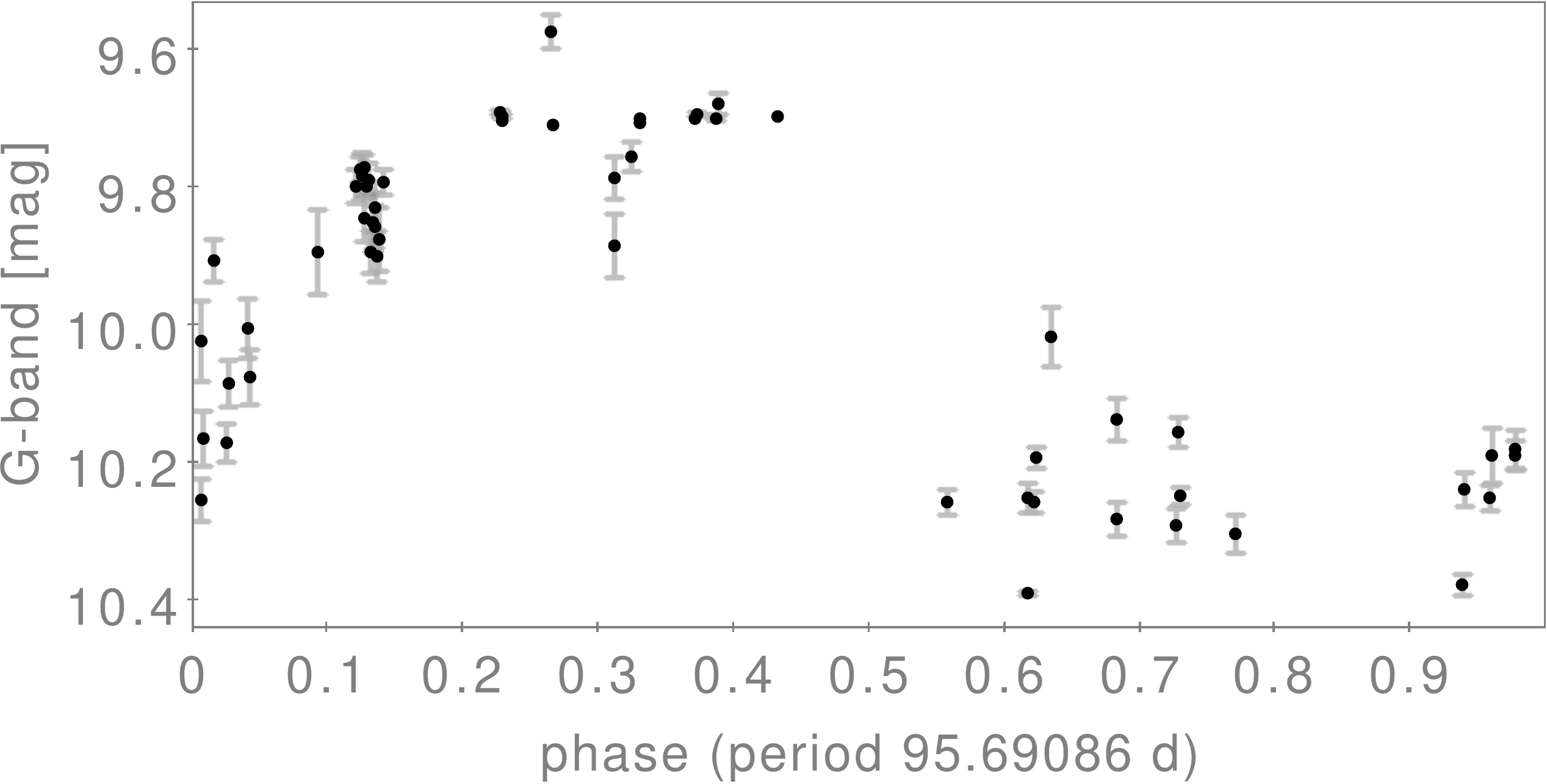}    
  & \includegraphics[trim=1.0cm 0 0 0, clip, height=0.15\textwidth]{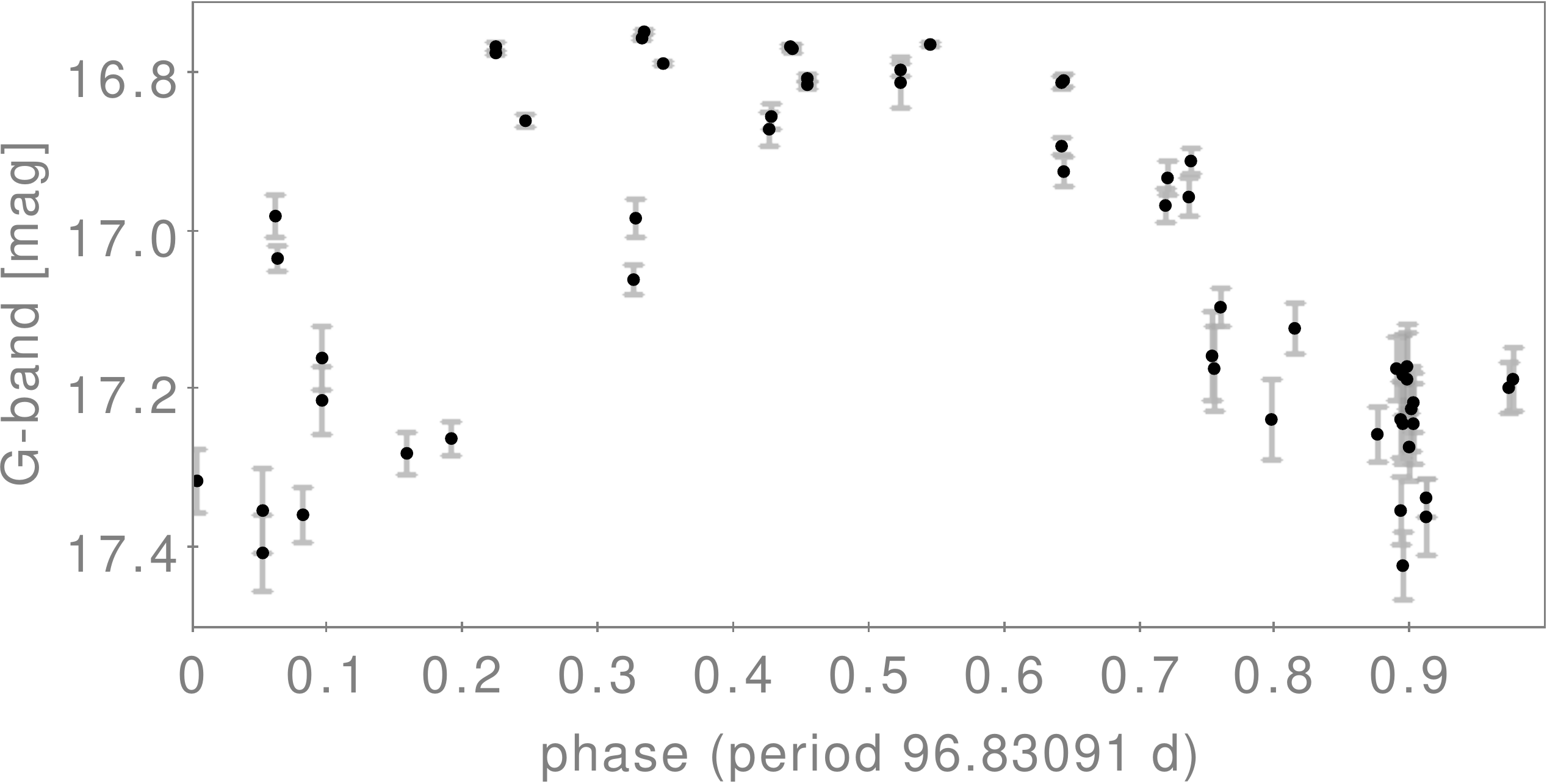}   
  & \includegraphics[trim=1.0cm 0 0 0, clip, height=0.15\textwidth]{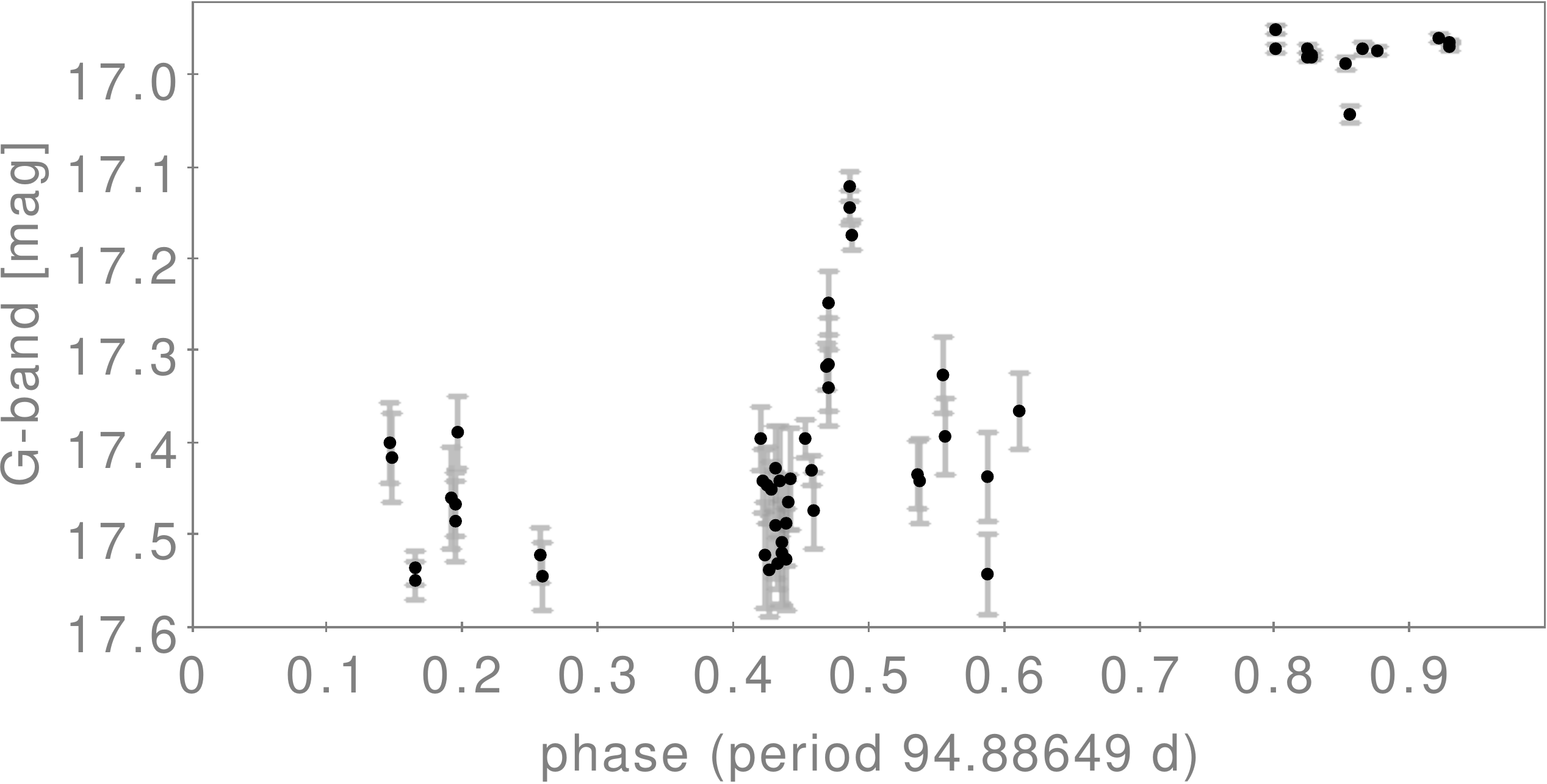} \\
& 6630570550225688320 & 2740120350948278528 & 5002124006200453888\\
 \rotatebox{0}{\normalsize $182.6 \pm  10$~d} 
  & \includegraphics[height=0.15\textwidth]{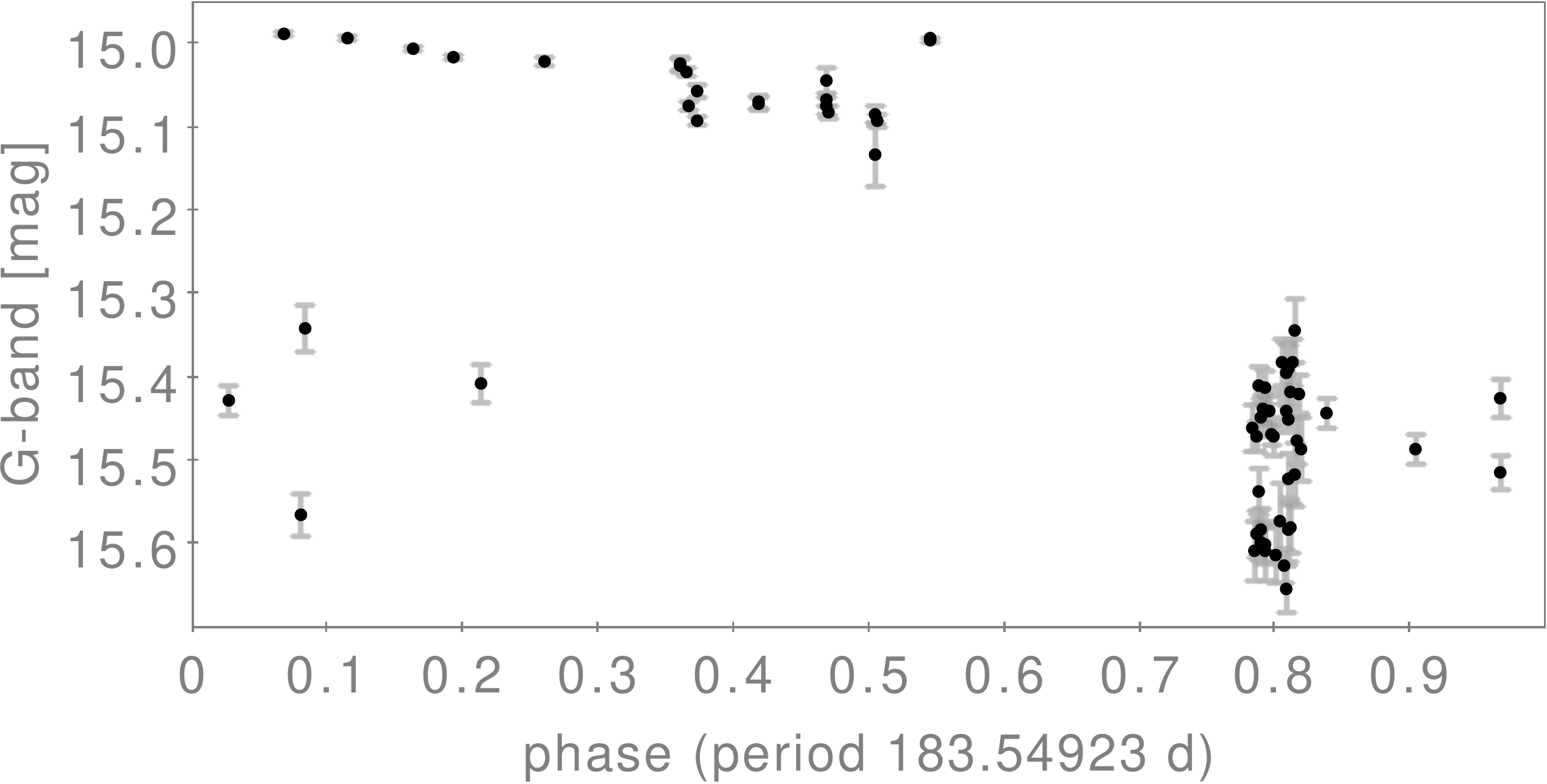}    
  & \includegraphics[trim=1.0cm 0 0 0, clip, height=0.15\textwidth]{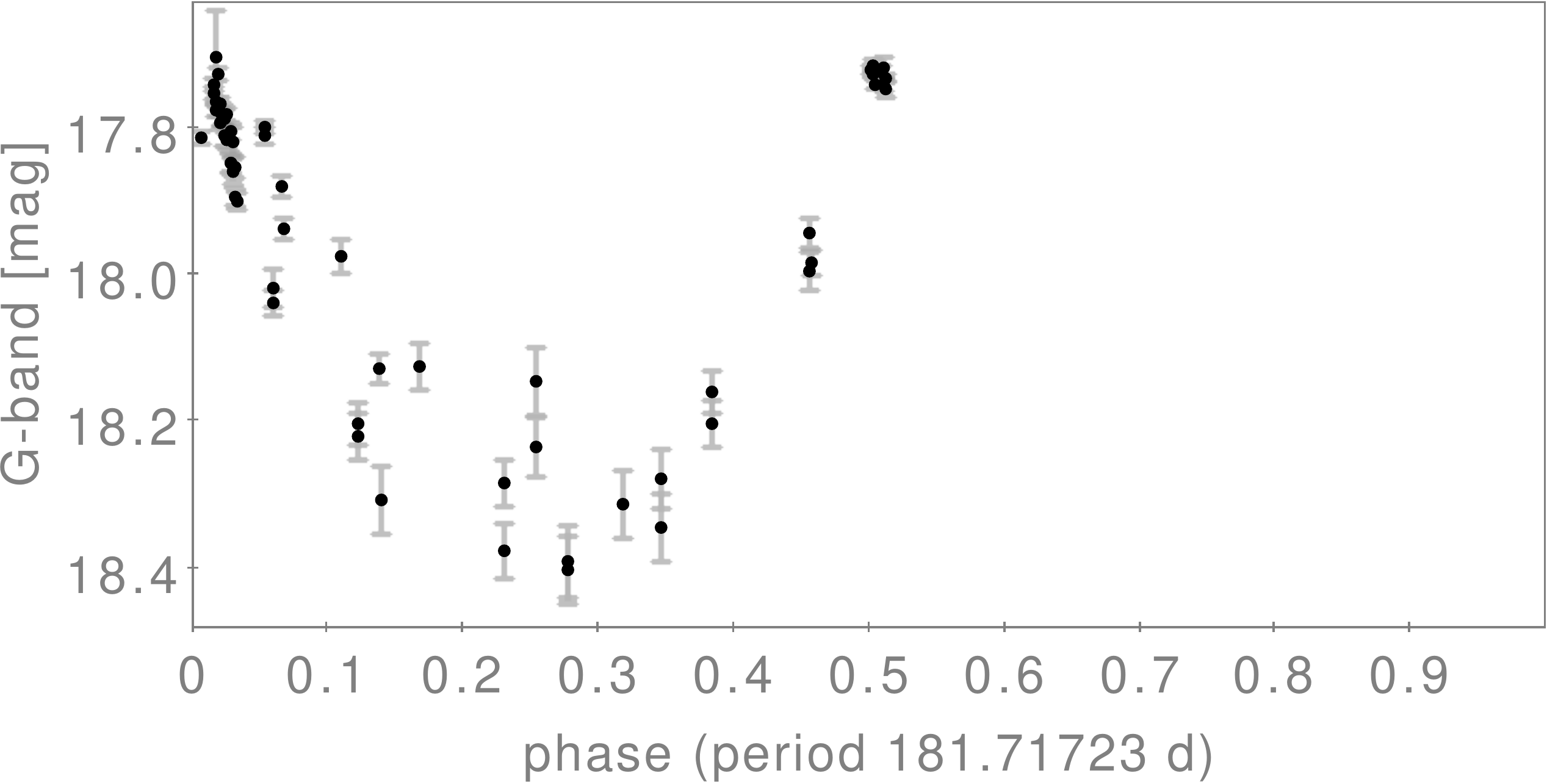}   
  & \includegraphics[trim=1.0cm 0 0 0, clip, height=0.15\textwidth]{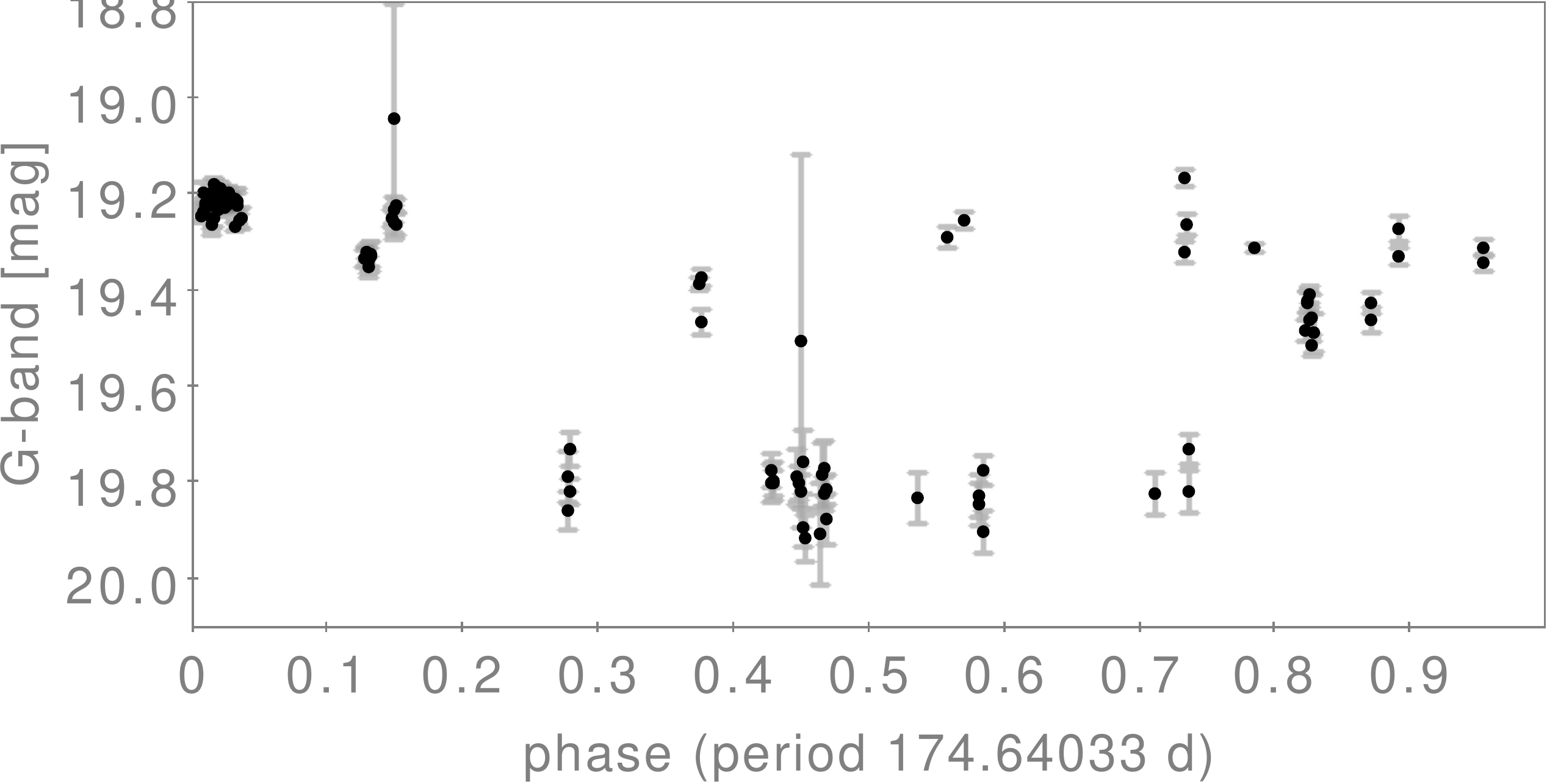}\\
\end{tabular}
 \vspace{-0.4cm}
\caption{Fig.~\ref{fig:foldedPeriodGExamples1} ctd. for longer periods of photometric data. 
}
\label{fig:foldedPeriodGExamples2}
\end{small}
\end{figure*}

\begin{figure*}[h]
\begin{tabular}{@{}lrr@{}}
\setlength{\tabcolsep}{0pt} 
\renewcommand{\arraystretch}{0} 
  \includegraphics[width=0.3\textwidth]{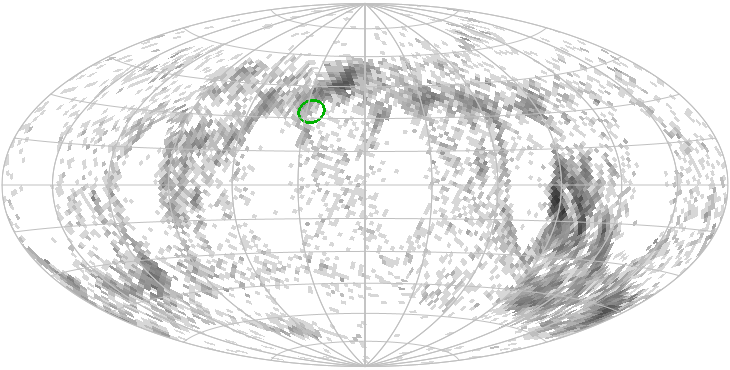}    
  & \includegraphics[width=0.3\textwidth]{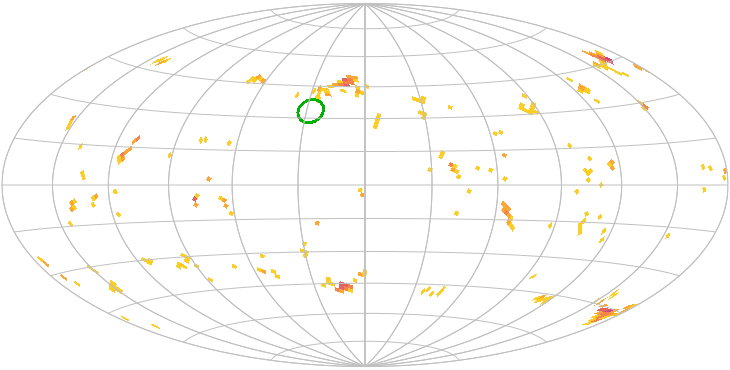}    
  & \includegraphics[width=0.3\textwidth]{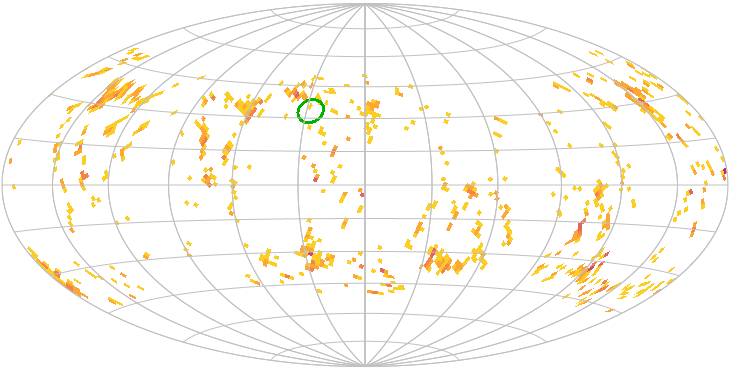}  \vspace{-3cm}\\
$13.95\pm 0.15$~d \quad \quad \quad \quad \ \ \  all-sky astro & sim. $k=2$ & sim. $k=4$  \vspace{2.4cm} \\[6pt]
  \includegraphics[width=0.3\textwidth]{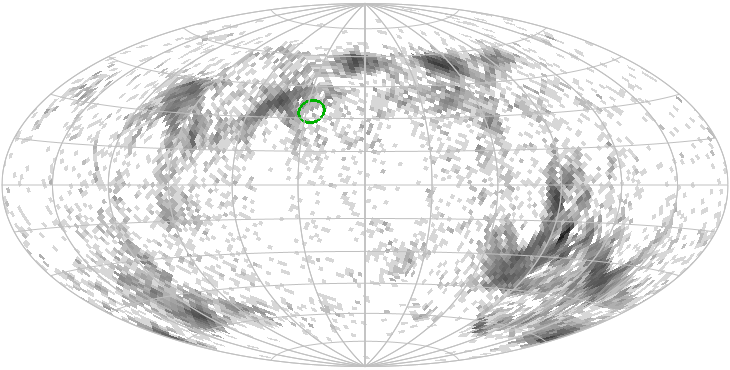}    
  & \includegraphics[width=0.3\textwidth]{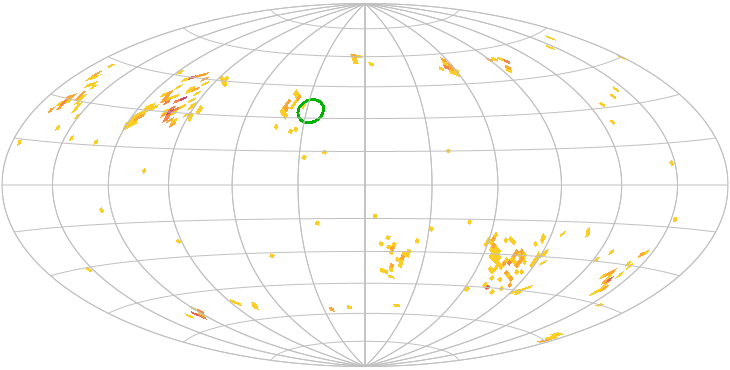}   
 & \includegraphics[width=0.3\textwidth]{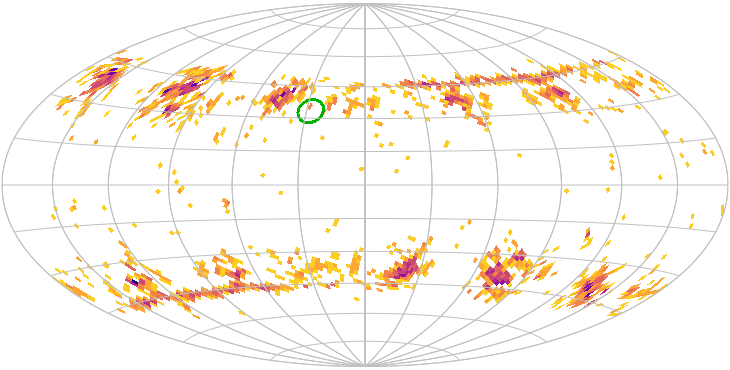} \vspace{-3cm}\\
$15.7  \pm  0.15$~d &  & \vspace{2.4cm}\\[6pt]
  \includegraphics[width=0.3\textwidth]{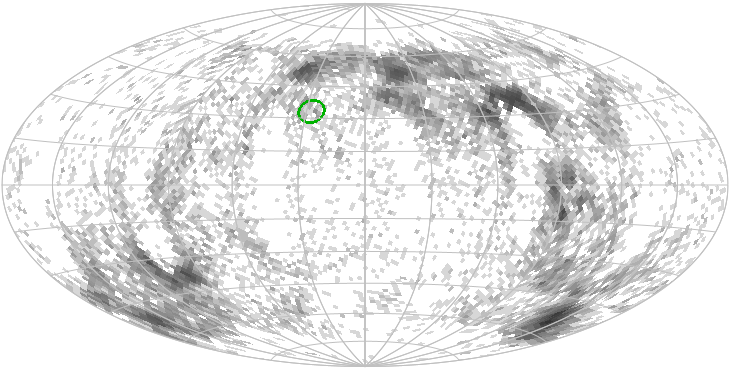}    
  & \includegraphics[width=0.3\textwidth]{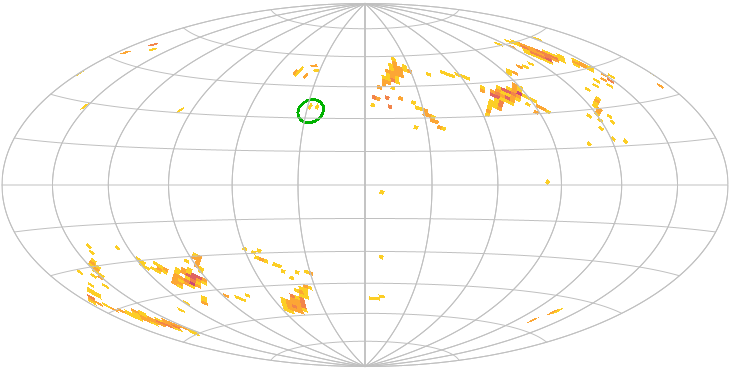}   
 & \includegraphics[width=0.3\textwidth]{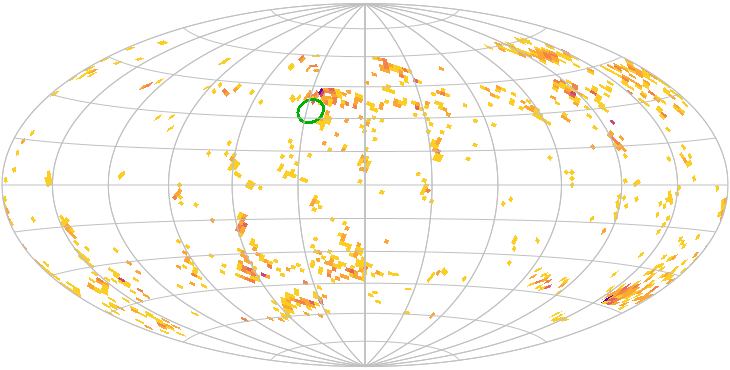}  \vspace{-3cm}\\
$16.35 \pm  0.2$~d &  &  \vspace{2.4cm}\\[6pt]
  \includegraphics[width=0.3\textwidth]{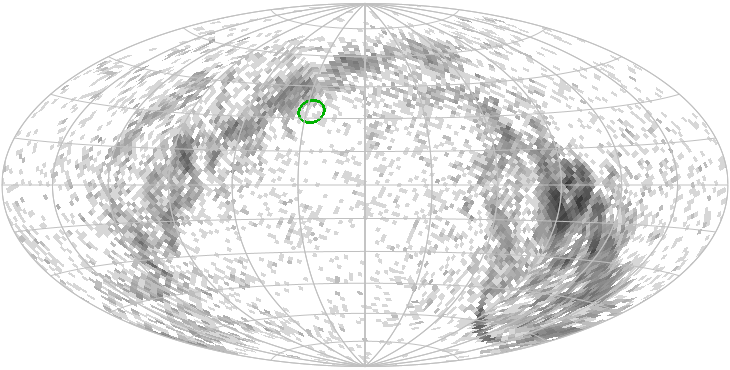}    
  &  \includegraphics[width=0.3\textwidth]{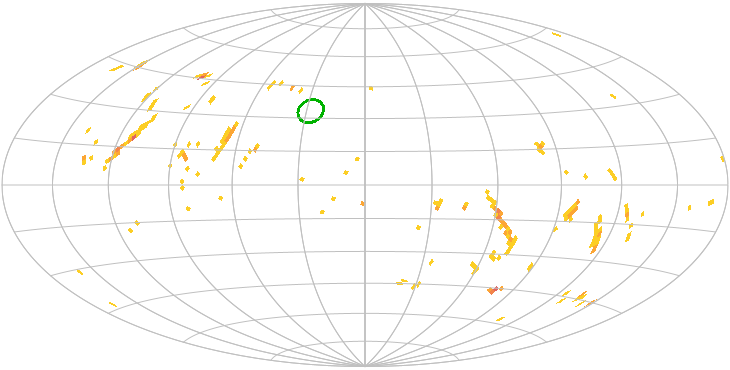}   
 & \includegraphics[width=0.3\textwidth]{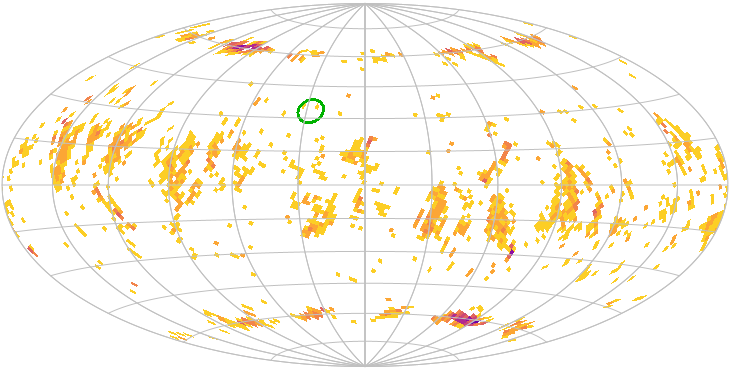}   \vspace{-3cm}\\
$18.8  \pm  0.2$~d &  &  \vspace{2.4cm}\\[6pt]
  \includegraphics[width=0.3\textwidth]{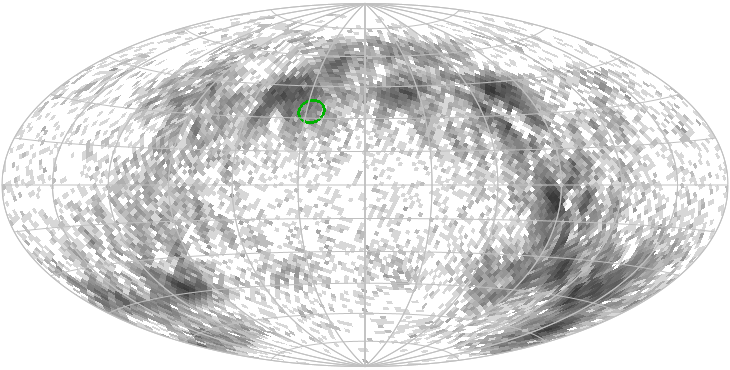}    
  &  \includegraphics[width=0.3\textwidth]{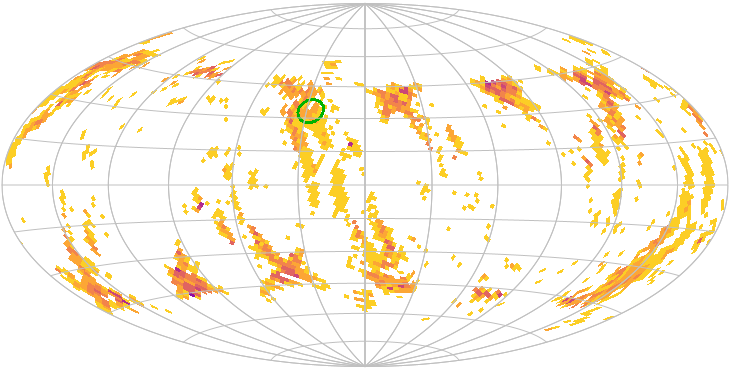}   
 & \includegraphics[width=0.3\textwidth]{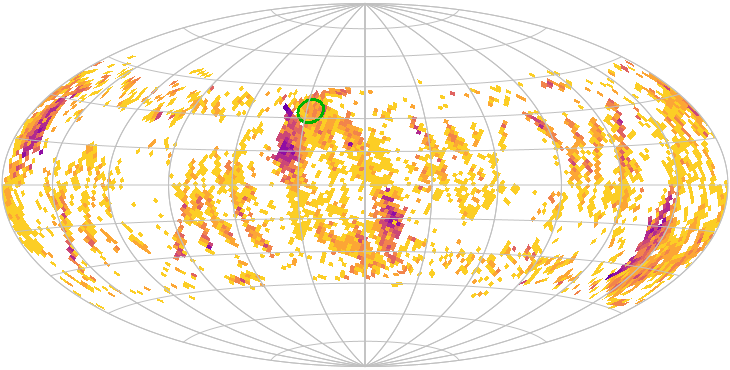}  \vspace{-3cm}\\
$19.9  \pm  0.2$~d &  &   \vspace{2.4cm}\\[6pt]
  \includegraphics[width=0.3\textwidth]{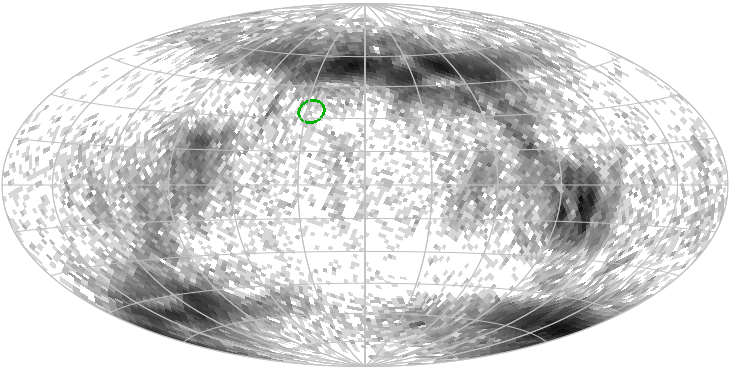}    
  &  \includegraphics[width=0.3\textwidth]{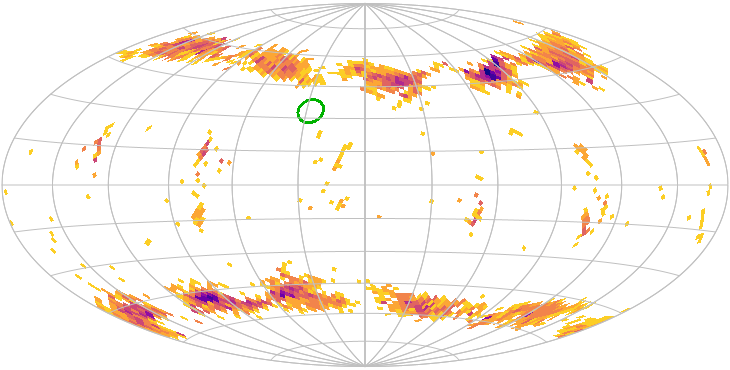}   
 & \includegraphics[width=0.3\textwidth]{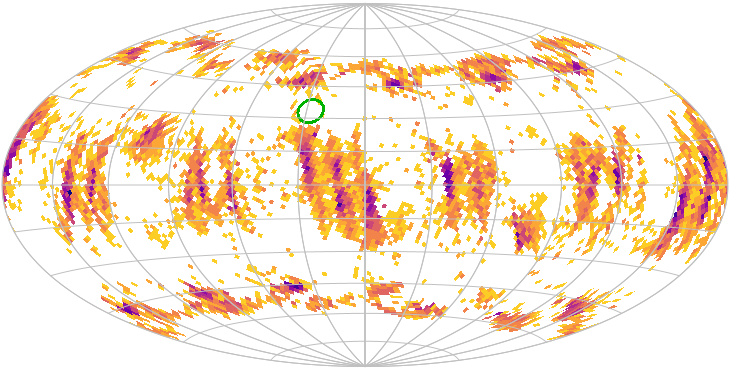}   \vspace{-3cm}\\
$25.1  \pm  0.3$~d &  & \vspace{2.4cm}\\[6pt]
   \includegraphics[width=0.3\textwidth]{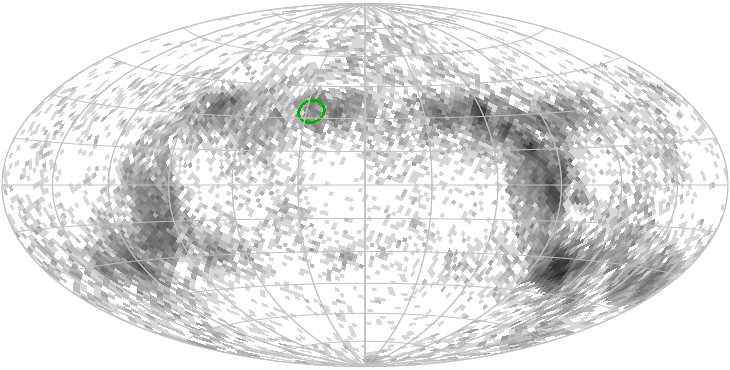}    
  & \includegraphics[width=0.3\textwidth]{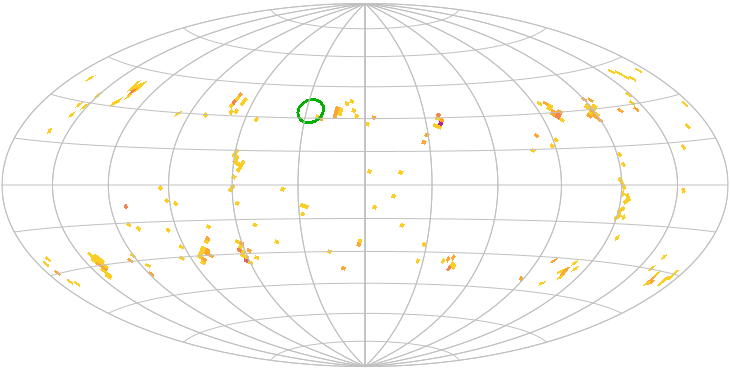}   
 & \includegraphics[width=0.3\textwidth]{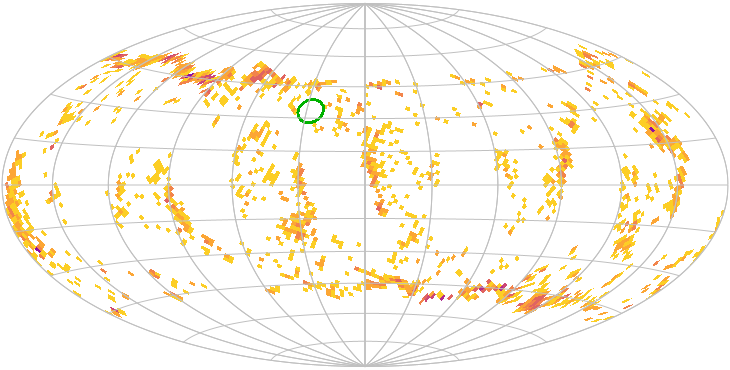}    \vspace{-3cm}\\
$26.9  \pm  0.4$~d &  &  \vspace{2.4cm}\\[6pt]
\multicolumn{3}{l}{\hspace{-0.32cm}\includegraphics[trim={0 3.3cm 0 0},clip,width=0.96\textwidth]{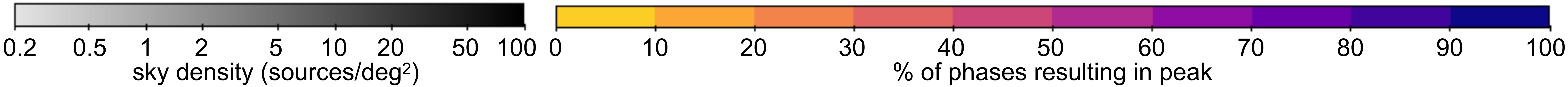} }  \vspace{-0.2cm}\\[6pt]
\end{tabular}
\caption{Ecliptic Aitoff projection of the astrometric period data presented in Fig.~\ref{fig:astrSimP}. Left: Source density of the astrometric peaks for the all-sky sample (top panel of Fig.~\ref{fig:astrSimP}). Right: Result of the  all-sky uniform simulations of our noiseless sampled bias model (panels 2 and 3 of Fig.~\ref{fig:astrSimP}). They are
colour-coded with the percentage of phases (position angles) that results in this peak: a low value means that only specific phasing of the scan-angle signal will result in a particular peak being observed at the given location. The green circle indicates the location of the GAPS catalogue. In all sky plots longitude zero is at the centre and increases to the left. }
\label{fig:skyplotDistrAstroAllsky1}
\end{figure*}

\begin{figure*}[h]
\begin{tabular}{@{}lrr@{}}
\setlength{\tabcolsep}{0pt} 
\renewcommand{\arraystretch}{0} 
  \includegraphics[width=0.3\textwidth]{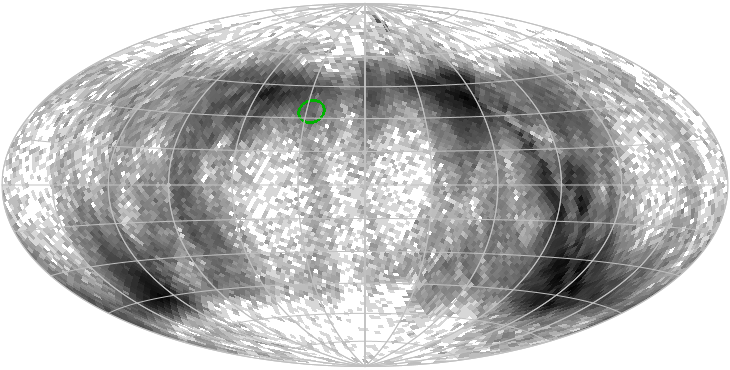}    
  & \includegraphics[width=0.3\textwidth]{images/simSpuriousPeriodSkyMaps/astr_ph5_k3_13p9d.png}    
  & \includegraphics[width=0.3\textwidth]{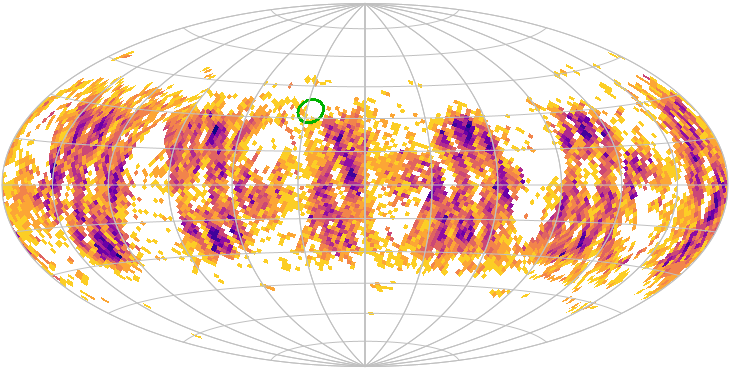}  \vspace{-3cm}\\
$31.5  \pm  0.6$~d \quad \quad \quad \quad \quad \ \ \ \  all-sky astro & sim. $k=2$ & sim. $k=4$   \vspace{2.4cm} \\[6pt]
  \includegraphics[width=0.3\textwidth]{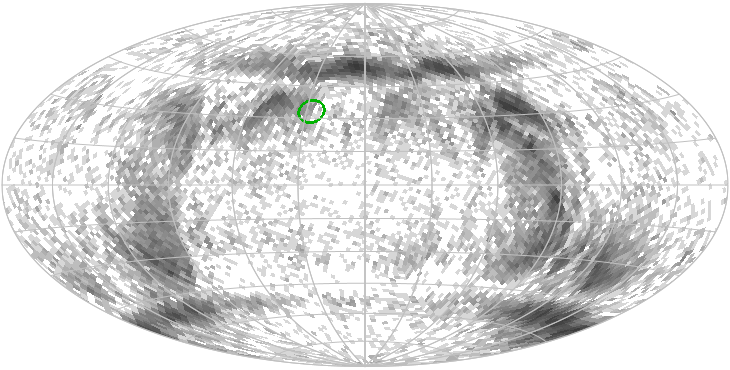}    
  & \includegraphics[width=0.3\textwidth]{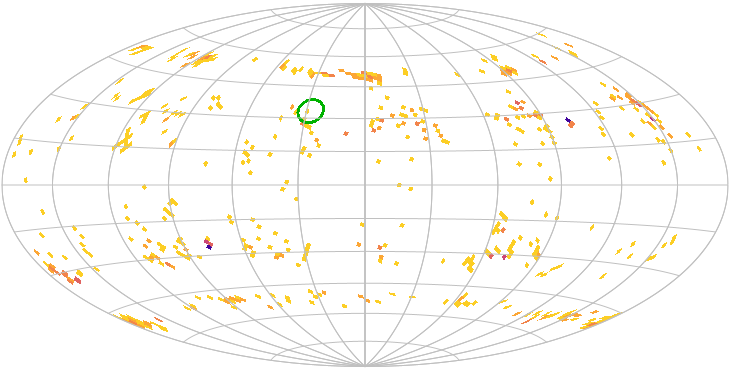}   
 & \includegraphics[width=0.3\textwidth]{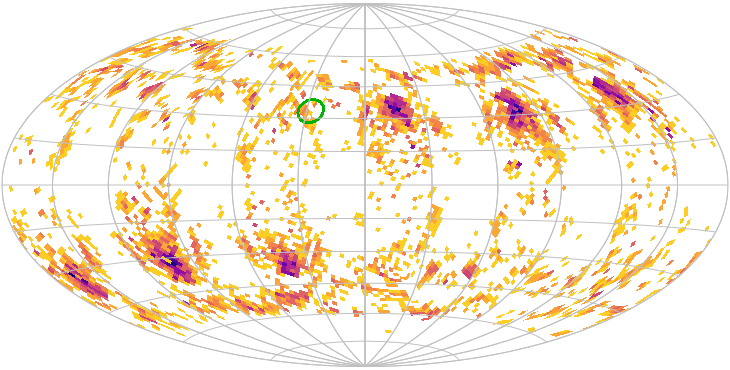} \vspace{-3cm}\\
$41.9  \pm  1.0$~d &  &  \vspace{2.4cm}\\[6pt]
  \includegraphics[width=0.3\textwidth]{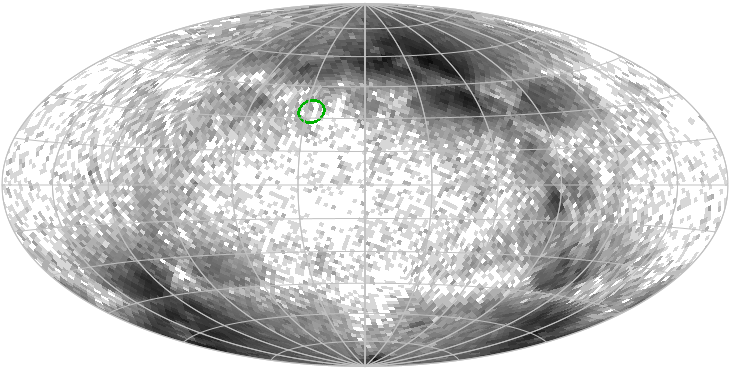}    
  & \includegraphics[width=0.3\textwidth]{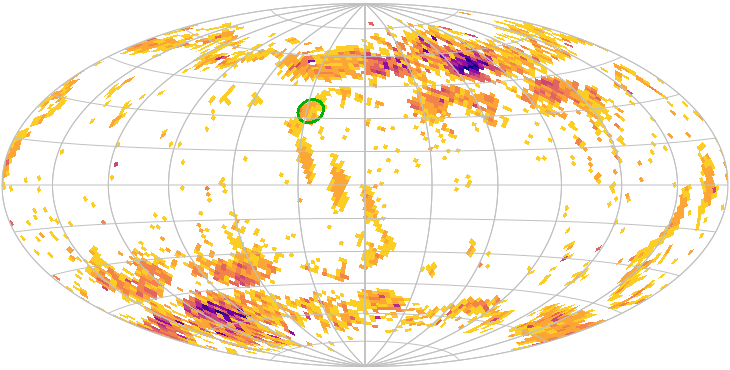}   
 & \includegraphics[width=0.3\textwidth]{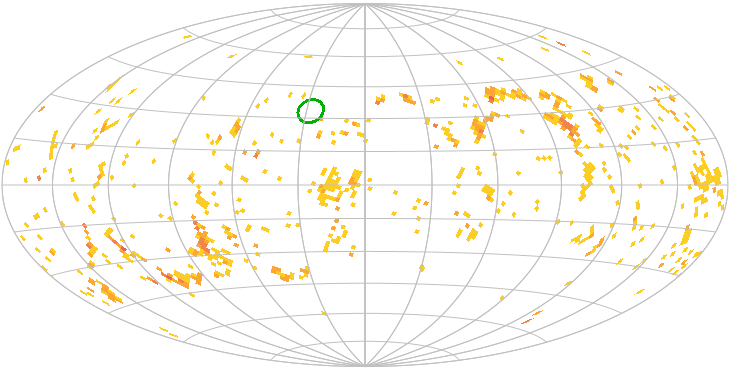}  \vspace{-3cm}\\
$46.8  \pm  1.1$~d &  &  \vspace{2.4cm}\\[6pt]
  \includegraphics[width=0.3\textwidth]{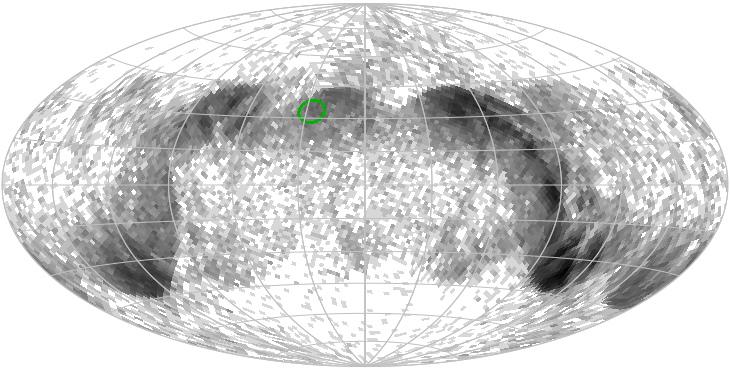}    
  &  \includegraphics[width=0.3\textwidth]{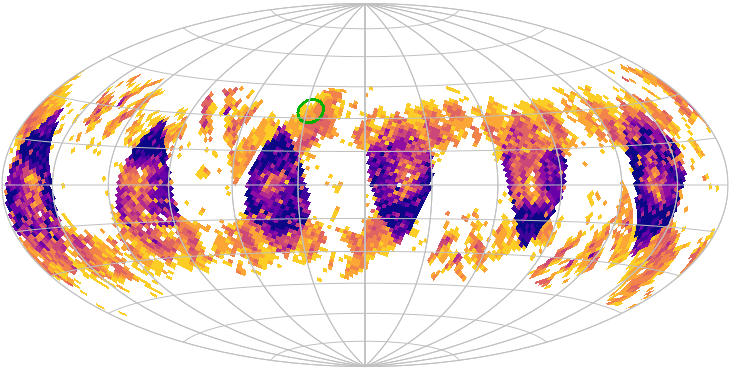}   
 & \includegraphics[width=0.3\textwidth]{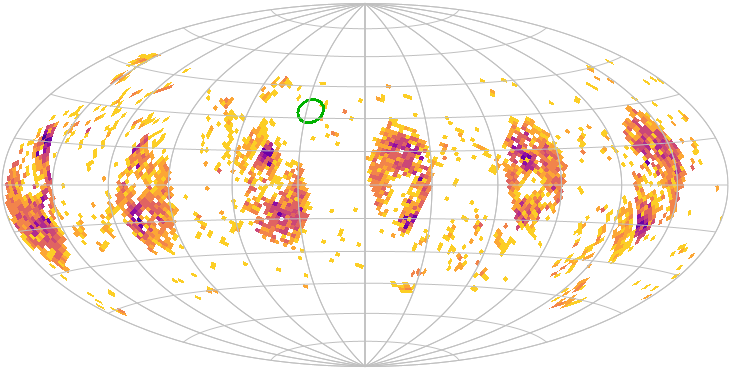}  \vspace{-3cm}\\
$53.7  \pm  1.1$~d &  &  \vspace{2.4cm}\\[6pt]
  \includegraphics[width=0.3\textwidth]{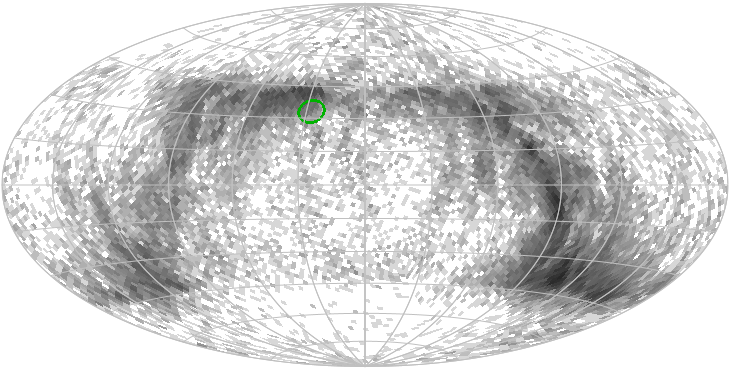}    
  &  \includegraphics[width=0.3\textwidth]{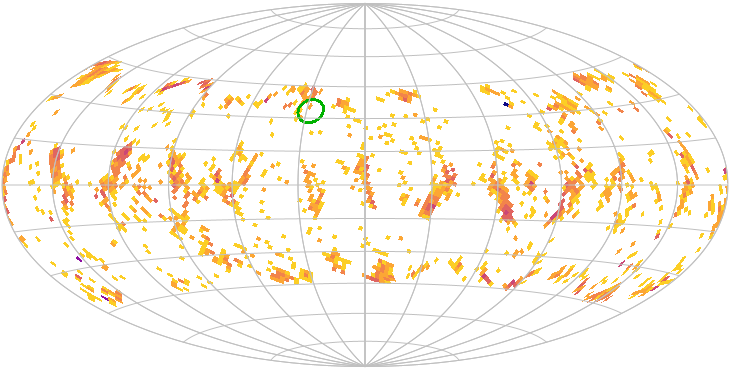}   
 & \includegraphics[width=0.3\textwidth]{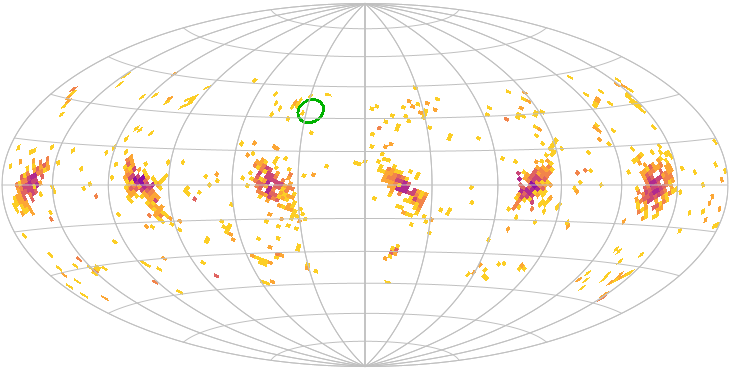}   \vspace{-3cm}\\
$76.1  \pm  1.7$~d &  & \vspace{2.4cm}\\[6pt]
  \includegraphics[width=0.3\textwidth]{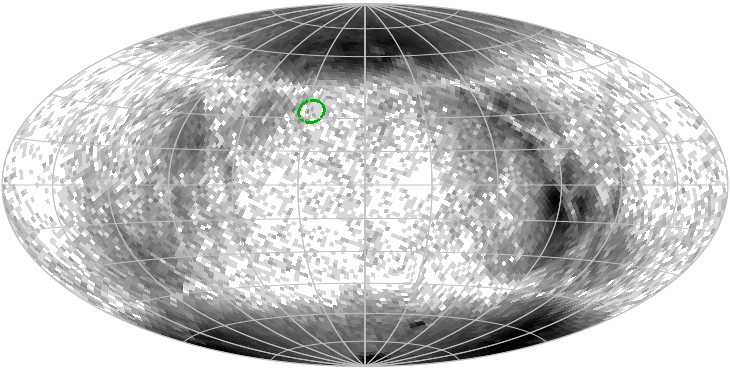}    
  &  \includegraphics[width=0.3\textwidth]{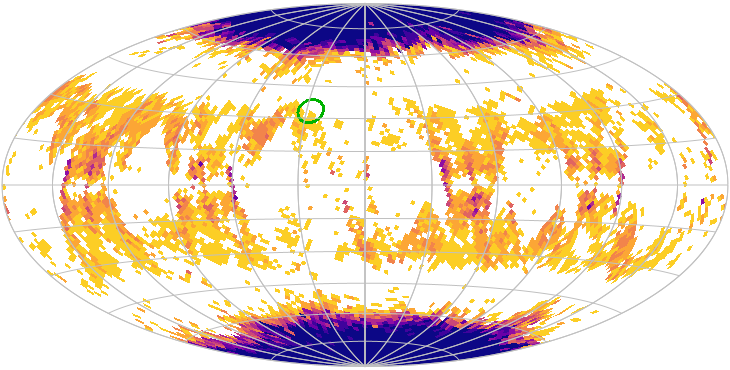}   
 & \includegraphics[width=0.3\textwidth]{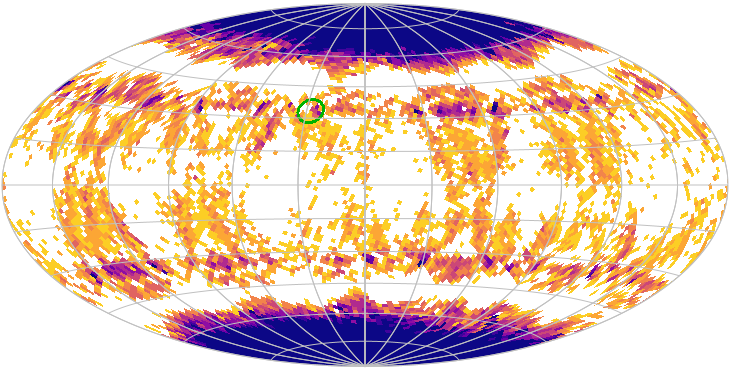}  \vspace{-3cm}\\
$91.3  \pm  2.4$~d &  &   \vspace{2.4cm}\\[6pt]
  \includegraphics[width=0.3\textwidth]{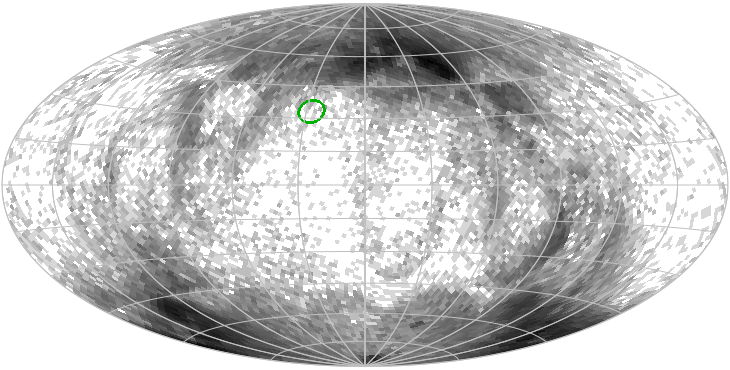}    
  &  \includegraphics[width=0.3\textwidth]{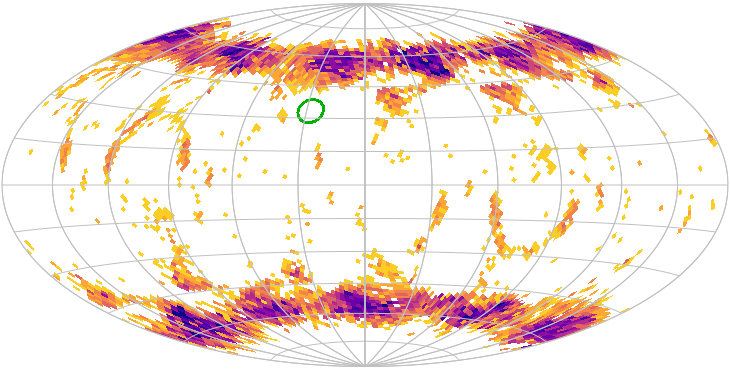}   
 & \includegraphics[width=0.3\textwidth]{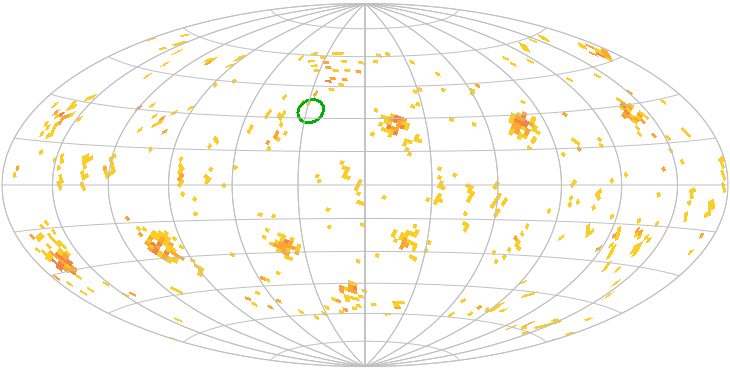}   \vspace{-3cm}\\
$96.1  \pm  2.4$~d &  & \vspace{2.4cm}\\[6pt]
   \includegraphics[width=0.3\textwidth]{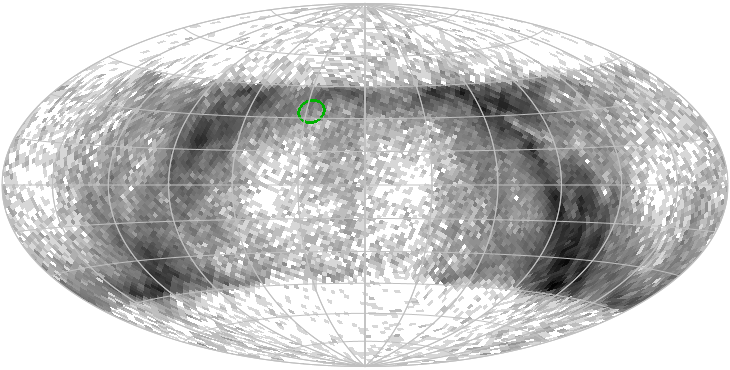}    
  & \includegraphics[width=0.3\textwidth]{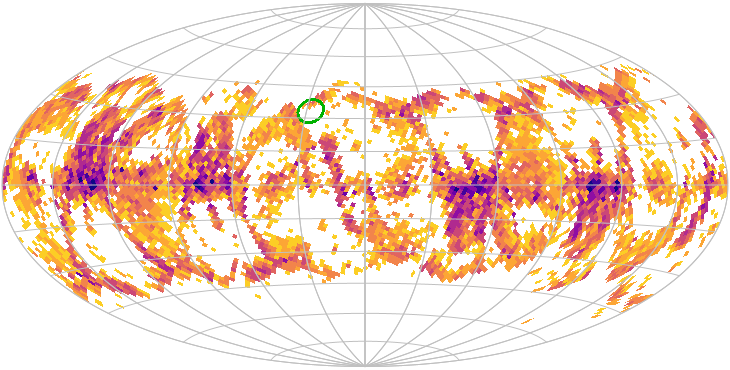}   
 & \includegraphics[width=0.3\textwidth]{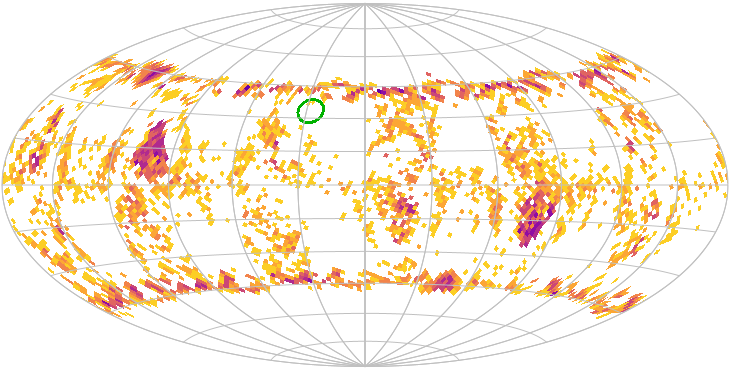}    \vspace{-3cm}\\
$182.6 \pm  10$~d &  &  \vspace{2.4cm}\\[6pt]
\multicolumn{3}{l}{\hspace{-0.32cm}\includegraphics[trim={0 3.3cm 0 0},clip,width=0.96\textwidth]{images/simSpuriousPeriodSkyMaps/FINAL_colourbar_astro.pdf} }  \vspace{-0.2cm}\\[6pt]
\end{tabular}
\caption{ Fig.~\ref{fig:skyplotDistrAstroAllsky1} ctd. for longer periods of astrometric data.}
\label{fig:skyplotDistrAstroAllsky2}
\end{figure*}

In addition to the sky maps, we also include example folded time-series of public photometry for the main spurious peaks in Figs.~\ref{fig:foldedPeriodGExamples1} and \ref{fig:foldedPeriodGExamples2}.


\FloatBarrier

\section{Scan position angle in ecliptic coordinates\label{sec:eclScanAngle}}

For use in Sect.~\ref{ssec:obsDistrSa}, we here derive an expression for the position angle of a scan relative to local {ecliptic} north,
$\psi_\text{ecl}$ (here generically labelled $\theta_\text{ecl}$), given the position of the source, $(\alpha,\delta)$, and the position angle of
the scan relative to local {equatorial} north, $\psi_\text{equ}$ (here generically labelled $\theta_\text{equ}$). Although we speak about the {scan}
position angle, the results apply to any position angle. We therefore use the more generic position-angle symbol $\theta$.

\begin{figure}[h]
\centering
  \centerline{\includegraphics[width=9cm]{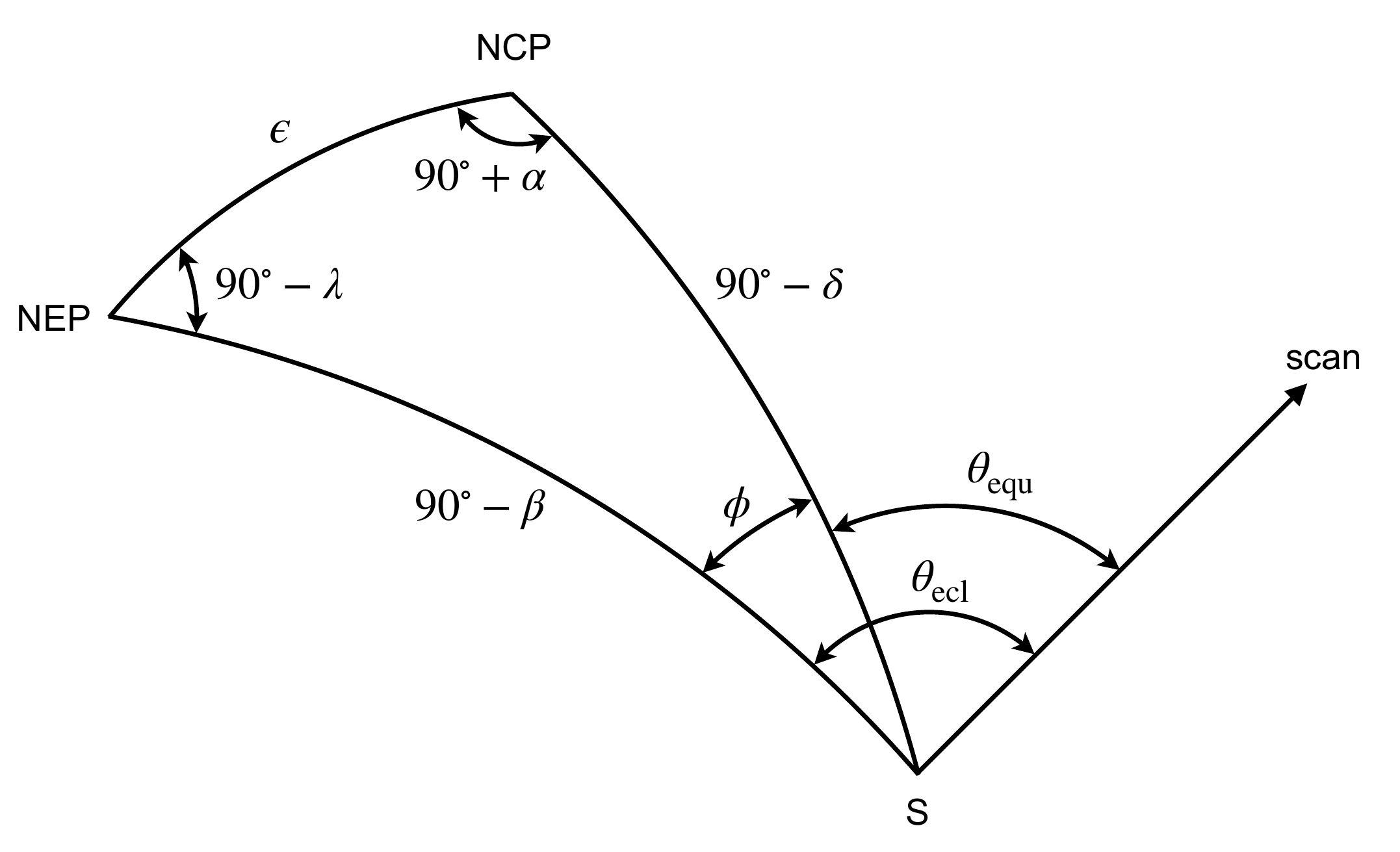}}
  \caption{Geometry of the scan across a source at S relative to the north celestial pole 
  and 
  north ecliptic pole, as viewed from outside the celestial sphere.}
  \label{fig1}
\end{figure}

The geometry of the scan is shown in Fig.~\ref{fig1}. The difference between
the two position angles, $\phi=\theta_\text{ecl}-\theta_\text{equ}$ , is the angle at the source
(S) in the spherical triangle it forms with the north celestial pole (NCP) and the north ecliptic pole (NEP), NCP--S--NEP, in which the other sides and angles depend on the
equatorial and ecliptic coordinates of the source, $(\alpha,\delta)$ and $(\lambda,\beta)$,
and the obliquity of the ecliptic, 
$\epsilon=84381.41100~\text{arcsec}$.
From the sine theorem, we have
\begin{equation}\label{e2}
\cos\beta\sin\phi = \cos\alpha\sin\epsilon
,\end{equation}
and from the cosine theorem,
\begin{equation}\label{e3}
\cos\epsilon = \sin\delta\sin\beta + \cos\delta\cos\beta\cos\phi \, 
\end{equation}
or
\begin{equation}\label{e4}
\cos\delta\cos\beta\cos\phi = \cos\epsilon - \sin\delta\sin\beta \, .
\end{equation}
From the last two equations, it is possible to solve $\phi$ without quadrant ambiguity, but the
equations involve not only $(\alpha,\delta),$ but also $\beta$, which may be inconvenient.
However, from the cosine theorem, we also have
\begin{equation}\label{e5}
\sin\beta = \cos\epsilon\sin\delta-\sin\epsilon\cos\delta\sin\alpha
,\end{equation}
which after insertion in Eq. (\ref{e4}) and division by $\cos\delta$ gives
\begin{equation}\label{e6}
\cos\beta\cos\phi=\cos\epsilon\cos\delta+\sin\epsilon\sin\delta\sin\alpha \, .
\end{equation}
Combination of Eqs. (\ref{e2}) and (\ref{e6}) now gives
\begin{equation}\label{e7}
\phi=\text{atan2}(\sin\epsilon\cos\alpha,\,\cos\epsilon\cos\delta+\sin\epsilon\sin\delta\sin\alpha)\, ,
\end{equation}
from which
\begin{equation}\label{eq:scanAngleE8}
\theta_\text{ecl}=\text{mod}(\theta_\text{equ}+\phi,\,2\pi)
\end{equation}
gives the ecliptic position angle in the interval $[0,2\pi)$.
This calculation is implemented in the MATLAB function {\tt scanPaEcl} reproduced below.
As a test example, $\alpha = 1.1$, $\delta = -0.5$, and $\theta_\text{equ} = 1.7$ gives
$\phi=0.276758777373696$ and $\theta_\text{ecl} = 1.9767587773737$.

{\tiny \begin{verbatim}
function [thetaEcl,phi] = scanPaEcl(alpha,delta,thetaEqu)
% Given the position of a source (alpha,delta) [rad] and the 
% position angle of the scan at the source in the local 
% equatorial system, thetaEqu [rad],  this function returns
% the position angle of the scan in the local ecliptic 
% system, thetaEcl [rad] and (optionally) the difference phi.
%
% Lennart Lindegren 2022-05-16

% obliquity of the ecliptic [rad]:
epsilon = 84381.41100 * pi/(180*3600);

ce = cos(epsilon);
se = sin(epsilon);
ca = cos(alpha);
sa = sin(alpha);
cd = cos(delta);
sd = sin(delta);
phi = atan2(se*ca, ce*cd+se*sd*sa);
thetaEcl = mod(thetaEqu + phi, 2*pi);

end
\end{verbatim}
}

\begin{figure}[h]
\centering
  \centerline{\includegraphics[width=9cm]{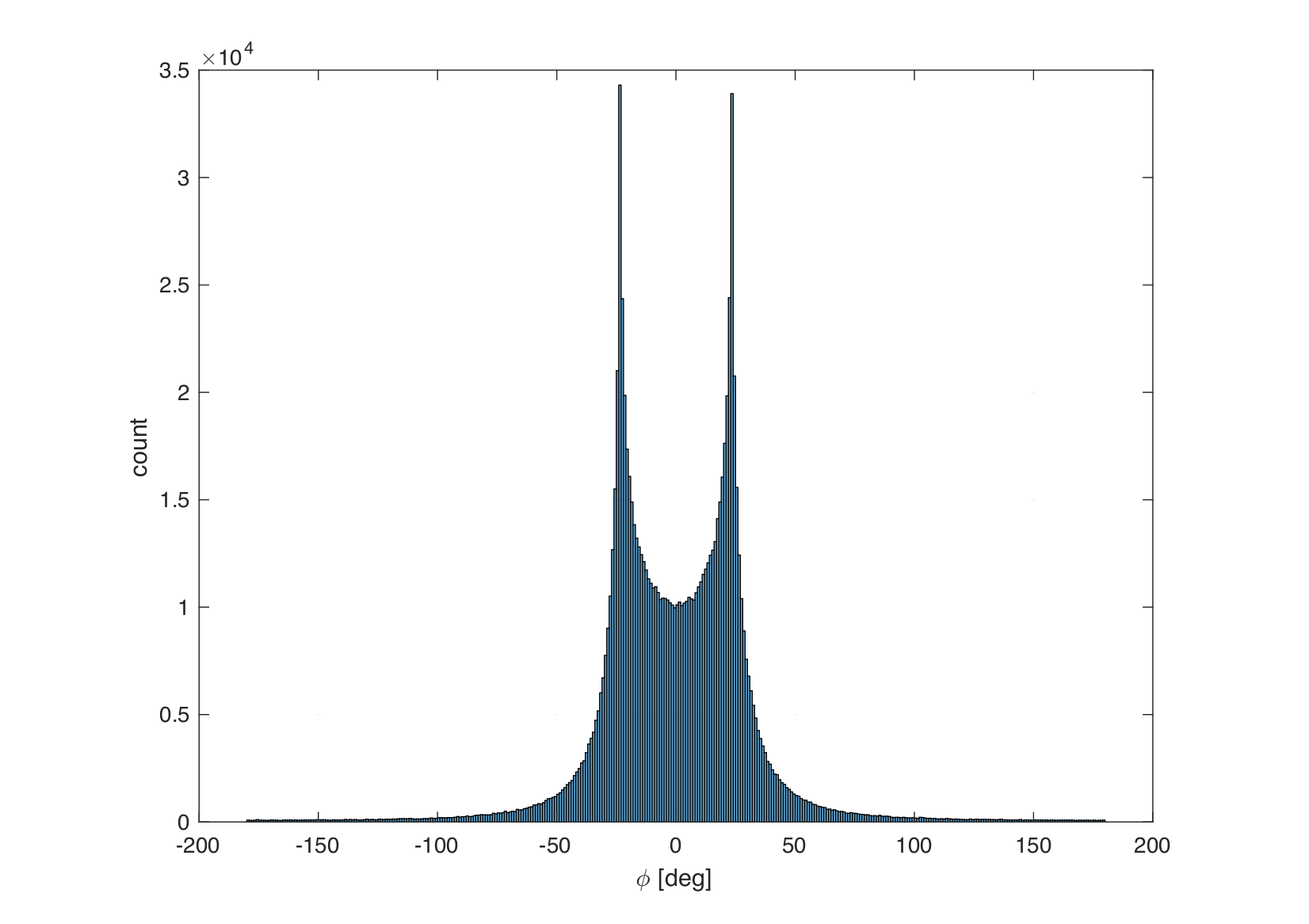}}
  \caption{Histogram of $\phi$ for $10^6$ random points of the sphere.}
  \label{fig2}
\end{figure}

\noindent Figure~\ref{fig1} was drawn for the case when $\phi>0$, but (\ref{e7})--(\ref{eq:scanAngleE8}) are
completely general and valid across the entire sky, except at the ecliptic pole ($\cos\beta=0$) ,
where $\theta_\text{ecl}$ is undefined. Figure~\ref{fig2} shows the distribution of $\phi$ for
random positions.

\section{\gaia acknowledgements\label{ssec:gaiaAcknowledgements}}

This work presents results from the European Space Agency (ESA) space mission \gaia. \gaia\ data are being processed by the \gaia\ Data Processing and Analysis Consortium (DPAC). Funding for the DPAC is provided by national institutions, in particular the institutions participating in the \gaia\ MultiLateral Agreement (MLA). The \gaia\ mission website is \url{https://www.cosmos.esa.int/gaia}. The \gaia\ archive website is \url{https://archives.esac.esa.int/gaia}.

The \gaia\ mission and data processing have financially been supported by, in alphabetical order by country:
\begin{itemize}
\item the Algerian Centre de Recherche en Astronomie, Astrophysique et G\'{e}ophysique of Bouzareah Observatory;
\item the Austrian Fonds zur F\"{o}rderung der wissenschaftlichen Forschung (FWF) Hertha Firnberg Programme through grants T359, P20046, and P23737;
\item the BELgian federal Science Policy Office (BELSPO) through various PROgramme de D\'{e}veloppement d'Exp\'{e}riences scientifiques (PRODEX) grants and the Polish Academy of Sciences - Fonds Wetenschappelijk Onderzoek through grant VS.091.16N, and the Fonds de la Recherche Scientifique (FNRS), and the Research Council of Katholieke Universiteit (KU) Leuven through grant C16/18/005 (Pushing AsteRoseismology to the next level with TESS, GaiA, and the Sloan DIgital Sky SurvEy -- PARADISE);  
\item the Brazil-France exchange programmes Funda\c{c}\~{a}o de Amparo \`{a} Pesquisa do Estado de S\~{a}o Paulo (FAPESP) and Coordena\c{c}\~{a}o de Aperfeicoamento de Pessoal de N\'{\i}vel Superior (CAPES) - Comit\'{e} Fran\c{c}ais d'Evaluation de la Coop\'{e}ration Universitaire et Scientifique avec le Br\'{e}sil (COFECUB);
\item the Chilean Agencia Nacional de Investigaci\'{o}n y Desarrollo (ANID) through Fondo Nacional de Desarrollo Cient\'{\i}fico y Tecnol\'{o}gico (FONDECYT) Regular Project 1210992 (L.~Chemin);
\item the National Natural Science Foundation of China (NSFC) through grants 11573054, 11703065, and 12173069, the China Scholarship Council through grant 201806040200, and the Natural Science Foundation of Shanghai through grant 21ZR1474100;  
\item the Tenure Track Pilot Programme of the Croatian Science Foundation and the \'{E}cole Polytechnique F\'{e}d\'{e}rale de Lausanne and the project TTP-2018-07-1171 `Mining the Variable Sky', with the funds of the Croatian-Swiss Research Programme;
\item the Czech-Republic Ministry of Education, Youth, and Sports through grant LG 15010 and INTER-EXCELLENCE grant LTAUSA18093, and the Czech Space Office through ESA PECS contract 98058;
\item the Danish Ministry of Science;
\item the Estonian Ministry of Education and Research through grant IUT40-1;
\item the European Commission’s Sixth Framework Programme through the European Leadership in Space Astrometry (\href{https://www.cosmos.esa.int/web/gaia/elsa-rtn-programme}{ELSA}) Marie Curie Research Training Network (MRTN-CT-2006-033481), through Marie Curie project PIOF-GA-2009-255267 (Space AsteroSeismology \& RR Lyrae stars, SAS-RRL), and through a Marie Curie Transfer-of-Knowledge (ToK) fellowship (MTKD-CT-2004-014188); the European Commission's Seventh Framework Programme through grant FP7-606740 (FP7-SPACE-2013-1) for the \gaia\ European Network for Improved data User Services (\href{https://gaia.ub.edu/twiki/do/view/GENIUS/}{GENIUS}) and through grant 264895 for the \gaia\ Research for European Astronomy Training (\href{https://www.cosmos.esa.int/web/gaia/great-programme}{GREAT-ITN}) network;
\item the European Cooperation in Science and Technology (COST) through COST Action CA18104 `Revealing the Milky Way with \gaia (MW-Gaia)';
\item the European Research Council (ERC) through grants 320360, 647208, and 834148 and through the European Union’s Horizon 2020 research and innovation and excellent science programmes through Marie Sk{\l}odowska-Curie grant 745617 (Our Galaxy at full HD -- Gal-HD) and 895174 (The build-up and fate of self-gravitating systems in the Universe) and grants 687378 (Small Bodies: Near and Far), 682115 (Using the Magellanic Clouds to Understand the Interaction of Galaxies), 695099 (A sub-percent distance scale from binaries and Cepheids -- CepBin), 716155 (Structured ACCREtion Disks -- SACCRED), 951549 (Sub-percent calibration of the extragalactic distance scale in the era of big surveys -- UniverScale), and 101004214 (Innovative Scientific Data Exploration and Exploitation Applications for Space Sciences -- EXPLORE);
\item the European Science Foundation (ESF), in the framework of the \gaia\ Research for European Astronomy Training Research Network Programme (\href{https://www.cosmos.esa.int/web/gaia/great-programme}{GREAT-ESF});
\item the European Space Agency (ESA) in the framework of the \gaia\ project, through the Plan for European Cooperating States (PECS) programme through contracts C98090 and 4000106398/12/NL/KML for Hungary, through contract 4000115263/15/NL/IB for Germany, and through PROgramme de D\'{e}veloppement d'Exp\'{e}riences scientifiques (PRODEX) grant 4000127986 for Slovenia;  
\item the Academy of Finland through grants 299543, 307157, 325805, 328654, 336546, and 345115 and the Magnus Ehrnrooth Foundation;
\item the French Centre National d’\'{E}tudes Spatiales (CNES), the Agence Nationale de la Recherche (ANR) through grant ANR-10-IDEX-0001-02 for the `Investissements d'avenir' programme, through grant ANR-15-CE31-0007 for project `Modelling the Milky Way in the \gaia era’ (MOD4Gaia), through grant ANR-14-CE33-0014-01 for project `The Milky Way disc formation in the \gaia era’ (ARCHEOGAL), through grant ANR-15-CE31-0012-01 for project `Unlocking the potential of Cepheids as primary distance calibrators’ (UnlockCepheids), through grant ANR-19-CE31-0017 for project `Secular evolution of galaxies' (SEGAL), and through grant ANR-18-CE31-0006 for project `Galactic Dark Matter' (GaDaMa), the Centre National de la Recherche Scientifique (CNRS) and its SNO \gaia of the Institut des Sciences de l’Univers (INSU), its Programmes Nationaux: Cosmologie et Galaxies (PNCG), Gravitation R\'{e}f\'{e}rences Astronomie M\'{e}trologie (PNGRAM), Plan\'{e}tologie (PNP), Physique et Chimie du Milieu Interstellaire (PCMI), and Physique Stellaire (PNPS), the `Action F\'{e}d\'{e}ratrice \gaia' of the Observatoire de Paris, the R\'{e}gion de Franche-Comt\'{e}, the Institut National Polytechnique (INP) and the Institut National de Physique nucl\'{e}aire et de Physique des Particules (IN2P3) co-funded by CNES;
\item the German Aerospace Agency (Deutsches Zentrum f\"{u}r Luft- und Raumfahrt e.V., DLR) through grants 50QG0501, 50QG0601, 50QG0602, 50QG0701, 50QG0901, 50QG1001, 50QG1101, 50\-QG1401, 50QG1402, 50QG1403, 50QG1404, 50QG1904, 50QG2101, 50QG2102, and 50QG2202, and the Centre for Information Services and High Performance Computing (ZIH) at the Technische Universit\"{a}t Dresden for generous allocations of computer time;
\item the Hungarian Academy of Sciences through the Lend\"{u}let Programme grants LP2014-17 and LP2018-7 and the Hungarian National Research, Development, and Innovation Office (NKFIH) through grant KKP-137523 (`SeismoLab');
\item the Science Foundation Ireland (SFI) through a Royal Society - SFI University Research Fellowship (M.~Fraser);
\item the Israel Ministry of Science and Technology through grant 3-18143 and the Tel Aviv University Center for Artificial Intelligence and Data Science (TAD) through a grant;
\item the Agenzia Spaziale Italiana (ASI) through contracts I/037/08/0, I/058/10/0, 2014-025-R.0, 2014-025-R.1.2015, and 2018-24-HH.0 to the Italian Istituto Nazionale di Astrofisica (INAF), contract 2014-049-R.0/1/2 to INAF for the Space Science Data Centre (SSDC, formerly known as the ASI Science Data Center, ASDC), contracts I/008/10/0, 2013/030/I.0, 2013-030-I.0.1-2015, and 2016-17-I.0 to the Aerospace Logistics Technology Engineering Company (ALTEC S.p.A.), INAF, and the Italian Ministry of Education, University, and Research (Ministero dell'Istruzione, dell'Universit\`{a} e della Ricerca) through the Premiale project `MIning The Cosmos Big Data and Innovative Italian Technology for Frontier Astrophysics and Cosmology' (MITiC);
\item the Netherlands Organisation for Scientific Research (NWO) through grant NWO-M-614.061.414, through a VICI grant (A.~Helmi), and through a Spinoza prize (A.~Helmi), and the Netherlands Research School for Astronomy (NOVA);
\item the Polish National Science Centre through HARMONIA grant 2018/30/M/ST9/00311 and DAINA grant 2017/27/L/ST9/03221 and the Ministry of Science and Higher Education (MNiSW) through grant DIR/WK/2018/12;
\item the Portuguese Funda\c{c}\~{a}o para a Ci\^{e}ncia e a Tecnologia (FCT) through national funds, grants SFRH/\-BD/128840/2017 and PTDC/FIS-AST/30389/2017, and work contract DL 57/2016/CP1364/CT0006, the Fundo Europeu de Desenvolvimento Regional (FEDER) through grant POCI-01-0145-FEDER-030389 and its Programa Operacional Competitividade e Internacionaliza\c{c}\~{a}o (COMPETE2020) through grants UIDB/04434/2020 and UIDP/04434/2020, and the Strategic Programme UIDB/\-00099/2020 for the Centro de Astrof\'{\i}sica e Gravita\c{c}\~{a}o (CENTRA);  
\item the Slovenian Research Agency through grant P1-0188;
\item the Spanish Ministry of Economy (MINECO/FEDER, UE), the Spanish Ministry of Science and Innovation (MICIN), the Spanish Ministry of Education, Culture, and Sports, and the Spanish Government through grants BES-2016-078499, BES-2017-083126, BES-C-2017-0085, ESP2016-80079-C2-1-R, ESP2016-80079-C2-2-R, FPU16/03827, PDC2021-121059-C22, RTI2018-095076-B-C22, and TIN2015-65316-P (`Computaci\'{o}n de Altas Prestaciones VII'), the Juan de la Cierva Incorporaci\'{o}n Programme (FJCI-2015-2671 and IJC2019-04862-I for F.~Anders), the Severo Ochoa Centre of Excellence Programme (SEV2015-0493), and MICIN/AEI/10.13039/501100011033 (and the European Union through European Regional Development Fund `A way of making Europe') through grant RTI2018-095076-B-C21, the Institute of Cosmos Sciences University of Barcelona (ICCUB, Unidad de Excelencia `Mar\'{\i}a de Maeztu’) through grant CEX2019-000918-M, the University of Barcelona's official doctoral programme for the development of an R+D+i project through an Ajuts de Personal Investigador en Formaci\'{o} (APIF) grant, the Spanish Virtual Observatory through project AyA2017-84089, the Galician Regional Government, Xunta de Galicia, through grants ED431B-2021/36, ED481A-2019/155, and ED481A-2021/296, the Centro de Investigaci\'{o}n en Tecnolog\'{\i}as de la Informaci\'{o}n y las Comunicaciones (CITIC), funded by the Xunta de Galicia and the European Union (European Regional Development Fund -- Galicia 2014-2020 Programme), through grant ED431G-2019/01, the Red Espa\~{n}ola de Supercomputaci\'{o}n (RES) computer resources at MareNostrum, the Barcelona Supercomputing Centre - Centro Nacional de Supercomputaci\'{o}n (BSC-CNS) through activities AECT-2017-2-0002, AECT-2017-3-0006, AECT-2018-1-0017, AECT-2018-2-0013, AECT-2018-3-0011, AECT-2019-1-0010, AECT-2019-2-0014, AECT-2019-3-0003, AECT-2020-1-0004, and DATA-2020-1-0010, the Departament d'Innovaci\'{o}, Universitats i Empresa de la Generalitat de Catalunya through grant 2014-SGR-1051 for project `Models de Programaci\'{o} i Entorns d'Execuci\'{o} Parallels' (MPEXPAR), and Ramon y Cajal Fellowship RYC2018-025968-I funded by MICIN/AEI/10.13039/501100011033 and the European Science Foundation (`Investing in your future');
\item the Swedish National Space Agency (SNSA/Rymdstyrelsen);
\item the Swiss State Secretariat for Education, Research, and Innovation through the Swiss Activit\'{e}s Nationales Compl\'{e}mentaires and the Swiss National Science Foundation through an Eccellenza Professorial Fellowship (award PCEFP2\_194638 for R.~Anderson);
\item the United Kingdom Particle Physics and Astronomy Research Council (PPARC), the United Kingdom Science and Technology Facilities Council (STFC), and the United Kingdom Space Agency (UKSA) through the following grants to the University of Bristol, the University of Cambridge, the University of Edinburgh, the University of Leicester, the Mullard Space Sciences Laboratory of University College London, and the United Kingdom Rutherford Appleton Laboratory (RAL): PP/D006511/1, PP/D006546/1, PP/D006570/1, ST/I000852/1, ST/J005045/1, ST/K00056X/1, ST/\-K000209/1, ST/K000756/1, ST/L006561/1, ST/N000595/1, ST/N000641/1, ST/N000978/1, ST/\-N001117/1, ST/S000089/1, ST/S000976/1, ST/S000984/1, ST/S001123/1, ST/S001948/1, ST/\-S001980/1, ST/S002103/1, ST/V000969/1, ST/W002469/1, ST/W002493/1, ST/W002671/1, ST/W002809/1, and EP/V520342/1.
\end{itemize}

The GBOT programme  uses observations collected at (i) the European Organisation for Astronomical Research in the Southern Hemisphere (ESO) with the VLT Survey Telescope (VST), under ESO programmes
092.B-0165,
093.B-0236,
094.B-0181,
095.B-0046,
096.B-0162,
097.B-0304,
098.B-0030,
099.B-0034,
0100.B-0131,
0101.B-0156,
0102.B-0174, and
0103.B-0165;
%
%
and (ii) the Liverpool Telescope, which is operated on the island of La Palma by Liverpool John Moores University in the Spanish Observatorio del Roque de los Muchachos of the Instituto de Astrof\'{\i}sica de Canarias with financial support from the United Kingdom Science and Technology Facilities Council, and (iii) telescopes of the Las Cumbres Observatory Global Telescope Network.

\end{appendix}

\end{document}